\documentclass[fleqn,usenatbib]{rasti}

\usepackage{newtxtext,newtxmath}

\usepackage[T1]{fontenc}

\DeclareRobustCommand{\VAN}[3]{#2}
\let\VANthebibliography\thebibliography
\def\thebibliography{\DeclareRobustCommand{\VAN}[3]{##3}\VANthebibliography}

\usepackage{graphicx}	
\usepackage{soul} 
\usepackage{mathrsfs}
\usepackage{bm}
\usepackage{bbm}
\usepackage[ruled,linesnumbered]{algorithm2e}
\usepackage{bbold}
\usepackage{url}
\usepackage{arydshln}
\usepackage{booktabs}
\usepackage{amsmath}	
\usepackage{xcolor}

\usepackage{multirow}
\usepackage{caption}
\usepackage{subcaption}

\usepackage{pifont}

\graphicspath{{./figs/}}

\newcommand{\cmark}{\ding{51}}%
\newcommand{\xmark}{\ding{55}}%

\DeclareMathOperator*{\argmax}{argmax}
\DeclareMathOperator*{\argmin}{argmin}

\SetKwRepeat{Do}{do}{while}

\newcommand{\tabL}{1.5mm}

\title[Scalable Bayesian UQ with data-driven priors]{Scalable Bayesian uncertainty quantification with data-driven priors for radio interferometric imaging}

\author[T.~I.~Liaudat et al.]{Tobías I. Liaudat$^{1,2,3}$\thanks{E-mail: \href{mailto:tobiasliaudat@gmail.com}{tobiasliaudat@gmail.com}},
Matthijs Mars$^{2}$,
Matthew A. Price$^{2}$,
Marcelo Pereyra$^{4,5}$,
Marta M. Betcke$^{1}$, and
\newauthor Jason D. McEwen$^{2,6 \thanks{E-mail: \href{mailto:jason.mcewen@ucl.ac.uk}{jason.mcewen@ucl.ac.uk}}}$
\\
$^{1}$Department of Computer Science, University College London (UCL), London WC1E 6BT, UK\\
$^{2}$Mullard Space Science Laboratory (MSSL), University College London (UCL), Holmbury St Mary, Dorking, Surrey RH5 6NT, UK\\
$^{3}$ IRFU, CEA, Université Paris-Saclay, F-91191 Gif-sur-Yvette, France \\
$^{4}$School of Mathematical and Computer Sciences, Heriot-Watt University, Edinburgh, EH14 4AS, Scotland, UK\\
$^{5}$Maxwell Institute for Mathematical Sciences, Bayes Centre, 47 Potterrow, Edinburgh, Scotland, UK\\
$^{6}$Alan Turing Institute, Euston Road, London NW1 2DB, UK
}

\date{Accepted ---. Received ---; in original form ---}

\pubyear{2023}

\begin{document}
\label{firstpage}
\pagerange{\pageref{firstpage}--\pageref{lastpage}}
\maketitle

\begin{abstract}
	Next-generation radio interferometers like the Square Kilometer Array have the potential to unlock scientific discoveries thanks to their unprecedented angular resolution and sensitivity. One key to unlocking their potential resides in handling the deluge and complexity of incoming data. This challenge requires building radio interferometric (RI) imaging methods that can cope with the massive data sizes and provide high-quality image reconstructions with uncertainty quantification (UQ). This work proposes a method coined \textsc{QuantifAI} to address UQ in RI imaging with data-driven (learned) priors for high-dimensional settings. Our model, rooted in the Bayesian framework, uses a physically motivated model for the likelihood. The model exploits a data-driven convex prior potential, which can encode complex information learned implicitly from simulations and guarantee the log-concavity of the posterior. We leverage probability concentration phenomena of high-dimensional log-concave posteriors to obtain information about the posterior, avoiding MCMC sampling techniques. We rely on convex optimisation methods to compute the MAP estimation, which is known to be faster and better scale with dimension than MCMC strategies. \textsc{QuantifAI} allows us to compute local credible intervals and perform hypothesis testing of structure on the reconstructed image. We propose a novel fast method to compute pixel-wise uncertainties at different scales, which uses $3$ and $6$ orders of magnitude less likelihood evaluations than other UQ methods like length of the credible intervals and Monte Carlo posterior sampling, respectively. We demonstrate our method by reconstructing RI images in a simulated setting and carrying out fast and scalable UQ, which we validate with MCMC sampling. Our method shows an improved image quality and more meaningful uncertainties than the benchmark method based on a sparsity-promoting prior. \textsc{QuantifAI}'s source code is available from \href{https://github.com/astro-informatics/QuantifAI}{github.com/astro-informatics/QuantifAI}.

\end{abstract}


\section{Introduction}

Radio astronomy plays a crucial role in expanding our understanding of the Universe, offering a unique perspective on astrophysical and cosmological phenomena. Among the transformative tools in an astronomer's toolkit, radio interferometric (RI) imaging stands out as an indispensable technique. Aperture synthesis and radio interferometry \citep{thompson2017} allow us to achieve high angular resolutions providing immense power to resolve objects. Furthermore, radio frequency signals are only weakly attenuated by our atmosphere, allowing for observations at the Earth's surface. The unparalleled angular resolution, high sensitivity and the different phenomena emitting in the radio wavelength regime make RI an ideal candidate to better help us understand our Universe.

The advent of the Square Kilometre Array \citep[SKA,][]{dewdney2009} heralds a new era in radio astronomy \citep{braun2015} spanning the study from the epoch of reionisation and fast radio bursts to galaxy evolution and dark energy. SKA's vast collecting area and sensitivity promise to provide a leap forward in our observational capabilities, opening doors to discoveries. However, this transformative potential comes with the formidable computational challenge of processing and making sense of the unprecedented volume of SKA-generated data. Developing and implementing algorithms that can efficiently handle SKA's data deluge is a challenge. In addition, achieving the high reconstruction performance required to unlock SKA's full potential is a significant obstacle in the SKA's data processing requirements.

The aperture synthesis techniques in RI probe the sky by acquiring specific Fourier measurements, which results in incomplete coverage of the Fourier domain of the sky's image of interest. The incomplete Fourier coverage makes the problem of estimating the underlying sky image, which we know as RI imaging, an ill-posed inverse problem, which is further complexified by the observational noise. Having a way to quantify the uncertainty in the image reconstructions becomes essential given the uncertainties involved in the RI imaging problem. To make scientifically sound inferences and informed decisions, we need the ability to quantify these uncertainties rigorously. This motivates the development of uncertainty quantification (UQ) methods tailored to the complexities of radio interferometric data, where scalability, i.e., the computational complexity with respect to the amount of data processed, and performance play a central role. We need to ensure that our reconstructions are not only insightful but also trustworthy.

In a nutshell, we want to develop RI imaging methods that can deliver precision with uncertainty quantification and that are highly scalable. Most existing methods only tackle some of these three requirements. The widely used CLEAN algorithm \citep{hogbom1974} built its success on scalability and fast inference. CLEAN and its extensions \citep{cornwell2008,offringa2014,offringa2017} have been continuously used in many RI imaging pipelines since its inception. Despite offering limited imaging quality and reconstruction artefacts compared to other approaches, CLEAN stands out due to its scalability. More recent approaches leverage compressed sensing theory, relying on sparse priors (often in wavelet representations) and convex optimisation techniques \citep{wiaux2009,mcewen2011,carrillo2012,carrillo2014,dabbech2015,dabbech2018,pratley2017}. These methods have been shown to improve the reconstruction quality at the expense of increased computational complexity. Considerable work has been directed to parallelisation and acceleration efforts for sparsity-based methods \citep{onose2016,pratley2019,pratley2019_2,thouvenin2022_a,thouvenin2022_b}.

The deep learning revolution has introduced a powerful way to encode complex image priors in neural networks, which may be used to solve complex high-dimensional inverse problems. This data-driven or learned paradigm has gained much traction across imaging landscape problems, including RI imaging \citep{allam2016,terris2022,aghabiglou2023,mars2023}. Learned methods can improve the reconstruction quality with respect to handcrafted priors, such as sparsity-based wavelet priors, as well as provide acceleration \citep{terris2022,mars2023,mars2024,aghabiglou2024} to convex optimisation-based methods.

Unfortunately, none of the RI imaging methods mentioned above, learned, sparsity-based or CLEAN-based, provide UQ tools for a given model. \citet{cai2018, cai2018_2} proposed methods for Bayesian UQ on RI imaging problems. \citet{cai2018} leverages proximal MCMC methods \citep{pereyra2016} to provide support for sparsity-promoting priors. The proposed method allows them to reconstruct the image and provide UQ by sampling the posterior probability distribution. The drawback of the method is the high computational cost suffered by all MCMC sampling techniques. The companion paper, \citet{cai2018_2}, overcomes the need for posterior sampling with maximum-a-posteriori (MAP) based UQ \citep{pereyra2017} relying on convex optimisation techniques. The second method \citep{cai2018_2} provides a significant speed-up with respect to the sampling-based method \citep{cai2018}, but its reconstruction quality is limited to sparsity-promoting priors.

Other RI imaging methods have addressed UQ for their reconstructions. \citet{dia2023} proposed to use score-based generative models as priors \citep{song2020,ho2020}, which by employing the convolved likelihood approximation \citep{remy2023,adam2022} are able to sample from the posterior. However, the method is computationally costly, and the sampling relies on MCMC methods. The Bayesian RI imaging method \texttt{comrade} \citep{tiede2022} was developed for very-long-baseline interferometry (VLBI) aiming to image black holes and active galactic nuclei. The \texttt{comrade} method relies on MCMC sampling methods like nested sampling \citep{ashton2022} and Hamiltonian Monte Carlo \citep{xu2020} to sample from the posterior. Posterior sampling based on MCMC methods can be enough for VLBI but cannot yet cope with the dimensions of SKA-like images. 

An alternative Bayesian RI imaging algorithm is the \texttt{resolve} method and its upgrades \citep{arras2018,arras2021,arras2022,roth2023,knollmuller2023}, introduced initially in \citet{junklewitz2016}, which can produce uncertainty maps from approximate posterior samples. The model is based on Gaussian random fields with different analytical priors used to promote certain aspects of the reconstruction, like positivity and spatial and temporal correlations. One novelty of the method is to approximate the true posterior probability distribution with variational inference and, therefore, be able to sample from the approximated distribution without resorting to MCMC methods that do not scale well in very high dimensions. The approximated posterior follows a multivariate Gaussian distribution, where their parameters are learnt by trying to maximise the information overlap between the true posterior and the model by minimising a Kullback-Leibler (KL-) divergence. Given that the full covariance matrix scales with the squared of the number of pixels, the authors exploit an approximation in the vicinity of the mean estimate. The method is based on the Metric Gaussian Variational Inference (MGVI) approach from \citet{knollmuller2019}. \texttt{resolve} has been used for reconstructions with VLBI observations of M87* \citep{arras2022} and Sagittarius A* \citep{knollmuller2023} from the Event Horizon Telescope (EHT) \citep{akiyama2019_1,akiyama2019_2}, as well as observations of Cygnus A from the Very Large Array (VLA) \citep{arras2018, arras2021, roth2023}.

In this article, we delve into the forefront of RI imaging and propose a method coined \textsc{QuantifAI}, based on learned convex priors, capable of delivering high-quality reconstructions with uncertainty quantification and being highly scalable. The method relies on the mathematically principled Bayesian framework to provide an understanding of the uncertainties through the posterior distribution. By restricting our model to log-concave posteriors, we can exploit recent MAP-based UQ techniques \citep{pereyra2017}, providing scalable optimisation-based UQ. We build upon recent advances in neural-network-based convex regularisers \citep{goujon2023}, allowing us to improve the reconstruction quality and obtain more meaningful uncertainties with respect to \citet{cai2018_2}. On top of the hypothesis tests of structure on the reconstructed image, we propose a novel fast method to estimate pixel-wise uncertainties as a function of scale.

The remainder of this article is organised as follows. In Section \ref{sc:RI_imaging}, we start by reviewing the RI imaging and techniques for the resulting inverse problem. Section \ref{sc:scalable_bayesian_method_uq} describes \textsc{QuantifAI}, the proposed method, and the RI image reconstruction algorithm. In Section \ref{sc:scalable_uq}, we introduce the core of our scalable UQ and the different UQ techniques it allows. The experimental results, including the performance of \textsc{QuantifAI} reconstruction and its UQ techniques, are presented in Section \ref{sc:experimental_results}. Section \ref{sc:realistic_exp} presents experimental results using more realistic observation based on simulated ungridded visibility patterns from the MeerKAT radio telescope. In Section \ref{sc:discussion} we provide a discussion on some limitations and possible extensions of the proposed methodology. We provide concluding remarks and present some future perspectives in Section \ref{sc:conclusions}.

\section{Radio interferometric imaging}
\label{sc:RI_imaging}

In this section, we start by reviewing the RI imaging inverse problem and discuss approaches to tackle it, including sparsity-based regularisation, the CLEAN method, and learned approaches. We then introduce the Bayesian framework elements needed such as MAP estimation and proximal MCMC sampling algorithms that will be later used as validation.

\subsection{Radio interferometry}
\label{sc:radio_interf_prob}

The interferometric measurement equation for a radio telescope \citep{thompson2017} in the monochromatic setting relates our observations represented by the visibility function $\mathcal{Y}$ to the sky brightness $\mathcal{X}$, which we want to reconstruct,
\begin{align}
	\label{eq:full_continuous_radio_eq}
	\mathcal{Y}( & u,v,w) = \iint \frac{\mathcal{X}(l,m) \, \mathcal{A}(l,m)}{\sqrt{1 - l^{2} - m^{2} }}                                                                                               \\
	             & \times \exp\left[ -2 \pi \text{i} w \left( \sqrt{1 - l^{2} - m^{2}} - 1 \right) \right] \exp \left[ -2 \pi \text{i} \left( lu + mv \right) \right] \text{d}l \text{d}m\,, \nonumber
\end{align}
where $\mathbf{u}=(u,v,w)$ are the interferometer baseline coordinates with units depending on the observation wavelength, $\mathbf{l}=(l,m,n)$ are cosine sky coordinates restricted to the unit sphere, and $\mathcal{A}$ includes direction-dependent effects (DDEs) like the primary beam of the dishes. The previous general model allows us to consider different DDEs through $\mathcal{A}$ and non-coplanar effects through the exponential term in $w$. These effects become considerable when considering wide fields of view and long baselines. There exists a rich body of literature incorporating such effects, e.g., \citet{smirnov2011_a,smirnov2011_b,smirnov2011_c,smirnov2011_d,thompson2017}, and there are scalable algorithms that take them into account, e.g., \citet{pratley2019}.

In this article, for the sake of simplicity but without loss of generality, we assume the coplanar setting, where the antennas are located in the same $w$ plane. We also assume that we observe a small field of view such that $1 - l^{2} - m^{2} \approx 1$. Consequently, we have $\exp\left[ -2 \pi \text{i} w \left( \sqrt{1 - l^{2} - m^{2}} - 1 \right) \right] \approx 1$, and Equation \ref{eq:full_continuous_radio_eq} reduces to
\begin{equation}
	\mathcal{Y}(u,v) \approx \iint \mathcal{X}(l,m) \, \mathcal{A}(l,m) \exp \left[ -2 \pi \text{i} \left( lu + mv \right) \right] \text{d}l \text{d}m\,.
	\label{eq:radio_eq_approx}
\end{equation}
From the previous equation, we can notice the remarkable result of $\mathcal{Y}(u,v) = \mathcal{F}(\mathcal{A} \mathcal{X})(u,v)$ where $\mathcal{F}$ is the two-dimensional Fourier transform.

To further simplify the problem, we will avoid using the continuous $\mathcal{Y}$ and $\mathcal{X}$ and work with their discrete counterparts, $\bm{x}$ and $\bm{y}$, respectively. The observational model we study for our RI imaging problem writes
\begin{equation}
	\bm{y} = \mathbf{\Phi} \bm{x} + \bm{n},
	\label{eq:linear_inv_prob}
\end{equation}
where $\bm{y} \in \mathbb{C}^{M}$ are the $M$ observed complex visibilities, $\bm{x} \in \mathbb{R}^{N}$ is the discrete sky brightness sampled on a $N$ point grid, and $\mathbf{\Phi} \in \mathbb{C}^{M \times N}$ is the linear measurement operator that models the acquisition process. Without loss of generality, the observational and instrumental noise $\bm{n} \in \mathbb{C}^{M}$ is assumed to be independent and identically distributed (iid) white Gaussian noise with zero mean and standard deviation $\sigma$. If the noise is not white, we can incorporate a noise whitening matrix in the $\mathbf{\Phi}$ operator such that the previous white noise assumption holds.

Each pair of antennas provides us with one visibility, which is a noisy Fourier component of the intensity image. Using an array of $n$ radio antennas allows us to sample $\binom{n}{2}=(n^2 - n)/2$ points in the $uv$-plane (or Fourier plane). The distribution of these points depends on the configuration of the radio antenna array. If different time intervals are considered, the Earth's rotation can be exploited to increase the number of $uv$ points. The $uv$ coverage is incomplete in all practical cases, and the measurements are noisy. Therefore, the linear operator $\mathbf{\Phi}$ is ill-posed. If we also consider a large number of measurements, recovering $\bm{x}$ from $\bm{y}$ becomes a challenging inverse problem.

The most basic reconstruction of $\bm{x}$ is often referred to as the naturally weighted dirty image that we will denote from now on as $\hat{\bm{x}}_{\text{dirty}}$ and call by dirty image as calibration weigths are not currently being taken into account in this work. This estimation is obtained by applying the adjoint of $\mathbf{\Phi}$ to the visibilities $\bm{y}$. To obtain a higher fidelity solution to the RI imaging inverse problem, we must regularise the problem by incorporating some prior information about the desired solutions $\bm{x}$. A broad range of methods can be characterized by what type of prior information is used to regularise the inverse problem and which algorithm is used to compute the reconstructed image $\hat{\bm{x}}$.

\subsection{Sparsity-based regularisation}
\label{sc:sparse_reg}

The last two decades have brought us a great number of RI imaging methods based on sparse representations. The prior information exploited is that the solution $\bm{x}$ is known to be sparsely represented in some bases or dictionaries. The bases are often built using multi-scale wavelets, or a dictionary is constructed with a collection of wavelets \citep{mallat2008}. We can represent our image $\bm{x}$ in a dictionary $\mathbf{\Psi} \in \mathbb{C}^{N \times L}$,
\begin{equation}
	\bm{x} = \mathbf{\Psi} \bm{a} = \sum_{i=1}^{L} \mathbf{\Psi}_i a_i \,,
\end{equation}
where $\bm{a} \in \mathbb{C}^{L}$ is a vector of coefficients of $\bm{x}$ weighting the corresponding dictionary atoms of $\mathbf{\Psi}$. The assumption made to regularise the inverse problem is that $\bm{a}$ is sparse or compressible, meaning that most of the coefficients are zero-valued or near zero, respectively. An array $\bm{a}$ is called $k$-sparse if it has only $k$ non-zero elements, which can be written as $\| \bm{a} \|_{0} = k$, where $\| \cdot \|_{0}$ denotes the $\ell_{0}$ pseudonorm.

Sparsity should be ideally enforced through the $\ell_{0}$ pseudonorm, which is non-convex. Consequently, a convex relaxation to the $\ell_1$ norm is used, which is a sparsity-promoting norm. The optimisation problem is formulated such that its solution coincides with the inverse problem solution. Therefore, the inverse problem can be tackled with an optimisation algorithm. The optimisation objective comprises two competing terms: (\textit{i}) a data-fidelity term $f(\cdot)$ that promotes consistency with the observed visibilities and depends on the statistics of the noise $\bm{n}$; and (\textit{ii}) a regularisation term $r(\cdot)$ that encodes our prior knowledge of $\bm{x}$. The optimisation problem reads
\begin{equation}
	\hat{\bm{x}} = \argmin_{\bm{x} \in \mathbb{R}^{N}} f(\bm{x}) + \sum_{k} \lambda_k r_k(\bm{x})\,,
\end{equation}
where we are using a sum of regularisation terms $r_k$, each with its corresponding regularisation strength parameter $\lambda_k$. Substituting the RI data fidelity and sparsity-enforcing regularisation in an overdetermined wavelet dictionary $\Psi$ terms into Equation \ref{eq:linear_inv_prob} we obtain
\begin{equation}
	\hat{\bm{x}} = \argmin_{\bm{x} \in \mathbb{R}^{N}} \frac{1}{2 \sigma^{2}} \left\| \bm{y} - \Phi \bm{x} \right\|_{2}^{2} + \lambda \big\| \Psi^{\dagger} \bm{x} \big\|_{1} \,,
	\label{eq:optim_l1_reg}
\end{equation}
where $\sigma$ is the noise standard deviation.

The previous formulation is referred to as unconstrained. Other works consider the constrained formulation, which minimises the $\ell_1$ term with respect to a hard $\ell_2$-ball constraint over $\bm{y}$ with a radius of $\epsilon$, which is related to the noise's $\sigma$ \citep{carrillo2012,pratley2017}. In this article, we will focus on the unconstrained formulation as it has a natural Bayesian interpretation. Obtaining the solution from Equation \ref{eq:optim_l1_reg} involves solving a convex optimisation problem, where we have the sum of a differentiable and a non-differentiable term. Proximal algorithms \citep{parikh2014} are well suited to tackle such optimisation problems. Recent developments brought us a wide collection of proximal optimisation algorithms, such as the forward-backward (FB) algorithm \citep{combettes2009}, the FISTA algorithm \citep{beck2009}, the alternating direction method of multipliers \citep[ADMM,][]{boyd2011}, and the primal-dual forward-backward algorithm \citep{chambolle2011,condat2013}, to mention a few. A rich literature exists exploiting the aforementioned concepts to tackle the RI imaging problem \citep{wiaux2009,mcewen2011,carrillo2012,carrillo2014,onose2016,pratley2017,pratley2019,pratley2019_2,pratley2019_3,cai2018_2}. For example, the Sparsity Averaging Reweighted Analysis (SARA) family of methods \citep{carrillo2012} use an over-complete dictionary composed of a concatenation of the Dirac basis and the first eight Daubechies wavelets \citep{daubechies1992} and has shown good performance in RI imaging.

\subsection{CLEAN}

Precursor of RI image reconstructions, the CLEAN algorithm \citep{hogbom1974} is a highly successful RI imaging method and it is still being used \citep{akiyama2019_1, akiyama2019_2} despite various negative characteristics. The CLEAN algorithm is a non-linear iterative method that assumes a sparse sky model. CLEAN iteratively removes the contribution of the brightest source convolved with the instrument's point spread function or dirty beam. This method can be interpreted as a matching pursuit algorithm \citep{wiaux2009}, or an $\ell_0$ regularisation with a basis composed of a sum of Dirac spikes. Several extensions of CLEAN have been developed over time \citep{bhatnagar2004, cornwell2008, stewart2011, offringa2014, offringa2017} achieving better reconstruction performance. See \citet{rau2009} for a review of CLEAN-based algorithms.

On top of a very early introduction, the success of CLEAN resides in its \textit{scalability}. However, CLEAN has been shown to produce artefacts when point sources do not well describe the underlying sky model limiting CLEAN's image quality and justifying the need for more advanced techniques, e.g. based on Section \ref{sc:sparse_reg}. CLEAN often requires manual intervention, making its use less practical. Furthermore, CLEAN and its extensions do not provide meaningful uncertainty quantification of its reconstruction.

\subsection{Learned approaches}
\label{sc:deep_learning_methods}

The advent of deep learning models has affected many imaging applications, and RI imaging is no exception. Handcrafted models and priors are limited in the information they can capture or represent with respect to recent, more expressive neural networks. Learned or data-driven methods can encode complex information existing in the data, e.g., astrophysical simulations, used in their training. In general, these approaches produce reconstructions with improved quality, a computational speed-up, or both. These reasons make learned approaches very relevant to the RI imaging reconstruction problem. However, there are issues regarding the robustness of learned methods to data distribution shifts \citep{hendrycks2020} and scalable methods for uncertainty quantification to the reconstruction.

\citet{allam2016} proposed a learned method for RI imaging based on convolutional neural networks \citep{dong2016} originally considered for super-resolution. The approach consists of learning to post-process dirty images with variants for both, known and unknown PSF. More recently, \citet{gheller2021} proposed to use a convolutional denoising autoencoder to learn to post-process radio images, e.g., the dirty image or CLEAN's output. \citet{connor2022} proposed a residual deep neural network (DNN) coined POLISH that works as a learned post-processing and super-resolution network. The DNN is based on the architecture proposed in \citet{yu2018} and takes as input dirty images at different wavelengths and resolutions. POLISH outputs a clean image at a higher resolution for each wavelength and shows a better reconstruction quality than CLEAN. The proposed method has been applied to simulations from the upcoming Deep Synoptic Array-2000 \citep{hallinan2019} and real data from the Very Large Array \citep[VLA,][]{perley2011}.

The Plug-and-Play (PnP) framework \citep{venkatakrishnan2013} provides a way to incorporate a deep learning model into a modern optimisation algorithm. The central idea is to replace a proximal regularisation term with a denoising deep neural network. \citet{ryu2019} studied conditions for the convergence of PnP algorithms. \citet{pesquet2021} proposed a new term for the denoiser's training loss that enforces the firm nonexpansiveness of the denoiser, which is usually deep learning-based. This training procedure allows the denoiser to suit a PnP framework with theoretical convergence conditions. The PnP framework with the nonexpansiveness enforced to the deep learning-based denoiser has been applied to the RI imaging problem in \citet{terris2022}, where the approach has been called AIRI for Artificial Intelligence for Regularisation in RI imaging. The approach achieved similar or better performance than competing prior-based approaches whilst providing a significant acceleration potential. The AIRI method was later validated on observations from the Australian Square Kilometre Array Pathfinder \citep[ASKAP,][]{wilber2023}.

Two learned approaches for an interferometric-based imager named Segmented Planar Imaging Detector for Electro-Optical Reconnaissance (SPIDER) were proposed by \citet{mars2023}. The first approach consists of a learned post-processing step from the dirty reconstruction based on a convolutional U-Net architecture \citep{ronneberger2015}. The second approach consists of a learned multiscale iterative method coined GU-Net, which incorporates the measurement operator to include measurement information at the different steps and scales of the method. GU-Net is more efficient than standard unrolling methods due to its multi-scale nature. The numerical results show an improved reconstruction quality and a faster convergence than proximal optimisation-based methods. In the following work \citep{mars2024}, the GU-Net was applied to the RI imaging problem. The variations of the $uv$-coverage are handled by training the neural network on a broad distribution of simulated $uv$-coverages and subsequently fine-tuning the network for a specific sampling distribution.

\citet{aghabiglou2023} recently proposed a series of DNNs that combines notions of PnP algorithms and unrolled optimisation methods \citep{adler2018,monga2019}. Each DNN is trained to transform a back-projected residual into an image residual, thus ideally improving the reconstruction of the previous iteration. The results show a significant speed-up with respect to AIRI or SARA-based methods while maintaining a similar reconstruction quality. Other recent approaches based on deep neural networks include:
\citet{wang2023} who proposed a denoising diffusion probabilistic model conditioned on the visibilities and the dirty reconstruction; and \citet{schmidt2022} who proposed a convolutional neural network based on residual blocks that intend to inpaint the measurements, or recover the entire $uv$ plane from an incomplete coverage.

\subsection{Bayesian framework}
\label{sc:bayesian_framework}

Bayesian inference provides a principled statistical framework to solve the inverse problem in Equation \ref{eq:linear_inv_prob} with statistical guarantees. This framework builds upon Bayes' famous theorem,
\begin{equation}
	\underbrace{p(\bm{x}|\bm{y})}_{\text{Posterior}} = \frac{\overbrace{p(\bm{y}|\bm{x})}^{\text{Likelihood}} \overbrace{p(\bm{x})}^{\text{Prior}}}{\underbrace{ \int_{\mathbb{R}^{N}} p(\bm{y} | \bm{x}) p(\bm{x}) {\rm d} \bm{x} }_{\text{Bayesian evidence}}} \,.
	\label{eq:bayes_theorem}
\end{equation}

Bayes' theorem relates the posterior distribution to the likelihood and prior terms that are the main constituents of a Bayesian model. The likelihood is associated with the data-fidelity term depending on the observational model and the noise statistics. The prior models expected properties of the solution $\bm{x}$, for example, smoothness, and piecewise regularity. This prior knowledge regularises the estimation problem.

The term in the denominator, commonly known as the Bayesian evidence, does not depend on $\bm{x}$ as we are marginalising over that variable, and it describes the likelihood of the observed data based on the modelling assumptions. The Bayesian evidence is crucial for making Bayesian model comparison \citep{robert2007}, which provides us with a consistent way to compare models. Such high dimensional integrals can be effectively estimated by, for example, nested sampling techniques \citep{skilling2006, ashton2022}, or the recently introduced learned harmonic mean estimator \citep{mcewen2021,spuriomancini2022,polanska2023}. Recent developments have focused on nested sampling to compute the model evidence in high-dimensional imaging problems with sparsity-based handcrafted priors \citep{cai2022} and deep learning-based priors \citep{mcewen2023}. Carrying out model selection is out of the scope of this work.

Under the Bayesian framework, we have the posterior distribution $p(\bm{x}|\bm{y})$ which assigns a probability to each possible solution $\bm{x}$ given some observations $\bm{y}$ and a model $\mathcal{M}$ consisting of the likelihood and prior terms. In imaging settings, explaining the information contained in the posterior distribution is not trivial due to its high-dimensional nature. The posterior distribution is generally characterised by samples computed by MCMC sampling. Efficiently sampling from high-dimensional posterior distributions is a current research topic, see e.g. \citet{klatzer2023}. Once $p(\bm{x}|\bm{y})$ is defined, we can say that the reconstruction method will be a point estimator of the posterior that will provide us with $\hat{\bm{x}}$. There are several choices for point estimators \citep{robert2007,arridge2019}, each with advantages and drawbacks. Some examples are a sample from the posterior, $\hat{\bm{x}} \sim p(\bm{x}|\bm{y})$, the maximum-a-posteriori estimator, $\hat{\bm{x}} = \argmax_{\bm{x}} p(\bm{x}|\bm{y})$, or the posterior mean, $\hat{\bm{x}} = \mathbb{E}[\bm{x}|\bm{y}]$.

The posterior also provides us with consistent ways of quantifying the uncertainty of the chosen point estimate or reconstruction \citep{robert2007}. For example, one way to represent uncertainty is to compute the posterior standard deviation. The pixels with a higher standard deviation are less constrained by the data and the prior allowing for more significant fluctuations.

One of the most significant drawbacks of Bayesian imaging methods is that they are known to be computationally expensive, even if there is a continuous effort targetting the scalability of these methods \citep{pereyra2016,durmus2018,pereyra2020,pereyra2022,klatzer2023}.

\subsubsection{Maximum-a-posteriori estimation}
\label{sc:map_estimation}

The MAP estimator is particularly interesting in high-dimensional problems like RI imaging as its formulation allows us to bypass the need for sampling from the posterior. Consequently, its computational footprint is significantly reduced. The likelihood and prior terms can be rewritten as $p(\bm{y}|\bm{x}) = \exp[-f(\bm{x},\bm{y})] $ and $p(\bm{x}) = \exp[-g(\bm{x})]$, respectively. The functions $f$ and $g$ are the likelihood and prior potentials. Using Bayes' theorem in Equation \ref{eq:bayes_theorem}, we can rewrite the MAP estimation as follows
\begin{equation}
	\hat{\bm{x}}_{\text{MAP}} = \argmax_{\bm{x} \in \mathbb{R}^{N}} p(\bm{x}|\bm{y}) = \argmax_{\bm{x} \in \mathbb{R}^{N}} p(\bm{y}|\bm{x}) p(\bm{x}) \,.
\end{equation}
The previous optimisation problem can be reformulated using the monotonicity of the logarithm as follows
\begin{align}
	\hat{\bm{x}}_{\text{MAP}} & =  \argmin_{\bm{x} \in \mathbb{R}^{N}} - \log p(\bm{y}|\bm{x}) - \log p(\bm{x}) \nonumber \\
	                          & = \argmin_{\bm{x} \in \mathbb{R}^{N}} f(\bm{x},\bm{y}) + g(\bm{x}) \,.
	\label{eq:map_estimation}
\end{align}

One advantage of the MAP estimator is that Equation \ref{eq:map_estimation} can be tackled efficiently with optimisation algorithms. We refer the reader to \citet{pereyra2019} for a deeper analysis of MAP estimation.

Coming back to the RI imaging inverse problem from Equation \ref{eq:linear_inv_prob}, we can define a (white) Gaussian likelihood,
\begin{equation}
	p(\bm{y}|\bm{x}) \propto \exp \left[ - \frac{1}{2 \sigma^{2}} \left\| \bm{y} - \Phi \bm{x} \right\|_{2}^{2} \right]\,
	\label{eq:gaussian_likelihood}
\end{equation}
and a sparsity-inducing Laplace-type prior defined as
\begin{equation}
	p(\bm{x}) \propto \exp \left[ - \lambda \big\| \Psi^{\dagger} \bm{x} \big\|_{1} \right] \,.
	\label{eq:sparsity_prior}
\end{equation}
Upon substitution of Equations \ref{eq:gaussian_likelihood} and \ref{eq:sparsity_prior} into Equation \ref{eq:map_estimation}, the MAP optimisation problem coincides with the one in Equation \ref{eq:optim_l1_reg}. Therefore, the MAP reconstruction, $\hat{\bm{x}}_{\text{MAP}}$, matches $\hat{\bm{x}}$ from Equation \ref{eq:optim_l1_reg}. Hence, sparsity-based approaches are MAP estimations with a prior based on the sparsity-promoting $\ell_1$ norm in a given dictionary, e.g., wavelets.

\subsubsection{Uncertainty quantification: more than a point estimate}

Computing a good reconstruction for an inverse problem in the form of Equation \ref{eq:linear_inv_prob} can itself be challenging. Moreover, the reconstruction is often insufficient for many scientific applications that require further quantification of the result. This demand opens the door to uncertainty quantification, which provides more than a point estimate. The Bayesian framework provides us with formidable tools to do uncertainty quantification. For example, if we choose the MAP estimator as our reconstruction following the model in Section \ref{sc:map_estimation}, we obtain the same reconstruction as in Section \ref{sc:sparse_reg}, which is the solution of Equation \ref{eq:optim_l1_reg}. However, with the Bayesian framework, we can sample from the posterior and estimate the posterior standard deviation, perform a Bayesian hypothesis test of some image structure \citep{cai2018, price2021}, or compute other pixel-wise uncertainty measurements like local credible intervals \citep[LCI,][]{cai2018, price2019}.

\subsubsection{Bayesian inference via MCMC sampling}
\label{sc:bayesian_inf}

Recent developments \citep{durmus2022} have considerably reduced the computational complexity of sampling high-dimensional posterior distributions in imaging inverse problems. Proximal MCMC sampling algorithms \citep{pereyra2016,durmus2018} extend the class of posterior distributions that can be studied by allowing the use of non-smooth terms. Sparse regularisers have been widely used in RI imaging \citep{carrillo2012,carrillo2014, pratley2017, cai2018}, and are usually enforced through a non-smooth $\ell_1$ term.

Let us note $\pi$ the target probability distribution that we are interested in sampling from, which in our case will be the posterior $p(\bm{x}|\bm{y})$. We consider a Langevin diffusion process on $\mathbb{R}^{N}$ such that its stationary distribution is $\pi$. Assuming that $\pi \in \mathcal{C}^{1}$ with Lipschitz gradients, we write the Langevin diffusion as the following stochastic process
\begin{equation}
	d L(t) = \frac{1}{2} \nabla \log \pi [ L(t) ] dt + d W(t), \quad L(0) = l_{0} \,,
\end{equation}
where $W$ is a $N$-dimensional Brownian motion. A usual discrete-time approximation of the Langevin diffusion consists of a forward Euler approximation with a step size $\delta$, known as the Euler-Maruyama approximation \citep{kloeden2011}. The resulting algorithm is known as the unadjusted Langevin algorithm (ULA),
\begin{equation}
	\bm{l}^{(m+1)} = \bm{l}^{(m)} + \delta \nabla \log \pi [\bm{l}^{(m)}] + \sqrt{2 \delta} \bm{w}^{(m+1)} \,,
\end{equation}
where $\bm{w}^{(m+1)} \sim \mathcal{N}(0,\mathbf{I}_{N})$ is the discrete counterpart of $W(t)$. The ULA-based Markov chain converges to $\pi$ with an asymptotic bias due to discretisation. The bias can be accounted for with a subsequent Metropolis-Hasting (MH) accept-reject step. Adding the MH step corrects the bias but increases the algorithm's computational complexity. The ULA algorithm with the subsequent MH step is known as the Metropolis-adjusted Langevin algorithm (MALA).

The ULA algorithm requires the target density $\pi$ to be continuously differentiable with Lipschitz gradients. Let us now consider $\pi(\bm{x}) \propto \exp [ -f(\bm{x}) - g(\bm{x}) ]$, where $f \in \mathcal{C}^{1}$ with Lipschitz gradient and $g$ is non-smooth but is a lower semicontinuous convex function that admits a proximal operator \citep{parikh2014}. Proximal MCMC algorithms \citep{pereyra2016} relax this assumption by approximating $g$, a non-smooth term in $\pi$, by its Moreau-Yosida envelope $g^{\gamma}$. The Moreau-Yosida approximation satisfies
\begin{align}
	\label{eq:MY_envelope_approx}
	 & \nabla g^{\gamma}(\bm{x}) = \frac{1}{\gamma}\left(\bm{x} - \text{prox}_{\gamma}^{g}\left(\bm{x}\right)\right)\,,                                                     \\
	 & \text{prox}_{\gamma}^{g}(\bm{x}) := \argmin_{\bm{u} \in \mathbb{R}^{N}} \left\{ g(\bm{u}) + \frac{1}{2 \gamma} \left\| \bm{u} - \bm{x} \right\|^{2}_{2} \right\} \,,
	\label{eq:prox_op_definition}
\end{align}
where $\gamma$ is the Moreau-Yosida approximation parameter and the proximal operator may or may not have a closed-form expression. Consequently, the non-smooth target density $\pi$ is approximated by the smooth $\pi_{\gamma}$, which replaces the $g$ term with its Moreau-Yosida approximation $g^{\gamma}$. The Markov chain targeting $\pi_{\gamma}$ writes
\begin{align}
	\bm{l}^{(m+1)} = \left(1 - \frac{\delta}{\gamma}\right) \bm{l}^{(m)} + \frac{\delta}{\gamma} \, \text{prox}_{\gamma}^{g} & \left(\bm{l}^{(m)}\right)                                                                \\
	                                                                                                                         & - \delta \nabla f\left(\bm{l}^{(m)}\right) + \sqrt{2 \delta} \bm{w}^{(m+1)} \,,\nonumber
\end{align}
and it is known as Moreau-Yosida regularised ULA (MYULA). If we add an MH step targetting the non-differentiable distribution $\pi$, the MCMC algorithm is known as Proximal MALA (Px-MALA). The proximal MCMC algorithms previously mentioned can be further accelerated by replacing the Euler-Maruyama approximation with the more involved Runge-Kutta-Chebyshev approximation \citep{abdullle2018}, giving rise to the SK-ROCK \citep{pereyra2020} algorithm.

\citet{cai2018} exploited the MYULA and Px-MALA algorithms to sample from the posterior in the RI imaging problem. The model is based on a Gaussian likelihood as in Equation \ref{eq:gaussian_likelihood} and a sparsity promoting prior akin to Equation \ref{eq:sparsity_prior}. However, the framework can be used with more complex noise models \citep{melidonis2023}, e.g. Poisson noise. In \citet{cai2018}, the RI image reconstruction relies on the minimum mean squared error (MMSE) estimator based on the posterior mean, while in \citet{cai2018_2}, the MAP is considered.

\section{Scalable Bayesian data-driven imaging with uncertainty quantification}
\label{sc:scalable_bayesian_method_uq}

\textsc{QuantifAI}\footnote{Code available at \href{https://github.com/astro-informatics/QuantifAI}{github.com/astro-informatics/QuantifAI}}, a scalable Bayesian data-driven method with uncertainty quantification is motivated by three principles:
\begin{itemize}
	\item[1.] \textit{Scalability:} The RI imaging inverse problem demands scalability for a method to be useful in real astronomical data scenarios such as SKA. The most time-consuming operation is evaluating the measurement operator $\Phi$ in the likelihood function. It is, therefore, essential to minimise the number of likelihood evaluations. For these reasons, we limit ourselves to the MAP estimator for our reconstruction corresponding to the solution of a convex optimization problem which converges quickly. We need to avoid sampling-based approaches as they are prohibitively expensive in terms of computations.
	\item[2.] \textit{High-quality reconstructions:} To improve the quality of our reconstruction, we consider data-driven or learned priors that can better encode the expected image structures. In Section \ref{sc:deep_learning_methods}, we have already seen that data-driven approaches can better represent complex imaging priors and provide reconstructions superior to handcrafted priors, such as sparsity-promoting priors based on wavelet dictionaries.
	\item[3.] \textit{Uncertainty quantification:} There are many ways to quantify uncertainty based on sampling the posterior distribution. However, using sampling-based methods is prohibitively expensive, and one of our key criteria is computational scalability. Therefore, we need to restrict ourselves to log-concave posteriors, which is equivalent to saying that the addition of our potentials $f + g$ has to be convex, and to explicit potentials. As we will later describe in more detail in Section \ref{sc:scalable_uq}, the first restriction enables the use of efficient methods relying on the concentration of probability for high-dimensional log-concave distribution \citep{pereyra2017}. Consequently, we can use approximate posterior information bypassing sampling methods. These methods are orders of magnitude faster resulting in a scalable Bayesian UQ method. In a nutshell, we require the posterior potential to be convex and explicit for scalable UQ. The likelihood is typically convex for RI imaging problems so we will enforce the prior potential $g$ to be convex and explicit. The requirement of explicit potentials will be explained in Section \ref{sc:scalable_uq}.
\end{itemize}

We continue by introducing the data-driven convex regularisers and the optimisation algorithm used to compute the MAP estimation for the proposed method.

\subsection{Learned convex regularisers}

As stated before, we need an expressive regulariser that is convex and has an explicit potential. Modern regularisers based on deep neural networks, like convolutional neural networks, used in RI imaging reconstruction methods satisfy neither of the two constraints. This last constraint, i.e., with an explicit potential required by the UQ approach, excludes a range of denoisers whose potentials are defined implicitly. PnP approaches \citep{terris2022} only require the denoising of the image without explicitly computing the regularisation potential. For example, a typical iteration from a PnP algorithm writes
\begin{equation}
	\bm{x}_{k+1} = \text{D}(\bm{x}_{k} - \gamma \nabla f(\bm{x}_{k}))\,, \quad (\forall k \in \mathbb{N})\,,
\end{equation}
where $\text{D}$ is the denoiser, $f$ is the data-fidelity term, $\gamma$ is the step size, and $k$ is the iteration number. The algorithm's convergence can be assured if $\text{D}$ and the stepsize satisfies some conditions \citep{pesquet2021,ryu2019}. Even if the denoiser $\text{D}$ is convex, we cannot use it for our approach as we must evaluate the potential.

\citet{mukherjee2020} proposed a learned convex regulariser parametrised by the architecture of a deep input-convex neural network \citep[ICNN,][]{amos2016}, which is convex by construction. The training of the regulariser is done with an adversarial framework introduced by \citet{lunz2018}.

Very recently, a learnable convex-ridge regulariser neural network \citep[CRR-NN\footnote{\href{https://github.com/axgoujon/convex_ridge_regularizers}{https://github.com/axgoujon/convex\_ridge\_regularizers}},][]{goujon2023} has been proposed, which comes with the required properties of being convex and having an explicit potential. In addition, the model focuses on being reliable and interpretable while still being expressive enough to provide excellent reconstruction quality. The CRR-NN regulariser, $R_{\bm{\theta}}$, has the form
\begin{equation}
	R_{\bm{\theta}} : \mathbb{R}^{N} \mapsto \mathbb{R},\quad R_{\bm{\theta}}(\bm{x}) = \sum_{i=1}^{N_{\text{C}}} \sum_{k} \psi_{i} \left( \left( \bm{h}_{i} \ast \bm{x} \right)[k] \right) ,
\end{equation}
where $\bm{h}_{n}$ are learnable 2D convolution kernels, $( \bm{h}_{i} \ast \bm{x} )[k]$ denotes the $k$-th pixel of the resulting convolution, $N_{\text{C}}$ is the number of channels or kernels, $\psi_i: \mathbb{R} \mapsto \mathbb{R}$ are learnable non-linear convex profile functions with a Lipschitz continuous derivative, i.e., $\psi_i \in C^{1,1}(\mathbb{R})$, and $\bm{\theta}$ in $R_{\bm{\theta}}$ represents all learnable parameters. The convexity constraint on the learnable activation functions, $\psi_i$, is enforced by making the pointwise $\sigma_i: \mathbb{R} \to \mathbb{R}$ monotonically increasing, with $\psi_{i}' = \sigma_{i}$, where $\sigma_{i} \in C^{0,1}_{\uparrow}(\mathbb{R})$, and $C^{0,1}_{\uparrow}(\mathbb{R})$ is the set of scalar Lipschitz continuous and increasing functions on $\mathbb{R}$. The $\sigma_i$ functions are chosen as learnable linear splines, which have been shown to be more expressive than ReLU functions in Propositions 3.3 and 3.5 from \citet[\S III.B]{goujon2023}. The main difference between a prior based on the CRR-NN and a wavelet dictionary is that the kernels (or filters) and the activation (or thresholding) functions are learnt in the first one. In the second one, they are fixed or handcrafted. We refer the reader to \citet[Figs. 5 and 6]{goujon2023} for examples of learned kernels $\bm{h}_{n}$ and activation functions for two trainings with two different noise levels. See \citet{goujon2023, bohra2020} for more information on learnable splines.

In the spirit of PnP approaches, the CRR-NN training is based on the denoising problem that reads
\begin{equation}
	\bm{x}^{\ast} = \argmin_{\bm{x} \in \mathbb{R}^{N}} \frac{1}{2} \| \bm{x} - \bm{y} \|_{2}^{2} + \lambda R_{\bm{\theta}}(\bm{x})\,,
	\label{eq:CRR-denoising}
\end{equation}
where $\bm{y}$ is a noisy version of $\bm{x}$, and $\lambda$ is a parameter controlling the regularisation strength. The denoising problem is addressed through the fixed point of the problem, which given the convexity assumptions, is unique. A gradient step of Equation \ref{eq:CRR-denoising} reads
\begin{equation}
	T_{R_{\bm{\theta}}, \lambda, \alpha}(\bm{x}) = \bm{x} - \alpha ((\bm{x} - \bm{y}) + \lambda \nabla R_{\bm{\theta}}(\bm{x})) \,,
	\label{eq:grad_step_CRR}
\end{equation}
where $\alpha$ is the stepsize. Convergence can be guaranteed if the stepsize satisfies $\alpha \in (0, 2/(1 + \lambda \,\text{Lip}(\nabla R_{\bm{\theta}})))$, where $\text{Lip}(\cdot)$ denotes the Lipschitz constant. By composing $t$ gradient descent updates of Equation \ref{eq:grad_step_CRR}, i.e., a $t$-fold composition, we obtain a multi-gradient step denoiser that we denote $T_{R_{\bm{\theta}}, \lambda, \alpha}^{t}$ following the notation of \citet{goujon2023}.

The denoising problem in Equation \ref{eq:CRR-denoising} can be formulated as a fix point problem for the $t$-step denoiser $T_{R_{\bm{\theta}}, \lambda, \alpha}^{t}$ as follows,
\begin{equation}
	T_{R_{\bm{\theta}}, \lambda, \alpha}^{t}(\bm{y}) \approx \bm{x} \,.
	\label{eq:fix_point_prob}
\end{equation}
We build the CRR-NN training by penalising the residual of the fix point problem in Equation \ref{eq:fix_point_prob} with a loss function $\mathcal{L}$, for a training set of pairs of noiseless and noisy images $\{\bm{x}^{(m)}, \bm{y}^{(m)}\}_{m=1}^{M}$, and reads
\begin{equation}
	\bm{\theta}_{t}^{\ast}, \lambda_{t}^{\ast} \in \argmin_{\bm{\theta}, \lambda} \sum_{m=1}^{M} \mathcal{L}\left( T_{R_{\bm{\theta}}, \lambda, \alpha}^{t}(\bm{y}^{(m)}), \bm{x}^{(m)} \right)\,.
	\label{eq:crr_training_loss}
\end{equation}

After having trained the denoiser, we define our prior potential as
\begin{equation}
	g(\bm{x})= \frac{\lambda}{\mu} R_{\bm{\theta}}(\mu \bm{x}) \,,
	\label{eq:crr_potential}
\end{equation}
where we have dropped the $\bm{\theta}_{t}^{\ast}, \lambda_{t}^{\ast}$ notation for $\bm{\theta}, \lambda$ and added a scaling parameter, $\mu$, to boost performance following \citet{goujon2023}. For the optimisation algorithm, we need the Lipschitz constant of the gradient of the potential in Equation \ref{eq:crr_potential}, which can be expressed as
\begin{equation}
	\text{Lip}(\nabla g) = \lambda \, \mu \, \text{Lip}(\nabla R_{\bm{\theta}}) \leq \lambda \, \mu \, \| \mathbf{W}^{T} \Sigma_{\infty} \mathbf{W} \| \,,
	\label{eq:crr_lipschitz}
\end{equation}
which is calculated in \citet[Prop. IV.1]{goujon2023}, and $\Sigma_{\infty} = \text{diag}(\| {\sigma}_{1}' \|_{\infty}, \ldots, \| \sigma_{N_C}' \|_{\infty} )$, and $\mathbf{W} = [\bm{w}_{1} \cdots \bm{w}_{N_C}]^{T}$ where $\bm{w}_{i}$ corresponds to the filter $\bm{h}_{i}$: $\bm{h}_{i} \ast \bm{x} = \bm{w}_{i}^{T} \bm{x}$.

	\subsection{Computing our reconstruction: the MAP}

	In our case, computing the MAP reduces to solving a convex optimisation problem. Following Equation \ref{eq:map_estimation}, the optimisation problem we address is the following one,
	\begin{equation}
		\hat{\bm{x}}_{\text{MAP}} = \argmin_{\bm{x} \in \mathbb{R}^{N}} \frac{1}{2 \sigma^{2}} \left\| \bm{y} - \Phi \bm{x} \right\|_{2}^{2} + \frac{\lambda}{\mu} R_{\bm{\theta}}(\mu \bm{x}) + \iota_{\mathbb{R}^{N}}(\bm{x}) \,,
		\label{eq:optim_target}
	\end{equation}
	where in addition we include $\iota_{\mathbb{R}^{N}}$, an indicator function enforcing the reconstructed image to be real. The proximal operator of the indicator function to a convex set is known and it amounts to a projection onto that convex set. In the the last term of Equation \ref{eq:optim_target} the proximal operator of the reality constraint is the projection of the vector to the real number, which is written as $\text{Re}(\cdot)$. We have assumed a (white) Gaussian likelihood and the prior term is based on a previously trained CRR-NN. The CRR-NN is smooth with Lipschitz continuous gradients. However, the non-smoothness of the reality enforcing constraint forces us to rely on proximal algorithms \citep{parikh2014} instead of an accelerated gradient descent method \citep{nesterov2018}. In this case, we use the FISTA algorithm \citep{beck2009}.

	For the optimisation, we need the gradient of the likelihood and prior terms
	\begin{align}
		\label{eq:grad_likelihood}
		 & \nabla_{\bm{x}} f(\bm{x},\bm{y}) = \frac{1}{\sigma^{2}} (\Phi^{\dagger} (\Phi \bm{x} - \bm{y})) \,, \\
		 & \nabla g(\bm{x}) = \lambda \nabla R_{\bm{\theta}}(\mu \bm{x}) \,,
		\label{eq:grad_prior_crr}
	\end{align}
	where, in our case, $(\cdot)^{\dagger}$ is the complex conjugate transpose.

	To ensure the algorithm's convergence we use the stepsize $\tau = 1/L$, where $L = \text{Lip}(\nabla_{\bm{x}} f(\bm{x},\bm{y}) + \nabla g(\bm{x}))$. We can estimate a simple bound for the Lipschitz constant as follows
	\begin{align}
		L & \leq \text{Lip}(\nabla_{\bm{x}} f(\bm{x},\bm{y})) + \text{Lip}(\nabla g(\bm{x})) = L_{\text{likelihood}} + L_{\text{prior-CRR-NN}}\,, \nonumber \\
		  & \leq \frac{\| \Phi^{\dagger} \Phi \|}{\sigma^{2}} + \lambda \, \mu \, \| \mathbf{W}^{T} \Sigma_{\infty} \mathbf{W} \| \,,
		\label{eq:lipschitz_optim}
	\end{align}
	where we have exploited the result from Equation \ref{eq:crr_lipschitz}, and $\| \Phi^{\dagger} \Phi \|$ denotes the spectral norm, which in the case of a linear operator coincides with its maximum singular value. In the simplified problem we are considering in Section \ref{sc:radio_interf_prob} with gridded visibilities, we have that $\| \Phi^{\dagger} \Phi \| = 1$. If a more realistic linear operator should be considered, the maximum singular value could be computed iteratively via the power method \citep{golub2013}.

	We initialise the optimisation with the naturally weighted dirty image, $\bm{x}_{(0)} = \text{Re}(\Phi^{\dagger}\bm{y})$. The optimisation procedure is summarised in Algorithm \ref{al:optim_alg}. We optimise for a fixed number of iterations $N_{\text{max}}$, or until a tolerance criterion of $\xi$ is reached. The stepsize is computed using the bound from Equation \ref{eq:lipschitz_optim}.

	\begin{algorithm}
		\caption{FISTA \citep{beck2009} tackling (\ref{eq:optim_target})}
		\label{al:optim_alg}
		\textbf{Input:}  $R_{\bm{\theta}}$, $\Phi$, $\sigma$, $\mu$, $\lambda$, $\xi$, $a_{(1)}=1$, $\bm{z}_{(1)} = \bm{x}_{(0)} = \text{Re}(\Phi^{\dagger}\bm{y})$,  $\tau = 0.98/L$.\\
		\textbf{Output:} $\hat{\bm{x}}_{\text{MAP}}$ \vspace{0.05in} \\
		\For{$n=1, \ldots, N_{\text{max}}$}{
			$\bm{x}_{(n)} = \bm{z}_{(n)} - \tau \left(\frac{1}{\sigma^{2}} \text{Re}(\Phi^{\dagger} (\Phi \bm{z}_{(n)} - \bm{y})) + \lambda \nabla R_{\bm{\theta}}(\mu \bm{z}_{(n)}) \right)$ \\
			$a_{(n+1)} = \frac{1}{2} (1 + \sqrt{4 a_{(n)}^{2} + 1})$  \\
			$\bm{z}_{(n+1)} = \bm{x}_{(n)} + \frac{a_{(n)} - 1}{a_{(n+1)}} (\bm{x}_{(n)} - \bm{x}_{(n-1)})$ \\
			\If{$\frac{\| \bm{x}_{(n)} - \bm{x}_{(n-1)} \|}{ \| \bm{x}_{(n-1)} \|} < \xi$}{
				break
			}
		} \vspace{0.05in}

		set $\hat{\bm{x}}_{\text{MAP}} = \bm{x}_{(n)}$
	\end{algorithm}

	\section{Scalable uncertainty quantification}
	\label{sc:scalable_uq}

	Enforcing the posterior's convexity and explicit potential opens the door to scalable UQ methodology that was unreachable otherwise. The restriction to log-concave posteriors is the price we pay to gain great scalability. Our approach is based on the work from \citet{pereyra2017}, which exploits concentration phenomena occurring in high-dimensional log-concave posteriors. The Bayesian high-posterior-density region can be approximated in log-concave models as the posterior probability mass tends to concentrate in particular regions on the parameter space. The approximation requires the MAP estimation, $\hat{\bm{x}}_{\text{MAP}}$, which we have already computed as it is the chosen point estimate for our reconstruction. This result allows us to estimate information from the posterior probability density function without MCMC sampling. In this Section, we introduce the main result we exploit for UQ. We then describe the proposed scalable UQ methods and how to validate our results with Langevin-based MCMC sampling algorithms.

	\subsection{Highest Posterior Density Regions}
	\label{sc:hpd_region_approx}

	Let us define a posterior credible region with a credible level of $100(1-\alpha)\%$ as a set $C_{\alpha} \in \mathbb{R}^{N}$ satisfying
	\begin{equation}
		p(\bm{x} \in C_{\alpha} | \bm{y}) = \int_{\bm{x} \in \mathbb{R}^{N}} p(\bm{x} | \bm{y}) \mathbbm{1}_{C_{\alpha}}(\bm{x}) {\rm d} \bm{x} = 1 - \alpha,
		\label{eq:posterior_credible_region}
	\end{equation}
	with $\mathbbm{1}_{C_{\alpha}}$ being being unity if $\bm{x} \in C_{\alpha}$ and zero otherwise. There are many regions satisfying the previous equation. We will focus on the highest posterior density region (HPD), which is defined as
	\begin{equation}
		C_{\alpha} := \left\{ \bm{x} \in \mathbb{R}^{N} : f(\bm{x}) + g(\bm{x}) \leq \gamma_{\alpha} \right\} \,,
		\label{eq:hpd_region}
	\end{equation}
	where $f$ and $g$ are the potentials of our likelihood and prior terms, and $\gamma_{\alpha}$ is a constant that defines a level-set of the log-posterior such that Equation \ref{eq:posterior_credible_region} holds. The HPD region has the property of having minimum volume \citep[\S 5.5]{robert2007}.

	Our posterior $p(\bm{x} | \bm{y}) = \exp[-f(\bm{x})-g(\bm{x})]/Z$ is log-concave on $\mathbb{R}^{N}$, where $Z$ is the Bayesian evidence. Then, following \citet[Theorem 3.1]{pereyra2017}, for any $\alpha \in (4 \exp[(-N/3)], 1)$, the HPD region $C_{\alpha}$ from Equation \ref{eq:hpd_region} is contained in
	\begin{equation}
		\hat{C}_{\alpha} = \left\{ \bm{x} \in \mathbb{R}^{N} : f(\bm{x}) + g(\bm{x}) \leq \hat{\gamma}_{\alpha} \right\},
		\label{eq:HPD_approx_region}
	\end{equation}
	where
	\begin{equation}
		\hat{\gamma}_{\alpha} = f(\hat{\bm{x}}_{\text{MAP}}) + g(\hat{\bm{x}}_{\text{MAP}}) + \sqrt{N} \tau_{\alpha} + N,
		\label{eq:HPD_thresh_approx}
	\end{equation}
	with a positive constant $\tau_{\alpha} = \sqrt{16 \log(3/\alpha)}$ independent of $p(\bm{x} | \bm{y})$.

	Theorem 3.2 in \citet{pereyra2017} gives the error analysis of the approximation, and we see that $0 \leq \hat{\gamma}_{\alpha} - \gamma_{\alpha} \leq \tau_{\alpha}\sqrt{N}+N$. Therefore, the error upper bound grows linearly with the dimension $N$, making $\hat{C}_{\alpha}$ a stable approximation of $C_{\alpha}$. The error lower bound along with the convexity of $f+g$ guarantees the inclusion $C \subseteq \hat{C}$ and consequently the resulting approximation is a conservative one where the errors are overestimated.

	After showing the main result allowing us to do UQ bypassing posterior sampling methods, it is clear from where the constraints of the prior come. The convexity is required to guarantee a log-concave posterior, as the likelihood potential is convex. The prior potential $g$ needs to be explicit to compute the approximate HPD region using Equation \ref{eq:HPD_thresh_approx}.

	\subsection{MAP-based UQ methods}

	We now explore different scalable UQ schemes based on the fast approximate implicit representation of the HPD region. For all the methods presented, we assume that we have already computed the $\hat{\bm{x}}_{\text{MAP}}$ estimation and the approximated HPD region threshold, $\hat{\gamma}_{\alpha}$.

	\subsubsection{Bayesian hypothesis testing of structure}

	A useful UQ tool is to perform a \textit{knock-out} hypothesis test to asses if a surrogate image still belongs to the HPD region \citep{cai2018,cai2018_2,price2021}. First, the surrogate image $\bm{x}_{\text{sgt}}$ is constructed by modifying the reconstruction, $\hat{\bm{x}}_{\text{MAP}}$. Then, it suffices to check if
	\begin{equation}
		f(\bm{x}_{\text{sgt}}) + g(\bm{x}_{\text{sgt}}) \leq \hat{\gamma}_{\alpha} \,.
		\label{eq:hyp_test_eq}
	\end{equation}
	If the inequality is satisfied, we cannot draw conclusions on the test we made, as $\bm{x}_{\text{sgt}}$ still belongs to the HPD region. However, if the inequality does not hold, we can conclude that $\bm{x}_{\text{sgt}}$ is out from the HDP region with a $100(1-\alpha)\%$ confidence level.

	This test can answer a variety of questions about our reconstructed image. One example is to interrogate some structure in the image to see if it is a reconstruction artefact or is physically motivated. For this question, the surrogate image would be composed of an image with the region of interest artificially inpainted with surrounding information. We need to take the inpainted image as our surrogate and evaluate Equation \ref{eq:hyp_test_eq} to see if the test is conclusive.

	The image inpainting algorithm is built similarly as in \citet{cai2018_2} but adapted to the CRR-NN-based prior. We start by selecting a region of interest $\Omega_{\text{D}}$, which is a subset of (typically contiguous) pixels from the image, where $\Omega_{\text{D}} \subseteq \Omega$, where $\Omega$ denotes the set of all the image pixels. The region $\Omega_{\text{D}}$ will be inpainted with background information. We then inpaint this region
	iteratively minimising $R_{\bm{\theta}}$ based on the following scheme
	\begin{align}
		\bm{x}_{\text{sgt}, (m+1)} = \hat{\bm{x}}_{\text{MAP}} & \mathbbm{1}_{\Omega - \Omega_{\text{D}}}                                                                                                                                    \\
		                                                       & + \left( \bm{x}_{\text{sgt}, (m)} - \alpha \lambda \nabla R_{\bm{\theta}}\left(\mu \, \bm{x}_{\text{sgt}, (m)}\right) \right) \mathbbm{1}_{\Omega_{\text{D}}} \,, \nonumber
	\end{align}
	where $\mathbbm{1}$ are indicator functions, and $\mathbbm{1}_{\Omega - \Omega_{\text{D}}}$ is a shorthand for $\mathbbm{1}_{\Omega} - \mathbbm{1}_{\Omega_{\text{D}}}$. We carry out a gradient step with the CRR-NN on the surrogate image and only update the region of interest. The hyperparameters, $\alpha$, $\lambda$, and $\mu$ are set as in Algorithm \ref{al:optim_alg}.

	Alternatively, \citet{repetti2019} presented a more sophisticated method to perform hypothesis testing of structure, which also exploits the approximations in Equations \ref{eq:HPD_approx_region}-\ref{eq:HPD_thresh_approx}. The method is dubbed Bayesian uncertainty quantification by optimisation (BUQO), and to answer the hypothesis test, it aims to study the intersection of two sets. The first one is defined in Equation \ref{eq:HPD_approx_region} that corresponds to the MAP estimate. The second one describes the set of feasible inpainted images given a region of interest and a set of constraints of desired properties. If the set intersection is empty, the structure of interest is considered present in the image with confidence $\alpha$ from Equation \ref{eq:HPD_approx_region}. \citet{tang2023} proposed an extension of the BUQO method to inpaint with data-driven models. However, these methods involve solving an expensive optimisation problem that does not scale with the high-dimensional settings we are considering in this work.

	Another example is to interrogate the reconstruction to see if the fine structure of the image is physical or likely an artefact. To construct the surrogate image we convolve the region of interest, $\Omega_{\text{D}}$, with a Gaussian smoothing kernel $G(0,\Sigma)$,
	\begin{equation}
		\bm{x}_{\text{sgt}} = \hat{\bm{x}}_{\text{MAP}} \mathbbm{1}_{\Omega - \Omega_{\text{D}}} + \left( \hat{\bm{x}}_{\text{MAP}} \ast G(0,\Sigma) \right) \mathbbm{1}_{\Omega_{\text{D}}} \,,
		\label{eq:blurring_op}
	\end{equation}
	where $\ast$ denotes the 2D convolution operation and test Equation \ref{eq:hyp_test_eq}.

	\subsubsection{Local credible intervals}
	\label{sc:pixel_UQ_def}

	Local credible intervals (LCIs) provide a tool to quantify spatial uncertainty per pixel at different resolutions. The LCIs are interpreted as Bayesian error bars for each pixel or superpixel, where with superpixel, we refer to a group of contiguous pixels. \citet{cai2018} computed LCIs using MCMC methods and then extended it in \citet{cai2018_2} to compute them based on the approximated HPD region based on the MAP. \citet{price2019} also exploited MAP-based LCIs in another imaging inverse problem, mass-mapping, for weak gravitational lensing convergence reconstruction.

	Let us write $\Omega = \{ \Omega_i \}_{i=1}^{M}$ the set of superpixels that partition the domain of $\bm{x}$. This partition is such that $\Omega_i \cap \Omega_j = \emptyset, \forall i \neq j$ and $\Omega = \cup_i \Omega_i$. We denote the indicator of the superpixel $\Omega_i$ as $\bm{\zeta}_{\Omega_i}$, that is one if the pixel belongs to the superpixel $\Omega_i$ and zero otherwise. The use of smaller or bigger superpixel sizes, i.e., $\| \bm{\zeta}_{\Omega_i} \|_{0}$, allows us to visualise the LCIs at different scales. The calculation of the LCIs is based on computing an upper and lower bound for each superpixel. Each bound is defined by the constant value we need to add or subtract to the superpixel region so that the modified image exits the approximate HPD credible region $\hat{C}_{\alpha}$. In other words, we compute the values that saturate the HPD region for each superpixel.

	We can isolate the superpixel region by defining the following surrogate image
	\begin{equation}
		\bm{x}_{i, \xi} = \hat{\bm{x}}_{\text{MAP}} (\bm{I} - \bm{\zeta}_{\Omega_i}) + (\xi + \bar{\bm{x}}_{\text{MAP}, \Omega_i}) \bm{\zeta}_{\Omega_i} \,,
	\end{equation}
	where $\bar{\bm{x}}_{\text{MAP}, \Omega_i}$ corresponds to the mean value of the pixels in the superpixel $\Omega_i$, and $\xi \in \mathbb{R}$. We vary the superpixel value from its mean by a uniform value $\xi$. The bounds for a superpixel $\Omega_i$ are computed as
	\begin{align}
		\label{eq:lci_bounds_1}
		\xi_{+, \Omega_i} = & \max_{\xi} \left\{ \xi \,|\, f(\bm{x}_{i, \xi}, \bm{y}) + g(\bm{x}_{i, \xi}) \leq \hat{\gamma}_{\alpha}, \, \xi \in [0, +\infty) \right\} \,, \\
		\xi_{-, \Omega_i} = & \min_{\xi} \left\{ \xi \,|\, f(\bm{x}_{i, \xi}, \bm{y}) + g(\bm{x}_{i, \xi}) \leq \hat{\gamma}_{\alpha}, \, \xi \in (-\infty, 0] \right\} \,,
		\label{eq:lci_bounds_2}
	\end{align}
	where we use the threshold $\hat{\gamma}_{\alpha}$ defined in Equation \ref{eq:HPD_thresh_approx}. The calculation of each bound can be recast as a problem of finding the zero of a function. The class of root-finding algorithms is well adapted for this root-finding problem, and, in practice, we use the bisection method \citep{burden1989}. \citet{price2021_2} proposed a faster way to compute the superpixel bounds by exploiting linearity that we could use to further accelerate the computation of $\xi_{+, \Omega_i}$ and $\xi_{-, \Omega_i}$.

	Once the bounds have been computed, we can collate the results for all superpixels and use the length of the LCIs to visualise the reconstruction uncertainty. The length of the LCI for superpixel $\Omega_i$ is defined as $l_{i} = \xi_{+, \Omega_i} - \xi_{-, \Omega_i}$, which we can visualise as an uncertainty image
	\begin{equation}
		\bm{\xi} = \sum_{i} \left(\xi_{+, \Omega_i} - \xi_{-, \Omega_i} \right) \bm{\zeta}_{\Omega_i} \,.
	\end{equation}

	The choice of using the mean on $\bar{\bm{x}}_{\text{MAP}, \Omega_i}$ for the region of the superpixel that will be studied constitutes a deviation from the original MAP reconstruction. We find it more physical to move the averaged superpixel rather than moving the original pixels belonging to the superpixel by a constant value. This choice constitutes another approximation to the proposed scheme that has already approximated the HPD region in Equation \ref{eq:HPD_approx_region}. Using superpixels allows us to gain sensibility and computing time at the expense of lowering the resolution of the LCI map, which can be a sensible trade-off for very large images.

	We will later validate the computed LCIs using the posterior samples obtained from computing the posterior standard deviation at different superpixel sizes. The method requires turning each posterior sample into an image with $M$ superpixels. We set the value of the superpixel to the mean of the values of belonging pixels.

	\subsubsection{Fast pixel uncertainty quantification at different scales}
	\label{sc:fast_pixel_UQ_new}

	The MAP-based LCIs described in the previous section are orders of magnitude faster than their MCMC-based counterparts \citep{cai2018,cai2018_2}. Nevertheless, to compute the LCIs, we still need to evaluate the likelihood potential, $f$, several times for each superpixel. As mentioned, evaluating the likelihood potential is by far the most time-consuming operation. The fact that we need to evaluate $f$ several times for each subpixel might make the LCIs impractical for SKA-scale problems.

	To overcome this issue, we propose a new approach relying on wavelet decomposition of the MAP reconstruction that reads
	\begin{equation}
		\hat{\bm{x}}_{\text{MAP}} = \mathbf{\Psi} \, \hat{\bm{a}}_{\text{MAP}} = \sum_{i=1}^{L} \mathbf{\Psi}_{i} \, \hat{a}_{\text{MAP}, i} \,,
		\label{eq:wav_transform_map}
	\end{equation}
	with a wavelet dictionary $\mathbf{\Psi}$. We define the hard thresholding operator for $\bm{a} \in \mathbb{C}^{L}$ with a threshold of $\xi_{\text{th}}$,
	\begin{equation}
		S_{\text{hard},\,\xi_{\text{th}}}(\bm{a}) = \left[ S_{\text{hard},\,\xi_{\text{th}}}(a_1), \ldots, S_{\text{hard},\,\xi_{\text{th}}}(a_L) \right]^{T}\,,
	\end{equation}
	as the point-wise application of the following hard-thresholding function
	\begin{equation}
		S_{\text{hard},\,\xi_{\text{th}}}(a_i) =
		\begin{cases}
			0,    & \text{if} \ |a_i| \le \xi_{\text{th}}, \\
			a_i , & \text{otherwise}.
		\end{cases}
	\end{equation}

	Let $\hat{\xi}_{\text{th}}$ be the thresholded value for which the thresholded image saturate the HPD region as follows
	\begin{align}
		\hat{\xi}_{\text{th}} = \max_{\xi_{\text{th}}} \{ \xi_{\text{th}} \, | & \, f(\hat{\bm{x}}_{\text{MAP}, \,\xi_{\text{th}}}, \bm{y}) + g(\hat{\bm{x}}_{\text{MAP}, \,\xi_{\text{th}}}) \leq \hat{\gamma}_{\alpha},                         \\
		                                                                       & \hat{\bm{x}}_{\text{MAP}, \,\xi_{\text{th}}} = \bm{\Psi} \, S_{\text{hard},\,\xi_{\text{th}}}(\hat{\bm{a}}_{\text{MAP}}),\, \xi \in [0, +\infty) \}\,. \nonumber
	\end{align}
	We can compute the threshold bound with a root-finding method, as was the case for the LCIs. The advantage of this formulation is that we are solving only one root-finding problem as opposed to one per superpixel in the LCIs calculation. This change considerably reduces the number of likelihood evaluations and, therefore, the computational complexity of the UQ method.

	By computing the difference between the MAP, $\hat{\bm{x}}_{\text{MAP}}$, and the thresholded surrogate, $\hat{\bm{x}}_{\text{MAP}, \, \hat{\xi}_{\text{th}}}$, we obtain an estimation of the solution's uncertainty and this can give us information about possible errors in the MAP. Furthermore, we can compare the MAP and the thresholded surrogate image to estimate errors as a function of scale, thus exposing the different structures of our reconstruction.

	Let us consider our wavelet transformation as a multiscale transform of level $J+1$ \citep{mallat2008,starck2010}. We can rewrite Equation \ref{eq:wav_transform_map} showcasing the multiscale coefficients as follows
	\begin{equation}
		\hat{\bm{x}}_{\text{MAP}} = \mathbf{\Psi} \, \hat{\bm{a}}_{\text{MAP}} = \sum_{l=0}^{J} \mathbf{\Psi}_{l} \, \hat{\bm{a}}_{\text{MAP}, l} \,,
		\label{eq:ml_wav_transform_map}
	\end{equation}
	where $\hat{\bm{a}}_{\text{MAP}, l}$ are the coefficients corresponding to the $l$-th level of decomposition, and the zeroth level corresponds to the coarse scale. We can now build thresholded surrogate images at different scales by replacing the MAP wavelet coefficients at level $l$ from Equation \ref{eq:ml_wav_transform_map} with the coefficients of the thresholded surrogate image $\hat{\bm{x}}_{\text{MAP}, \, \hat{\xi}_{\text{th}}}$. Let us write the thresholded surrogate image at level $j$ as follows
	\begin{equation}
		\hat{\bm{x}}_{\text{MAP}, \, \hat{\xi}_{\text{th}}, \, j} = \sum_{\substack{l=0,\\ l \neq j}}^{J} \mathbf{\Psi}_{l} \, \hat{\bm{a}}_{\text{MAP}, l} + \mathbf{\Psi}_{j} \hat{\bm{a}}_{\text{MAP}, \, \hat{\xi}_{\text{th}}, \, j} \,,
		\label{eq:coefficient_replacement}
	\end{equation}
	where $\hat{\bm{a}}_{\text{MAP}, \, \hat{\xi}_{\text{th}}, \, j}$ corresponds to the wavelet coefficients of the thesholded surrogate image $\hat{\bm{x}}_{\text{MAP}, \, \hat{\xi}_{\text{th}}}$ at level $j$. The errors at level $j$ can be approximated by the difference between $\hat{\bm{x}}_{\text{MAP}}$ and $\hat{\bm{x}}_{\text{MAP}, \, \hat{\xi}_{\text{th}}, \, j}$.

	There are two main advantages of this approach to pixel-based UQ with respect to the LCIs described in Section \ref{sc:pixel_UQ_def}. The first one is the reduced computational complexity, as we only need to solve a single root-finding problem, significantly reducing the number of likelihood evaluations. The second is that when we saturate the HPD region, we consider the entire image simultaneously. In the LCI computation, we only change a small group of pixels until it saturates the HDP region that corresponds to the entire image. This behaviour can be problematic as, for example, the LCI error bounds will be larger if the size of the image grows and the superpixel size is kept the same, which is an undesirable property. Consequently, the predicted errors from the new pixel UQ approach are faster to compute and considerably tighter than the LCIs.

	\subsection{MCMC sampling and UQ validation}
	\label{sc:bayes_inf_for_uq_val}

	As stated before, MCMC sampling is not yet scalable to the high dimensions of the RI imaging problems we target. However, sampling is still helpful in validating the UQ approaches of this paper. If we sample from the posterior distribution, we can compute the posterior standard deviation, providing a visual representation of the posterior, including the learned convex regulariser. Sampling from the posterior also allows us to compare the MAP estimator with another widely known estimator, the posterior mean \citep{arridge2019}, which coincides with the minimum mean squared error (MMSE) estimator under some assumptions.

	The log-posterior distribution of the \textsc{QuantifAI} model with the CRR-NN reads
	\begin{equation}
		-\log p_{\text{\scriptsize CRR-NN}}(\bm{x}|\bm{y}) \propto \frac{1}{2 \sigma^{2}} \left\| \bm{y} - \Phi \bm{x} \right\|_{2}^{2} + \frac{\lambda}{\mu} R_{\bm{\theta}}(\mu \bm{x}) + \iota_{\mathbb{R}^{N}}(\bm{x}) \,,
	\end{equation}
	with the first two terms belonging to $\mathcal{C}^{1}$ with Lipschitz gradients, we do not need to use any approximation, e.g., the MY envelope, to sample from it. The reality constraint is enforced directly when evaluating the gradient of the log-likelihood. The Langevin diffusion sampling algorithms reviewed in Section \ref{sc:bayesian_inf} require the gradient of the log-posterior distribution, which have been computed in Equation \ref{eq:grad_likelihood} and Equation \ref{eq:grad_prior_crr}. In practice, we will use the SK-ROCK algorithm \citep{pereyra2020} as it provides a faster convergence than the ULA algorithm. We omit the subsequent MH step to accelerate the calculations motivated by the consistent results from \citet{cai2018}.

	The log-posterior distribution of the analysis formulation of the model from \citet{cai2018} with a wavelet-based sparsity enforcing prior reads
	\begin{equation}
		-\log p_{\text{\scriptsize WAV}}(\bm{x}|\bm{y}) \propto \frac{1}{2 \sigma^{2}} \left\| \bm{y} - \Phi \bm{x} \right\|_{2}^{2} + \lambda \big\| \Psi^{\dagger} \bm{x} \big\|_{1} + \iota_{\mathbb{R}^{N}}(\bm{x}) \,,
	\end{equation}
	which includes the non-smooth prior term with the $\ell_1$ norm, and the reality constraint which we again apply to the gradient of the log-likelihood. We resort to the MY envelope $\gamma$-approximation of the sparsity-inducing prior term as shown in Equation \ref{eq:MY_envelope_approx}. The proximal operator of the prior term has a closed-form solution that reads
	\begin{equation}
		S_{\text{soft},\,\beta_{\text{th}}}(\bm{a}) = \left[ S_{\text{soft},\,\beta_{\text{th}}}(a_1), \ldots, S_{\text{soft},\,\beta_{\text{th}}}(a_L) \right]^{T}\,,
	\end{equation}
	where $\bm{a} = \Psi^{\dagger} \bm{x}$ and we have applied element-wise the soft-thresholding function
	\begin{equation}
		S_{\text{soft},\,\beta_{\text{th}}}(a_i) =
		\begin{cases}
			0,                                                          & \text{if} \ |a_i| \le \beta_{\text{th}}, \\
			\frac{a_i}{|a_i|} \left(|a_i| - \beta_{\text{th}} \right) , & \text{otherwise}.
		\end{cases}
		\label{eq:soft_thresh_op}
	\end{equation}
	The threshold $\beta_{\text{th}}$ used in practice is $\lambda \gamma$, the product of the regularisation strength and the parameter of the MY approximation. See \citet{cai2018} for more details on sampling the model with a wavelet-based regularisation. In practice, we again rely on the SK-ROCK algorithm for sampling and avoid using an MH step for the reasons mentioned above.

	\section{Experimental results}
	\label{sc:experimental_results}

	In this section, we demonstrate the \textsc{QuantifAI} model against the wavelet-based model presented in \citet{cai2018,cai2018_2} as it is one of the few RI imaging methods providing UQ capabilities. We use a simulated setup with real reconstructed RI images considered as the ground truth. We focus on the UQ capabilities of the methods, while also considering reconstruction performance.

	\subsection{Dataset}
	\label{sc:simulations}

	The base images used in our experiment are the ones from \citet{cai2018}: (\textit{i}) the HI region of the M31 galaxy in Figure \ref{fi:M31_reconstruction}~(a), (\textit{ii}) the W28 supernova remnant in Figure \ref{fi:W28_reconstruction}~(a), (\textit{iii}) the 3C288 radio galaxy in Figure \ref{fi:3c288_reconstruction}~(a), and (\textit{iv}) the Cygnus A radio galaxy in Figure \ref{fi:CYN_reconstruction}~(a). All the images are $256 \times 256$ pixels, except for the Cygnus A galaxy, which is $256 \times 512$. As the ground truth images are reconstructed from real observations, some minor original residuals and backgrounds are more noticeable in the log scale images; for example, see Figure \ref{fi:3c288_reconstruction}~(b). The ground truth images' values have been normalized to a unitless range between $0$ and $1$, and therefore, the colour bars in the reconstruction figures follow this range.  

	The previous images correspond to $\bm{x}$ in our observational model from Equation \ref{eq:linear_inv_prob}. The complex Gaussian noise $\bm{n} \in \mathbb{C}^{M}$ is composed of independent and identically distributed (\textit{i.i.d.}) samples. Each sample is simulated following a complex Gaussian distribution, $n_i \sim \mathcal{N}_{\mathcal{C}}(0, \sigma^2)$, which implies that $\text{Re}(\bm{n}), \text{Im}(\bm{n}) \sim \mathcal{N}(0, \sigma/\sqrt{2})$ \citep{tse2005}. The noise is set such that we get a specific input signal-to-noise ratio (ISNR) on each image. For all the experiments, we set the ISNR to $30$dB, and the noise standard deviation is computed as follows
	\begin{equation}
		\sigma = \frac{\| \bm{\Phi} \bm{x} \|_2}{\sqrt{M}}\, 10^{-\text{ISNR}/20} \,.
	\end{equation}

	To mimic the $uv$ plane coverage, we reuse the Fourier mask from \citet[Fig. 2]{cai2018} and use it to generate the visibilities from $\bm{y}$. The variable sampling density profile was taken from \citet{puy2011} and represents a $10\%$ coverage of the Fourier plane. In the experiments in the current section, we work with gridded visibilities where we have around $1.3 \times 10^{4}$ visibilities for Cygnus A and $6.5 \times 10^{3}$ visibilities for the rest of the images. The validation of the UQ techniques through MCMC sampling requires a large amount of iterations. The use of gridded visibilities allows us to base the forward operator $\mathbf{\Phi}$ from Equation \ref{eq:linear_inv_prob} on the FFT \citep{fft1965}, helping to alleviate the computational burden of the validation. Section \ref{sc:realistic_exp} presents results with ungridded visibilities.

	\subsection{Models and experiment settings}
	\label{sc:models}

	\subsubsection{RI imaging models}

	The CRR-NN in the \textsc{QuantifAI} model is a pretrained model with $t=5$, Gaussian white noise with standard deviation $\sigma=5$, and parameters $\mu=20$ , $\lambda=5 \times 10^{4}$. The model was trained on a set of natural images from the BSD500 dataset \citep{arbelaez2011} cointaining images of landscapes, people, animals and objects among others. The images are scaled to the $[0,255]$ range, using $\ell_1$ norm as the loss function in Equation \ref{eq:crr_training_loss} with the Adam optimiser \citep{kingma2017}. The training parameters followed \citet[\S VI.A]{goujon2023}.

	The wavelet dictionary $\Psi$ used in the wavelet-based model is composed of Daubechies $8$ wavelets \citep{daubechies1992} with a multiresolution level $J=4$ following the setup from \citet{cai2018,cai2018_2}. The regularisation parameter $\lambda_{\text{\scriptsize WAV}}$ was set to $1 \times 10^{2}$.

	The regularisation strengths of both models, $\lambda$ and $\lambda_{\text{\scriptsize WAV}}$, were set to maximise the MAP reconstruction SNR using a grid search. We observed that \textsc{QuantifAI}'s reconstruction SNR is significantly more robust with respect to the choice of regularisation strength than the wavelet-based models.

	\subsubsection{Optimisation settings}

	For \textsc{QuantifAI}, we used the optimisation algorithm shown in Algorithm \ref{al:optim_alg}. The wavelet-based model also requires a proximal algorithm due to its non-smooth component and to provide a fair comparison we used the FISTA algorithm \citep{beck2009} presented with more detail in the Appendix \ref{ap:wavelet_map}. In these experiments, we assumed that the noise level $\sigma$ is known. If the noise level is unknown, it may be estimated by standard techniques \citep{price2021}. Both algorithms' tolerance criterion $\xi$ was set to $10^{-5}$, and the total number of iterations to $1.5 \times 10^{4}$. Nevertheless, both optimisation algorithms converged before the total number of iterations was reached.

	\subsubsection{MCMC sampling settings}

	We generate $5 \times 10^{4}$ samples from each posterior distribution, with $5 \times 10^{4}$ burn-in iterations and a thinning factor of $10$. The burn-in iterations consist of generating several samples that will be discarded so that the chain passes its transient period. The thinning factor corresponds to the number of samples that need to be generated between two samples so that they can be considered independently drawn from the target distribution. The sampling algorithm produced a total of $5.5 \times 10^{5}$ samples for each model. We have set to $10$ the number of stages for the SK-ROCK algorithm \citep{pereyra2020}, which is one of its main hyperparameters. The sampling of the posterior probability distributions is used as a validation, and therefore we set the sampling parameters focusing on good reconstructions and posterior samples rather than speed.

	The wavelet-based model requires the MY envelope approximation to guarantee the chain's convergence, as described in Section \ref{sc:bayesian_inf} and Section \ref{sc:bayes_inf_for_uq_val}. The MY approximation parameter $\gamma$ was set to the inverse of the likelihood gradient's Lipschitz constant, c.f.~the first term of Equation \ref{eq:lipschitz_optim}.

	The choice of the step sizes is critical to ensure the chains' convergence to the target distribution in a reasonable amount of time. The step size is chosen as a function of each posterior gradient's Lipschitz constant. The step sizes $\delta_{\textsc{Q}}$ and $\delta_{\text{W}}$, corresponding to the \textsc{QuantifAI} and wavelet-based models, respectively, are computed as follows
	\begin{equation}
		\delta_{\textsc{Q}} = \frac{\kappa_{\textsc{Q}}}{L_{\text{likelihood}} + L_{\text{prior-CRR-NN}}} \,,\quad \delta_{\text{W}} = \frac{\kappa_{\text{W}}}{L_{\text{likelihood}} + \gamma^{-1}} \,,
	\end{equation}
	where the Lipschitz constant bounds are shown in Equation \ref{eq:lipschitz_optim}, and $\kappa_{\textsc{Q}}$ and $\kappa_{\text{W}}$, are two positive constants smaller than one, here set to $0.98$. We have followed the advise from \citet{durmus2018,cai2018} to set the sampling parameters.

	\subsubsection{UQ settings}

	We set $\alpha = 0.01$ in all the UQ methods, so the confidence level is $99\%$. We used the bisection algorithm to compute the LCIs and the fast pixel UQ at different scales, with tolerance $10^{-4}$ and maximum number of iterations $200$, for both models. We used the same wavelet dictionary as in the wavelet-based model for the fast pixel UQ at different scales.

	The inpainting algorithm uses the same stopping criterion as Algorithm \ref{al:optim_alg}. In this case, the tolerance is set to $5 \times 10^{-6}$, and the total number of iterations to $1.5 \times 10^{4}$. The CRR-NN used for the inpainting is the same one used in the \textsc{QuantifAI} model.

	The Gaussian blurring kernel $G(0,\Sigma)$ from Equation \ref{eq:blurring_op} is set using $\Sigma=\sigma_{G}^{2} I_{2 \times 2}$, with $\sigma_{G}$ being $3.5$ pixels and a truncation radius of $7$ pixels, giving a kernel $G \in \mathbb{R}^{15 \times 15}$.

	\subsection{Image reconstruction}

	\begin{table}
		\begin{center}
			\caption{Reconstruction performance of the different point estimates for the dataset images in terms of SNR with respect to the ground truth. We compare the MAP and the MMSE reconstruction of the wavelet-based and the \textsc{QuantifAI} model. We include the dirty reconstruction as a reference. We observe that the MAP estimation from \textsc{QuantifAI} outperforms the other reconstructions from the wavelet-based prior and all the MMSE estimations.}
			\label{tb:reconstruction_SNRs}
			\begin{tabular}{cccccc}
				\toprule
				\multirow{3}{*}{Images}                 & \multicolumn{5}{c}{Reconstruction SNR [dB]} \vspace{0.01in}                                                                              \\
				\cmidrule{2-6}
				                                        &
				\multirow{2}{*}{Dirty}                  &
				\multicolumn{2}{c}{Wavelet-based prior} &
				\multicolumn{2}{c}{\textsc{QuantifAI}}                                                                                                                                             \\
				                                        &                                                             & \multicolumn{1}{c}{MMSE} & \multicolumn{1}{c}{MAP} &
				\multicolumn{1}{c}{MMSE}                & \multicolumn{1}{c}{MAP}                                                                                                                  \\
				\hline \hline
				W28                                     & $3.39$                                                      & $18.17$                  & $23.04$                 & $23.38$ & $\bf 26.85$ \\
				M31                                     & $5.01$                                                      & $23.78$                  & $25.52$                 & $24.61$ & $\bf 27.48$ \\
				3C288                                   & $7.02$                                                      & $14.31$                  & $14.15$                 & $23.23$ & $\bf 24.10$ \\
				Cygnus A                                & $4.60$                                                      & $20.52$                  & $17.53$                 & $25.36$ & $\bf 30.25$ \\
				\bottomrule
			\end{tabular}
		\end{center}
	\end{table}

\begin{figure*}
		\centering
		\captionsetup[subfigure]{skip=-2pt}
		\begin{subfigure}[b]{0.23\textwidth}
			\centering
			\includegraphics[height=3.6cm]{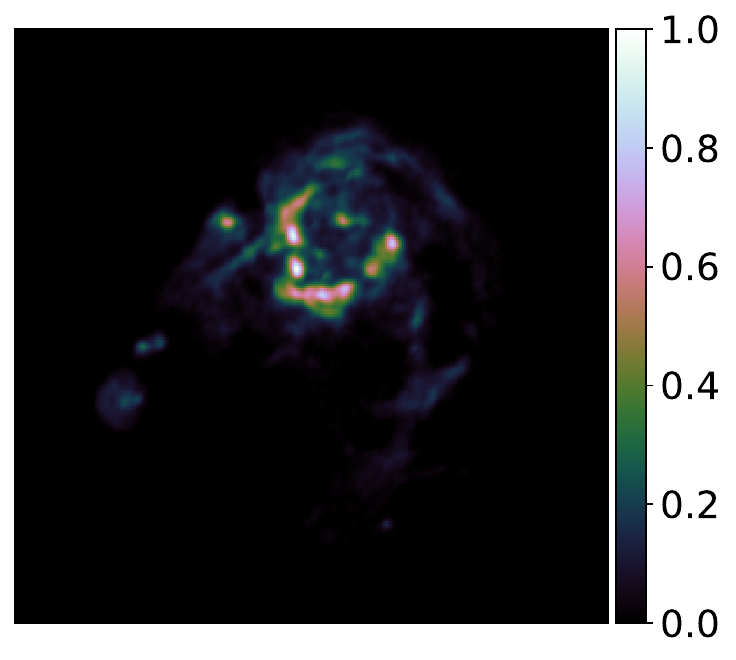}
			\caption{Ground truth (linear scale)}
			\label{fi:M31_reconstruction_r1_c1}
		\end{subfigure}
		\hfill
		\begin{subfigure}[b]{0.23\textwidth}
			\centering
			\includegraphics[height=3.6cm]{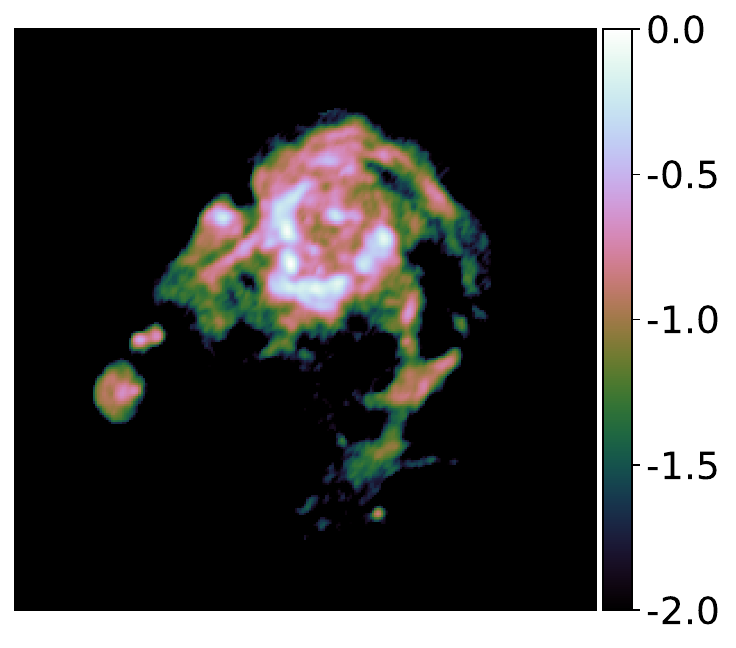}
			\caption{Ground truth}
			\label{fi:M31_reconstruction_r1_c2}
		\end{subfigure}
		\hfill
		\begin{subfigure}[b]{0.23\textwidth}
			\centering
			\includegraphics[height=3.6cm]{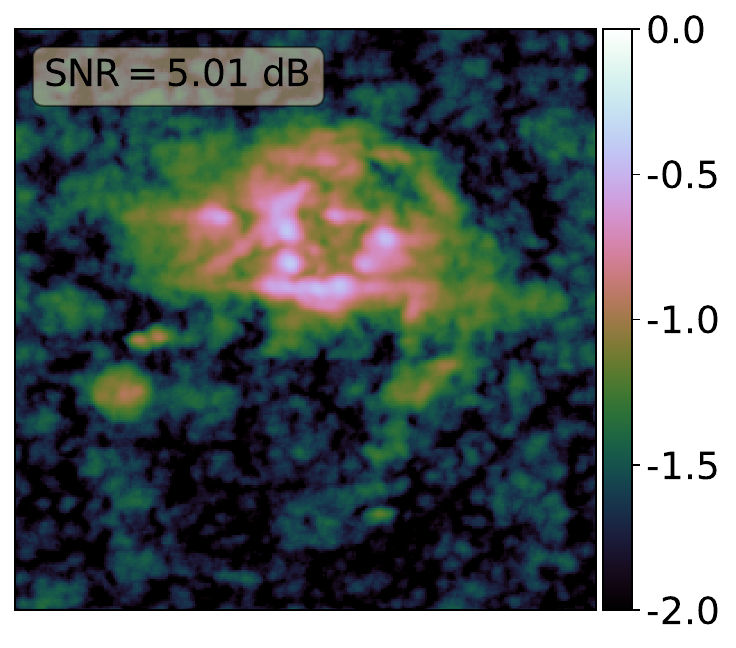}
			\caption{Dirty reconstruction}
			\label{fi:M31_reconstruction_r1_c3}
		\end{subfigure}
		\hfill
		\begin{subfigure}[b]{0.23\textwidth}
			\centering
			\includegraphics[height=3.6cm]{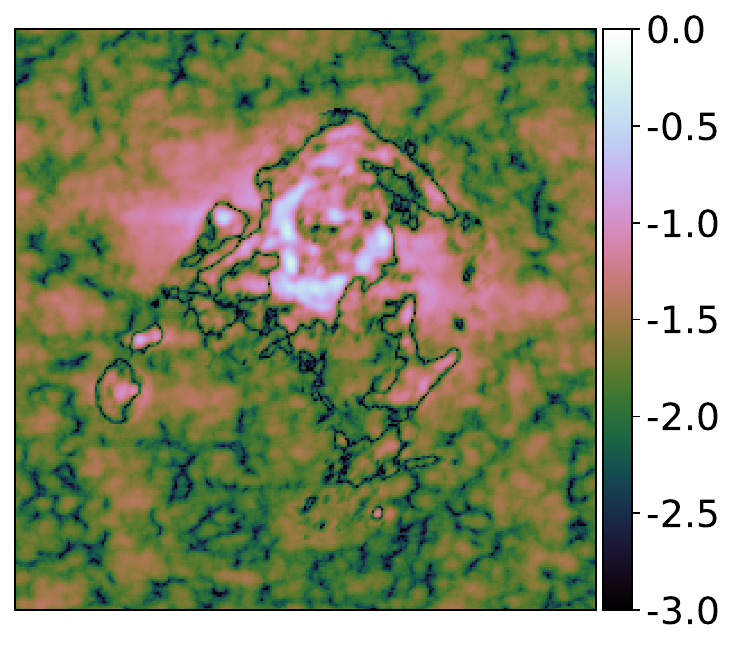}
			\caption{Dirty reconstruction error}
			\label{fi:M31_reconstruction_r1_c4}
		\end{subfigure}\\
		\begin{subfigure}[b]{0.23\textwidth}
			\centering
			\includegraphics[height=3.6cm]{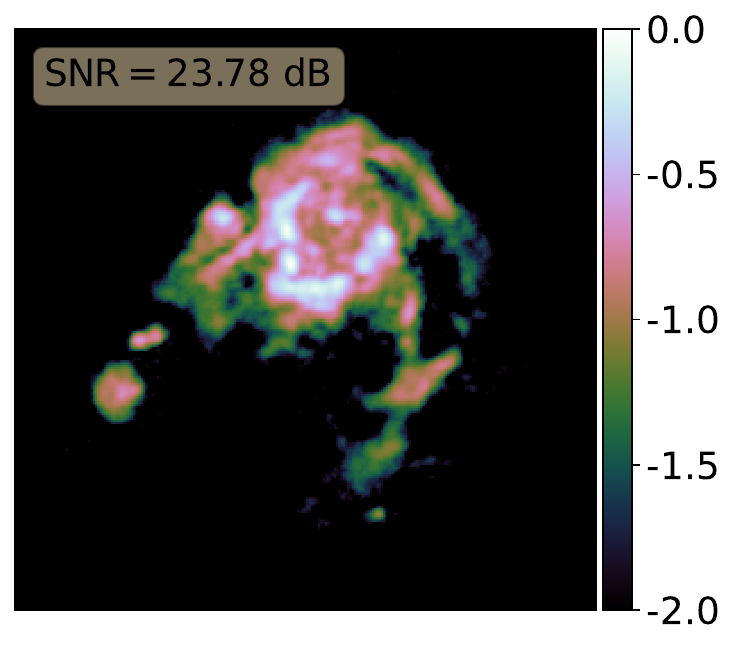} \put(-125,15){\rotatebox{90}{Wavelet-based prior}}
			\caption{MMSE reconstruction}
			\label{fi:M31_reconstruction_r2_c1}
		\end{subfigure}
		\hfill
		\begin{subfigure}[b]{0.23\textwidth}
			\centering
			\includegraphics[height=3.5cm]{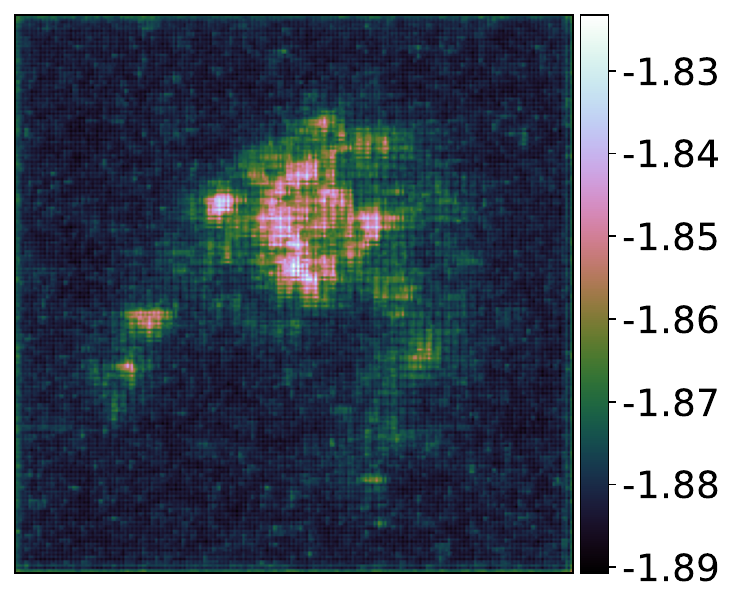}
			\caption{Posterior standard deviation}
			\label{fi:M31_reconstruction_r2_c2}
		\end{subfigure}
		\hfill
		\begin{subfigure}[b]{0.23\textwidth}
			\centering
			\includegraphics[height=3.6cm]{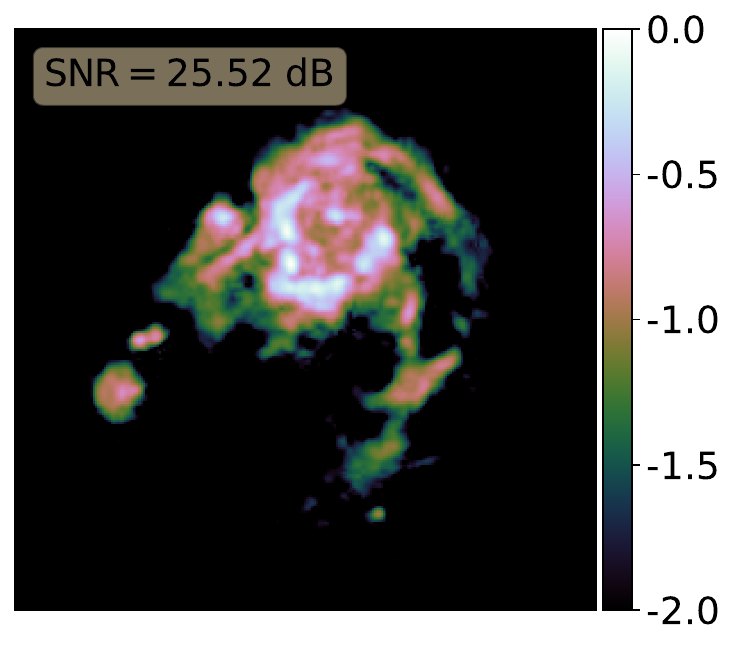}
			\caption{MAP reconstruction}
			\label{fi:M31_reconstruction_r2_c3}
		\end{subfigure}
		\hfill
		\begin{subfigure}[b]{0.23\textwidth}
			\centering
			\includegraphics[height=3.6cm]{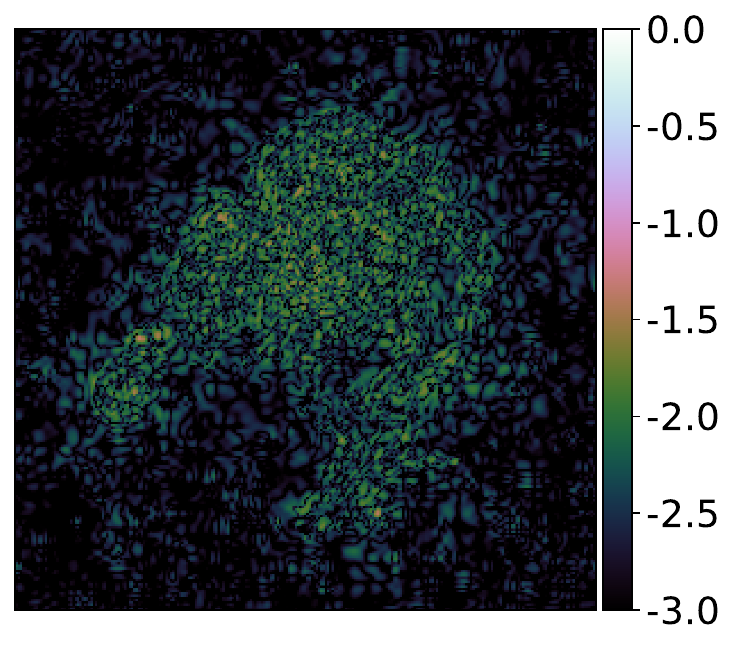}
			\caption{MAP reconstruction error}
			\label{fi:M31_reconstruction_r2_c4}
		\end{subfigure}\\
		\begin{subfigure}[b]{0.23\textwidth}
			\centering
			\includegraphics[height=3.6cm]{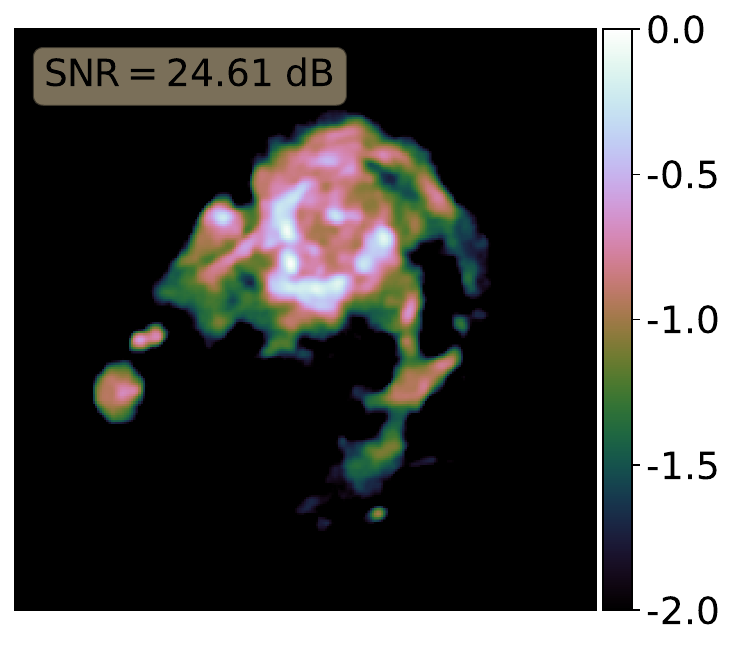} \put(-125,38){\rotatebox{90}{\textsc{QuantifAI}}}
			\caption{MMSE reconstruction}
			\label{fi:M31_reconstruction_r3_c1}
		\end{subfigure}
		\hfill
		\begin{subfigure}[b]{0.23\textwidth}
			\centering
			\includegraphics[height=3.5cm]{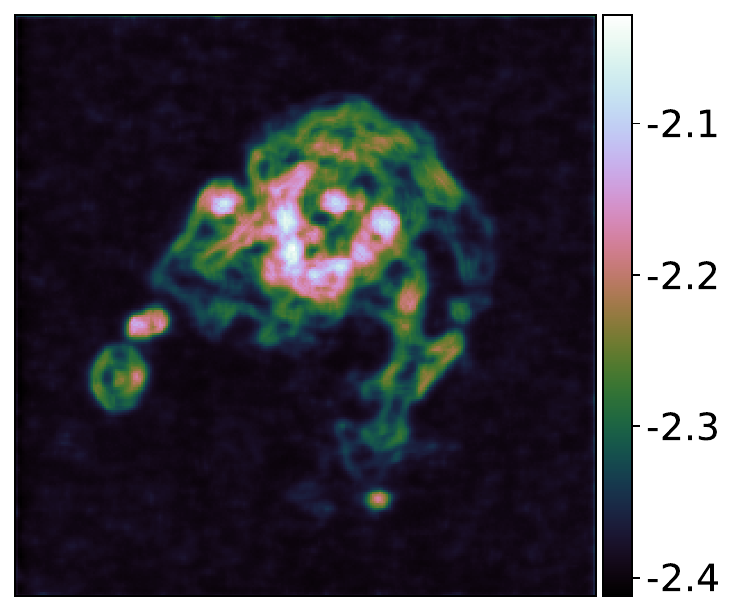}
			\caption{Posterior standard deviation}
			\label{fi:M31_reconstruction_r3_c2}
		\end{subfigure}
		\hfill
		\begin{subfigure}[b]{0.23\textwidth}
			\centering
			\includegraphics[height=3.6cm]{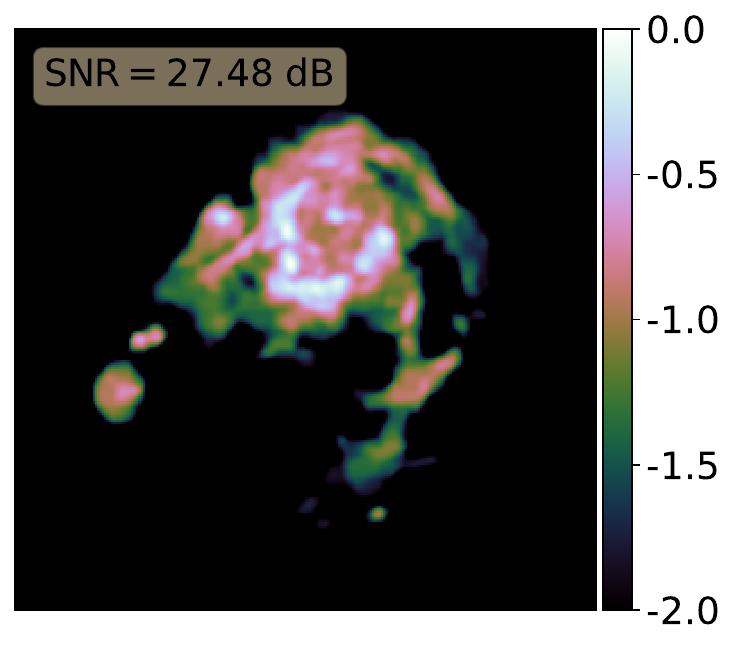}
			\caption{MAP reconstruction}
			\label{fi:M31_reconstruction_r3_c3}
		\end{subfigure}
		\hfill
		\begin{subfigure}[b]{0.23\textwidth}
			\centering
			\includegraphics[height=3.6cm]{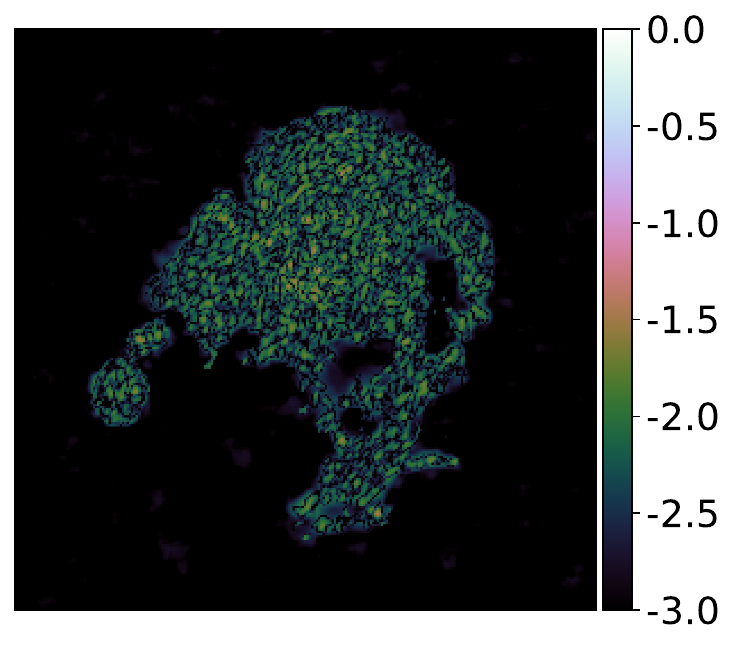}
			\caption{MAP reconstruction error}
			\label{fi:M31_reconstruction_r3_c4}
		\end{subfigure}
		\caption{RI image reconstructions for M31. The images are shown in a $\log_{10}$ scale except for subfigure (a).
			\textbf{Top row:} The first two images show the ground truth intensity image in linear and $\log_{10}$ scales, respectively. The third image shows the dirty reconstruction, computed by applying the pseudo-inverse of the measurement operator on the observations. The fourth image shows the error of the dirty reconstruction with respect to the ground truth.
			\textbf{Middle row:} We show the results of the wavelet-based model. The first and second columns show the minimum mean squared error (MMSE) estimator and the posterior standard deviation, respectively. Both images are computed using the $5 \times 10^{4}$ generated posterior samples. The third column shows the MAP reconstruction obtained through an optimisation algorithm. The fourth column depicts the error of the MAP reconstruction with respect to the ground truth.
			\textbf{Bottom row:} We present the results of the \textsc{QuantifAI} model. The columns are presented in the same order as for the Wavelet reconstructions in the middle row. For every reconstruction, we display the SNR with respect to the ground truth in the top left corner. Compared with the wavelet-based model, \textsc{QuantifAI} recovers a reconstruction with a higher SNR and shows more meaningful uncertainties, which can be seen by comparing the posterior standard deviation and the MAP reconstruction error.}
		\label{fi:M31_reconstruction}
	\end{figure*}

\begin{figure*}
		\centering
		\captionsetup[subfigure]{skip=-2pt}
		\begin{subfigure}[b]{0.23\textwidth}
			\centering
			\includegraphics[height=3.6cm]{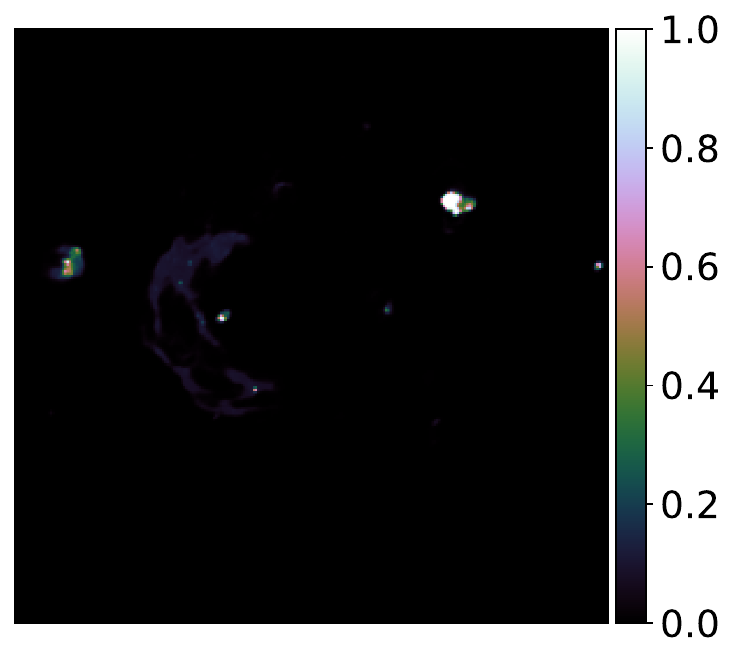}
			\caption{Ground truth (linear scale)}
			\label{fi:W28_reconstruction_r1_c1}
		\end{subfigure}
		\hfill
		\begin{subfigure}[b]{0.23\textwidth}
			\centering
			\includegraphics[height=3.6cm]{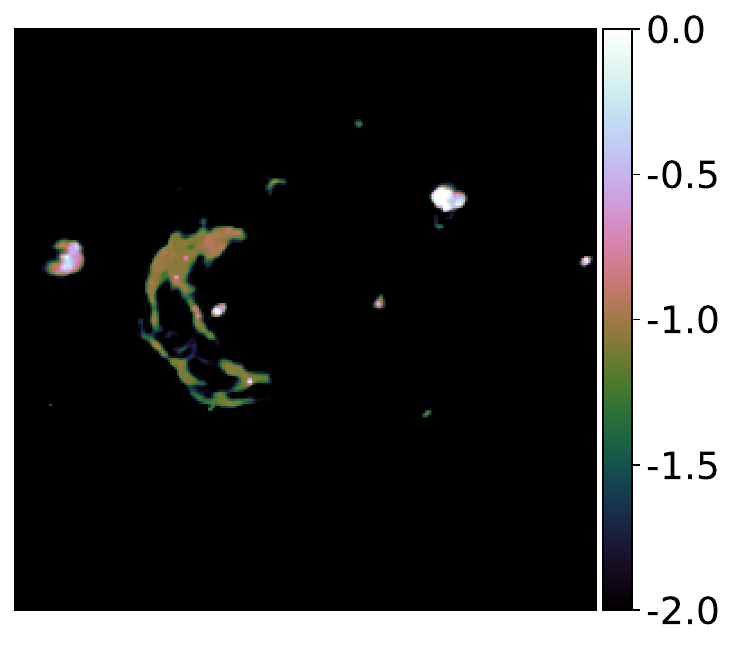}
			\caption{Ground truth}
			\label{fi:W28_reconstruction_r1_c2}
		\end{subfigure}
		\hfill
		\begin{subfigure}[b]{0.23\textwidth}
			\centering
			\includegraphics[height=3.6cm]{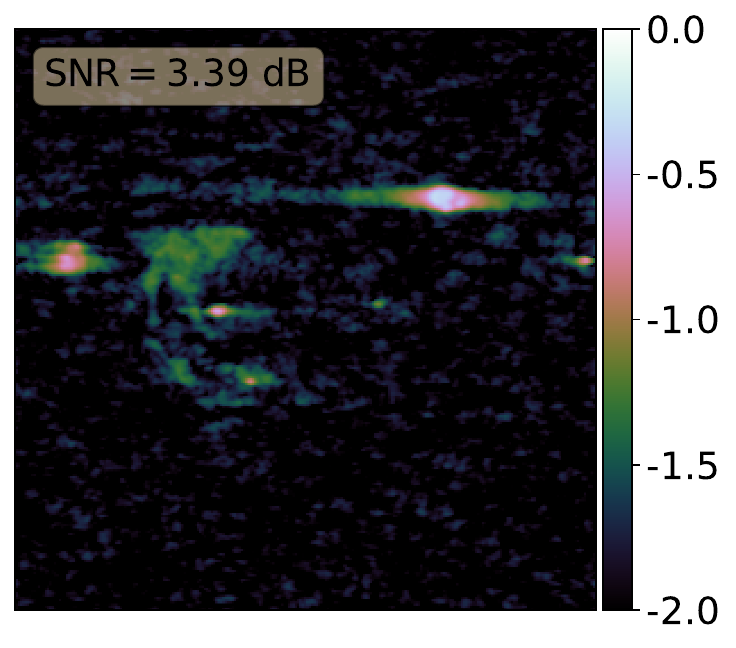}
			\caption{Dirty reconstruction}
			\label{fi:W28_reconstruction_r1_c3}
		\end{subfigure}
		\hfill
		\begin{subfigure}[b]{0.23\textwidth}
			\centering
			\includegraphics[height=3.6cm]{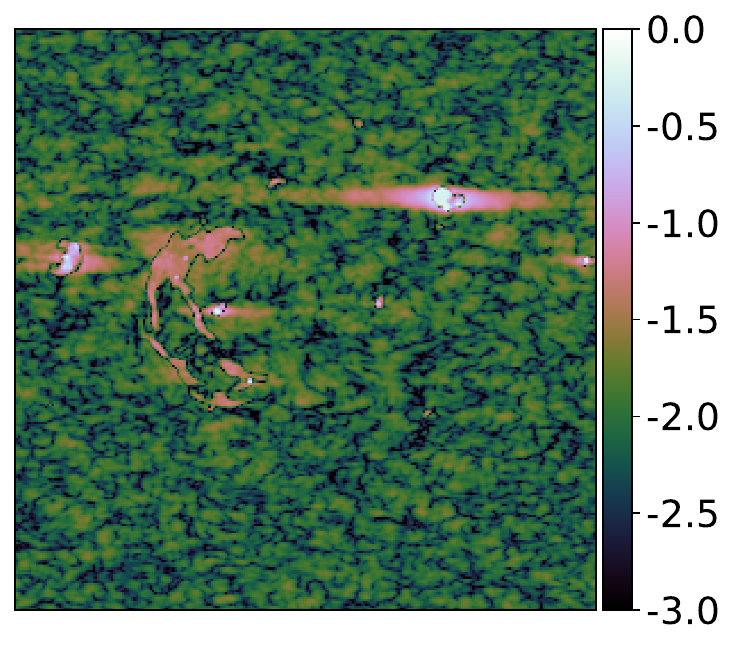}
			\caption{Dirty reconstruction error}
			\label{fi:W28_reconstruction_r1_c4}
		\end{subfigure}\\
		\begin{subfigure}[b]{0.23\textwidth}
			\centering
			\includegraphics[height=3.6cm]{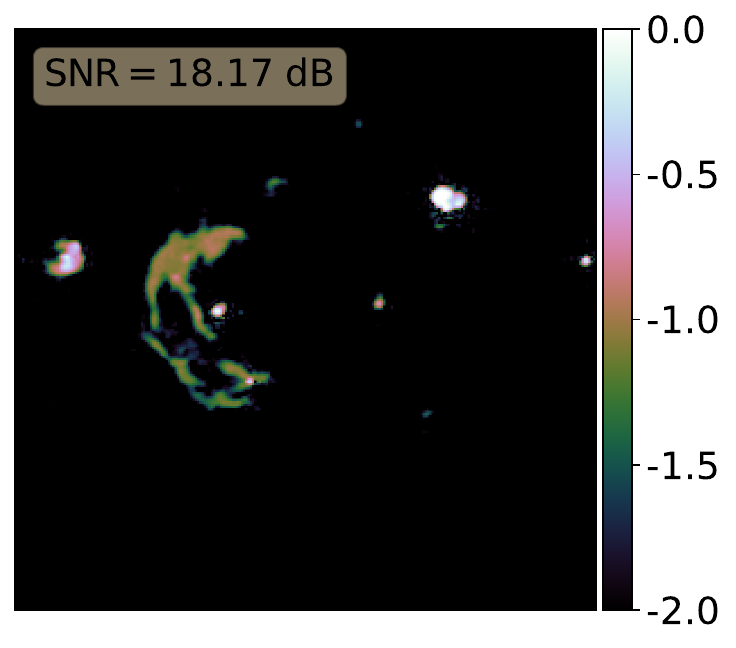} \put(-125,15){\rotatebox{90}{Wavelet-based prior}}
			\caption{MMSE reconstruction}
			\label{fi:W28_reconstruction_r2_c1}
		\end{subfigure}
		\hfill
		\begin{subfigure}[b]{0.23\textwidth}
			\centering
			\includegraphics[height=3.4cm]{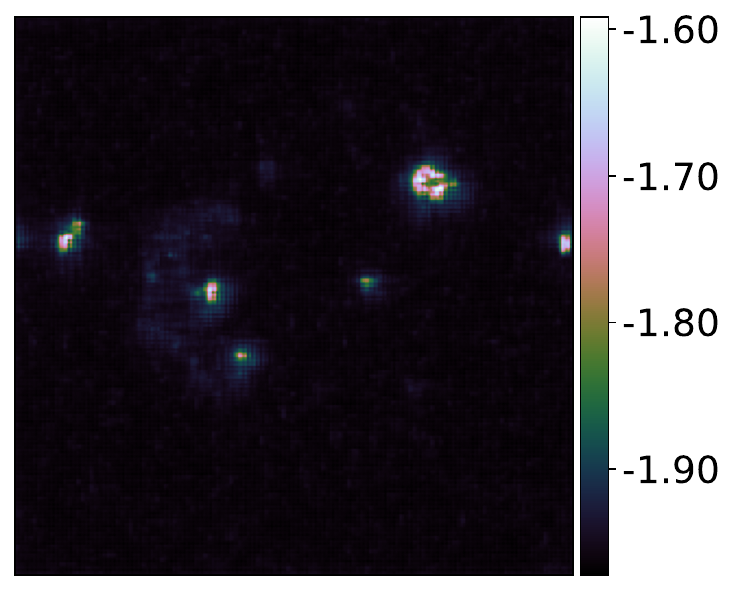}
			\vspace*{-0.7\baselineskip}
			\caption{Posterior standard deviation}
			\label{fi:W28_reconstruction_r2_c2}
		\end{subfigure}
		\hfill
		\begin{subfigure}[b]{0.23\textwidth}
			\centering
			\includegraphics[height=3.6cm]{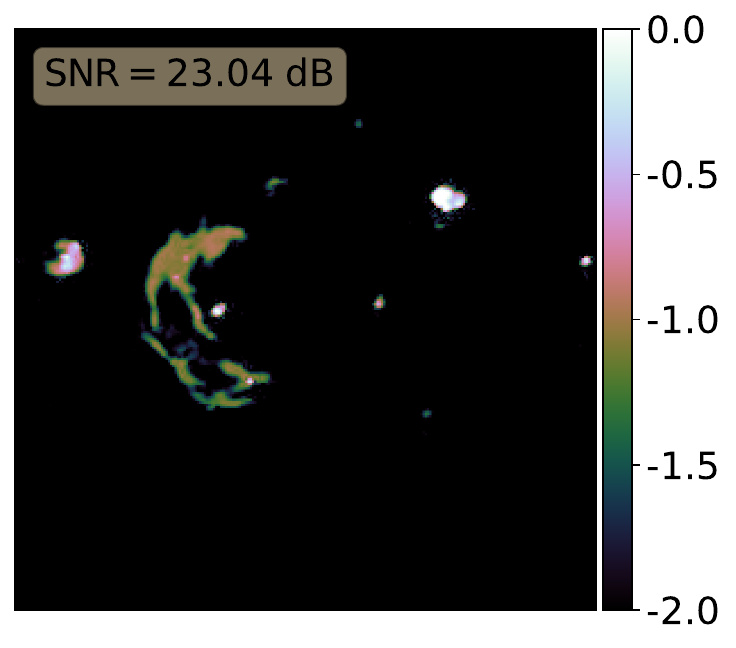}
			\caption{MAP reconstruction}
			\label{fi:W28_reconstruction_r2_c3}
		\end{subfigure}
		\hfill
		\begin{subfigure}[b]{0.23\textwidth}
			\centering
			\includegraphics[height=3.6cm]{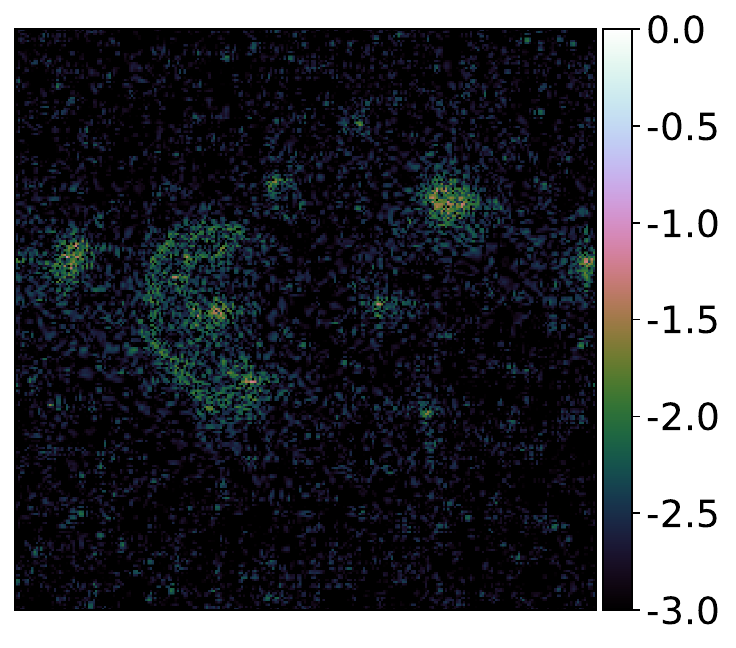}
			\caption{MAP reconstruction error}
			\label{fi:W28_reconstruction_r2_c4}
		\end{subfigure}\\
		\begin{subfigure}[b]{0.23\textwidth}
			\centering
			\includegraphics[height=3.6cm]{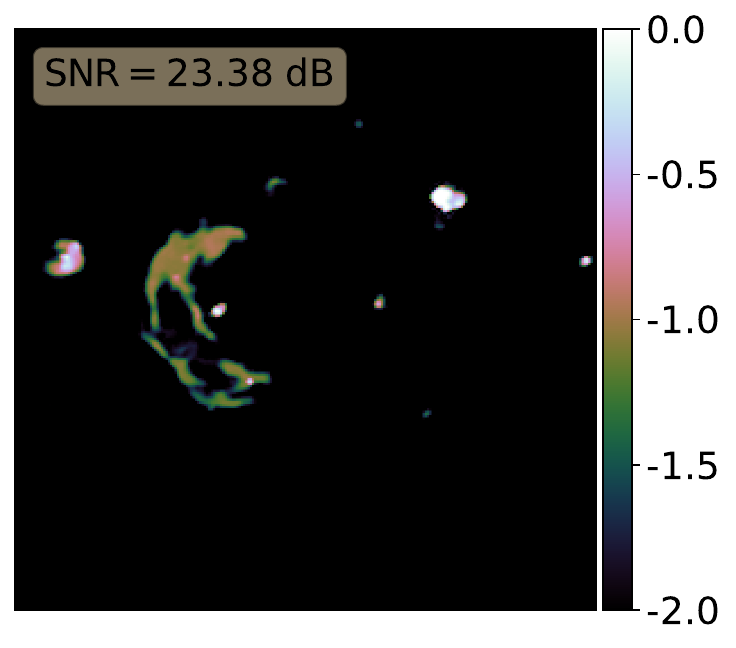} \put(-125,38){\rotatebox{90}{\textsc{QuantifAI}}}
			\caption{MMSE reconstruction}
			\label{fi:W28_reconstruction_r3_c1}
		\end{subfigure}
		\hfill
		\begin{subfigure}[b]{0.23\textwidth}
			\centering
			\includegraphics[height=3.4cm]{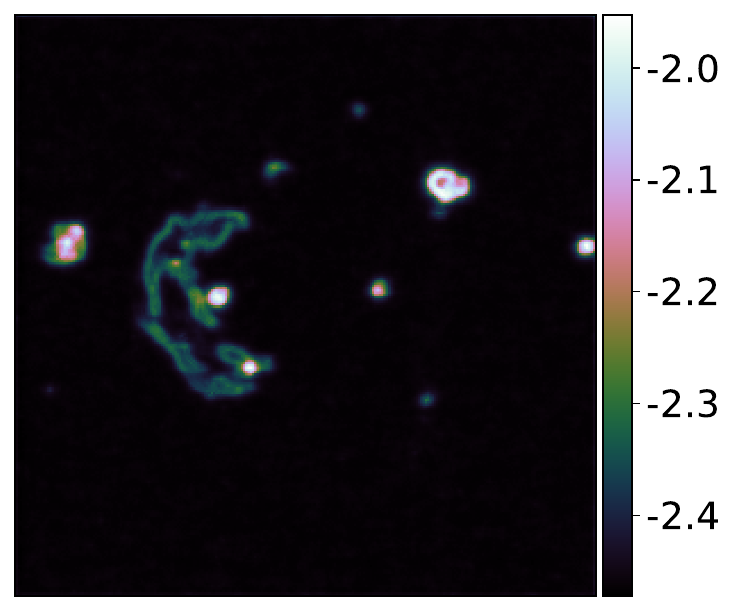}
			\vspace*{-0.7\baselineskip}
			\caption{Posterior standard deviation}
			\label{fi:W28_reconstruction_r3_c2}
		\end{subfigure}
		\hfill
		\begin{subfigure}[b]{0.23\textwidth}
			\centering
			\includegraphics[height=3.6cm]{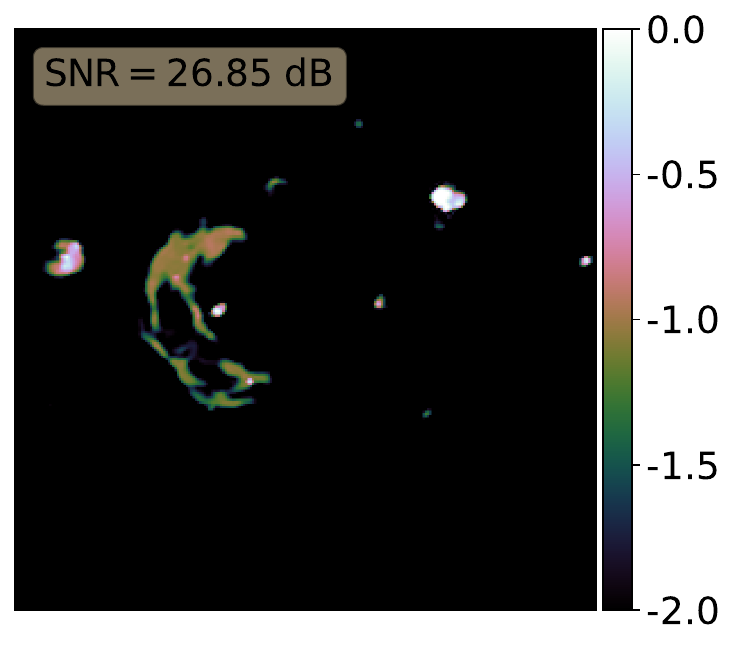}
			\caption{MAP reconstruction}
			\label{fi:W28_reconstruction_r3_c3}
		\end{subfigure}
		\hfill
		\begin{subfigure}[b]{0.23\textwidth}
			\centering
			\includegraphics[height=3.6cm]{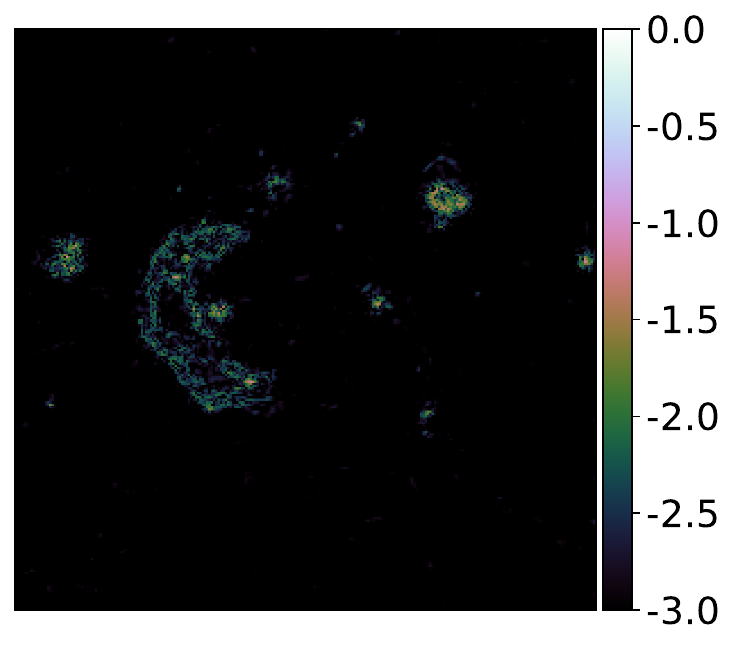}
			\caption{MAP reconstruction error}
			\label{fi:W28_reconstruction_r3_c4}
		\end{subfigure}
		\caption{RI image reconstructions for W28. The order of the images follows the M31 results presented in Figure \ref{fi:M31_reconstruction}.
			Compared with the wavelet-based model, \textsc{QuantifAI} recovers a reconstruction with a higher SNR and shows more meaningful uncertainties, which can be seen by comparing the posterior standard deviation and the MAP reconstruction error.}
		\label{fi:W28_reconstruction}
	\end{figure*}

\begin{figure*}
		\centering
		\captionsetup[subfigure]{skip=-2pt}
		\begin{subfigure}[b]{0.23\textwidth}
			\centering
			\includegraphics[height=3.6cm]{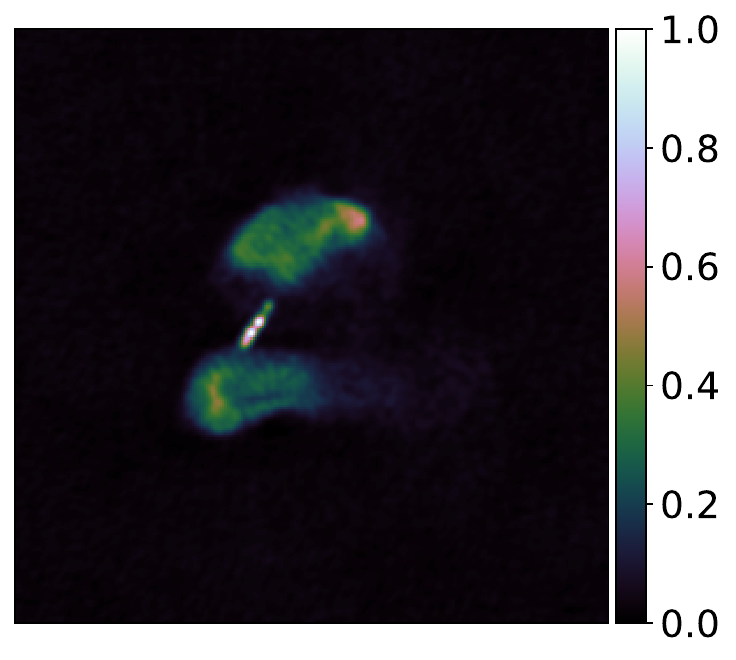}
			\caption{Ground truth (linear scale)}
			\label{fi:3c288_reconstruction_r1_c1}
		\end{subfigure}
		\hfill
		\begin{subfigure}[b]{0.23\textwidth}
			\centering
			\includegraphics[height=3.6cm]{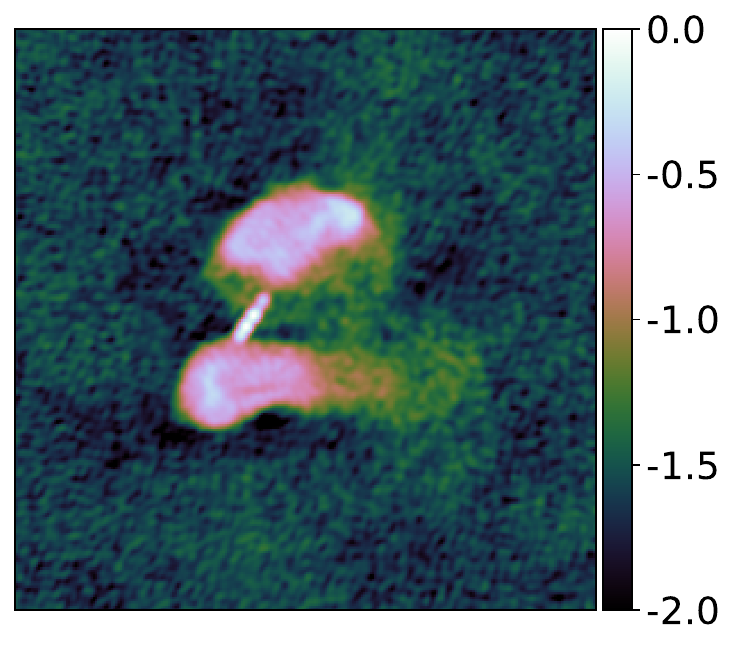}
			\caption{Ground truth}
			\label{fi:3c288_reconstruction_r1_c2}
		\end{subfigure}
		\hfill
		\begin{subfigure}[b]{0.23\textwidth}
			\centering
			\includegraphics[height=3.6cm]{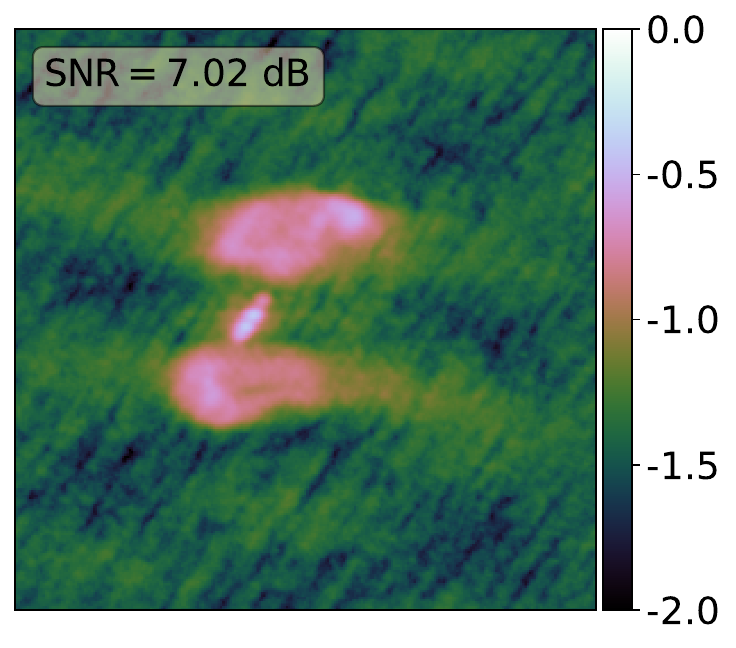}
			\caption{Dirty reconstruction}
			\label{fi:3c288_reconstruction_r1_c3}
		\end{subfigure}
		\hfill
		\begin{subfigure}[b]{0.23\textwidth}
			\centering
			\includegraphics[height=3.6cm]{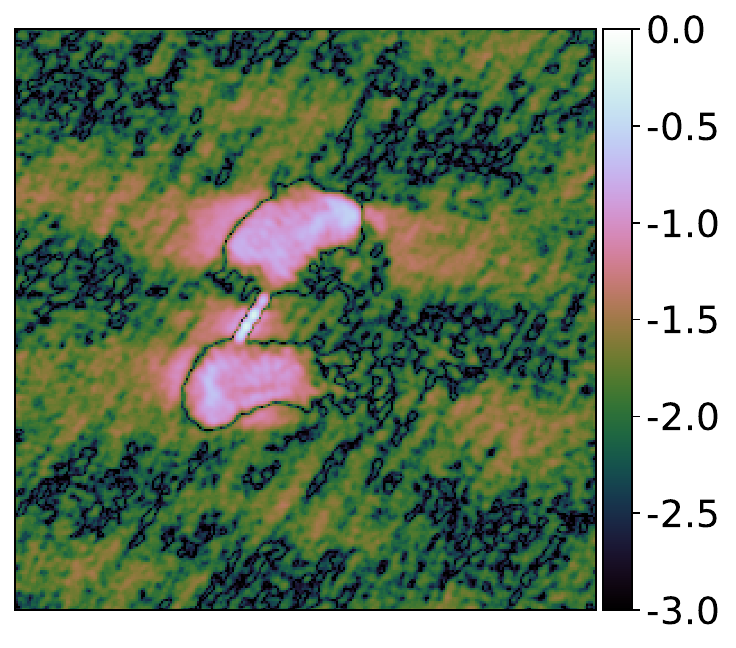}
			\caption{Dirty reconstruction error}
			\label{fi:3c288_reconstruction_r1_c4}
		\end{subfigure}\\
		\begin{subfigure}[b]{0.23\textwidth}
			\centering
			\includegraphics[height=3.6cm]{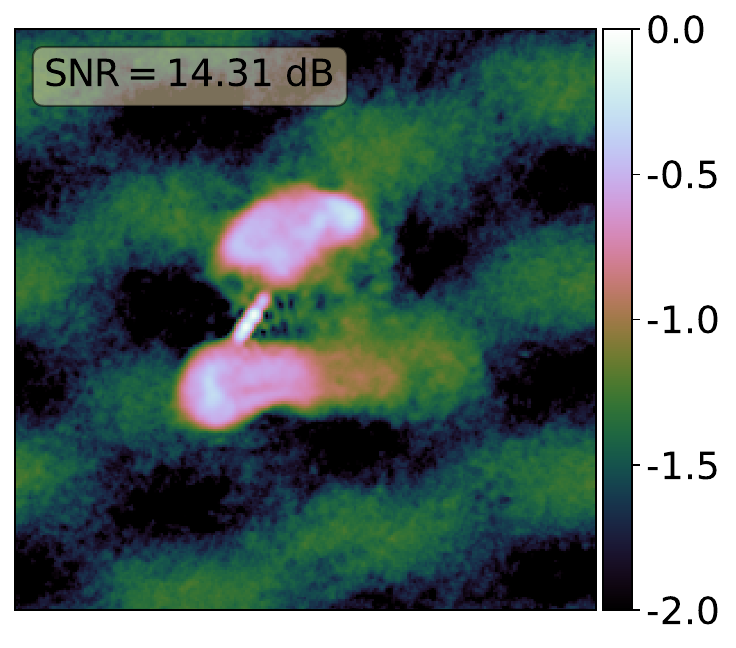} \put(-125,15){\rotatebox{90}{Wavelet-based prior}}
			\caption{MMSE reconstruction}
			\label{fi:3c288_reconstruction_r2_c1}
		\end{subfigure}
		\hfill
		\begin{subfigure}[b]{0.23\textwidth}
			\centering
			\includegraphics[height=3.4cm]{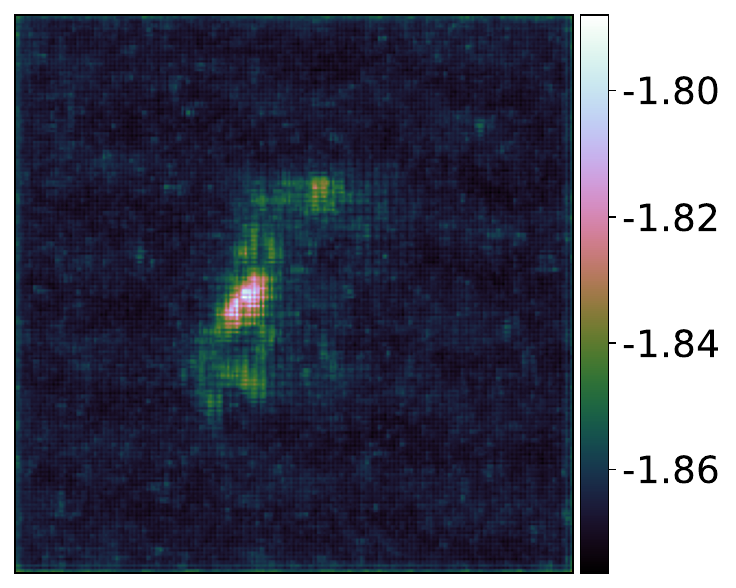}
			\vspace*{-0.7\baselineskip}
			\caption{Posterior standard deviation}
			\label{fi:3c288_reconstruction_r2_c2}
		\end{subfigure}
		\hfill
		\begin{subfigure}[b]{0.23\textwidth}
			\centering
			\includegraphics[height=3.6cm]{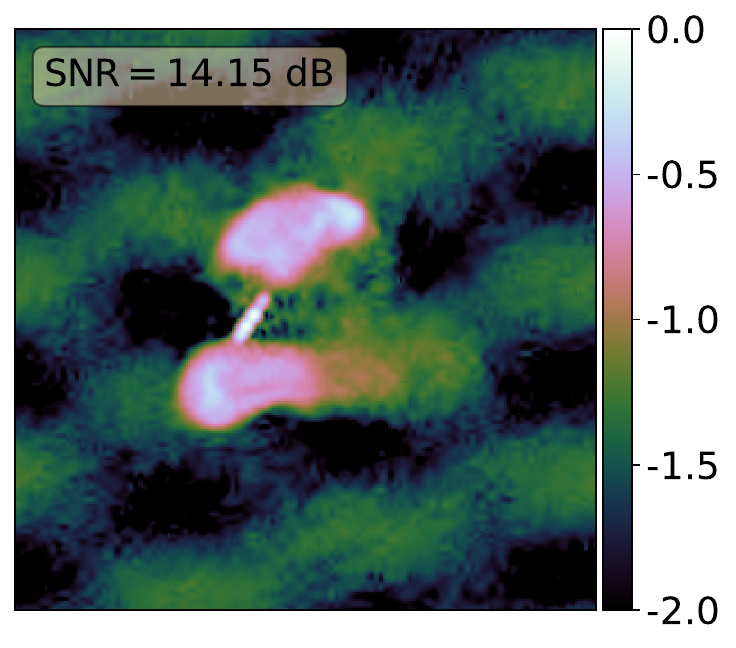}
			\caption{MAP reconstruction}
			\label{fi:3c288_reconstruction_r2_c3}
		\end{subfigure}
		\hfill
		\begin{subfigure}[b]{0.23\textwidth}
			\centering
			\includegraphics[height=3.6cm]{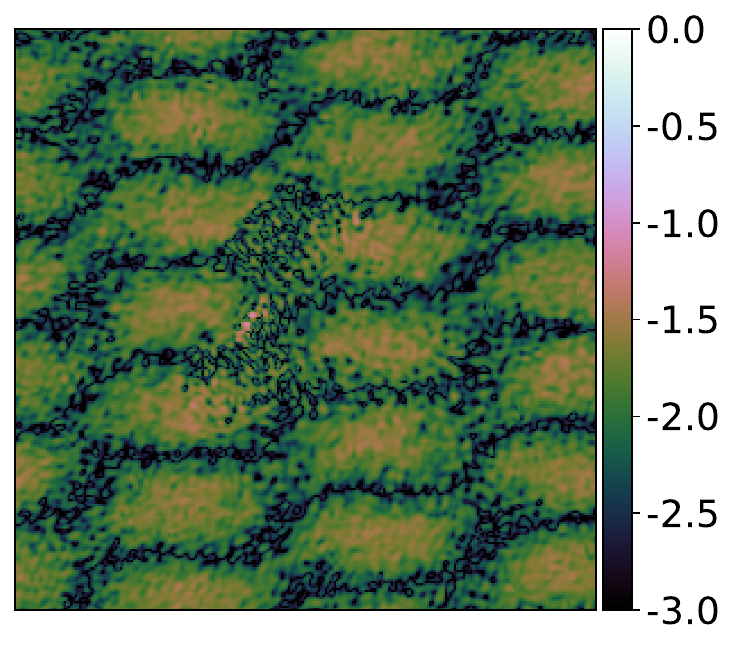}
			\caption{MAP reconstruction error}
			\label{fi:3c288_reconstruction_r2_c4}
		\end{subfigure}\\
		\begin{subfigure}[b]{0.23\textwidth}
			\centering
			\includegraphics[height=3.6cm]{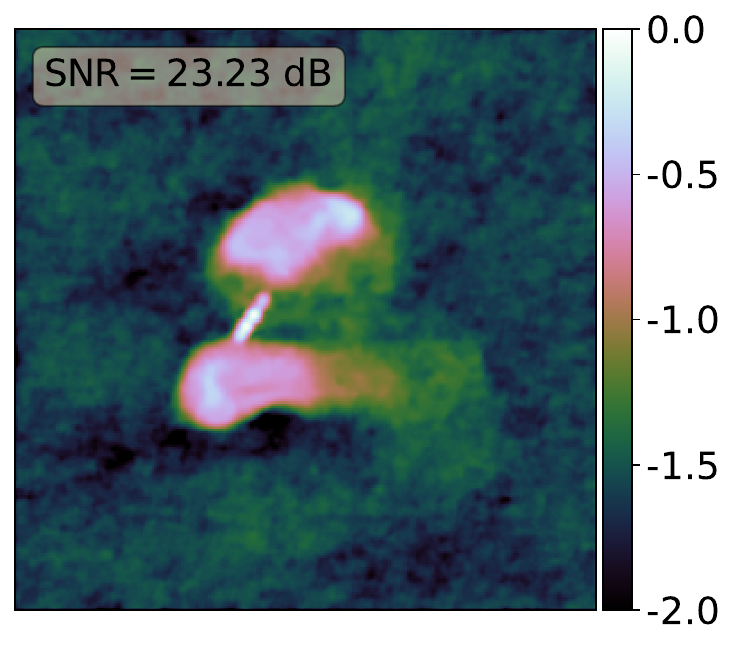} \put(-125,38){\rotatebox{90}{\textsc{QuantifAI}}}
			\caption{MMSE reconstruction}
			\label{fi:3c288_reconstruction_r3_c1}
		\end{subfigure}
		\hfill
		\begin{subfigure}[b]{0.23\textwidth}
			\centering
			\includegraphics[height=3.4cm]{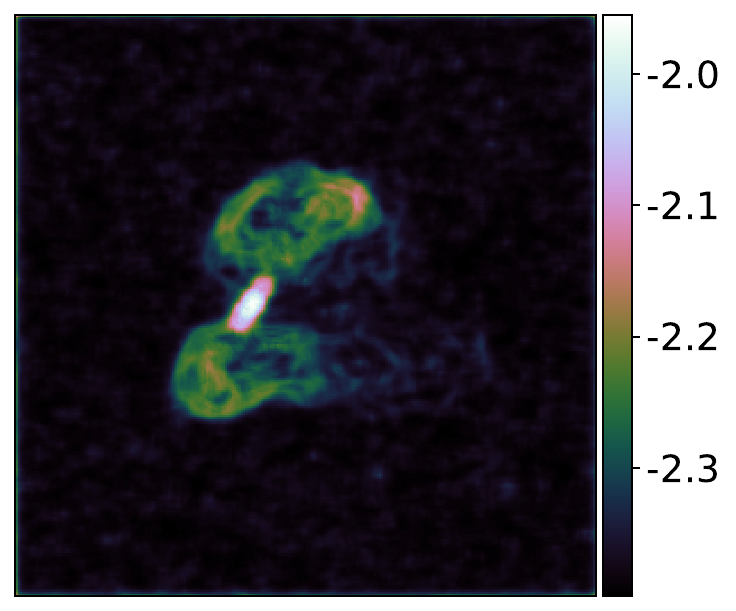}
			\vspace*{-0.7\baselineskip}
			\caption{Posterior standard deviation}
			\label{fi:3c288_reconstruction_r3_c2}
		\end{subfigure}
		\hfill
		\begin{subfigure}[b]{0.23\textwidth}
			\centering
			\includegraphics[height=3.6cm]{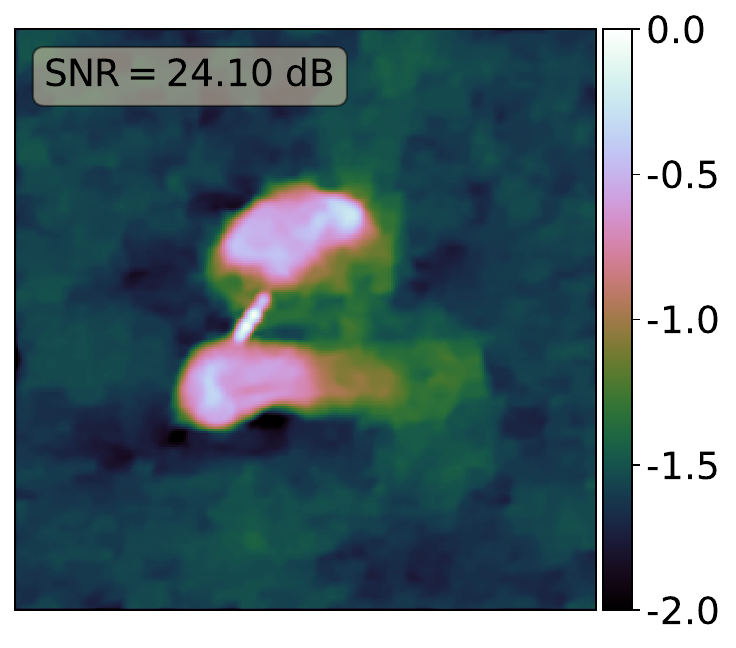}
			\caption{MAP reconstruction}
			\label{fi:3c288_reconstruction_r3_c3}
		\end{subfigure}
		\hfill
		\begin{subfigure}[b]{0.23\textwidth}
			\centering
			\includegraphics[height=3.6cm]{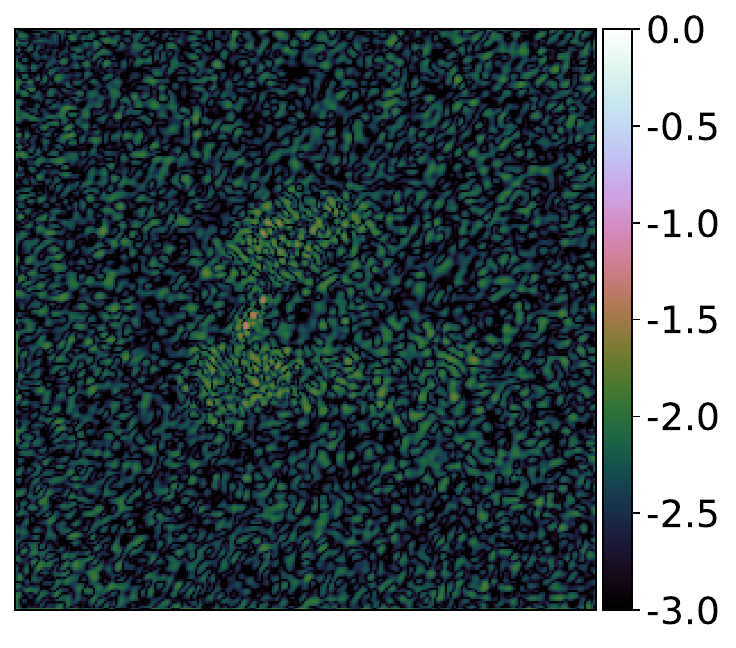}
			\caption{MAP reconstruction error}
			\label{fi:3c288_reconstruction_r3_c4}
		\end{subfigure}
		\caption{RI image reconstructions for 3C288. The order of the images follows the M31 results presented in Figure \ref{fi:M31_reconstruction}. Compared with the wavelet-based model, \textsc{QuantifAI} recovers a reconstruction with a higher SNR and shows more meaningful uncertainties, which can be seen by comparing the posterior standard deviation and the MAP reconstruction error.}
		\label{fi:3c288_reconstruction}
	\end{figure*}

\begin{figure*}
		\centering
		\captionsetup[subfigure]{skip=-2pt}
		\begin{subfigure}[b]{0.23\textwidth}
			\centering
			\includegraphics[height=2.0cm]{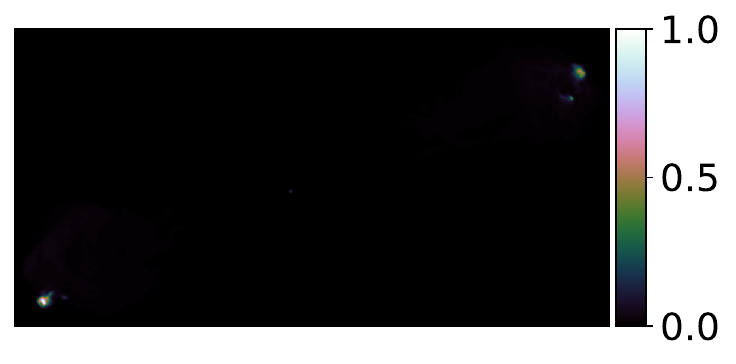}
			\caption{Ground truth (linear scale)}
			\label{fi:CYN_reconstruction_r1_c1}
		\end{subfigure}
		\hfill
		\begin{subfigure}[b]{0.23\textwidth}
			\centering
			\includegraphics[height=2.0cm]{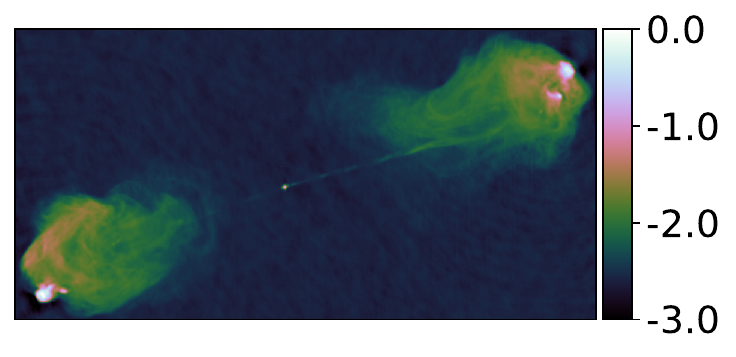}
			\caption{Ground truth}
			\label{fi:CYN_reconstruction_r1_c2}
		\end{subfigure}
		\hfill
		\begin{subfigure}[b]{0.23\textwidth}
			\centering
			\includegraphics[height=2.0cm]{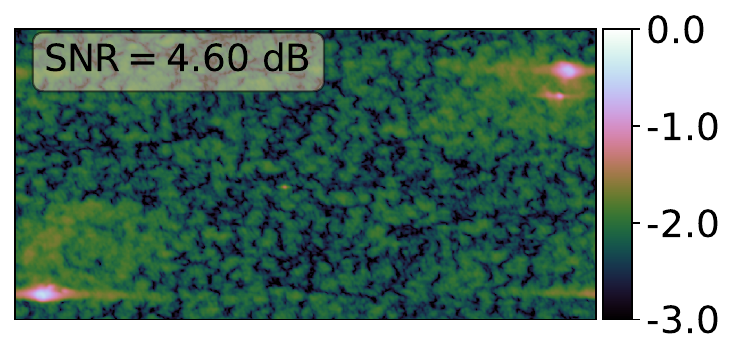}
			\caption{Dirty reconstruction}
			\label{fi:CYN_reconstruction_r1_c3}
		\end{subfigure}
		\hfill
		\begin{subfigure}[b]{0.23\textwidth}
			\centering
			\includegraphics[height=2.0cm]{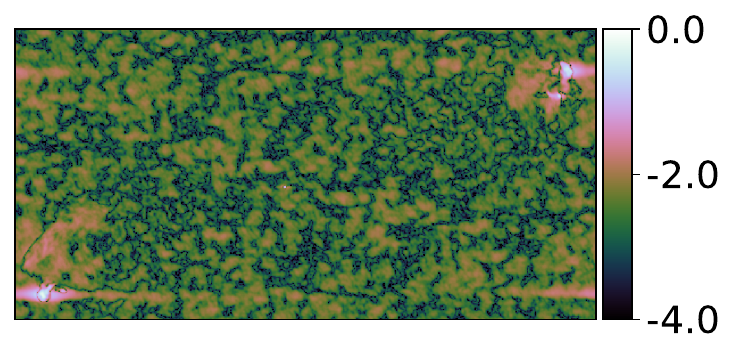}
			\caption{Dirty reconstruction error}
			\label{fi:CYN_reconstruction_r1_c4}
		\end{subfigure}\\
		\begin{subfigure}[b]{0.23\textwidth}
			\centering
			\includegraphics[height=2.0cm]{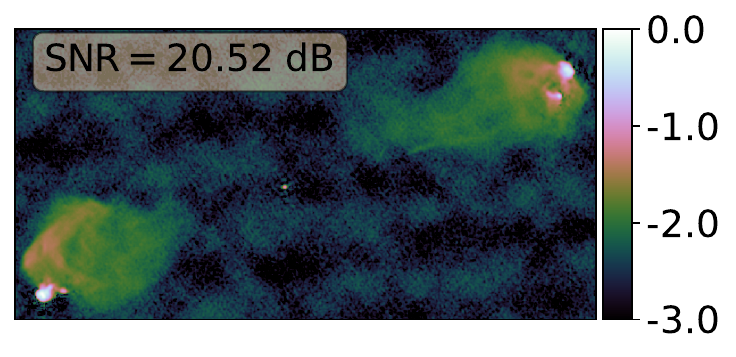} \put(-130,8){\rotatebox{90}{Wavelet prior}}
			\caption{MMSE reconstruction}
			\label{fi:CYN_reconstruction_r2_c1}
		\end{subfigure}
		\hfill
		\begin{subfigure}[b]{0.23\textwidth}
			\centering
			\includegraphics[height=1.8cm]{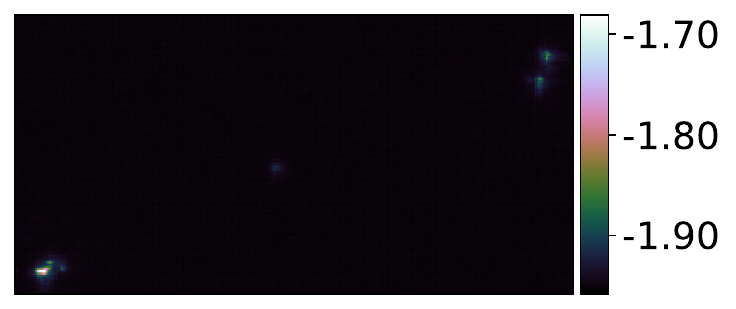}
			\vspace*{-0.7\baselineskip}
			\caption{Posterior standard deviation}
			\label{fi:CYN_reconstruction_r2_c2}
		\end{subfigure}
		\hfill
		\begin{subfigure}[b]{0.23\textwidth}
			\centering
			\includegraphics[height=2.0cm]{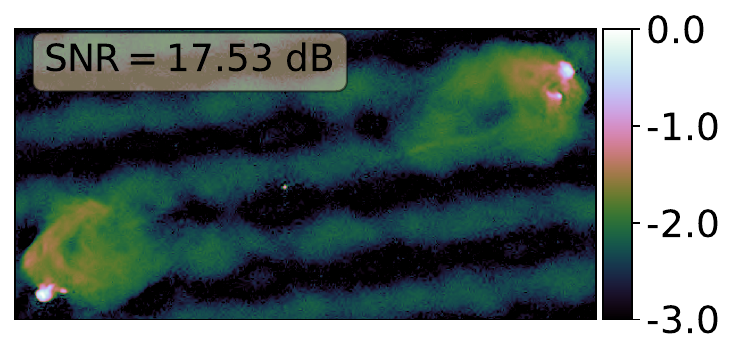}
			\caption{MAP reconstruction}
			\label{fi:CYN_reconstruction_r2_c3}
		\end{subfigure}
		\hfill
		\begin{subfigure}[b]{0.23\textwidth}
			\centering
			\includegraphics[height=2.0cm]{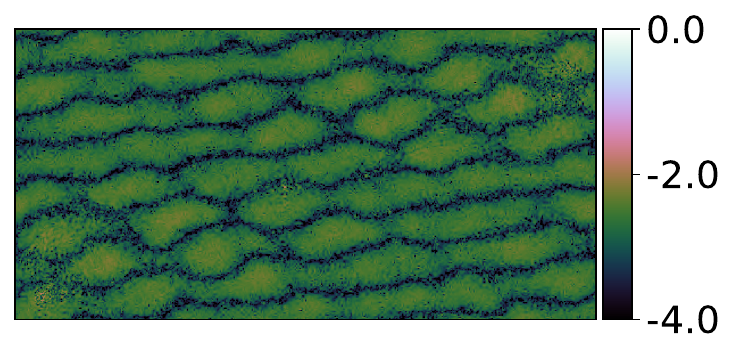}
			\caption{MAP reconstruction error}
			\label{fi:CYN_reconstruction_r2_c4}
		\end{subfigure}\\
		\begin{subfigure}[b]{0.23\textwidth}
			\centering
			\includegraphics[height=2.0cm]{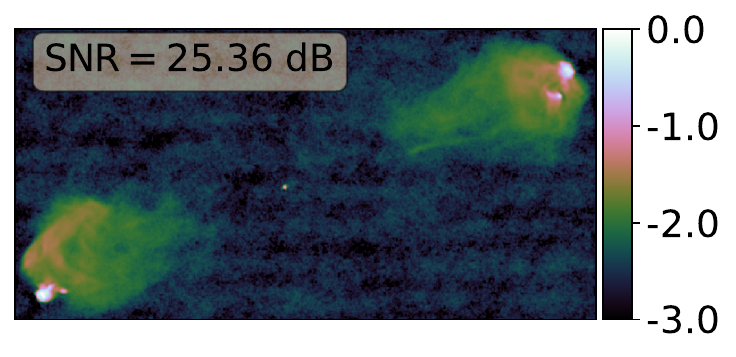} \put(-130,12){\rotatebox{90}{\textsc{QuantifAI}}}
			\caption{MMSE reconstruction}
			\label{fi:CYN_reconstruction_r3_c1}
		\end{subfigure}
		\hfill
		\begin{subfigure}[b]{0.23\textwidth}
			\centering
			\includegraphics[height=1.8cm]{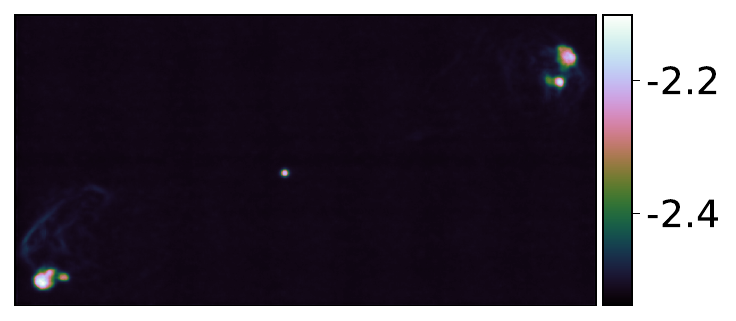}
			\vspace*{-0.7\baselineskip}
			\caption{Posterior standard deviation}
			\label{fi:CYN_reconstruction_r3_c2}
		\end{subfigure}
		\hfill
		\begin{subfigure}[b]{0.23\textwidth}
			\centering
			\includegraphics[height=2.0cm]{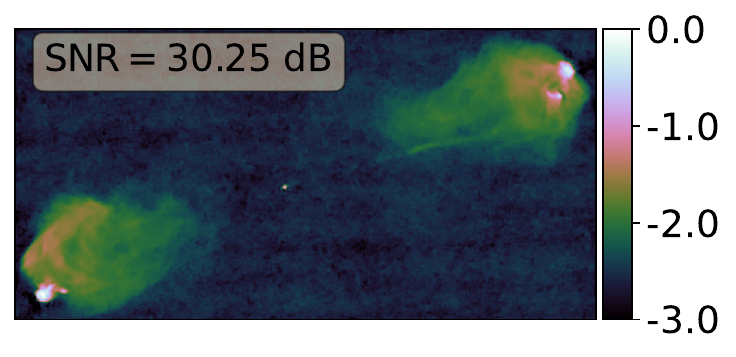}
			\caption{MAP reconstruction}
			\label{fi:CYN_reconstruction_r3_c3}
		\end{subfigure}
		\hfill
		\begin{subfigure}[b]{0.23\textwidth}
			\centering
			\includegraphics[height=2.0cm]{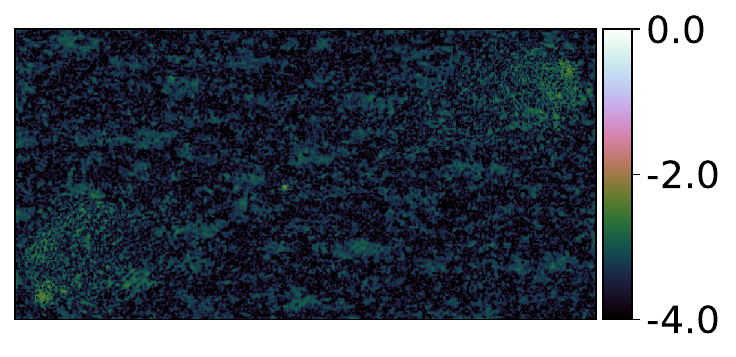}
			\caption{MAP reconstruction error}
			\label{fi:CYN_reconstruction_r3_c4}
		\end{subfigure}
		\caption{RI image reconstructions for Cygnus A. The order of the images follows the M31 results presented in Figure \ref{fi:M31_reconstruction}. Compared with the wavelet-based model, \textsc{QuantifAI} recovers a reconstruction with a higher SNR and shows more meaningful uncertainties, which can be seen by comparing the posterior standard deviation and the MAP reconstruction error.}
		\label{fi:CYN_reconstruction}
	\end{figure*}

	We present the RI image reconstructions of the four ground truth test images in Figure \ref{fi:M31_reconstruction}, Figure \ref{fi:W28_reconstruction}, Figure \ref{fi:3c288_reconstruction} and Figure \ref{fi:CYN_reconstruction}. In each figure, we compare the wavelet-based and \textsc{QuantifAI} models, and we include the dirty reconstruction as a reference. The metric used to compare the RI image reconstruction is the SNR expressed in dB defined as follows
	\begin{equation}
		\text{SNR}(\bm{x}, \bm{x}_{\text{gt}}) = - 20 \log_{10}\left( \frac{\| \bm{x}_{\text{gt}} - \bm{x} \|_{2}}{\|\bm{x}_{\text{gt}} \|_{2}} \right) \,,
		\label{eq:snr_def}
	\end{equation}
	where $\bm{x}_{\text{gt}}$ corresponds to the reference or ground truth, and $\bm{x}$ to the estimation, and $\| \cdot \|_{2}$ is the usual $\ell_2$ norm.

	The quantitative reconstruction performance results are presented in Table \ref{tb:reconstruction_SNRs}. The MAP reconstruction from \textsc{QuantifAI} performs significantly better than the wavelet-based counterpart in every image from our dataset. The performance gain lies between $1.9$dB and $12.7$dB, with an average gain of $7$dB. It is difficult to see the \textsc{QuantifAI} improvements by eye when inspecting reconstructed images. However, when observing the errors in the fourth column, the improved quality of \textsc{QuantifAI}'s reconstructions becomes evident. Shifting towards the sampling results, we observe a similar behaviour of the MMSE reconstruction in favour of \textsc{QuantifAI}'s images. The MAP is considerably faster than the MMSE, relying on optimisation rather than posterior sampling. Recall that the MMSE is built as averaging posterior samples. In addition, the MAP consistently provides improved reconstruction performance with respect to the MMSE.

	The posterior standard deviation provides a qualitative way to validate the posterior model and its uncertainties. The comparison of the posterior standard deviation with the MAP reconstruction error shows a higher correlation for the \textsc{QuantifAI} model than the wavelet-based model. In addition, the posterior standard deviation of \textsc{QuantifAI} shows lower variance than its wavelet-based counterpart, which is in agreement with \textsc{QuantifAI}'s smaller reconstruction error. For example, in image W28 in Figure \ref{fi:W28_reconstruction}, we observe in subfigure (\ref{fi:W28_reconstruction_r3_c2}) that the posterior standard deviation value is large near the edges of the ground truth image. It is reassuring that \textsc{QuantifAI}'s reconstruction error also shows the same behaviour.

	The performance results showcase the expressive power of the CRR-NN-based prior even if the regulariser is constrained to be convex. The results also confirm the generalisation power of the CRR-NN-based prior. Even if trained on natural images, the CRR-NN can provide remarkable reconstruction performances for astronomical images and meaningful posterior standard deviations.

	The reconstructions using the wavelet-based prior model do exhibit some low-intensity artefacts in the Cygnus A and 3C288 images, as shown in Figures \ref{fi:CYN_reconstruction_r2_c4} and \ref{fi:3c288_reconstruction_r2_c4}. These artefacts are due to the patterns in the ground truth images, which originate from real observations, and the thresholding of the orthogonal wavelet basis. Such patterns are absent in the M31 and W28 images because the noise was removed in a preprocessing step, as seen in Figures \ref{fi:M31_reconstruction_r1_c2} and \ref{fi:W28_reconstruction_r1_c2} in comparison to Figures \ref{fi:CYN_reconstruction_r1_c2} and \ref{fi:3c288_reconstruction_r1_c2}. The regularisation strength for the wavelet-prior model was selected to maximise the reconstruction SNR. This chosen value is lower than in \citet{cai2018_2}, explaining the observed patterns and the finer details in our reconstructions. Employing a wavelet dictionary instead of an orthogonal wavelet basis and adding a positivity-enforcing constraint could mitigate the appearance of these artefacts.

	\subsection{Hypothesis testing of image structure}

	We start by carrying out hypothesis tests of image structure, which are the most scalable UQ techniques we will study. First, a surrogate image is created by modifying one region of interest. It only takes one further evaluation of the likelihood and prior potential to carry out the hypothesis test. The test helps to quantitatively answer a scientific question with a $100(1 - \alpha)\%$ confidence level. The scientific question targeted depends on the constructed surrogate image, and in this work, we consider two scenarios.

	\addtolength{\tabcolsep}{-\tabL}
	\begin{figure}
		\centering
		\begin{tabular}{cc}
			\includegraphics[width=0.48\linewidth]{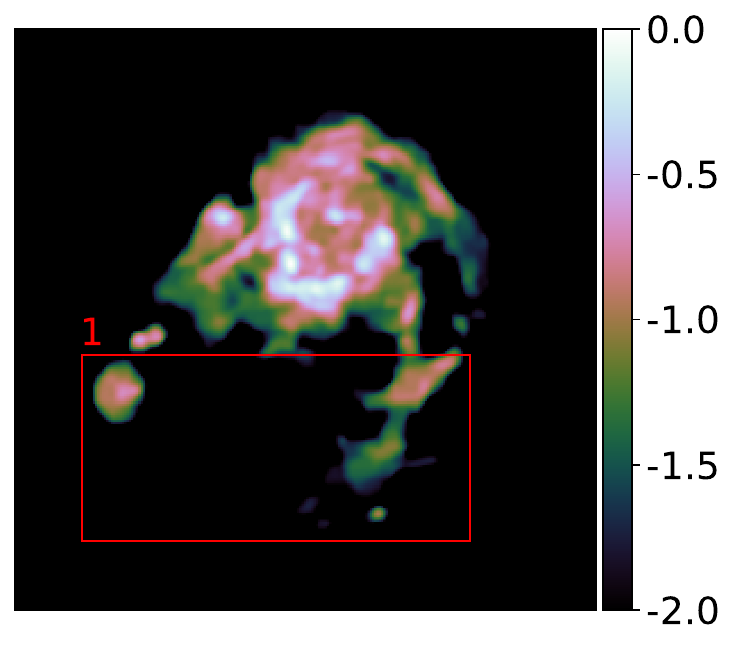} \put(-124,42){\rotatebox{90}{M31}}      &
			\includegraphics[width=0.48\linewidth]{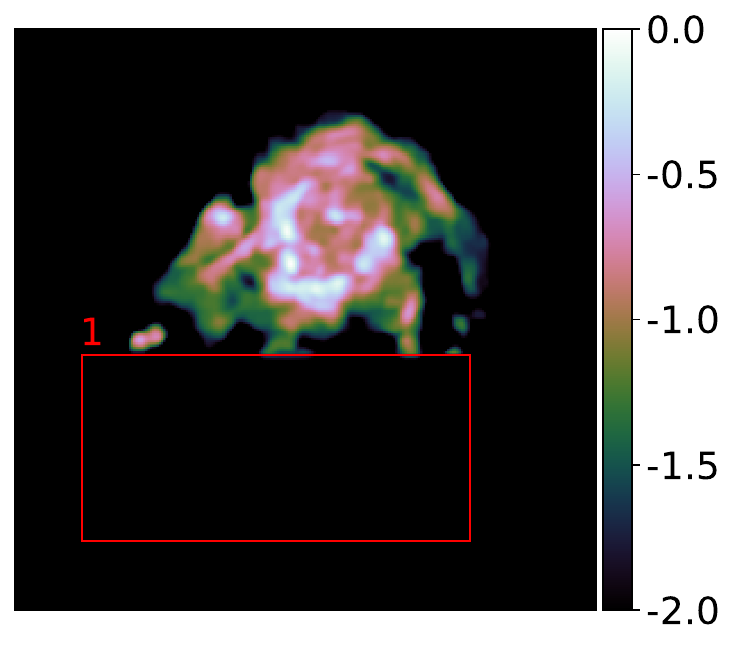}                                                               \\
			\includegraphics[width=0.48\linewidth]{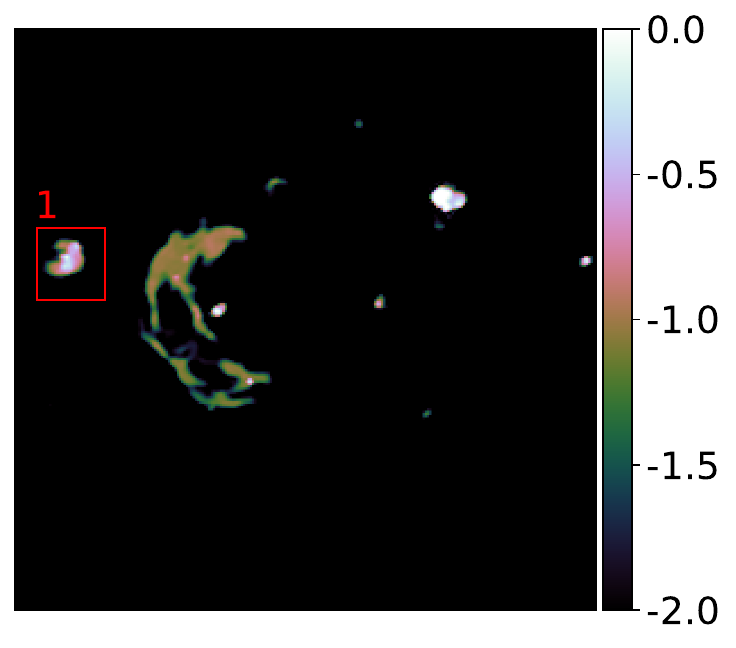} \put(-124,42){\rotatebox{90}{W28}}      &
			\includegraphics[width=0.48\linewidth]{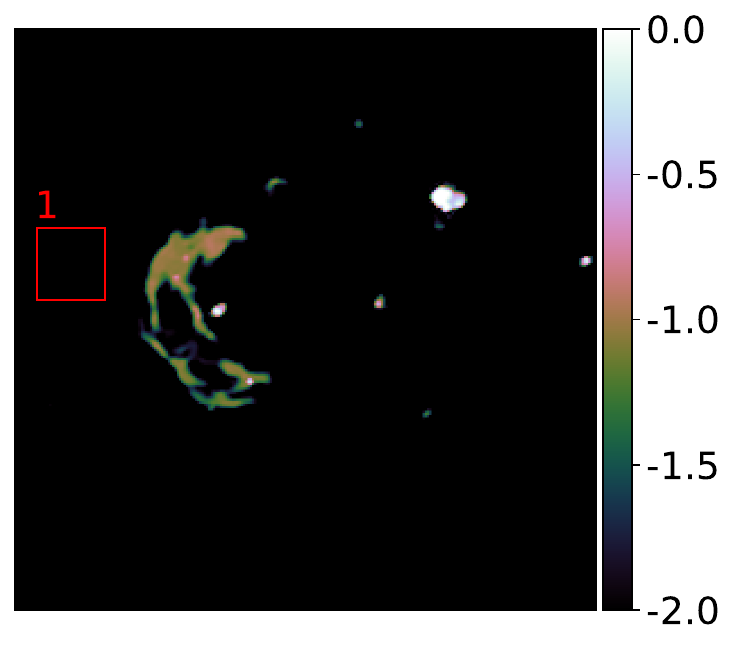}                                                               \\
			\includegraphics[width=0.48\linewidth]{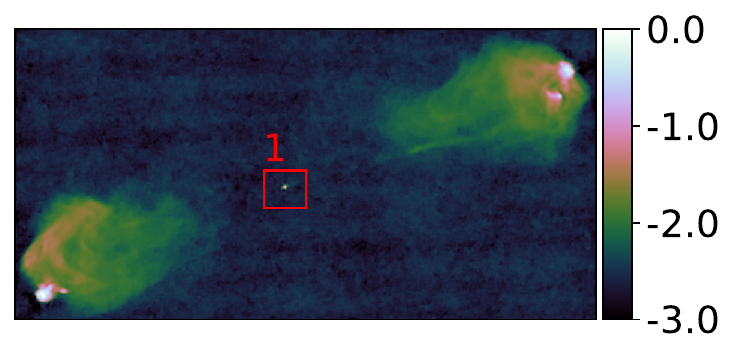} \put(-124,12){\rotatebox{90}{Cygnus A}} &
			\includegraphics[width=0.48\linewidth]{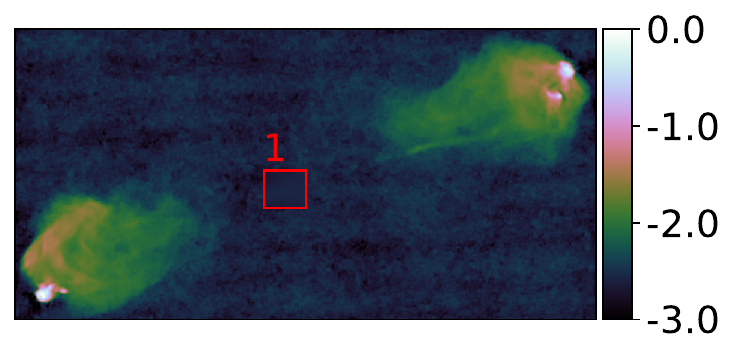}                                                               \\
			\includegraphics[width=0.48\linewidth]{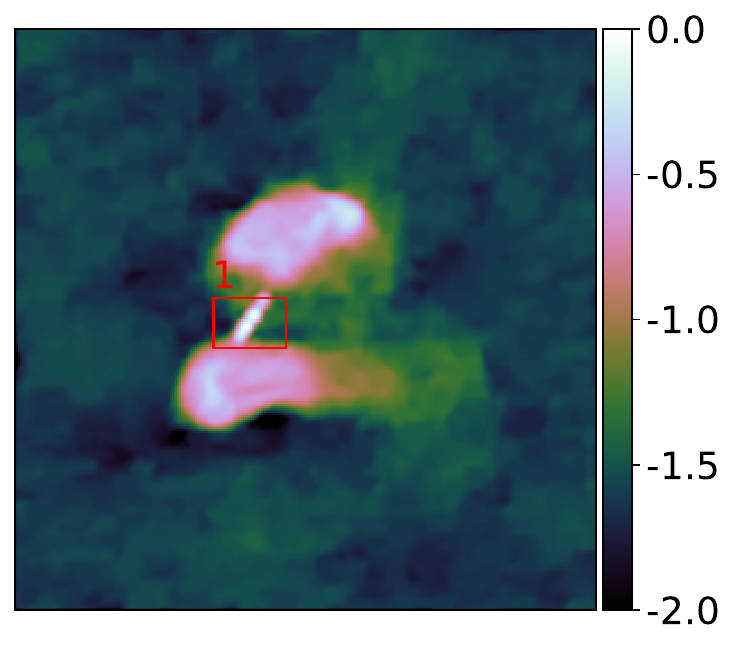} \put(-124,42){\rotatebox{90}{3C288}}  &
			\includegraphics[width=0.48\linewidth]{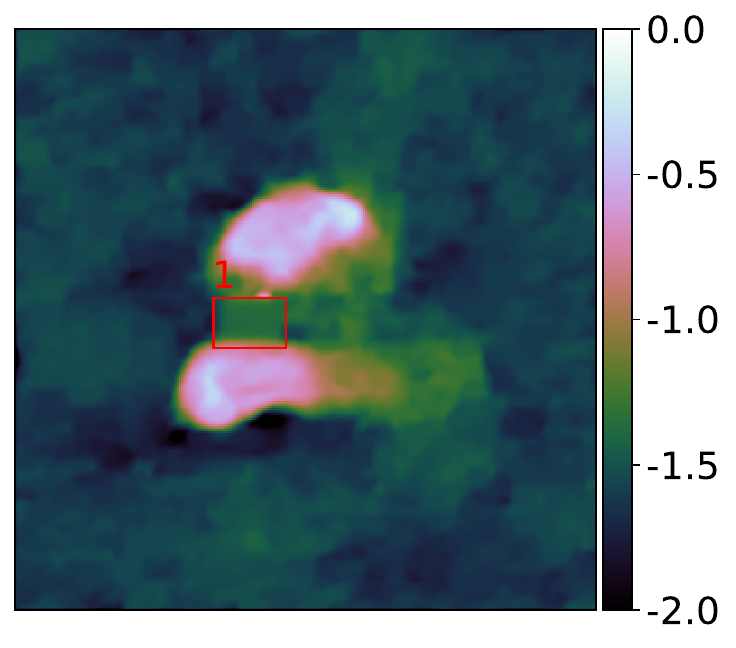}                                                             \\
			\includegraphics[width=0.48\linewidth]{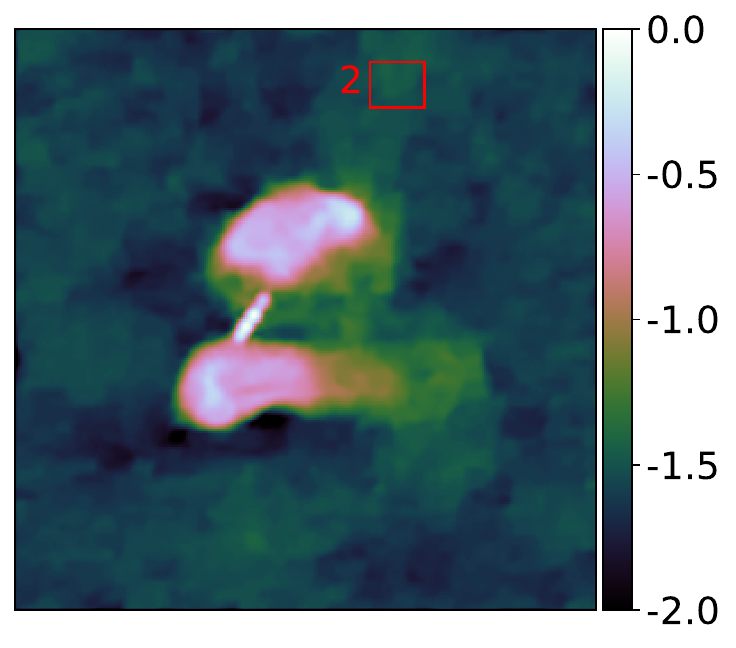} \put(-124,42){\rotatebox{90}{3C288}}  &
			\includegraphics[width=0.48\linewidth]{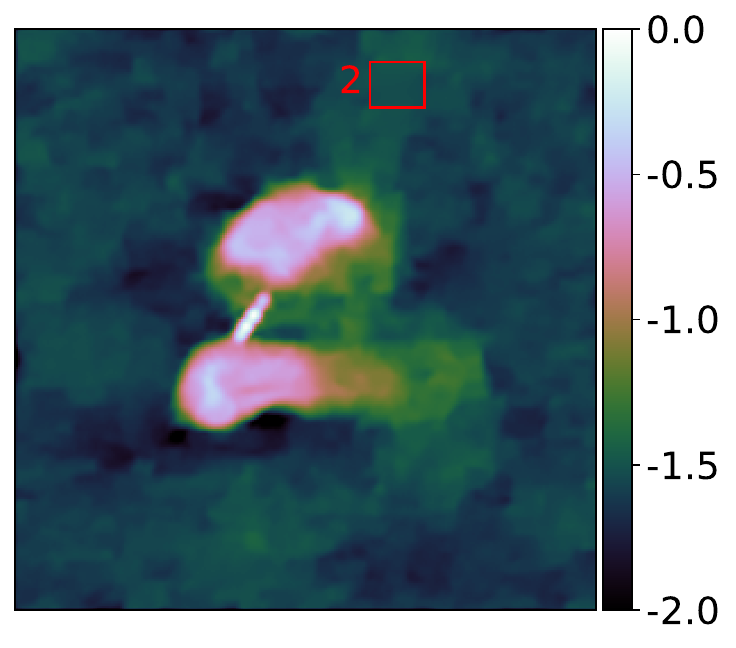}                                                             \\
			(a) MAP reconstruction                                                                                                                  & (b) Inpainted surrogate
		\end{tabular}
		\caption{Hypothesis test of different regions of the four \textsc{QuantifAI} MAP reconstructions for M31, W28, Cygnus A, and 3C288. All the images are shown in $\log_{10}$ scale. The left column shows the respective MAP reconstruction with the region of interest framed in a red rectangle. The right column shows the surrogate images inpainted using the \textsc{QuantifAI} prior. The first four rows show regions corresponding to physical structures present in the ground truth images. The last row corresponds to a non-physical region. Results of the hypothesis tests are summarised in Table \ref{tb:hp_test_inpainting_CRR}.}
		\label{fi:hp_test_inpainting_CRR}
	\end{figure}
	\addtolength{\tabcolsep}{\tabL}

	In the first scenario, we consider a particular structure in the reconstructed intensity image. We can query whether the structure's origin is physical or not. For example, the structure could be a reconstruction artefact or a physical process. Figure \ref{fi:hp_test_inpainting_CRR} shows this option, where we have analysed different regions of the four images.  The first four inpainted regions correspond to physical structures, and the fifth region, i.e., region number $2$ of image 3C288, does not correspond to a physical structure. The surrogate images are produced with an inpainting algorithm using \textsc{QuantifAI}'s prior so that the inpainted region agrees with the prior.

	\addtolength{\tabcolsep}{-\tabL}
	\begin{figure}
		\centering
		\begin{tabular}{cc}
			\includegraphics[width=0.48\linewidth]{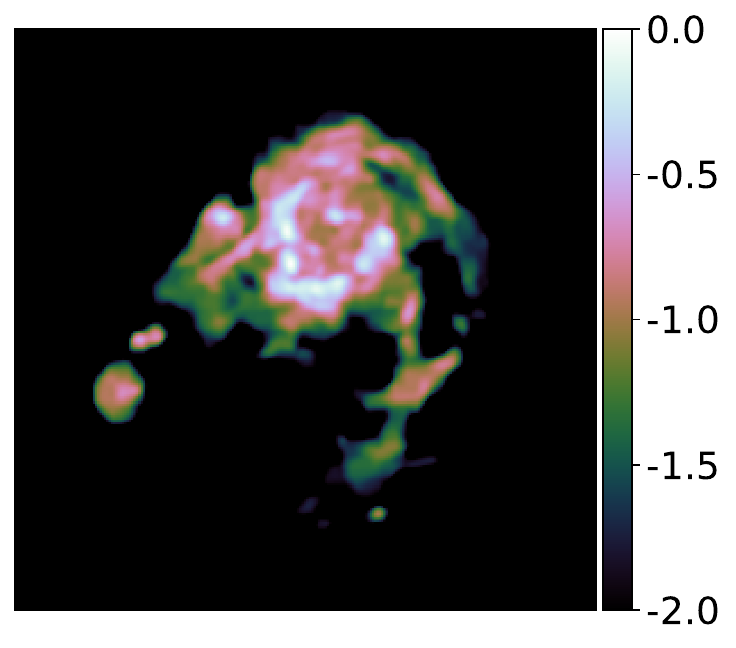} \put(-124,42){\rotatebox{90}{M31}}      &
			\includegraphics[width=0.48\linewidth]{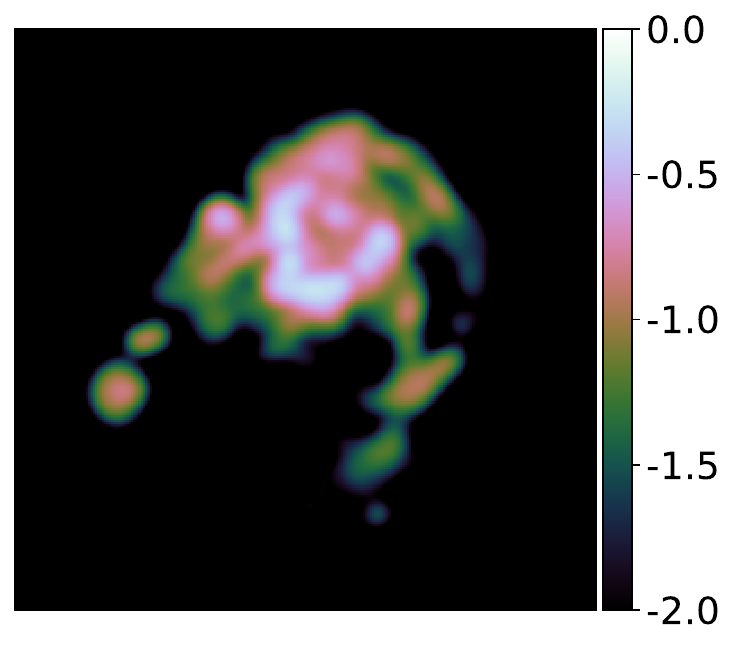}                                                \\
			\includegraphics[width=0.48\linewidth]{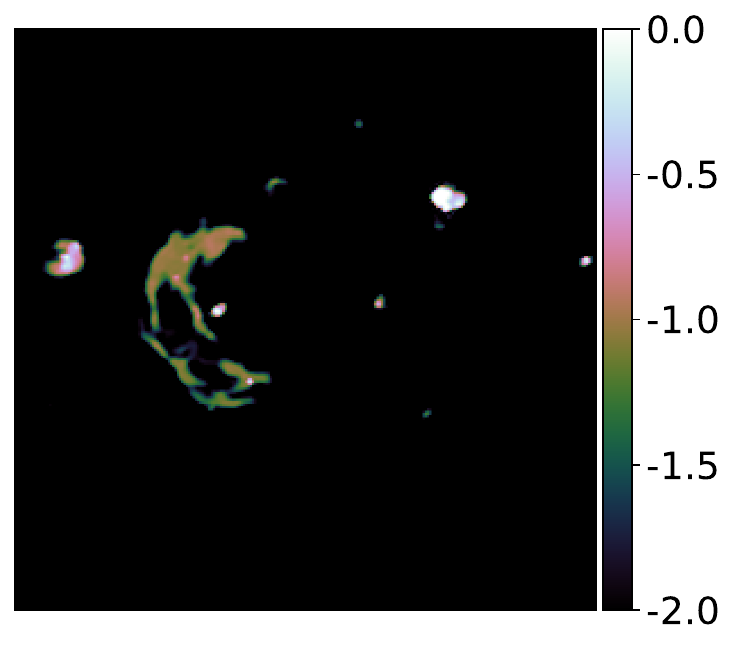} \put(-124,42){\rotatebox{90}{W28}}      &
			\includegraphics[width=0.48\linewidth]{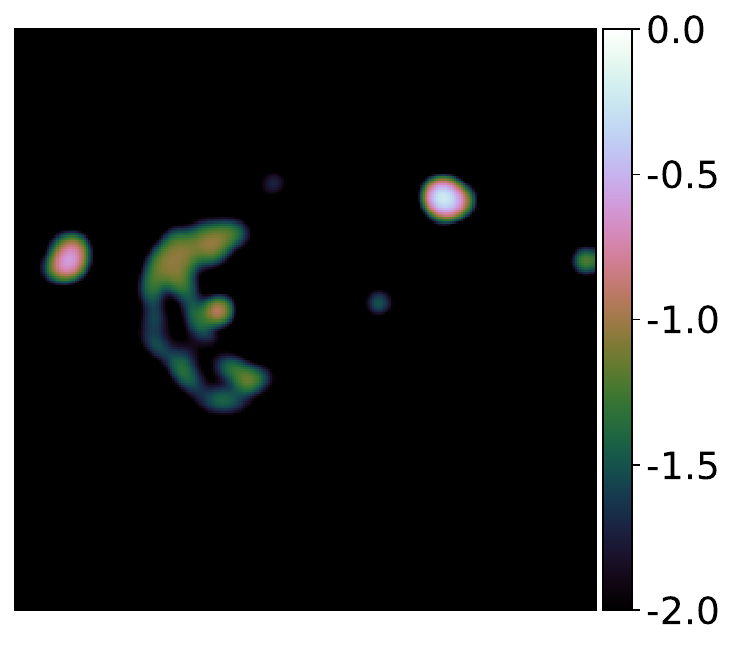}                                                \\
			\includegraphics[width=0.48\linewidth]{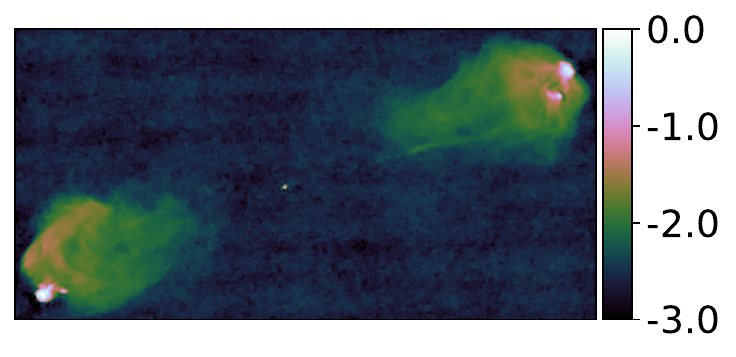} \put(-124,12){\rotatebox{90}{Cygnus A}} &
			\includegraphics[width=0.48\linewidth]{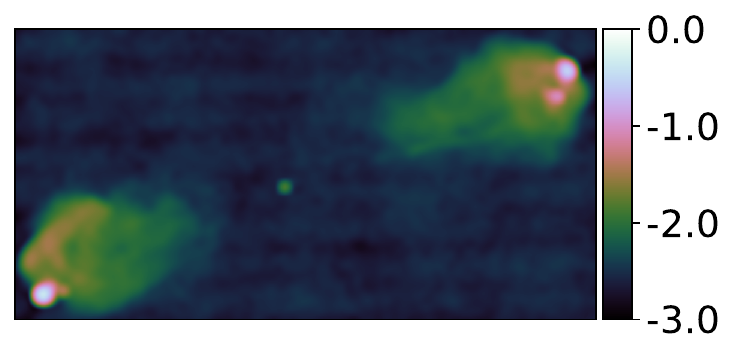}                                                \\
			\includegraphics[width=0.48\linewidth]{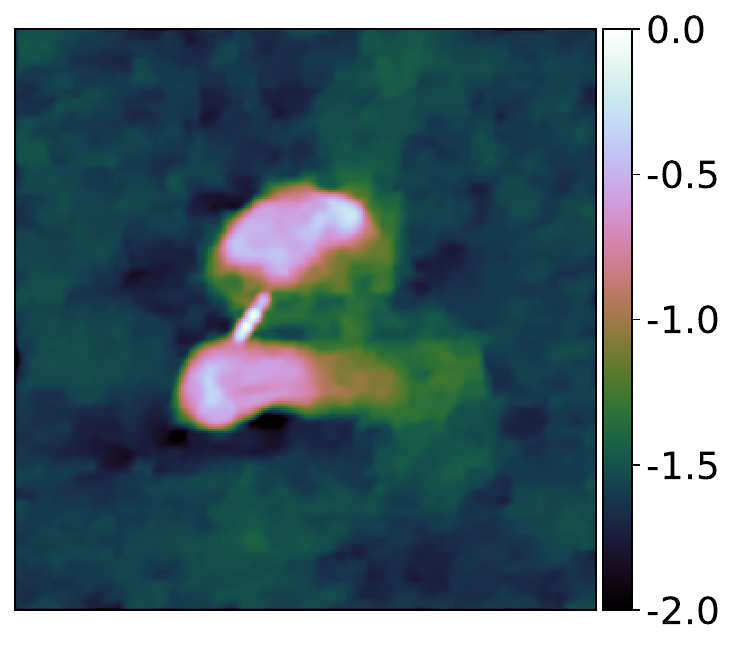} \put(-124,42){\rotatebox{90}{3C288}}  &
			\includegraphics[width=0.48\linewidth]{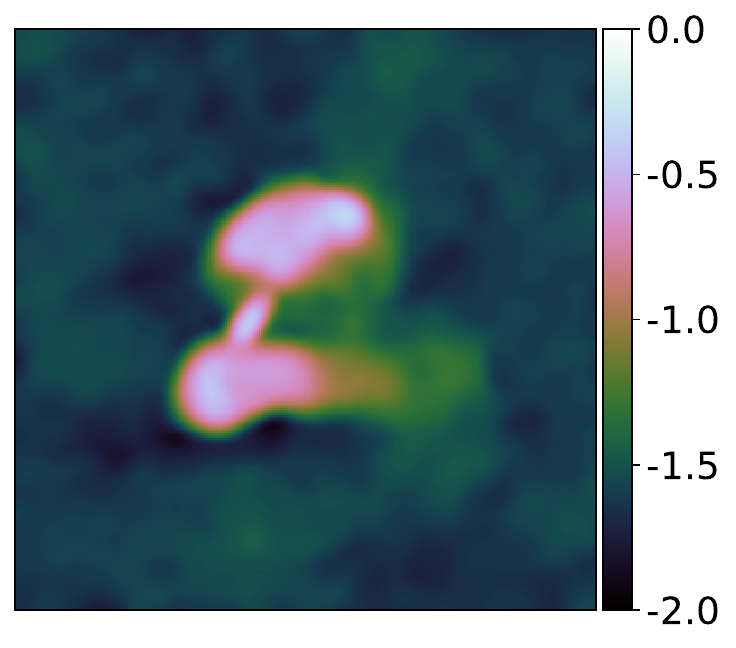}                                              \\
			(a) MAP reconstruction                                                                                                  & (b) Blurred surrogate
		\end{tabular}
		\caption{Hypothesis test of the fine structure in the four \textsc{QuantifAI} MAP reconstructions for M31, W28, Cygnus A, and 3C288. All the images are shown in $\log_{10}$ scale. The fine structure is blurred using a Gaussian kernel with a standard deviation of $3.5$ pixels and a radius of $7$ pixels. The blurred surrogate images in the second column are constructed by blurring the MAP reconstruction shown in the first column. The hypothesis tests are done with the \textsc{QuantifAI} model.}
		\label{fi:hp_test_blurring_CRR}
	\end{figure}
	\addtolength{\tabcolsep}{\tabL}

	The second scenario is to blur the finer structure in the reconstructed image and perform a hypothesis test to elucidate the question of whether the blurred structure is physical or not. The test is illustrated in Figure \ref{fi:hp_test_blurring_CRR}. In this case, all four blurred images represent physical structures.

	In both cases, we compare the hypothesis test using a MAP-based approach described in this work and a sampling-based approach for validation. In the MAP-based approach, we build the HPD region in Equation \ref{eq:HPD_approx_region} with the approximation in Equation \ref{eq:HPD_thresh_approx} and use the MAP estimation as our reconstruction. In the sampling-based approach, we use the MMSE as the reconstruction, i.e., the mean of the posterior samples, and compute the threshold defining the HPD region using the quantile function on the potentials of the posterior samples following \citet[\S 5.2]{cai2018}.

	\begin{table*}
		\begin{center}
			\caption{Hypothesis test results for the inpainted surrogates in Figure \ref{fi:hp_test_inpainting_CRR} using the \textsc{QuantifAI} model. The function $(f+g)(\cdot)$ corresponds to the combined potential of the likelihood and the prior. The reconstruction $\hat{\bm{x}}^{*}$ represents the point estimate used in the sampling or optimisation scenarios, which are the MMSE and the MAP, respectively. The SK-ROCK method corresponds to the posterior sampling techniques. The image $\hat{\bm{x}}^{*,{\text{sgt}}}$ corresponds to the surrogate image, where the areas of interest shown in Figure \ref{fi:hp_test_inpainting_CRR} have been inpainted. The isocontours, $\hat{\gamma}_{0.01}$, or thresholds, are calculated with an $\alpha$ of $0.01$ giving a credible set of $99\%$. In the MAP row, the threshold is computed following the approximation in Equation \ref{eq:HPD_thresh_approx}. In the SK-ROCK row, the threshold is computed from the posterior samples following \citet{cai2018}. The symbols ~\cmark~ and ~\xmark~ in the Ground truth column indicate if the inpainted region contains a physical structure from the ground truth or not, respectively. In the last column, the ~\cmark~ indicates that the hypothesis test is conclusive. All values are scaled with $10^{5}$. \textsc{QuantifAI} is able to correctly reject the hypothesis for all images where the tested structure is physical except for the Cygnus A image. In all cases, the MAP-based and MCMC sampling-based results agree with each other.}
			\label{tb:hp_test_inpainting_CRR}
			\begin{tabular}{cccccccc}
				\toprule
				\multirow{2}{*}{Images}   & Test               & Ground                   & \multirow{2}{*}{Method} & \multicolumn{1}{c}{Point estimate}              & \multicolumn{1}{c}{Surrogate}                                & \multicolumn{1}{c}{Isocontour}            & Hypothesis \\
				                          & area               & truth                    &                         & \multicolumn{1}{c}{$(f + g)(\hat{\bm{x}}^{*})$} & \multicolumn{1}{c}{$(f + g)(\hat{\bm{x}}^{*,{\text{sgt}}})$} & \multicolumn{1}{c}{$\hat{\gamma}_{0.01}$} & test       \\
				\hline \hline
				\multirow{2}{*}{M31}      & \multirow{2}{*}{1} & \multirow{2}{*}{ \cmark} &
				SK-ROCK                   & $0.340$            & $\bf 1.159$              & $0.742$                 & \cmark                                                                                                                                                                  \\
				                          &                    &                          & MAP                     & $0.310$                                         & $\bf 1.161$                                                  & $0.990$                                   & \cmark     \\
				\midrule
				\multirow{2}{*}{Cygnus A} & \multirow{2}{*}{1} & \multirow{2}{*}{ \cmark} &
				SK-ROCK                   & $0.125$            & $0.182$                  & $\bf 0.848$             & \xmark                                                                                                                                                                  \\
				                          &                    &                          & MAP                     & $0.105$                                         & $0.169$                                                      & $\bf 1.450$                               & \xmark     \\
				\midrule
				\multirow{2}{*}{W28}      & \multirow{2}{*}{1} & \multirow{2}{*}{ \cmark} &
				SK-ROCK                   & $0.222$            & $\bf 4.649$              & $0.612$                 & \cmark                                                                                                                                                                  \\
				                          &                    &                          & MAP                     & $0.196$                                         & $\bf 4.699$                                                  & $0.876$                                   & \cmark     \\
				\midrule
				\multirow{4}{*}{3C288}    & \multirow{2}{*}{1} & \multirow{2}{*}{\cmark}  &
				SK-ROCK                   & $0.257$            & $\bf 1.918$              & $0.659$                 & \cmark                                                                                                                                                                  \\
				                          &                    &                          & MAP                     & $0.229$                                         & $\bf 1.908$                                                  & $0.908$                                   & \cmark     \\
				\cdashline{2-8}
				                          & \multirow{2}{*}{2} & \multirow{2}{*}{ \xmark} &
				SK-ROCK                   & $0.257$            & $0.257$                  & $\bf 0.659$             & \xmark                                                                                                                                                                  \\
				                          &                    &                          & MAP                     & $0.229$                                         & $0.229$                                                      & $\bf 0.908$                               & \xmark     \\
				\bottomrule
			\end{tabular}
		\end{center}
	\end{table*}

	Table \ref{tb:hp_test_inpainting_CRR} presents the results for the inpainting hypothesis test, where the inpainted surrogates are shown in Figure \ref{fi:hp_test_inpainting_CRR}. The MAP- and sampling-based results are consistent in all the images studied, where the threshold computed with the posterior samples is slightly tighter than the MAP-based approximation. The hypothesis tests correctly classify the structure in images M31, W28 and 3C288, including the two cases of the latter image. The UQ methods cannot make a strong statistical statement about the structures in the Cygnus A image. In this image, where the inpainted region has a tiny physical structure, the potentials of the inpainted surrogate image rest close to the MAP and MMSE estimators. We include the hypothesis test results of the same inpainting experiment for the wavelet-based model in Appendix \ref{ap:hyp_test_structure} to provide a comparison between the models. We used the wavelet prior to inpaint the region of interest to allow for a fair comparison. All results from the wavelet-based model are in agreement with \textsc{QuantifAI}.

	\begin{table*}
		\begin{center}
			\caption{Hypothesis test results for the blurred surrogates of Figure \ref{fi:hp_test_blurring_CRR} using the \textsc{QuantifAI} model. The description of Figure \ref{fi:hp_test_blurring_CRR} holds in this table. All values are scaled with $10^{5}$. \textsc{QuantifAI} is able to correctly reject the hypothesis in all cases, and the MAP-based outcome agrees with its MCMC sampling-based counterpart.}
			\label{tb:hp_test_blurring_CRR}
			\begin{tabular}{cccccccc}
				\toprule
				\multirow{2}{*}{Images}   & \multirow{2}{*}{Method} & \multicolumn{1}{c}{Initial}                     & \multicolumn{1}{c}{Surrogate}                                & \multicolumn{1}{c}{Isocontour}            & Hypothesis \\
				                          &                         & \multicolumn{1}{c}{$(f + g)(\hat{\bm{x}}^{*})$} & \multicolumn{1}{c}{$(f + g)(\hat{\bm{x}}^{*,{\text{sgt}}})$} & \multicolumn{1}{c}{$\hat{\gamma}_{0.01}$} & test       \\
				\hline \hline
				\multirow{2}{*}{M31}      &
				SK-ROCK                   & $0.340$                 & $\bf 1.905$                                     & $0.742$                                                      & \cmark                                                 \\
				                          & MAP                     & $0.310$                                         & $\bf 1.906$                                                  & $0.990$                                   & \cmark     \\
				\midrule
				\multirow{2}{*}{Cygnus A} &
				SK-ROCK                   & $0.125$                 & $\bf 9.642$                                     & $0.848$                                                      & \cmark                                                 \\
				                          & MAP                     & $0.105$                                         & $\bf 9.643$                                                  & $1.450$                                   & \cmark     \\
				\midrule
				\multirow{2}{*}{W28}      &
				SK-ROCK                   & $0.222$                 & $\bf 8.389$                                     & $0.612$                                                      & \cmark                                                 \\
				                          & MAP                     & $0.196$                                         & $\bf 8.387$                                                  & $0.876$                                   & \cmark     \\
				\midrule
				\multirow{2}{*}{3C288}    &
				SK-ROCK                   & $0.257$                 & $\bf 0.920$                                     & $0.659$                                                      & \cmark                                                 \\
				                          & MAP                     & $0.229$                                         & $\bf 0.922$                                                  & $0.908$                                   & \cmark     \\
				\bottomrule
			\end{tabular}
		\end{center}
	\end{table*}

	The results from the blurred surrogates of Figure \ref{fi:hp_test_blurring_CRR} are presented in Table \ref{tb:hp_test_blurring_CRR}. In all the images, the hypothesis test concludes that the blurred fine structure is physical as the potential falls out of the HPD region. The MAP- and sampling-based results are consistent with each other.

	The different hypothesis tests have shown consistent results between the sampling-based and highly scalable MAP-based results. In addition, the results from the hypothesis tests are coherent between the \textsc{QuantifAI} and wavelet-based model. We remark that the approach based on the MAP requires one further measurement operator evaluation to carry out the hypothesis test. The test provides a highly scalable way to answer scientific questions about the uncertainty of the RI imaging reconstructions.

	\subsection{Local credible intervals}

\begin{figure*}
		\centering
		\begin{tabular}{cccc}
			{\large M31}                                                                                                                                     & {\large W28} & {\large 3C288} & {\large Cygnus A}
			\\
			\includegraphics[width=0.24\linewidth]{M31/M31-CRR-MAP_image.pdf} \put(-130,25){\rotatebox{90}{MAP estimation}}                                  &
			\includegraphics[width=0.24\linewidth]{W28/W28-CRR-MAP_image.pdf}                                                                                &
			\includegraphics[width=0.24\linewidth]{3c288/3c288-CRR-MAP_image.pdf}                                                                            &
			\raisebox{.5\height}{\includegraphics[width=0.24\linewidth]{CYN/CYN-CRR-MAP_image.pdf}}
			\\
			\multicolumn{4}{c}{\large Superpixel size: $4 \times 4$}
			\\
			\includegraphics[width=0.24\linewidth]{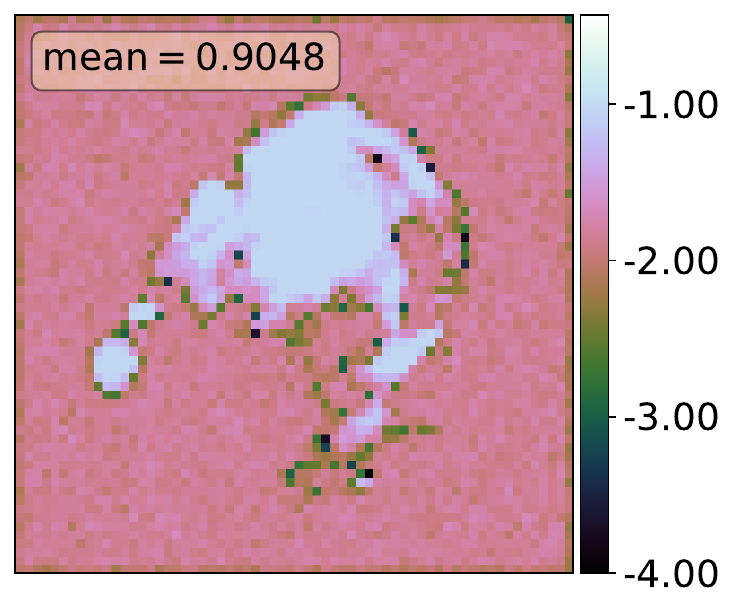} \put(-130,30){\rotatebox{90}{$\mathrm{LCI} - <\mathrm{LCI}>$}} &
			\includegraphics[width=0.24\linewidth]{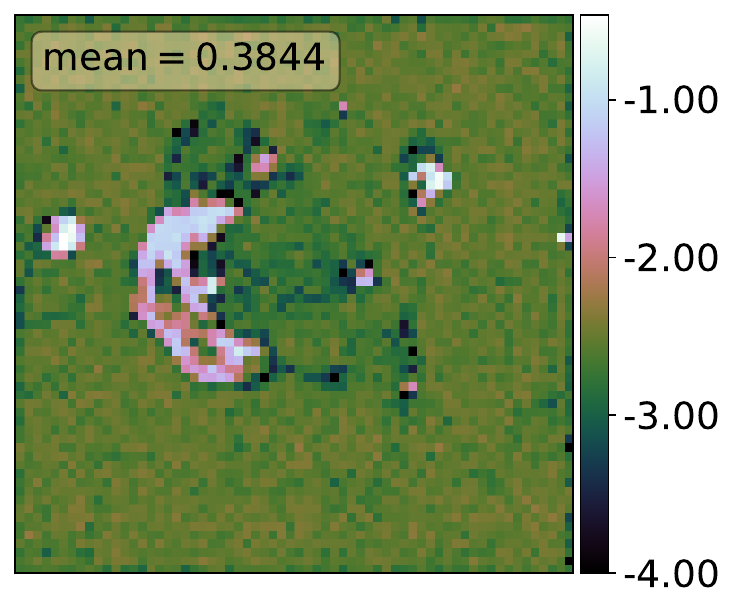}                                                                &
			\includegraphics[width=0.24\linewidth]{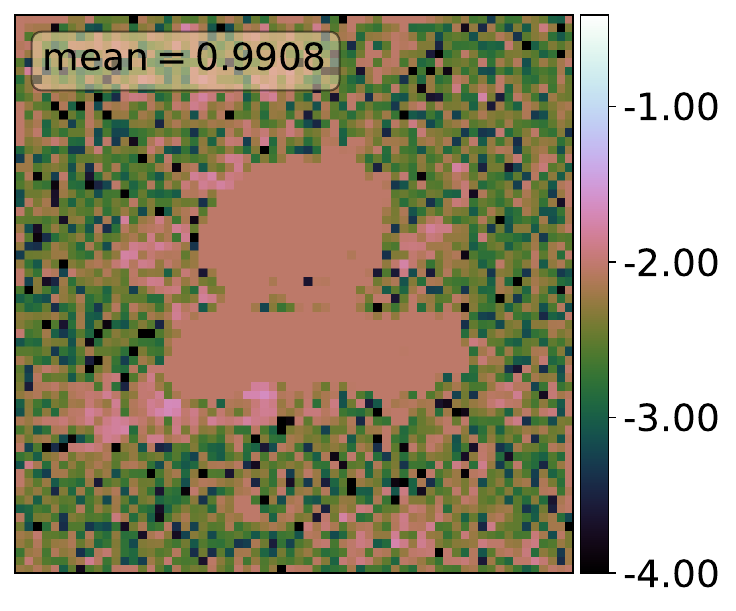}                                                            &
			\raisebox{.5\height}{\includegraphics[width=0.24\linewidth]{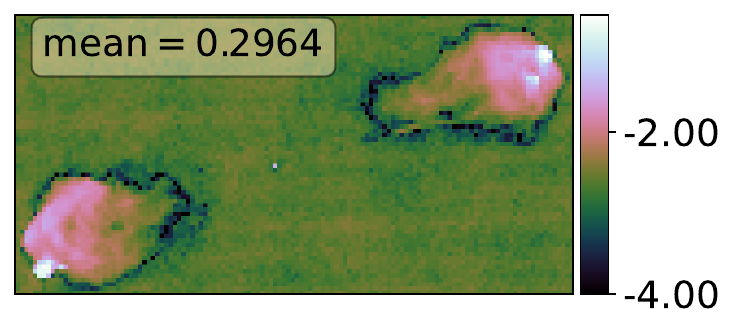}}
			\\
			\includegraphics[width=0.24\linewidth]{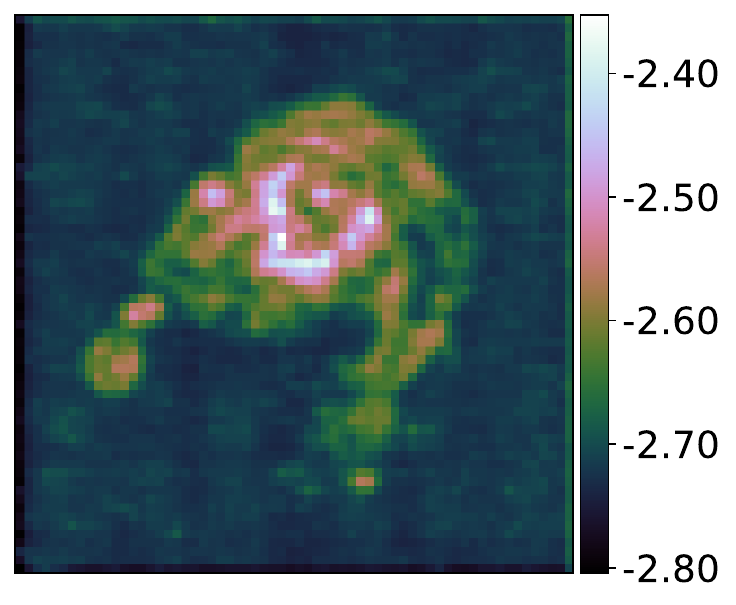} \put(-130,7){\rotatebox{90}{Posterior standard deviation}}  &
			\includegraphics[width=0.24\linewidth]{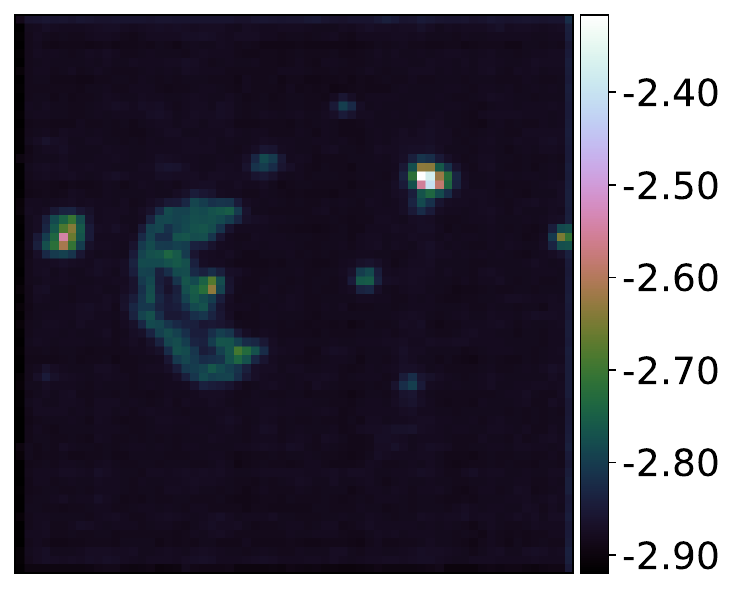}                                                             &
			\includegraphics[width=0.24\linewidth]{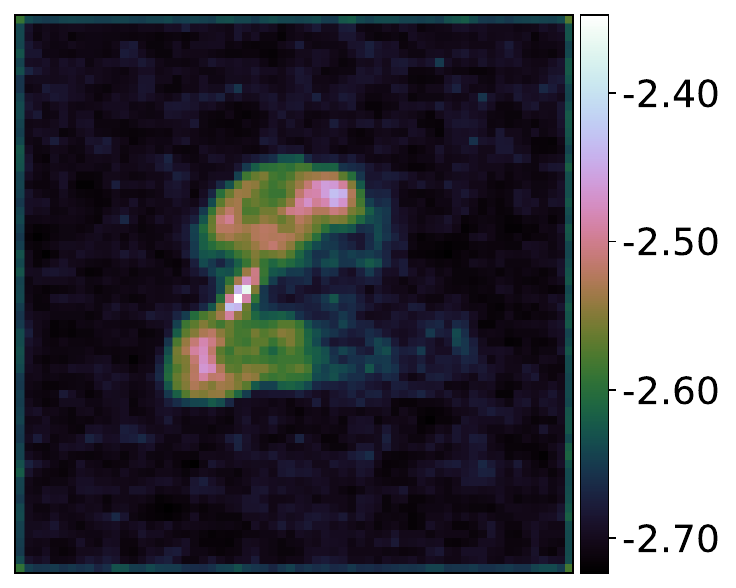}                                                         &
			\raisebox{.5\height}{\includegraphics[width=0.24\linewidth]{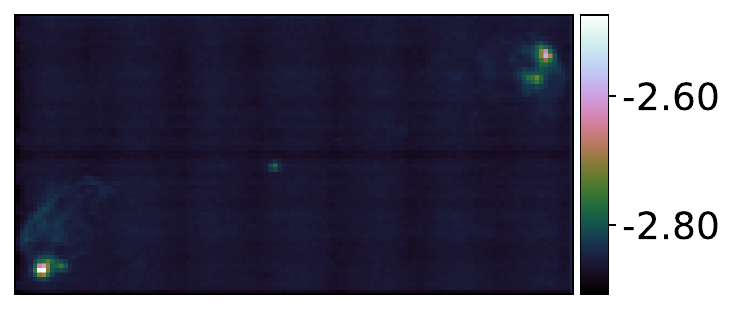}}
			\\
			\multicolumn{4}{c}{\large Superpixel size: $8 \times 8$}
			\\
			\includegraphics[width=0.24\linewidth]{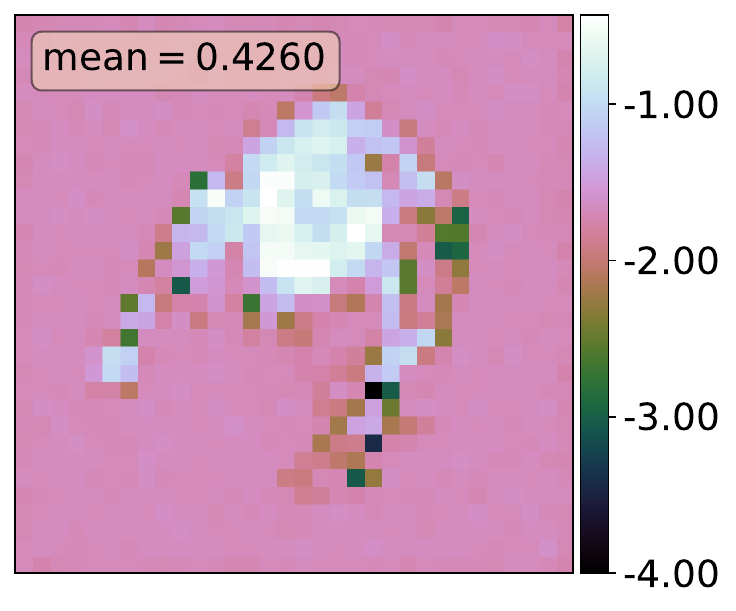} \put(-130,30){\rotatebox{90}{$\mathrm{LCI} - <\mathrm{LCI}>$}} &
			\includegraphics[width=0.24\linewidth]{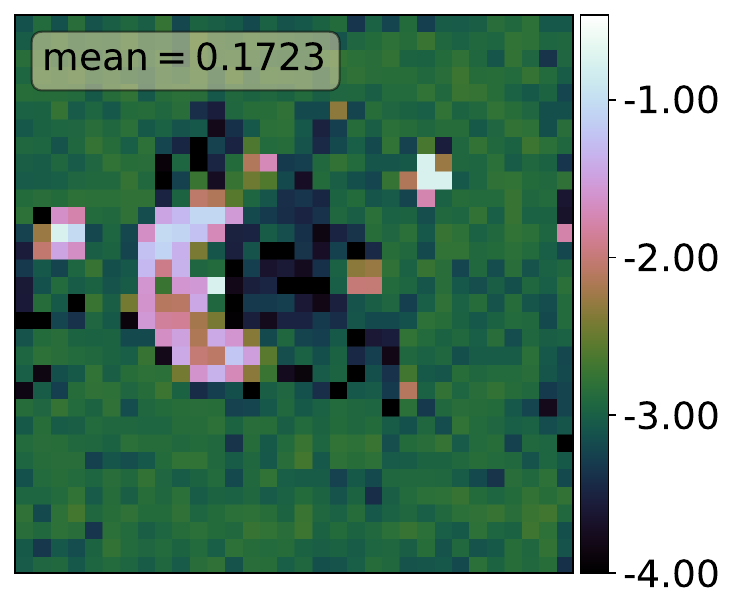}                                                                &
			\includegraphics[width=0.24\linewidth]{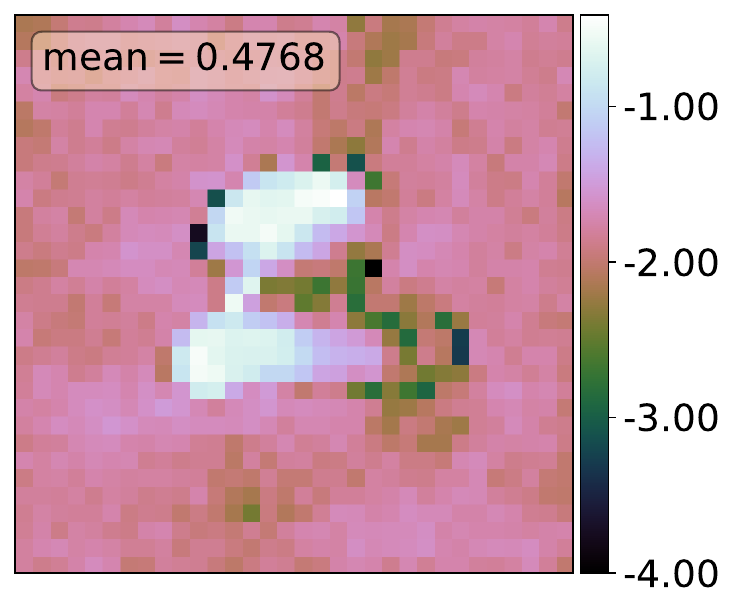}                                                            &
			\raisebox{.5\height}{\includegraphics[width=0.24\linewidth]{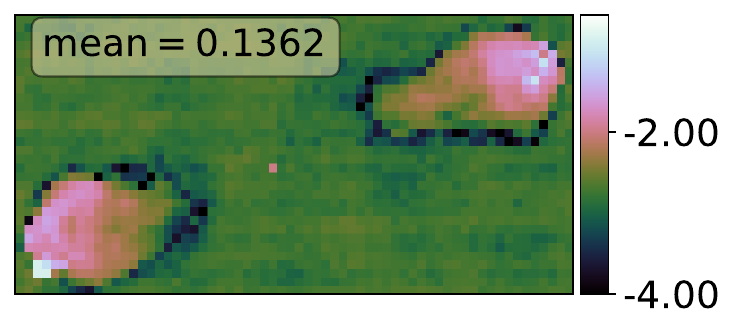}}
			\\
			\includegraphics[width=0.24\linewidth]{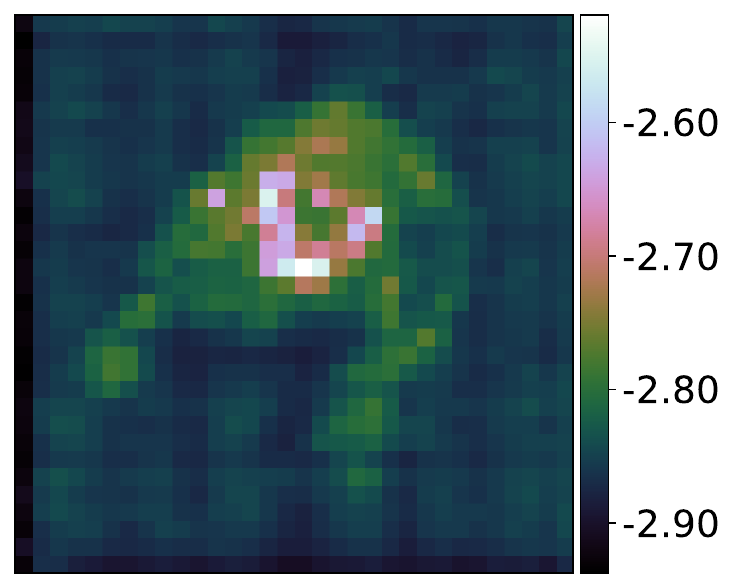} \put(-130,7){\rotatebox{90}{Posterior standard deviation}}  &
			\includegraphics[width=0.24\linewidth]{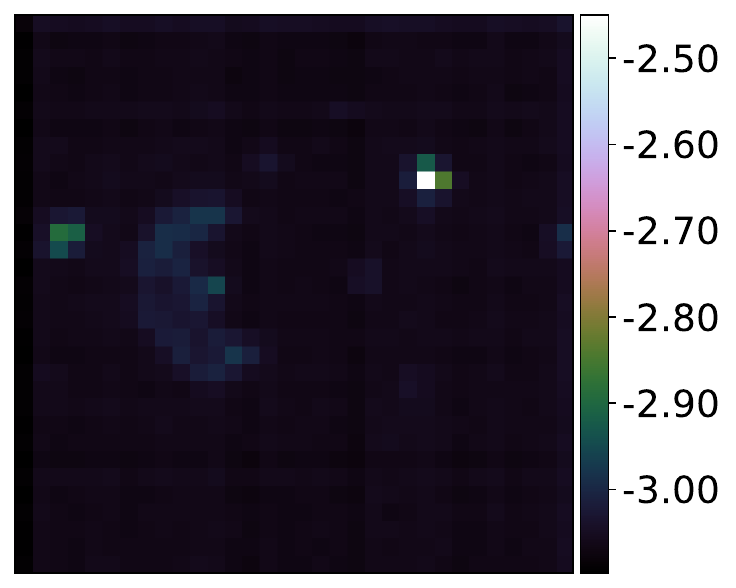}                                                             &
			\includegraphics[width=0.24\linewidth]{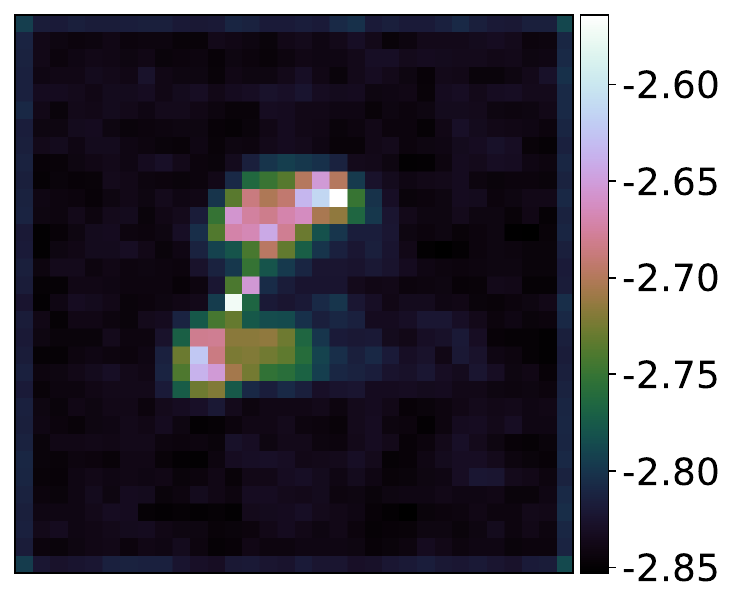}                                                         &
			\raisebox{.5\height}{\includegraphics[width=0.24\linewidth]{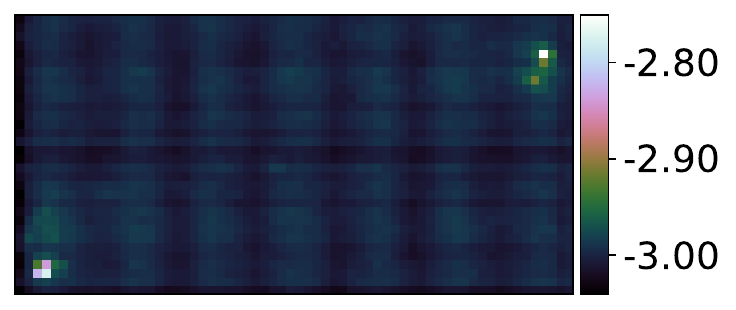}}
		\end{tabular}
		\caption{Length of the local credible intervals (LCIs), \textit{cf.} Bayesian error bars, computed with a $99\%$ credible level using superpixel sizes of $4 \times 4$ and $8 \times 8$. Each column represents one of the four images in our dataset. The first row shows the MAP estimation of each image at its original resolution.
			The second row displays the variation of the LCIs around their mean, recorded in a box in the upper left corners. This display choice allows us to visualise the structure of the LCIs better while keeping the LCIs mean information. The third row presents the posterior standard deviation computed with the same superpixel size. The fourth and fifth rows present the equivalent information for the superpixel size of $8 \times 8$. There is reasonable agreement between the uncertainty captured by the LCI and the posterior standard deviation.}
		\label{fi:pixel_uq_all}
	\end{figure*}

	We have exploited the approximation of the HPD region from Section \ref{sc:hpd_region_approx} based on the MAP estimations and a credible level of $99\%$. The approximate HPD regions were then used to compute the LCIs, whose lengths per pixel are visualized as an image, c.f.~Figure \ref{fi:pixel_uq_all}. The LCI lengths are displayed after subtracting the mean LCI length overall superpixels in the image, which is shown in the top left corner of the image. The UQ results for \textsc{QuantifAI} are presented for two superpixel sizes, $4 \times 4$ and $8 \times 8$. We have omitted LCIs from the wavelet-based prior for conciseness. The posterior standard deviations at the two superpixel sizes are included for comparison with the significantly faster MAP-based UQ technique of the LCIs. We find a reasonable agreement between the structure in the LCI plots and the posterior standard deviation. For example, the 3C288 image with superpixel size $8 \times 8$ yields tighter LCIs in the two elliptical regions and in the small connecting structure in the centre of the image. The corresponding posterior standard deviation is smaller in the aforementioned regions, which is expected as most of the observed signal concentrates there. The LCIs and the posterior standard deviation represent different quantiles, so we would not expect an exact agreement even without any approximation in the computation of the LCIs.

	We observe, as expected, that the larger superpixels have tighter LCIs, as seen in the mean LCIs shown on the top left corner of the subfigures in Figure \ref{fi:pixel_uq_all}. The reconstructions are naturally less uncertain on the larger scales due to the properties of our measurement operator, as the visibilities are generally concentrated towards the low frequencies. In addition, varying the value of a larger superpixel saturates the HPD region faster than for a small superpixel. We have also computed the LCIs for the superpixels of size $16 \times 16$, which we have not included for conciseness. The corresponding mean LCI values are $0.20$, $0.08$, $0.24$, and $0.07$ for the images in the same order as in Figure \ref{fi:pixel_uq_all}.

	When comparing the mean value of the LCIs from the four reconstructions from Figure \ref{fi:pixel_uq_all} we notice that two of them, M31 and 3C288, have higher uncertainty than the rest. The higher the uncertainty, the larger the mean value of the LCI gets, as the superpixel values need larger changes before they saturate the HPD region. Image 3C288, with a superpixel size of $4 \times 4$, is an example where the LCIs have saturated as the mean is close to unity\footnote{n.b. The images are scaled in the range $[0,1]$.}; therefore, the LCI image's detailed structure is lost due to the saturation. This saturation highlights the need for superpixel sizes to be selected appropriately, depending on the case at hand.

	\addtolength{\tabcolsep}{-\tabL}
	\addtolength{\tabcolsep}{\tabL}

	\subsection{Fast pixel uncertainty quantification at different scales}

\begin{figure*}
		\centering
		\begin{tabular}{cccc}
			\multicolumn{2}{c}{\large M31}                                                                                                                                       & \multicolumn{2}{c}{\large W28}                                                                 \\
			\includegraphics[width=0.2\linewidth]{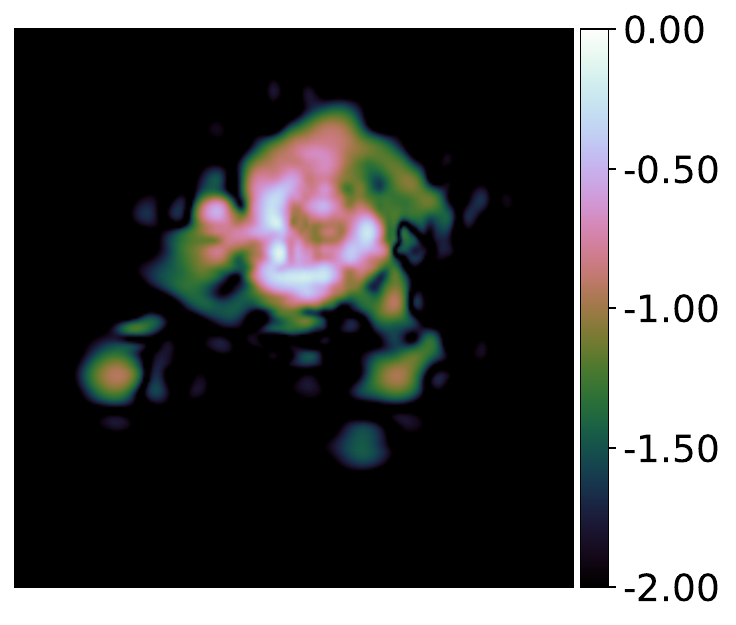}                                                              &
			\includegraphics[width=0.2\linewidth]{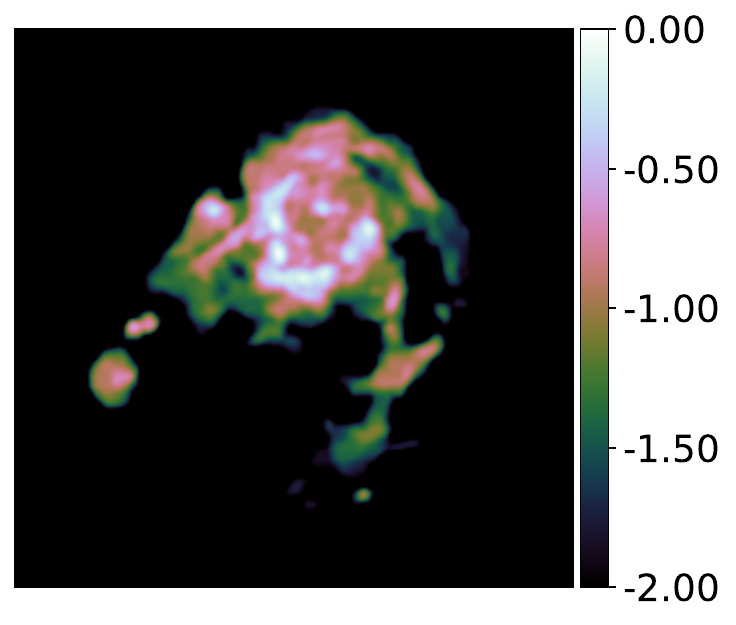}                                                                           &
			\includegraphics[width=0.2\linewidth]{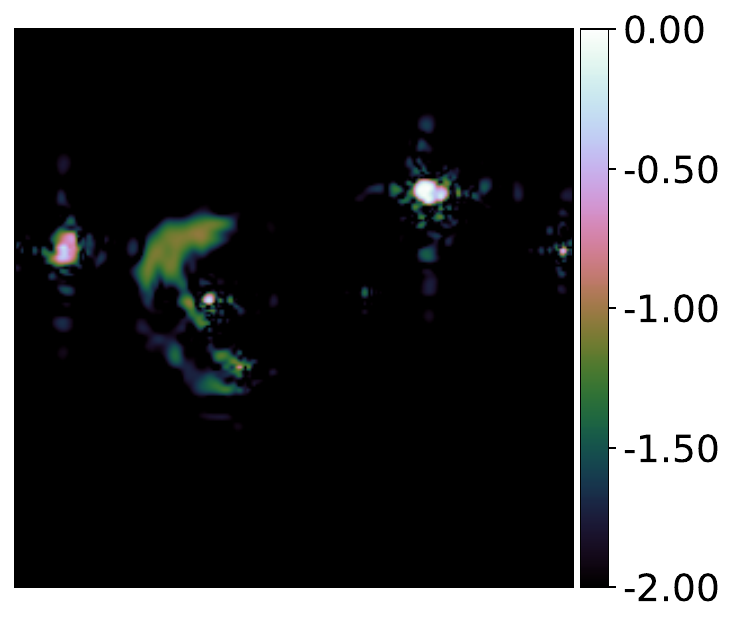}                                                              &
			\includegraphics[width=0.2\linewidth]{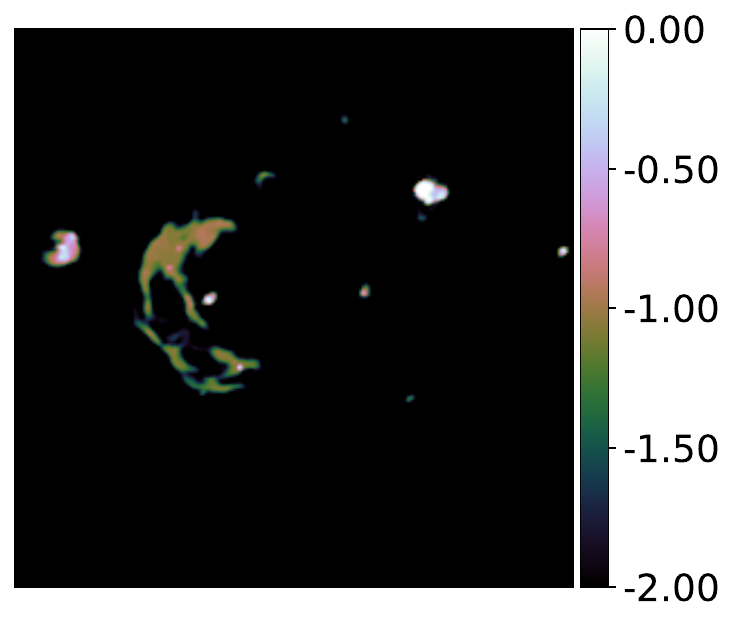}
			\\
			{\small (a) Thresholded MAP}                                                                                                                                         & {\small (b) MAP estimation}       & {\small (c) Thresholded MAP} & {\small (d) MAP estimation}
			\\
			\\
			\multicolumn{2}{c}{\large Errors}                                                                                                                                    & \multicolumn{2}{c}{\large Errors}                                                              \\
			{\small Predicted}                                                                                                                                                   & {\small True errors}              & {\small Predicted}           & {\small True errors}
			\\
			\includegraphics[width=0.2\linewidth]{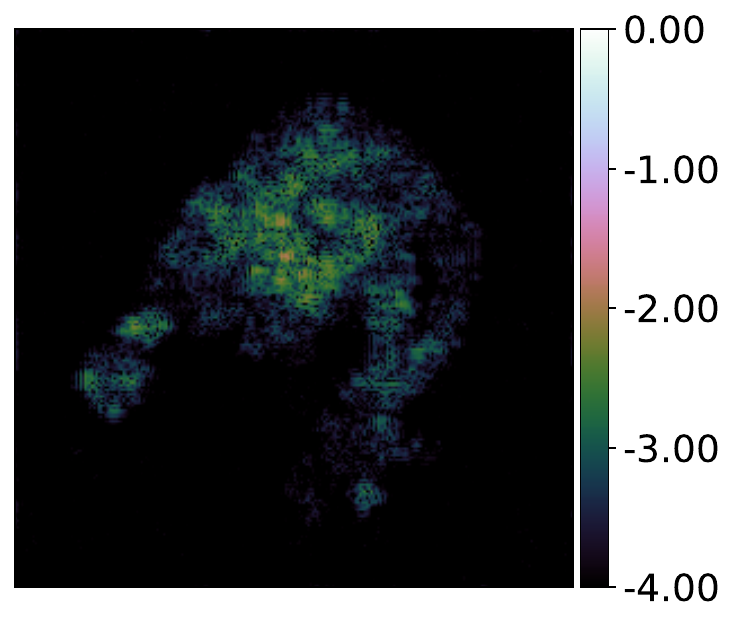} \put(-110,30){\rotatebox{90}{\large Level $4$}} &
			\includegraphics[width=0.2\linewidth]{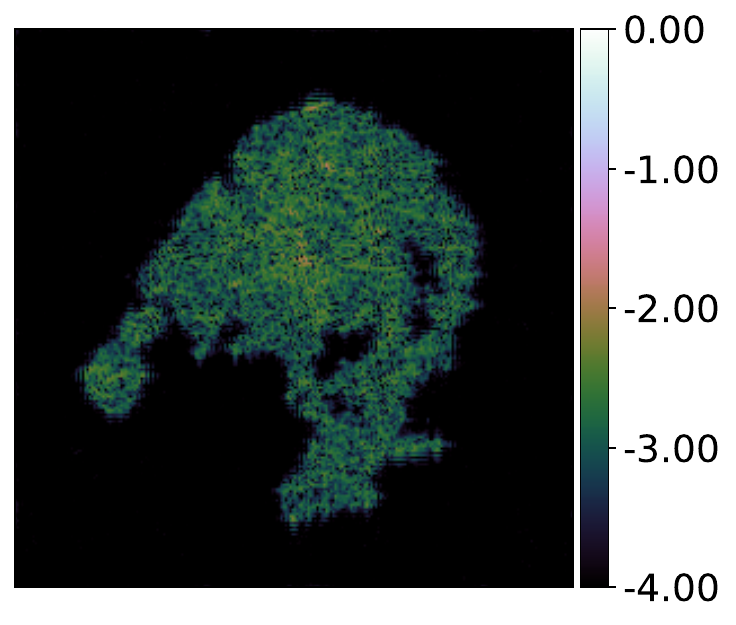}                                                          &
			\includegraphics[width=0.2\linewidth]{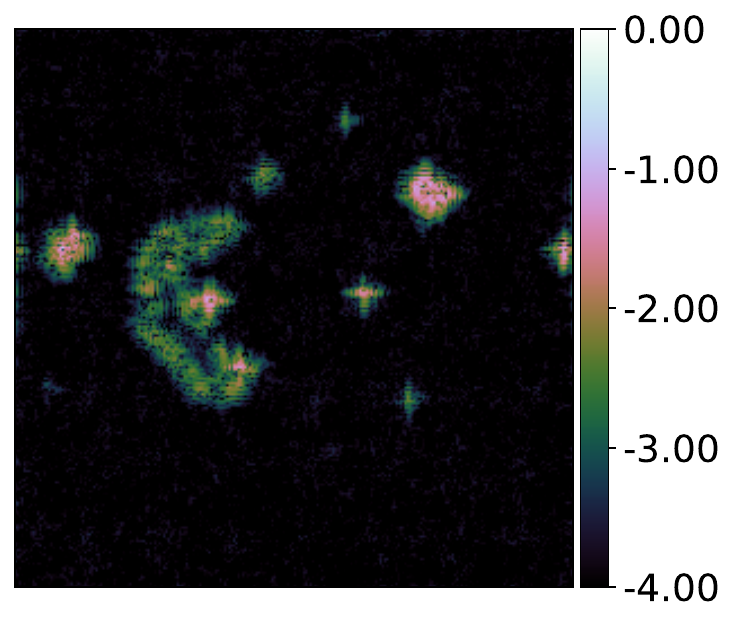}                                                 &
			\includegraphics[width=0.2\linewidth]{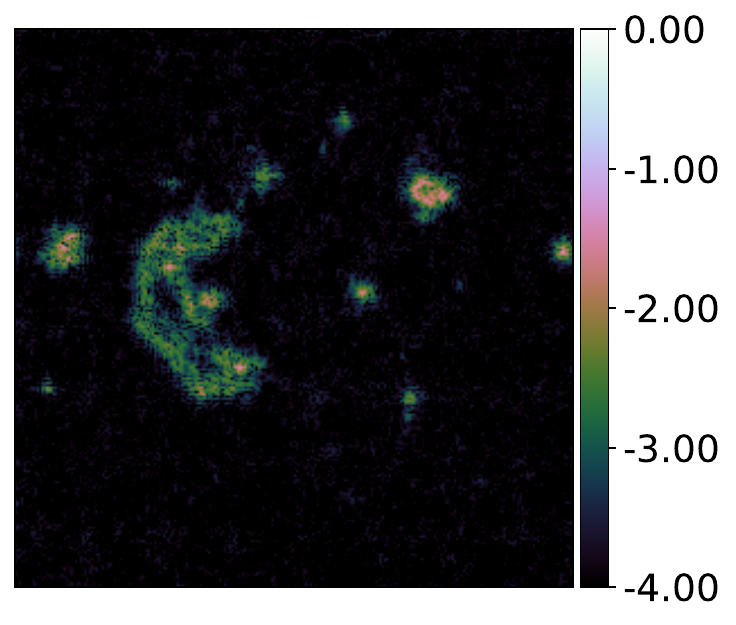}
			\\
			\includegraphics[width=0.2\linewidth]{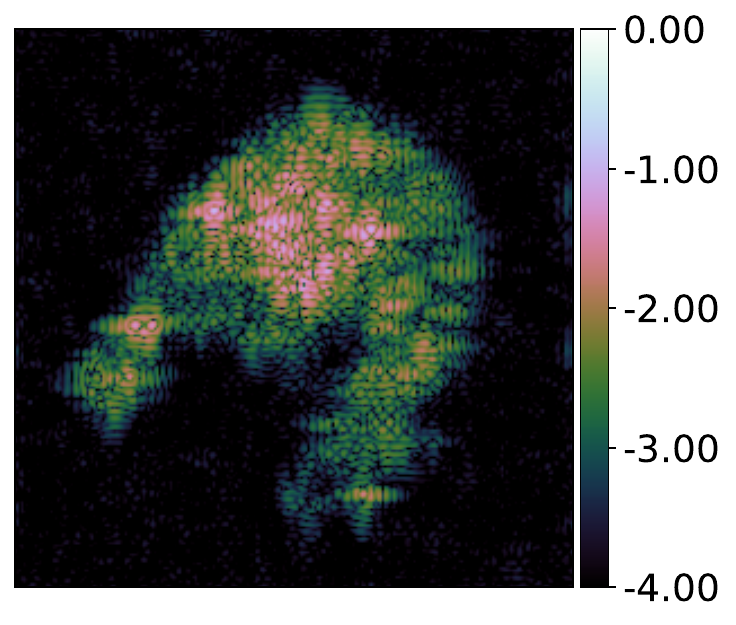} \put(-110,30){\rotatebox{90}{\large Level $3$}} &
			\includegraphics[width=0.2\linewidth]{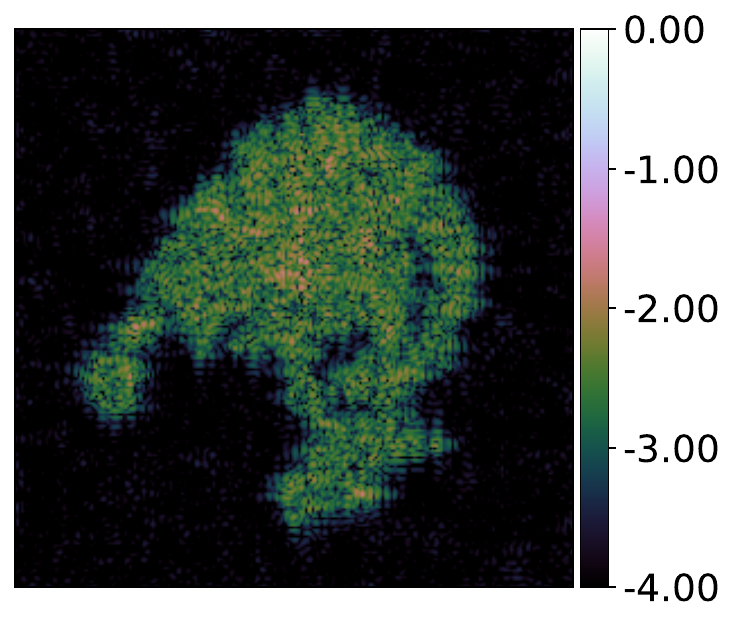}                                                          &
			\includegraphics[width=0.2\linewidth]{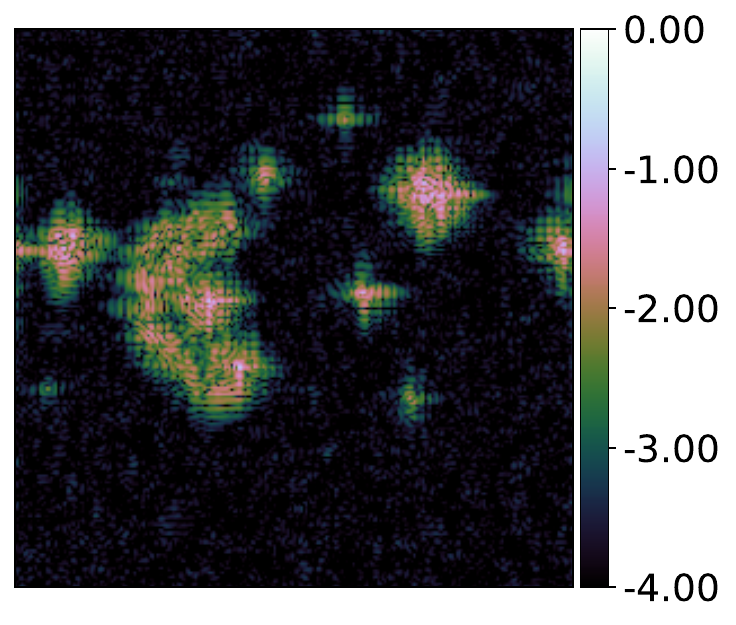}                                                 &
			\includegraphics[width=0.2\linewidth]{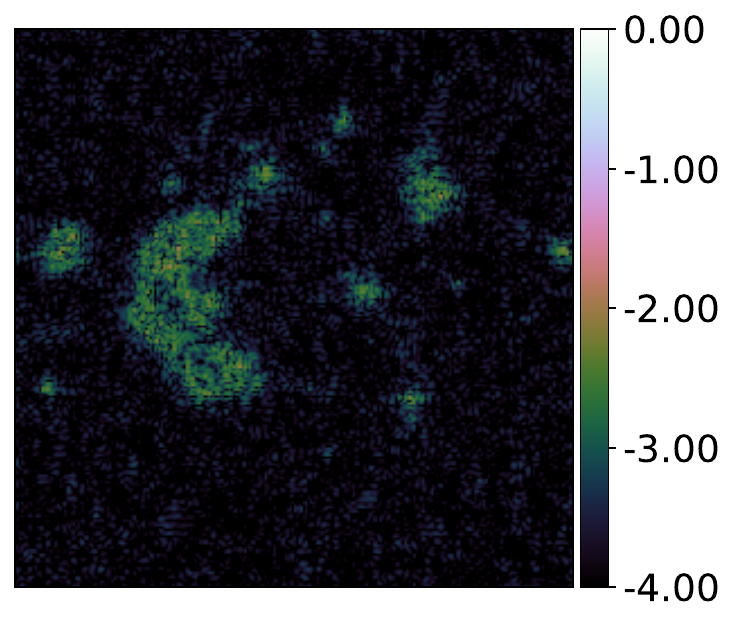}
			\\
			\includegraphics[width=0.2\linewidth]{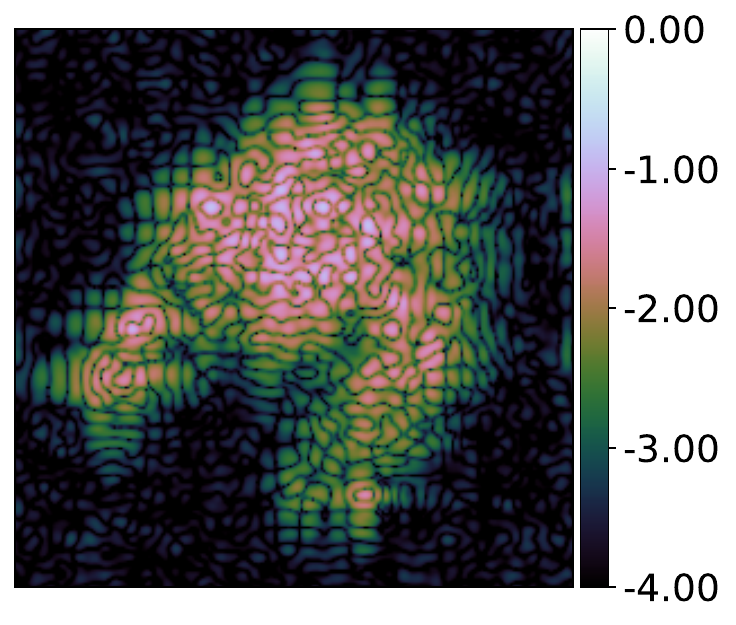} \put(-110,30){\rotatebox{90}{\large Level $2$}} &
			\includegraphics[width=0.2\linewidth]{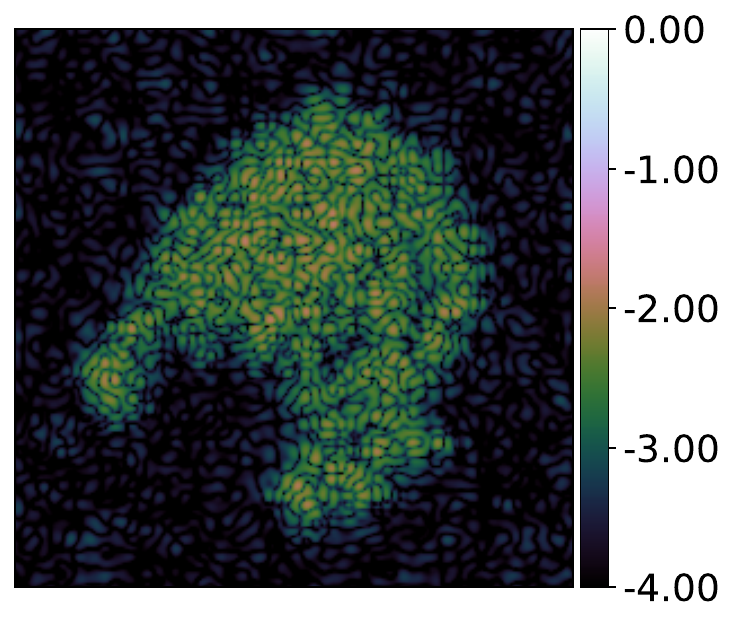}                                                          &
			\includegraphics[width=0.2\linewidth]{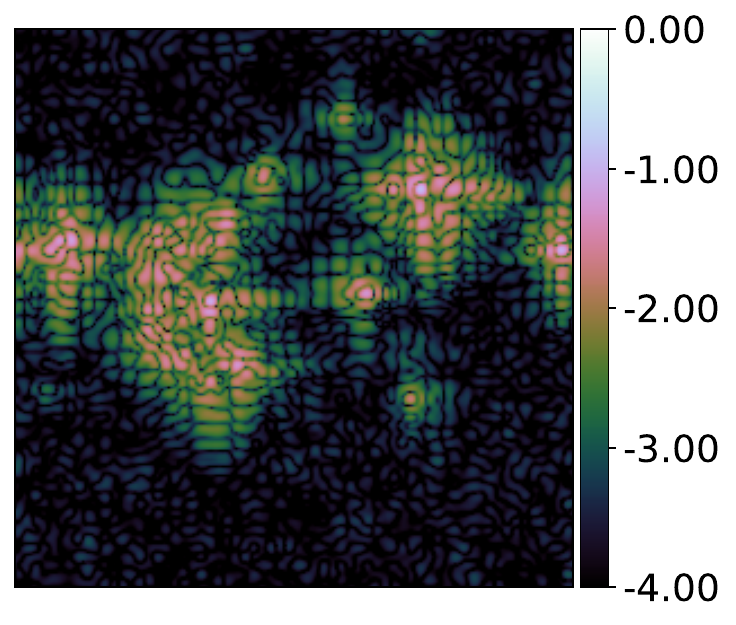}                                                 &
			\includegraphics[width=0.2\linewidth]{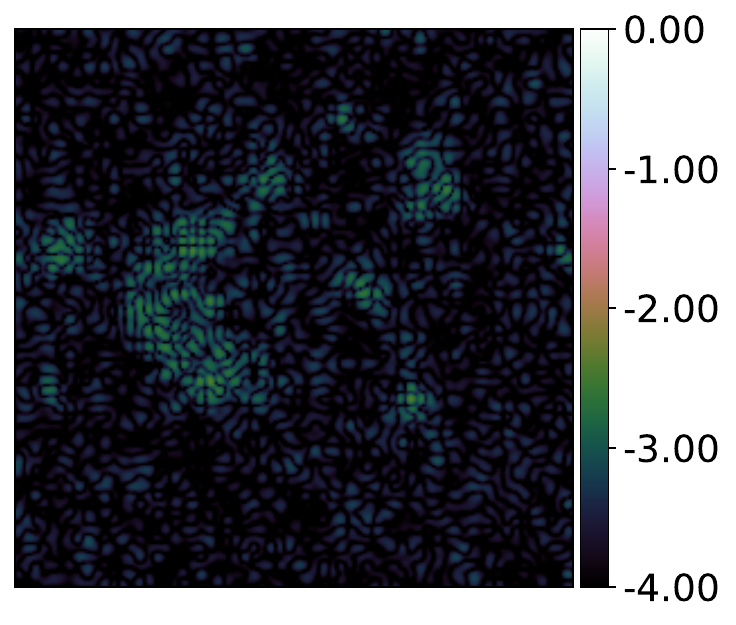}
			\\
			\includegraphics[width=0.2\linewidth]{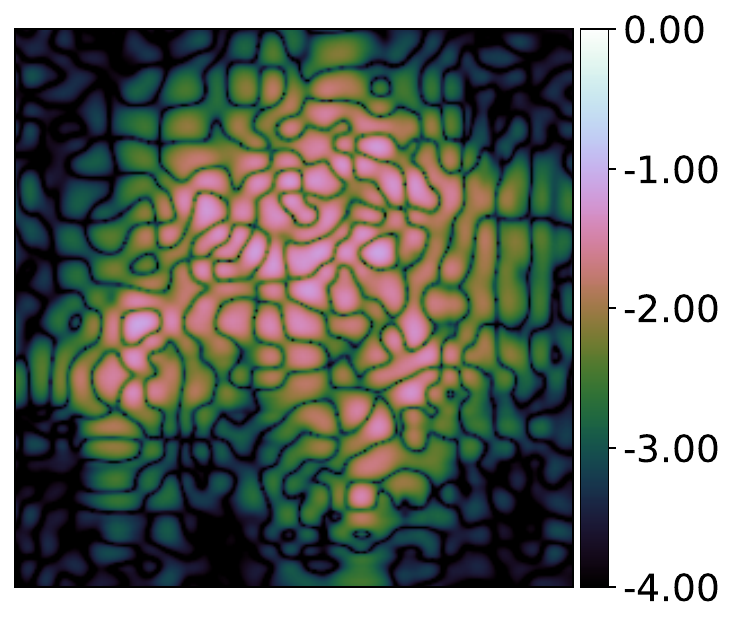} \put(-110,30){\rotatebox{90}{\large Level $1$}} &
			\includegraphics[width=0.2\linewidth]{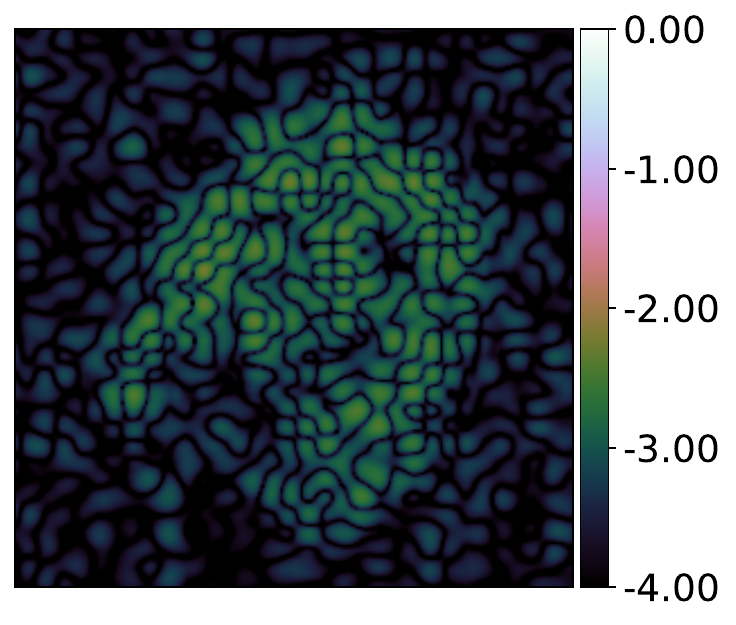}                                                          &
			\includegraphics[width=0.2\linewidth]{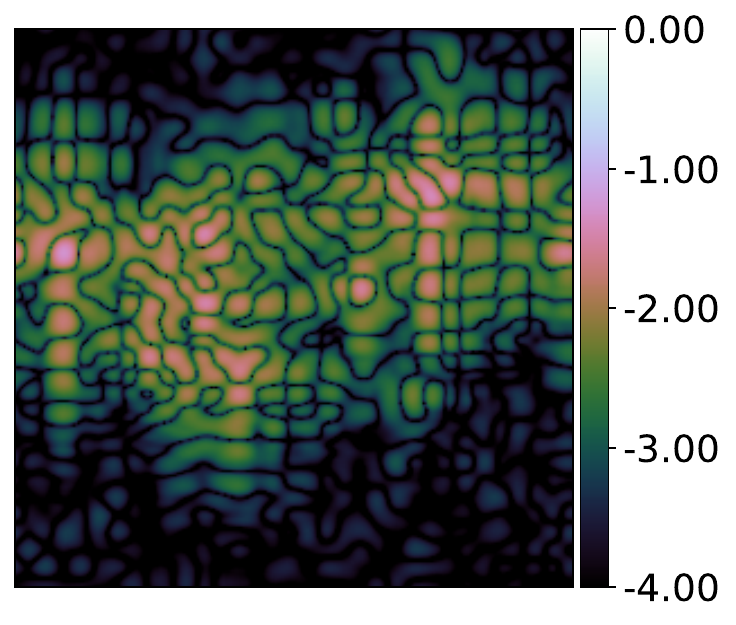}                                                 &
			\includegraphics[width=0.2\linewidth]{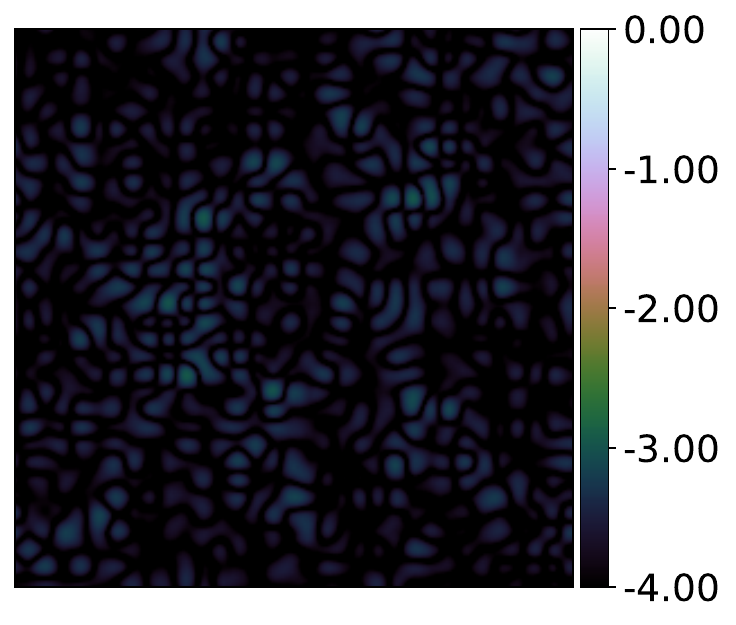}
			\\
			\includegraphics[width=0.2\linewidth]{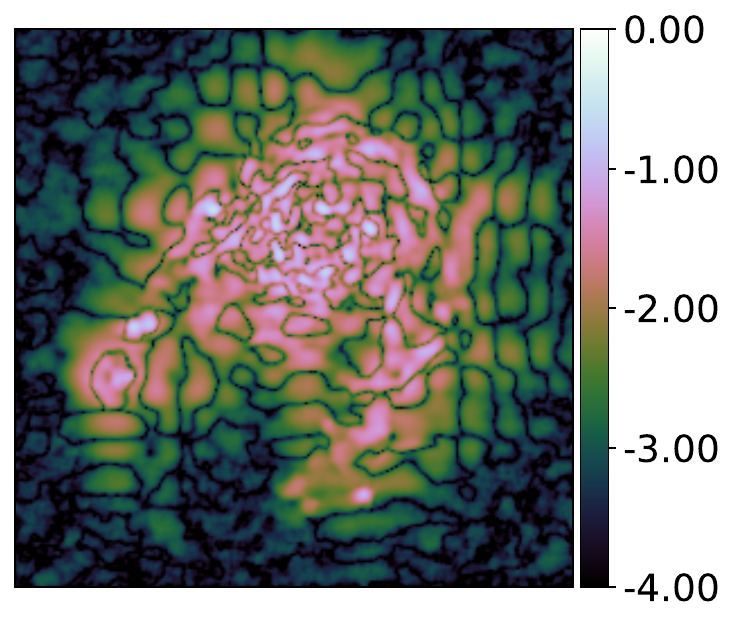} \put(-110,30){\rotatebox{90}{\large All levels}}        &
			\includegraphics[width=0.2\linewidth]{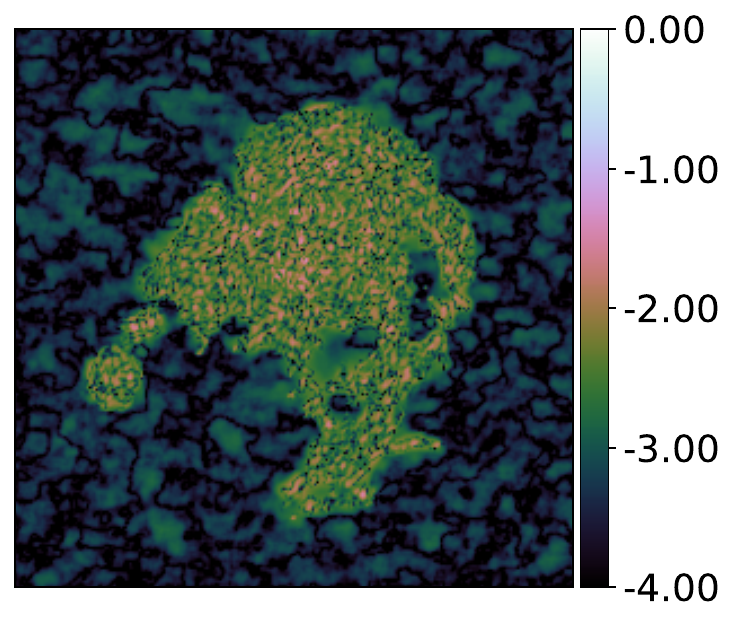}                                                                  &
			\includegraphics[width=0.2\linewidth]{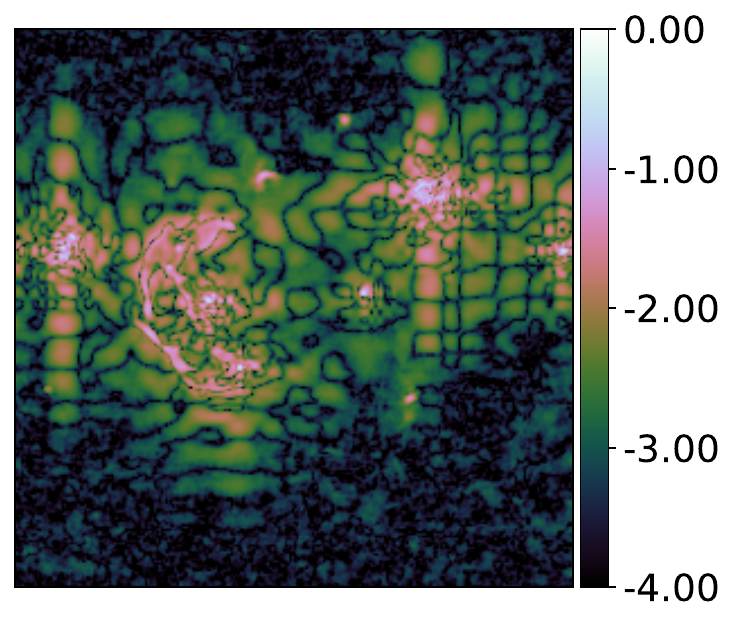}                                                         &
			\includegraphics[width=0.2\linewidth]{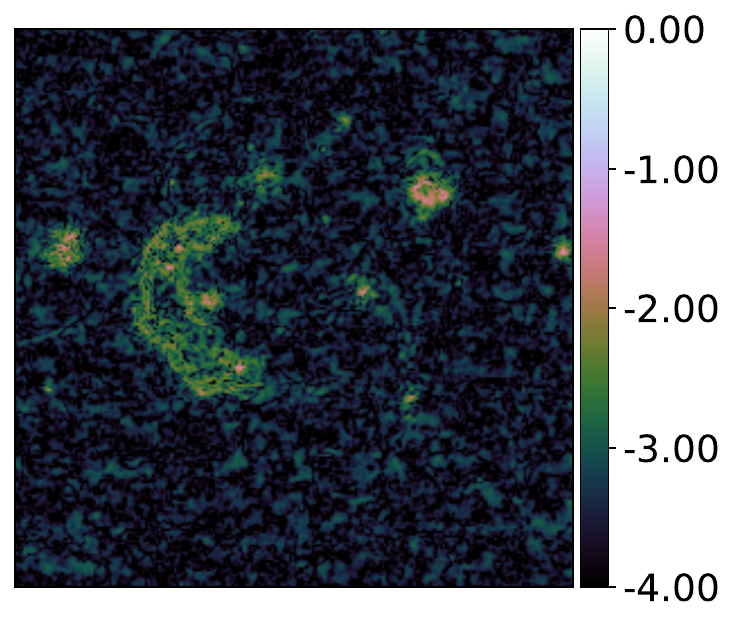}
			\\
		\end{tabular}
		\caption{Fast pixel uncertainty quantification (UQ) with the \textsc{QuantifAI} model on the images M31 and W28. The first two columns correspond to M31, while the last two columns to W28. The first row displays the pairs of the thresholded MAP reconstruction that saturates the HPD region versus the original MAP reconstruction. The following rows compare the predicted error of the thresholded MAP computed with the fast pixel UQ method against the MAP reconstruction error using ground truth images at each wavelet scale. The last row shows the cumulative error when considering all scales.}
		\label{fi:new_pixel_UQ_hard}
	\end{figure*}

	The fast pixel UQ method results for the images M31 and W28 are reported in Figure \ref{fi:new_pixel_UQ_hard}. We use the error between the MAP estimation and the ground truth image, i.e., true error, to validate the predicted uncertainty of the fast UQ method. The true error at different scales can be computed following Equation \ref{eq:coefficient_replacement},
	\begin{equation}
		\hat{\bm{x}}_{\text{GT}, \, j} = \sum_{\substack{l=0,\\ l \neq j}}^{J} \mathbf{\Psi}_{l} \, \bm{a}_{\text{GT}, l} + \mathbf{\Psi}_{j} \hat{\bm{a}}_{\text{MAP}, \, j} \,,
		\label{eq:coefficient_replacement_GT}
	\end{equation}
	where $\bm{a}_{\text{GT}, l}$ are the wavelet decomposition coefficients of the ground truth image at multi-resolution level $l$. We have replaced the ground truth image's wavelet coefficient at a single level with the coefficients from the MAP reconstruction.

	We observe a good agreement between the predicted and ground truth errors at the different multi-resolution levels. There is an overestimation of the errors, which can come from two sources. First, the approximation of the HPD region is conservative, as it has been discussed in \citet{pereyra2017}. Second, the MAP estimation is already missing some of the fine or high-frequency structures in the ground truth images. This fact can be seen in the MAP reconstruction errors in subfigures \ref{fi:M31_reconstruction_r2_c4} and \ref{fi:W28_reconstruction_r2_c4}. The missing high-frequency structure is expected due to the properties of the measurement operator discussed in Section \ref{sc:radio_interf_prob}.

	The structure from the chosen wavelet representation, $\mathbf{\Psi}$, underpinning the UQ method can be observed in the predicted errors. This structure is visible mainly in the higher frequencies of the W28, where point sources are in the image. The wavelet structure should be taken into account when analysing the reconstruction errors.

	This fast pixel UQ method allows us to approximate the reconstruction errors made at different scales for a fraction of the computational cost of the LCI pixel UQ method. The evaluations of the measurement operators are reduced by three orders of magnitude, resulting in an ultra-fast and truly scalable pixel UQ method. Furthermore, a single nonlinear equation solve, e.g. root finding problem, of the new pixel UQ method suffices to predict the errors at all scales, while with LCIs, we are required to repeat the process for each superpixel size.

	\addtolength{\tabcolsep}{-\tabL}
	\addtolength{\tabcolsep}{\tabL}

	\subsection{Computation time}

	The computation wall-clock time for both models, \textsc{QuantifAI} and the wavelet-based, are summarised in Table \ref{tb:computation_time}. We include the results for only the W28 image in Table \ref{tb:computation_time} and \ref{tb:measurement_op_evaluations} as they are representative of the other images. All the computations for both models were done using an Nvidia-A100 40GB GPU using Pytorch \citep{pytorch2019}. We observe a lower computation time for the \textsc{QuantifAI} model with respect to its wavelet-based counterpart. One reason is the lightweight CRR-NN model that quickly evaluates its gradient and potential. Note that the regularisation strength has an impact on the number of iterations and it could be changed to favour a faster convergence. The regularisation strength was chosen to optimize MAP reconstruction quality.

	The results shown in Table \ref{tb:computation_time} highlight the importance of relying on optimisation-based rather than sampling-based reconstructions when focusing on the scalability of the method. There is a difference of approximately four orders of magnitude in the computation time of the MAP and the MMSE which relies on MCMC sampling techniques. Focusing on UQ, the posterior sampling is $60$ times slower than the computation of the LCIs with $8 \times 8$ superpixels and more than $37500$ times slower than the fast pixel UQ proposed in this work. The new fast pixel UQ provides an extremely rapid approach to providing pixel-based UQ, over $630$ times faster than the $8 \times 8$ LCIs.

	The evaluation of the measurement operator is the most time-consuming operation in a real large-scale RI imaging scenario. If we target scalability, we need to monitor the number of measurement operator evaluations. Table \ref{tb:measurement_op_evaluations} summarises the number of measurement operator evaluations required for the UQ techniques. The results are only shown for the \textsc{QuantifAI} model as they are representative of the wavelet-based model. As mentioned before, we note the reduction of evaluations between optimisation and sampling-based reconstructions. We remark on the reduction in the number of evaluations for the UQ tasks, approximately $3$ orders of magnitude between the sampling and the LCIs, and $3$ subsequent orders of magnitude between the LCIs and the fast pixel UQ. These results make the fast pixel UQ $6$ orders of magnitude faster than MCMC sampling. The MAP estimation for the CRR required $1082$ measurement operator evaluations. However, the algorithm's settings were chosen to maximise the reconstruction SNR. By modifying the regularisation parameter of the CRR-based prior, we can reduce the number of evaluations by an order of magnitude. Recent developments in code parallelisation for RI imaging reconstruction algorithms\footnote{\href{https://github.com/astro-informatics/purify}{https://github.com/astro-informatics/purify} and \href{https://github.com/astro-informatics/sopt}{https://github.com/astro-informatics/sopt}} \citep{pratley2019_2,pratley2019_3} could be integrated to push the scalability of the method further.

	\begin{table}
		\begin{center}
			\caption{Computation wall-clock times for the W28 image in seconds for both models being compared.}
			\label{tb:computation_time}
			\begin{tabular}{ccccc}
				\toprule
				\multirow{2}{*}{Models}       &
				\multirow{1}{*}{MAP}          &
				\multicolumn{1}{c}{Posterior} &
				\multicolumn{1}{c}{LCIs}      &
				\multicolumn{1}{c}{Fast}                                                                                                           \\
				                              & \multirow{1}{*}{optim.} & \multicolumn{1}{c}{sampling} & \multicolumn{1}{c}{$8 \times 8$} &
				\multicolumn{1}{c}{pixel UQ}                                                                                                       \\
				\hline \hline
				Wavelet-based                 & $0.94$                  & $36.0 \times 10^{3}$         & $149.7$                          & ---    \\
				\textsc{QuantifAI}            & $0.64$                  & $6.44 \times 10^{3}$         & $108.2$                          & $0.17$ \\
				\bottomrule
			\end{tabular}
		\end{center}
	\end{table}

	\begin{table}
		\begin{center}
			\caption{The number of measurement operator evaluations used by the \textsc{QuantifAI} for the W28 image. We do not distinguish between the measurement operator and its adjoint. Therefore, evaluating the log-likelihood gradient counts as two evaluations of the measurement operator. The fast pixel UQ is three and six orders of magnitude faster than the MCMC sampling and LCIs, respectively.}
			\label{tb:measurement_op_evaluations}
			\begin{tabular}{cccc}
				\toprule
				\multicolumn{1}{c}{MCMC}     &
				\multicolumn{1}{c}{LCIs}     &
				\multicolumn{1}{c}{LCIs}     &
				\multicolumn{1}{c}{Fast}                                                                                    \\
				\multicolumn{1}{c}{sampling} & \multicolumn{1}{c}{$8 \times 8$} & \multicolumn{1}{c}{$16 \times 16$} &
				\multicolumn{1}{c}{pixel UQ}                                                                                \\
				\hline \hline
				$11 \times 10^{6}$           & $81.5 \times 10^{3}$             & $21.2 \times 10^{3}$               & $28$ \\
				\bottomrule
			\end{tabular}
		\end{center}
	\end{table}

\section{Experimental results with ungridded visibilities}
\label{sc:realistic_exp}

In the previous section, we have validated the proposed UQ methods from \textsc{QuantifAI} in a simple setting with gridded visibilities. This choice lets us use the FFT algorithm for the forward model, which allows us to run MCMC algorithms for UQ validation in a sensible amount of time. In this section, we showcase \textsc{QuantifAI} in an experiment using simulated visibility patterns from the MeerKAT radio telescope \citep{meerkat2016}. The main difference is that the visibility patterns are ungridded following a realistic distribution, which obliges us to rely on the Non-Uniform FFT (NUFFT) for the forward model. We remark that in this experiment, we are not addressing many of the challenges of dealing with real data, which are beyond the scope of this article.

\subsection{Dataset and experiment settings}

\begin{figure*}
	\centering
	\captionsetup[subfigure]{skip=-2pt}
	\begin{subfigure}[b]{0.24\linewidth}
		\centering
		\includegraphics[height=3.3cm]{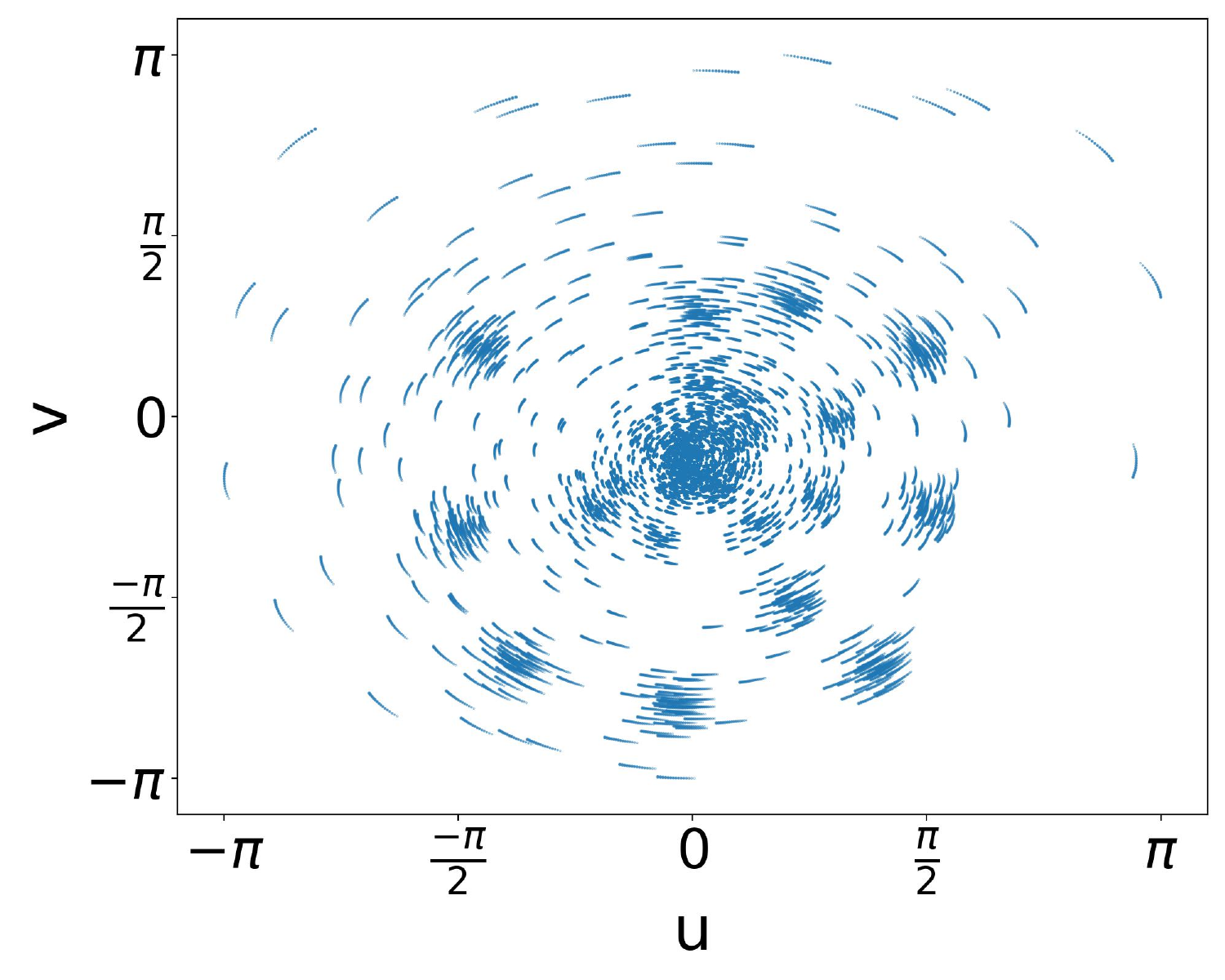}
		\caption{1 hour of synthesis time}
		\label{fi:meerkat_vis_1h}
	\end{subfigure}
	\hfill
	\begin{subfigure}[b]{0.24\linewidth}
		\centering
		\includegraphics[height=3.3cm]{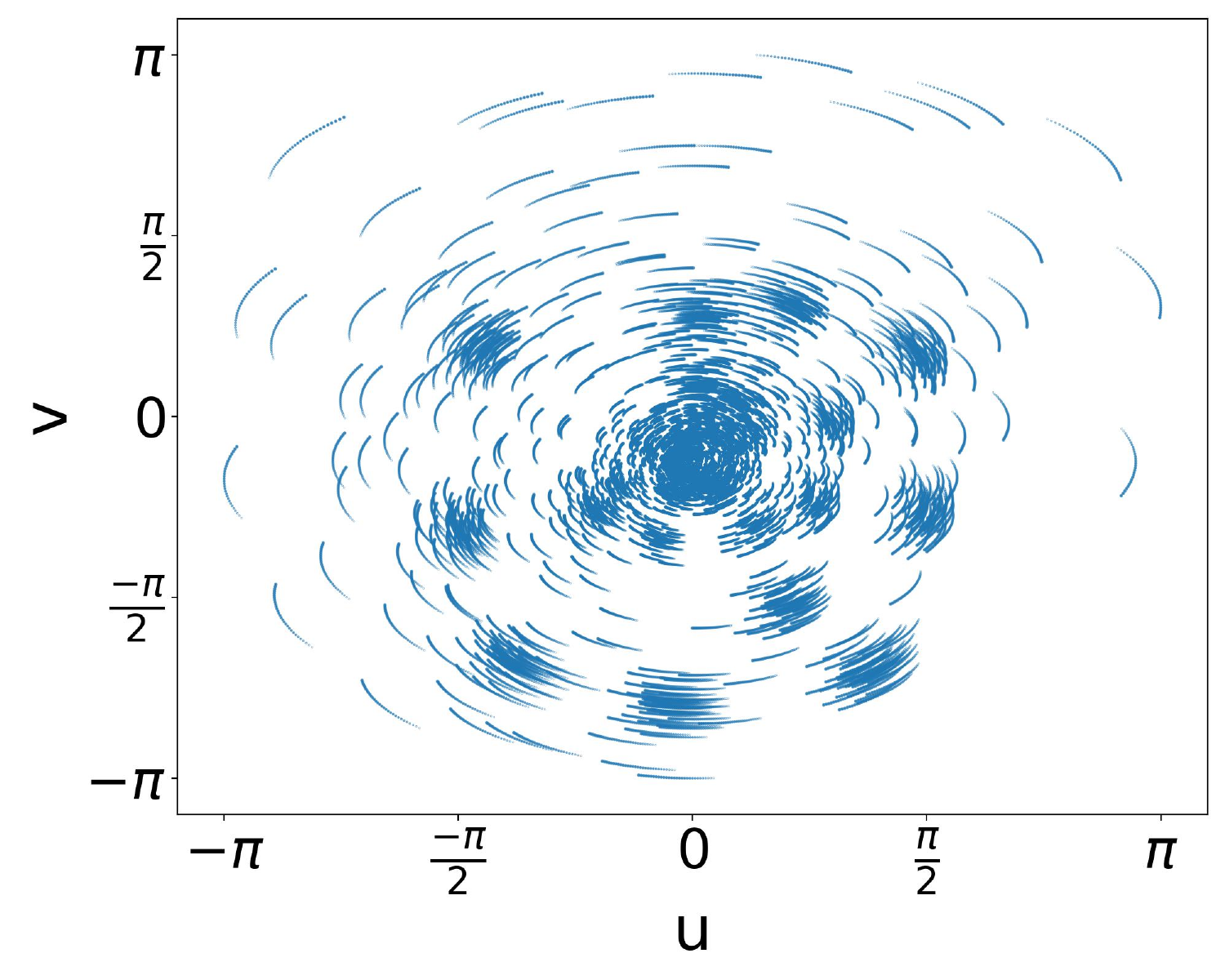}
		\caption{2 hours of synthesis time}
		\label{fi:meerkat_vis_2h}
	\end{subfigure}
	\hfill
	\begin{subfigure}[b]{0.24\linewidth}
		\centering
		\includegraphics[height=3.3cm]{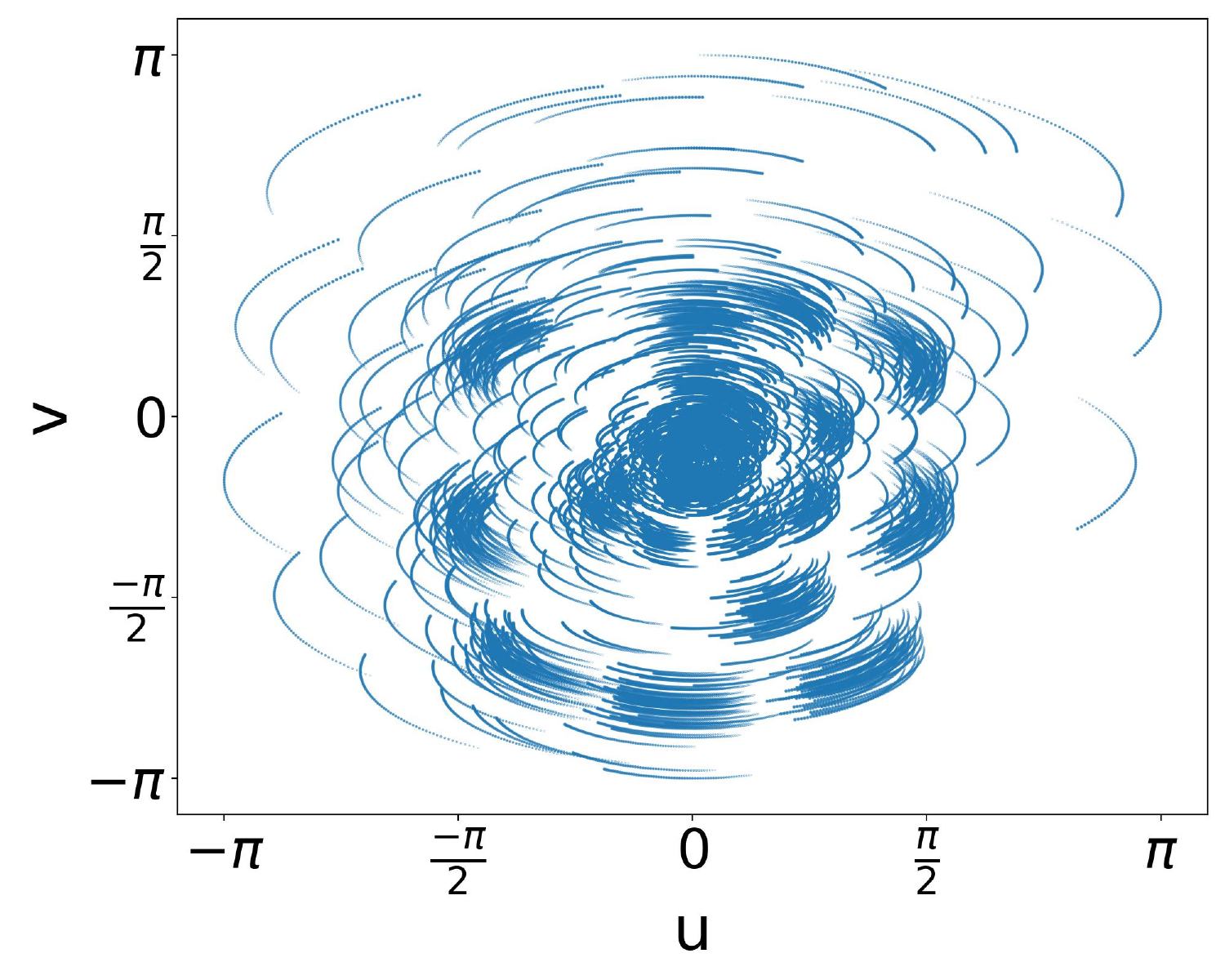}
		\caption{4 hours of synthesis time}
		\label{fi:meerkat_vis_4h}
	\end{subfigure}
	\hfill
	\begin{subfigure}[b]{0.24\linewidth}
		\centering
		\includegraphics[height=3.3cm]{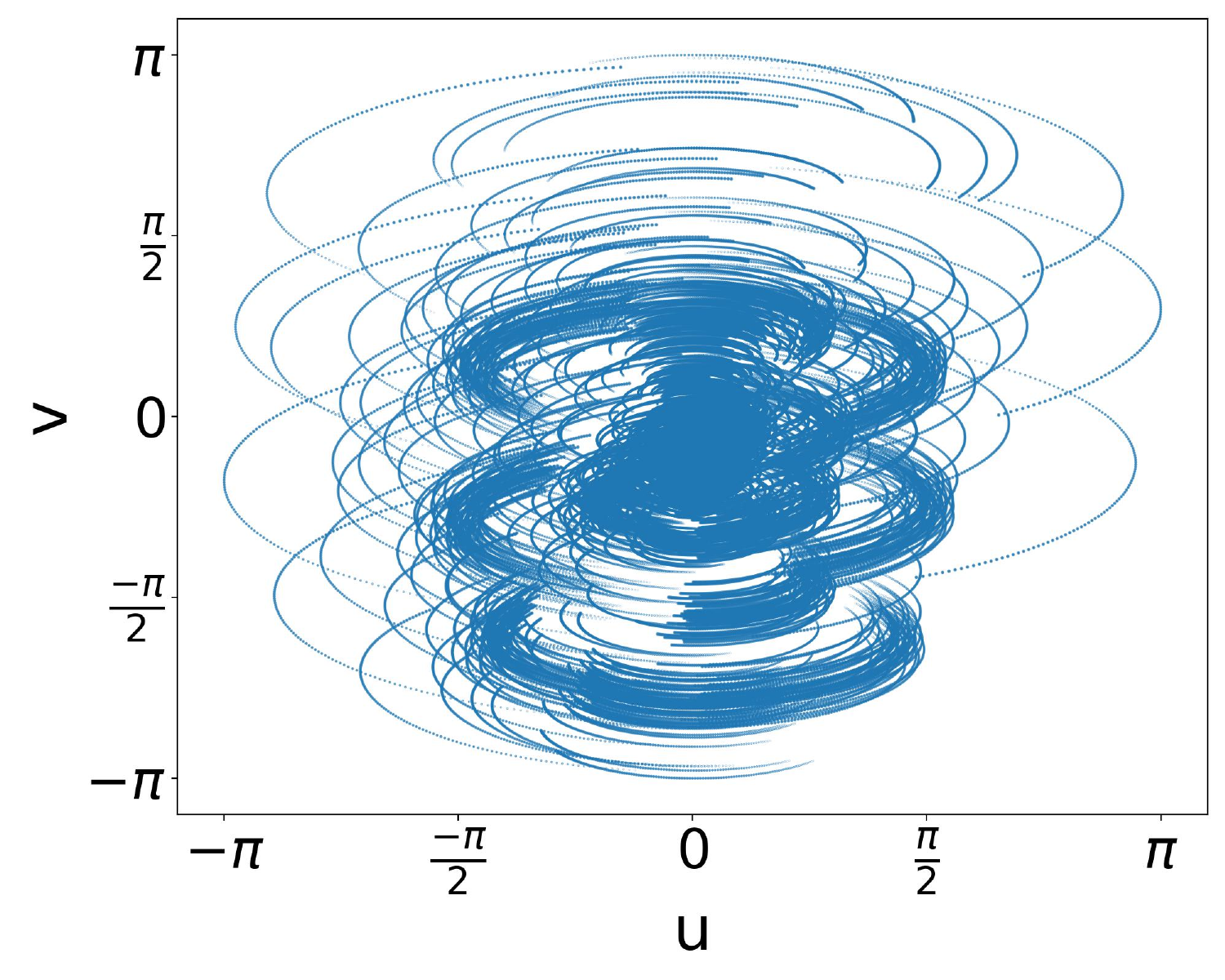}
		\caption{8 hours of synthesis time}
		\label{fi:meerkat_vis_8h}
	\end{subfigure}\\
	\caption{The four sets of simulated ungridded visibilities for the MeerKAT radio telescope with synthesis times of $1$, $2$, $4$ and $8$ hours.}
	\label{fi:meerkat_vis}
\end{figure*}

We have simulated four single-frequency ungridded visibility patterns of differing synthesis times for MeerKAT. The start frequency is set to $1400$MHz with a channel width of $10$MHz. The pointing position has been randomly selected and set to $(13$h$18$m$54.86$s, $-15$d$36$m$04.25$s$)$ in the J$2000$ reference, and it was maintained for the four generated datasets. We have used a publicly available code\footnote{\href{https://github.com/ratt-ru/simms}{https://github.com/ratt-ru/simms}} that is based on the CASA simulation software \citep{CASA_software}. The synthesis times used are $1$, $2$, $4$, and $8$ hours, with a constant integration time of $240$s. Each dataset has a field of view of approximately $1$ deg$^2$. The number of visibilities of each dataset is $3\times10^{4}$, $6\times10^{4}$, $1.2\times10^{5}$, and $2.4\times10^{5}$, correspondingly. Figure \ref{fi:meerkat_vis} presents the simulated visibility patterns. We reuse the images described in Section \ref{sc:simulations} as the ground truth, keeping their original dimensions, which are unrealistically small to be representative of a MeerKAT observation grid size.

To cope with the ungridded visibilities, we have to change the forward operator $\mathbf{\Phi}$ from Equation \ref{eq:linear_inv_prob} that before was based on the FFT. We rely on the pytorch-based NUFFT implementation from \citet{muckley2020}\footnote{\href{https://github.com/mmuckley/torchkbnufft}{https://github.com/mmuckley/torchkbnufft}} based on Kaiser-Bessel gridding \citep{fessler2003}. We have used the same images, Gaussian noise model presented in Section \ref{sc:simulations}, and trained CRR-NN in \textsc{QuantifAI}. We have reused the previously introduced hyperparameter values of \textsc{QuantifAI} except for $\lambda$, which we have tuned to maximise the reconstruction's SNR for the four datasets to $10^{4}$, $1.4 \times 10^{4}$, $1.9 \times 10^{4}$, and $2.25 \times 10^{4}$, respectively.

\subsection{Results}

\addtolength{\tabcolsep}{-\tabL}
\begin{figure*}
	\centering
	\begin{tabular}{cccc}
		{\Large 1h} 	& {\Large 2h}	& {\Large 4h} & {\Large 8h} \\
		\includegraphics[width=0.18\linewidth]{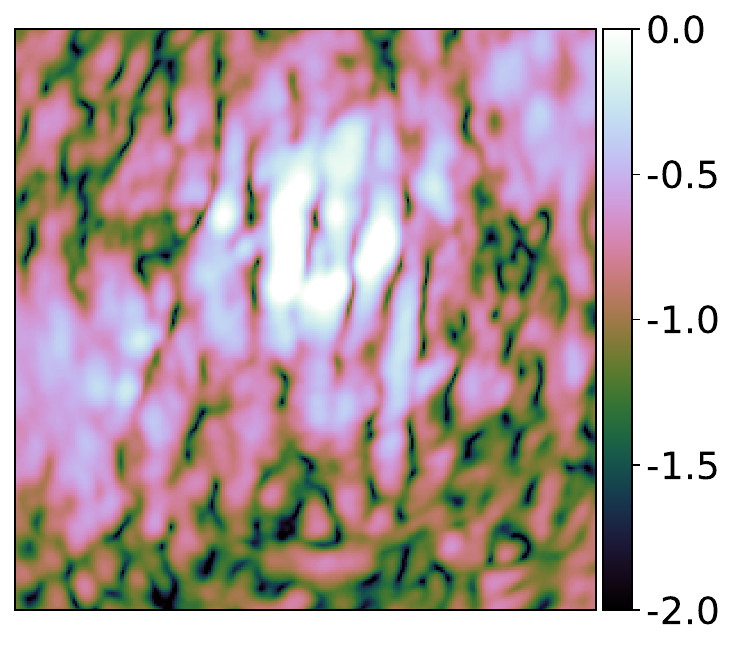} \put(-110,22){\rotatebox{90}{\large Dirty rec.}}  &
		\includegraphics[width=0.18\linewidth]{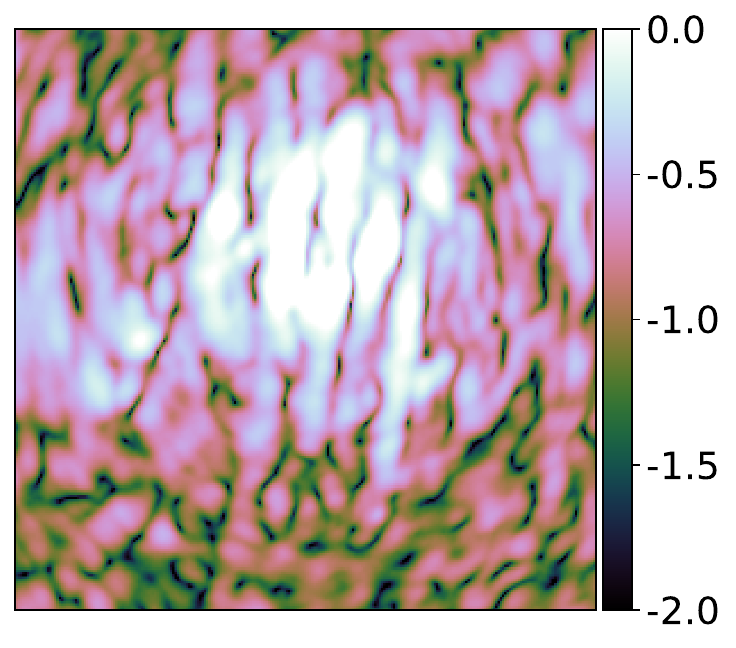}                                                                           &
		\includegraphics[width=0.18\linewidth]{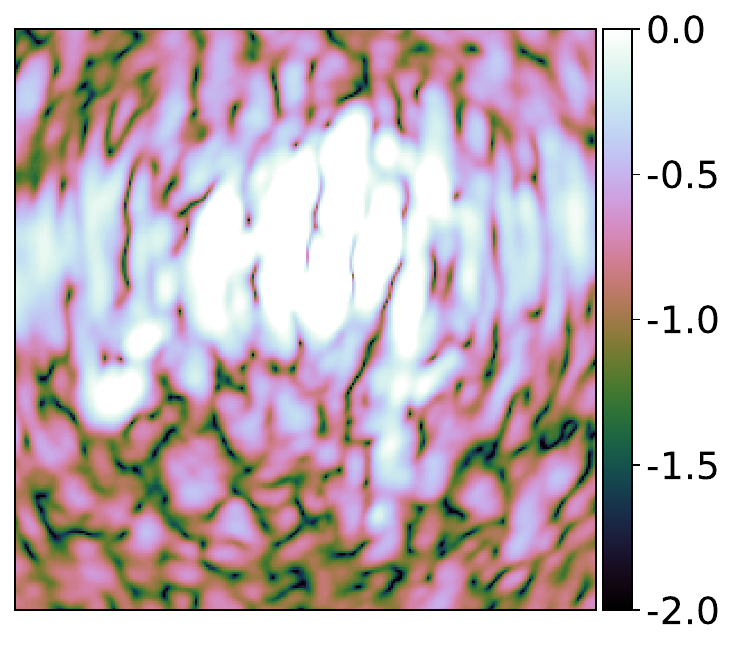}                                                              &
		\includegraphics[width=0.18\linewidth]{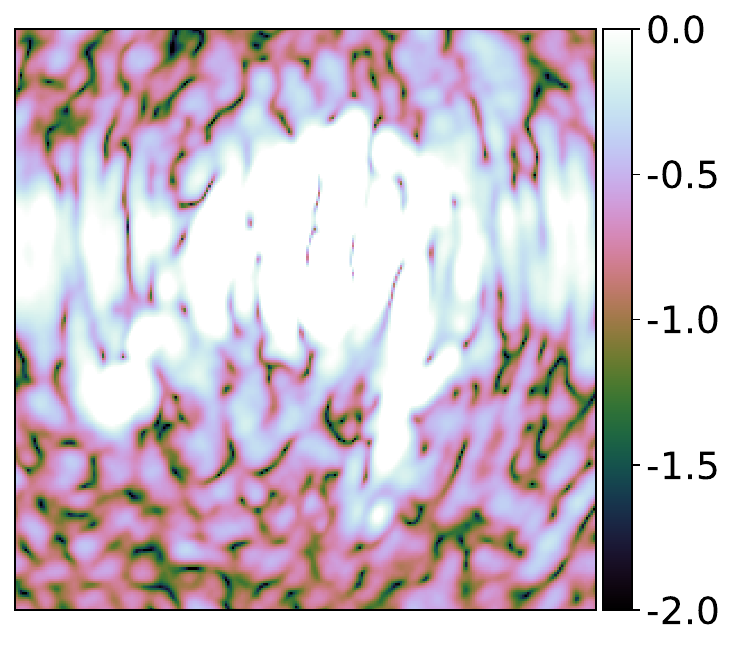}
		\\
		\includegraphics[width=0.18\linewidth]{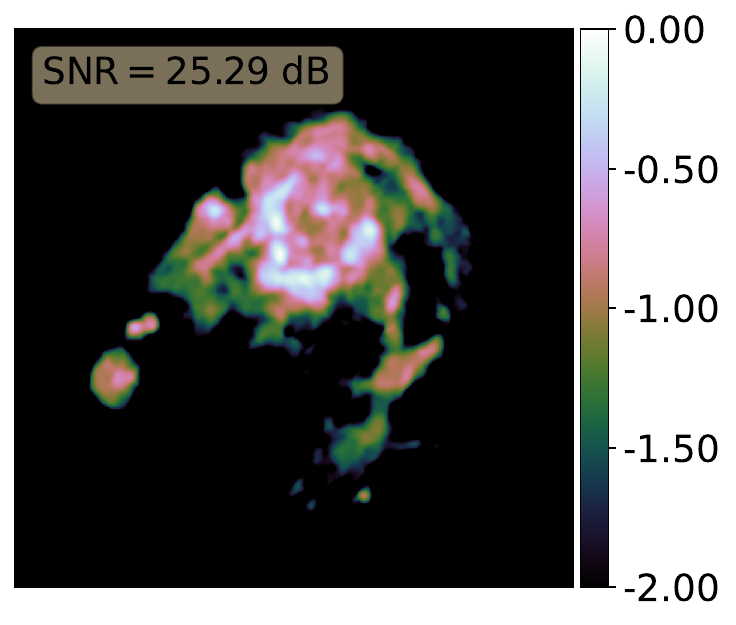} \put(-110,22){\rotatebox{90}{\large MAP rec.}} &
		\includegraphics[width=0.18\linewidth]{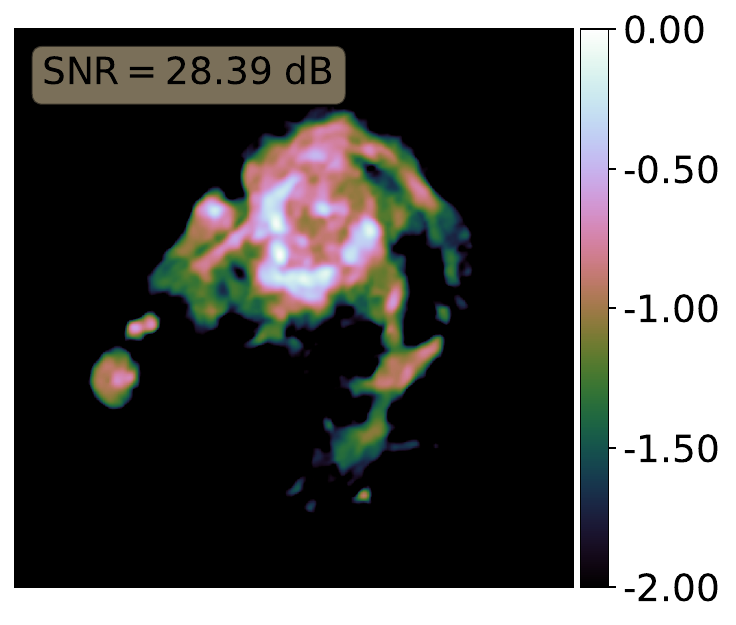}                                                                           &
		\includegraphics[width=0.18\linewidth]{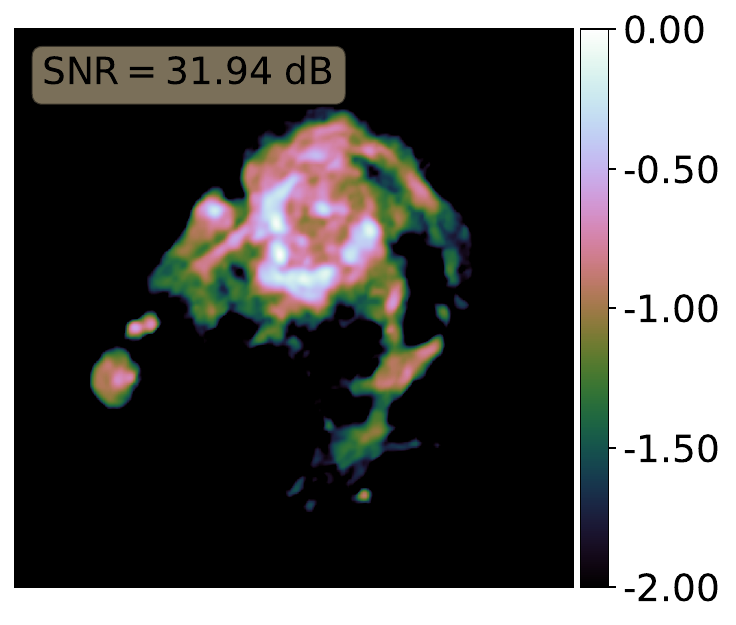}                                                              &
		\includegraphics[width=0.18\linewidth]{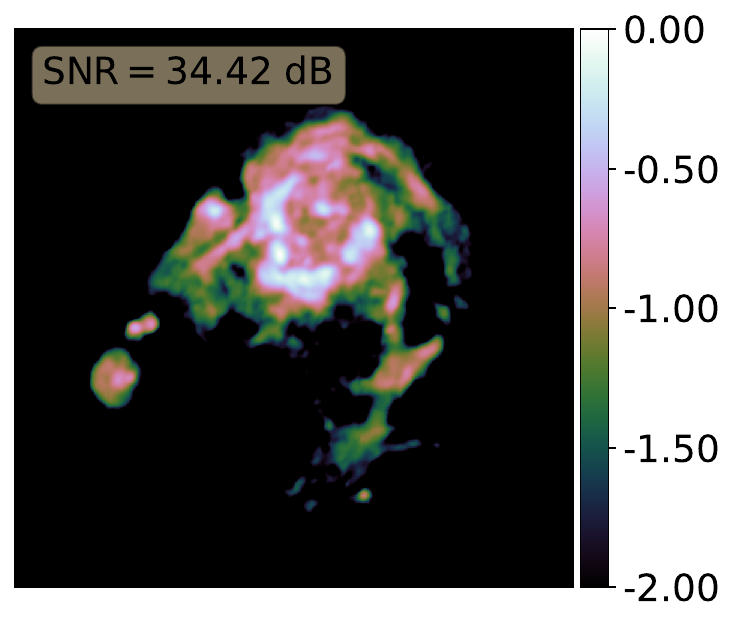}
		\\
		\includegraphics[width=0.18\linewidth]{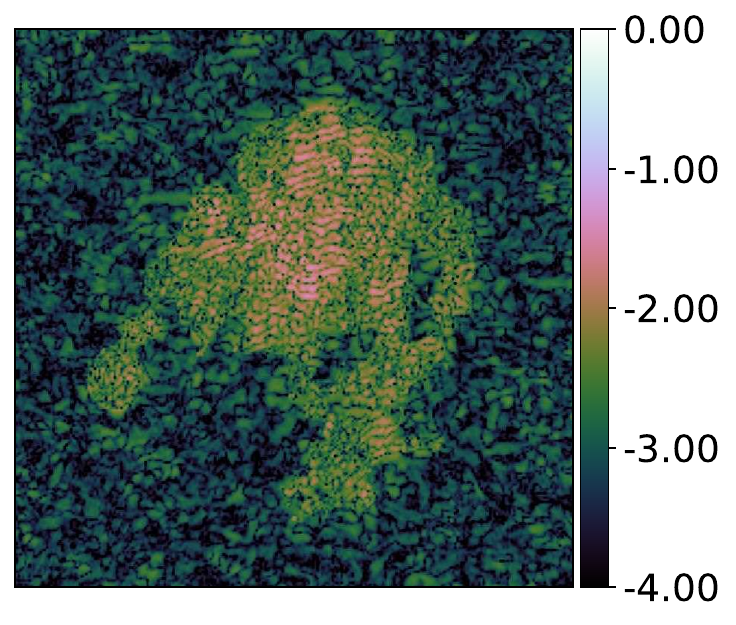} \put(-110,20){\rotatebox{90}{\large Oracle error}} &
		\includegraphics[width=0.18\linewidth]{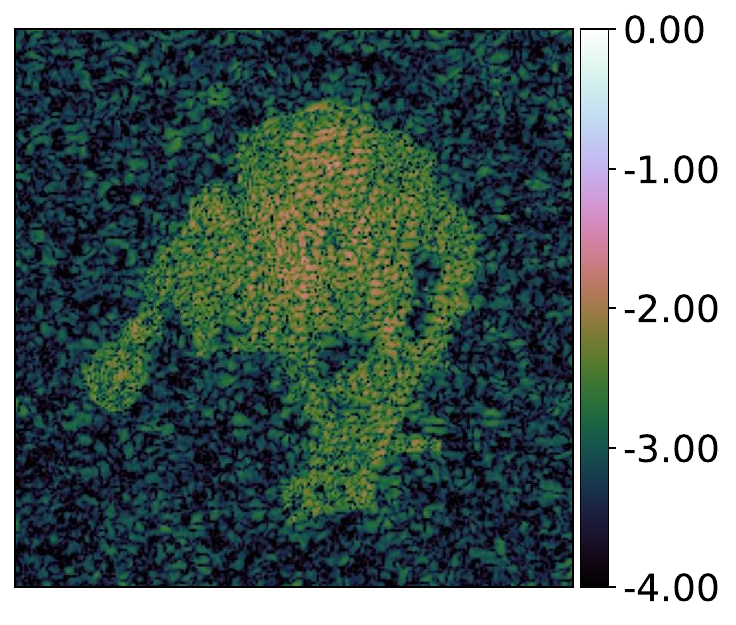}                                                                           &
		\includegraphics[width=0.18\linewidth]{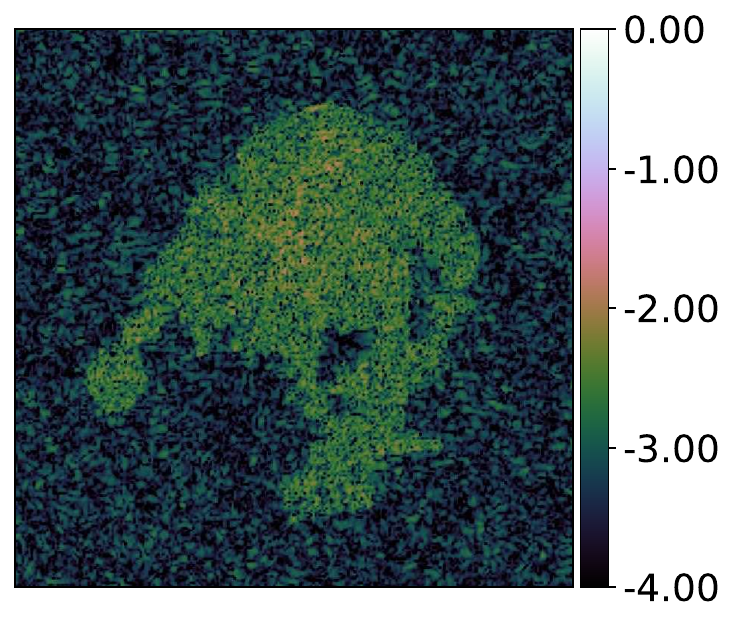}                                                              &
		\includegraphics[width=0.18\linewidth]{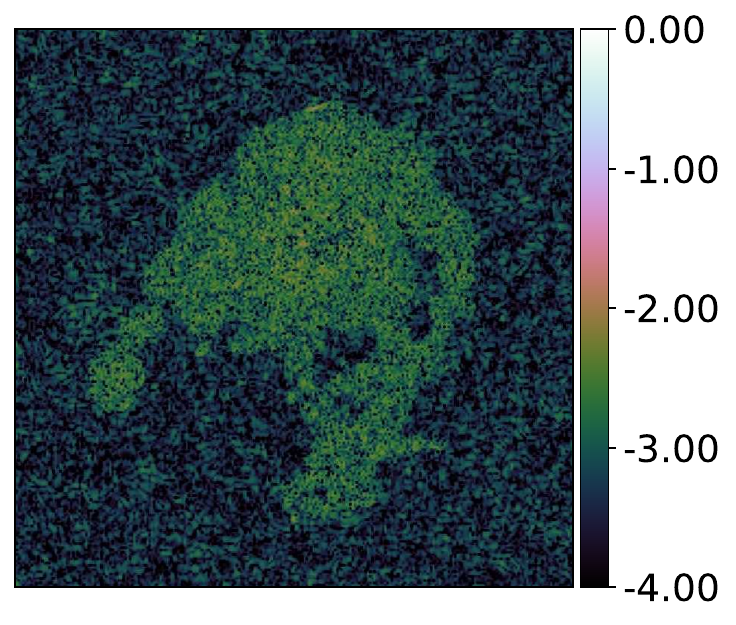}
		\\
		\multicolumn{4}{c}{\large Predicted error}  \\
		\includegraphics[width=0.18\linewidth]{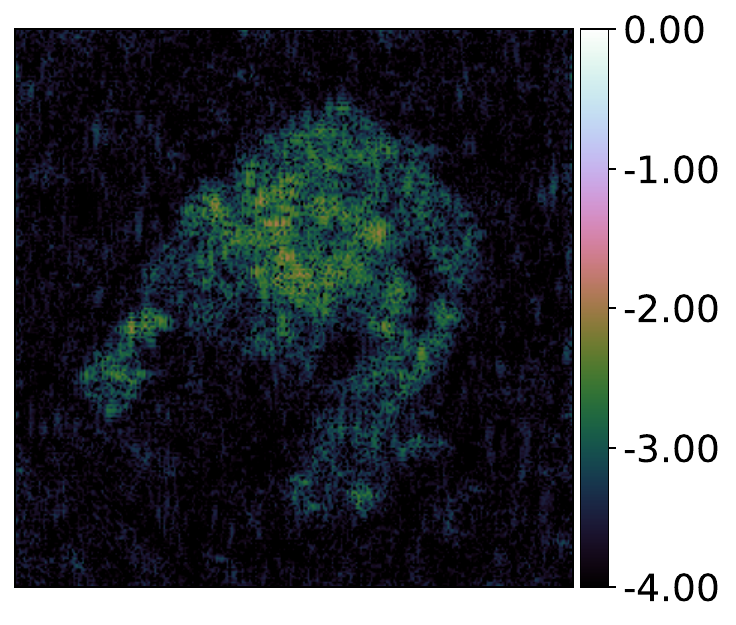} \put(-110,30){\rotatebox{90}{\large Level $4$}} &
		\includegraphics[width=0.18\linewidth]{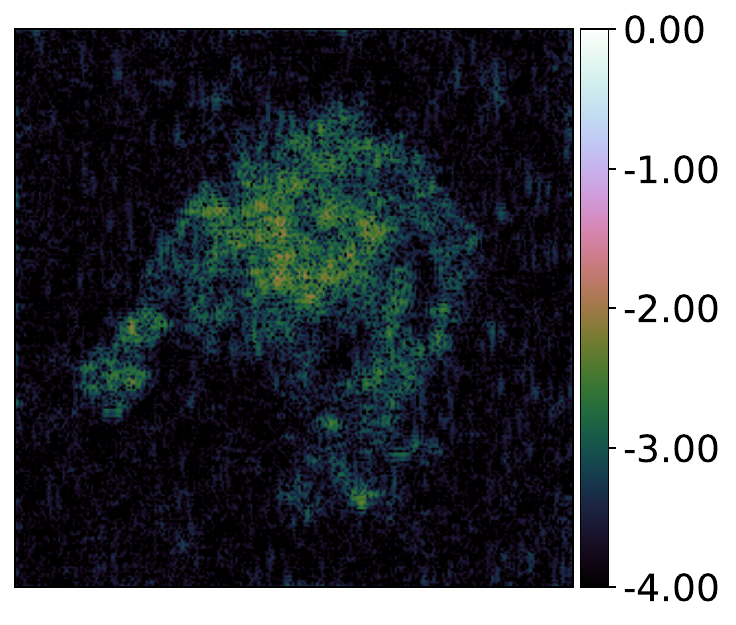}    &
		\includegraphics[width=0.18\linewidth]{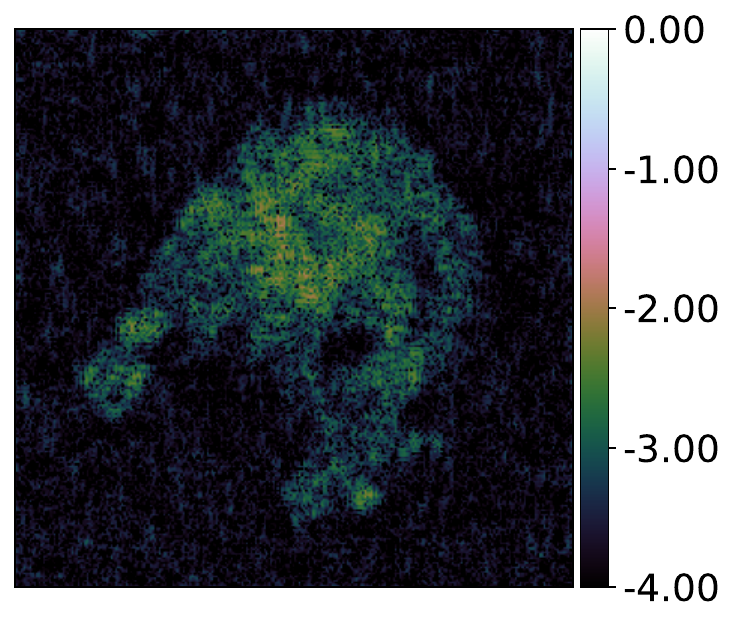}                                                 &
		\includegraphics[width=0.18\linewidth]{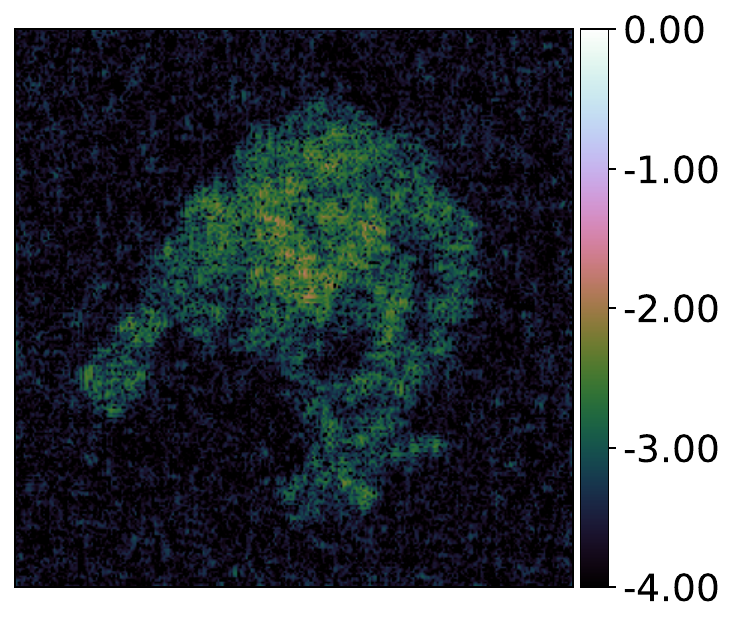}
		\\
		\includegraphics[width=0.18\linewidth]{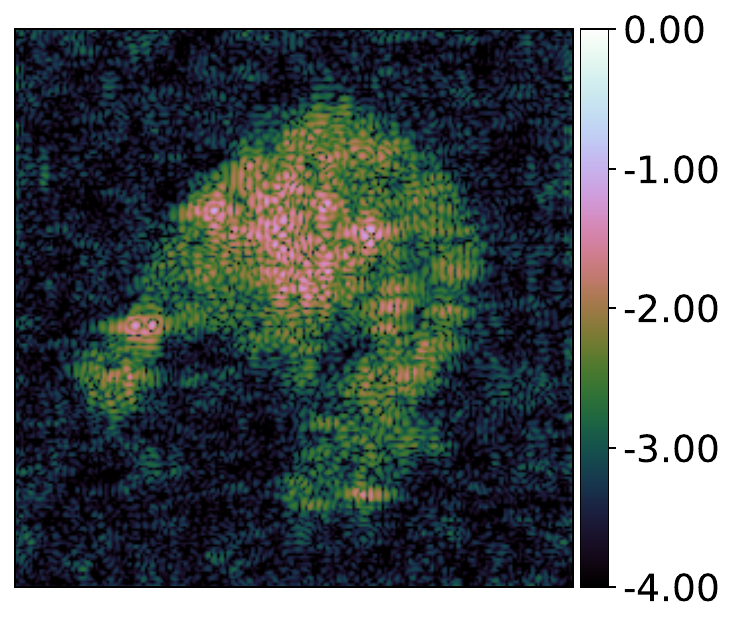} \put(-110,30){\rotatebox{90}{\large Level $3$}} &
		\includegraphics[width=0.18\linewidth]{fastUQ_ungridded/M31-CRR1h-newPixelUQ-MAP_thresholded_error_level_3.pdf}                                                          &
		\includegraphics[width=0.18\linewidth]{fastUQ_ungridded/M31-CRR1h-newPixelUQ-MAP_thresholded_error_level_3.pdf}                                                 &
		\includegraphics[width=0.18\linewidth]{fastUQ_ungridded/M31-CRR1h-newPixelUQ-MAP_thresholded_error_level_3.pdf}
		\\
		\includegraphics[width=0.18\linewidth]{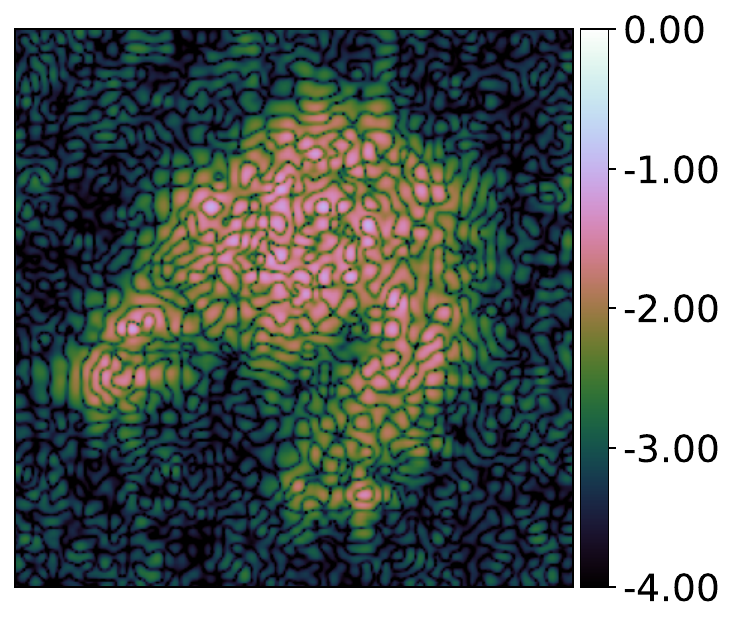} \put(-110,30){\rotatebox{90}{\large Level $2$}} &
		\includegraphics[width=0.18\linewidth]{fastUQ_ungridded/M31-CRR1h-newPixelUQ-MAP_thresholded_error_level_2.pdf}                                                          &
		\includegraphics[width=0.18\linewidth]{fastUQ_ungridded/M31-CRR1h-newPixelUQ-MAP_thresholded_error_level_2.pdf}                                                 &
		\includegraphics[width=0.18\linewidth]{fastUQ_ungridded/M31-CRR1h-newPixelUQ-MAP_thresholded_error_level_2.pdf}
		\\
		\includegraphics[width=0.18\linewidth]{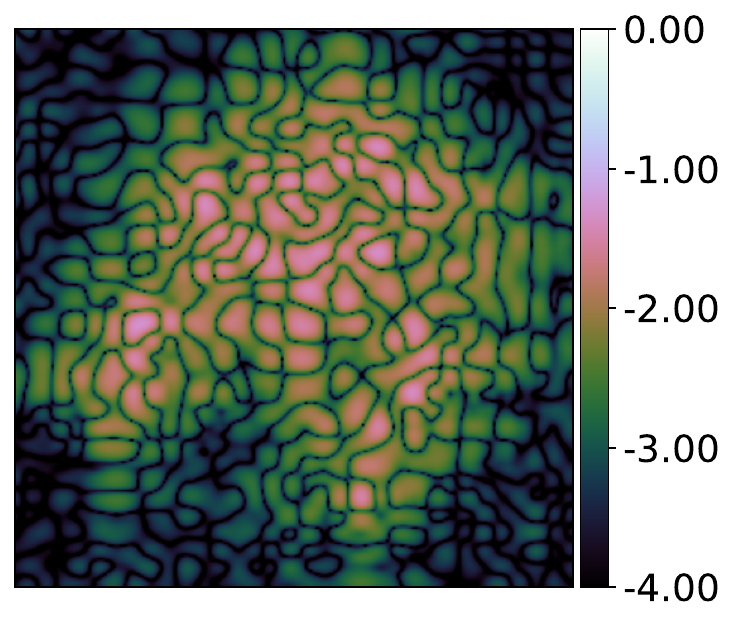} \put(-110,30){\rotatebox{90}{\large Level $1$}} &
		\includegraphics[width=0.18\linewidth]{fastUQ_ungridded/M31-CRR1h-newPixelUQ-MAP_thresholded_error_level_1.pdf}                                                          &
		\includegraphics[width=0.18\linewidth]{fastUQ_ungridded/M31-CRR1h-newPixelUQ-MAP_thresholded_error_level_1.pdf}                                                 &
		\includegraphics[width=0.18\linewidth]{fastUQ_ungridded/M31-CRR1h-newPixelUQ-MAP_thresholded_error_level_1.pdf}
		\\
	\end{tabular}
	\caption{Reconstructions and fast pixel uncertainty quantification (UQ) with the \textsc{QuantifAI} model for the M31 image with the four sets of simulated MeerKAT ungridded visibilities. Each column corresponds to the four datasets with synthesis times of $1$, $2$, $4$ and $8$ hours for a field of view of approximately $1$ deg$^2$. The first row represents the dirty reconstruction. The MAP reconstruction is presented in the second row, while the oracle error, which we do not have access to with real data, is shown in the third row. The different decomposition levels of pixel UQ are shown in the last four rows.}
	\label{fi:ungridded_results_M31}
\end{figure*}
\addtolength{\tabcolsep}{\tabL}

\begin{table*}
	\begin{center}
		\caption{Main results of \textsc{QuantifAI} for the M31 image with the realistic ungridded MeerKAT visibility patterns with differing synthesis times. As the number of visibilities grows with the synthesis timer, so does the reconstruction SNR. The number of visibilities increases proportionally to the synthesis times.}
		\label{tb:realistic_results_M31}
		\begin{tabular}{cccccc}
			\toprule
			&\multirow{2}{*}{Metrics} & \multicolumn{4}{c}{Datasets} \\
			\cmidrule{3-6}
			& & 1h	& 2h	& 4h	& 8h \\
			\hline \hline
			& Number of visibilities			& $3 \times 10^{4}$  	& $6 \times 10^{4}$ 	& $1.2 \times 10^{5}$   	& $2.4 \times 10^{5}$  	\\
			& MAP reconstrucion SNR [dB]		& $25.29$  				& $28.39$ 				& $31.94$   				& $34.42$    			\\
			\cmidrule{3-6}
			\multirow{2}{*}{Reconstruction} 	& Measurement op. evaluations		& $3288$  	& $2916$  	& $3006$   		& $3114$ 	\\
												& Wall-clock time [s]				& $17.01$  	& $28.19$ 	& $53.24$   	& $105.25$ 	\\
			\cmidrule{3-6}
			\multirow{2}{*}{UQ} 				& Measurement op. evaluations		& $26$  	& $28$  	& $30$   		& $30$ 		\\
												& Wall-clock time [s]				& $0.28$  	& $0.44$ 	& $0.73$   		& $1.26$ 	\\
			\bottomrule
		\end{tabular}
	\end{center}
\end{table*}

The reconstructions and the fast pixel UQ maps for the image M31 are shown in Figure \ref{fi:ungridded_results_M31}. Each column corresponds to each of the four datasets. The results for the other images are postponed to Appendix \ref{ap:more_realistic_results}. Table \ref{tb:realistic_results_M31} presents quantitative results regarding reconstruction SNR, measurement operator evaluations and wall-clock computing time. The reconstructions present good quality in terms of SNR, which increases with longer synthesis times, which was expected as the Fourier coverage increases, as seen in Figure \ref{fi:meerkat_vis}. The wall-clock reconstruction time increases with the longer synthesis times as the time for each evaluation of the measurement operator becomes longer due to the larger number of visibilities. 

Even in this more challenging setting, we find a good correlation between the predicted and the oracle error. The pixel UQ maps still show some patterns characteristic of the multiscale wavelet representation used and should, therefore, be considered in the analysis of the pixel UQ maps. The fast pixel UQ represents a fraction of the time and number of measurement operator evaluations required for the MAP reconstruction.

\section{Discussion}
\label{sc:discussion}

In this work, we have worked with synthetic datasets, starting with gridded visibilities and ending with realistic ungridded visibility patterns. The handling of real or realistic data is beyond the scope of this work, which focuses on the methodology presented and the validation of the UQ techniques. Some problems faced while handling large-scale observation include the calibration of direction-independent and direction-dependent effects. Further studies of the performance of \textsc{QuantifAI} to more realistic images with differing large dynamic ranges and bright point sources are left for the future. The extension of \textsc{QuantifAI} to incorporate a frequency axis in the reconstruction to cope with multi-frequency observations is also left for further study. Even if \textsc{QuantifAI} could reconstruct images with UQ maps with $\sim 10^{5}$ visibilities in less than two minutes, there is ongoing work to exploit existing C++ parallelisation capabilities to scale the method further. A detailed performance and computing time benchmark is expected for the future implementation.

The current approach to set the regularisation strength of the CRR-NN, $\lambda$, with a grid search is not compatible with real data as we require access to the ground truth. There are several ways to circumvent this problem in the large-scale setting. One way forward is to consider a subset of the observations to alleviate the computational burden and rely on the empirical Bayesian approach from \citep{vidal2020,debortoli2020} to estimate the regularisation parameter directly from the observed data. Another way forward is to follow a heuristic approach similar to \citet{terris2022}. The study of the best strategy for \textsc{QuantifAI} is left for further study. 

We have not included a positivity constraint in the \textsc{QuantifAI} model, as is the case for \citet{cai2018_2}, which we use for comparison. However, there is no impediment to adding a positivity constraint to \textsc{QuantifAI}; it amounts to changing the indicator function from Equation \ref{eq:optim_target} to an indicator function of the real positive orthant. The proximal operator associated with this indicator function has a closed form: the projection to the positive orthant. Consequently, we can compute the corresponding MAP solution and evaluate the potential using the modified indicator function.

The fast pixel UQ proposed in this article improved the tightness of the UQ bounds with respect to the LCIs by considering the entire image when saturating the HPD region. Nevertheless, these pixel-level UQ maps are intended to be shared as visual aids accompanying the reconstructions. The pixel UQ maps can drive the astronomer's attention to a specific region in the image where a consequent hypothesis test can be carried out using the techniques described in this work. The choice of wavelet basis impacts the fast pixel UQ maps produced, so it should be considered when analysing the maps. A way to improve the fast pixel UQ maps would be to replace the orthogonal wavelet basis used with a more performant dictionary as in SARA \citep{carrillo2012}, which has shown to be well adapted for RI images.

\section{Conclusions}
\label{sc:conclusions}

In this work, we propose a new method coined \textsc{QuantifAI} that addresses uncertainty quantification in radio-interferometric (RI) imaging with data-driven (learned) priors in very high-dimensional settings. We have focused on three fundamental points in the RI imaging pipeline: scalability, estimation performance, and uncertainty quantification (UQ).

Our model builds upon a principled Bayesian framework for the UQ analysis, which is known to be computationally expensive when exploiting MCMC sampling methods. However, in this work, we leverage convex optimisation techniques to estimate the maximum-a-posteriori (MAP), the point estimate of the posterior distribution we use as reconstruction. We restrict our model to a log-concave posterior distribution to remain highly scalable and have Bayesian UQ techniques. This restriction is equivalent to having convex potentials for our likelihood and prior. In this scenario, we can exploit an approximation of the high posterior density (HPD) region, which only requires the MAP estimation \citep{pereyra2017} and bypasses expensive sampling techniques.

We want to include data-driven priors that can encode complex information learned implicitly from training data making them more expressive. Consequently, the learned priors allow us to improve performance with respect to previous models based on handcrafted priors \citep{cai2018_2}, e.g., wavelet-based sparsity-promoting priors. To support fast UQ techniques, our models must be convex, hence we adopt the recently introduced learnable convex-ridge regulariser neural network \citep[CRR-NN,][]{goujon2023}. The CRR-NN-based prior is performant, reliable and has shown to be robust to data distribution shifts. The \textsc{QuantifAI} model uses an analytic physically motivated model for the likelihood and the learned CRR-NN-based prior. In this work, we are focusing on the methodology, which is why we have only considered small problems, i.e., images of $256 \times 256$. Nevertheless, \textsc{QuantifAI} can be integrated into the distributed frameworks \citep{pratley2019_2,pratley2019_3}, which is the focus of ongoing work.

Numerical experiments are conducted with four images representative of RI imaging. We compare the \textsc{QuantifAI} model with the model containing a wavelet-based prior of \citet{cai2018_2}. Our results show a considerable improvement in the reconstruction performance for \textsc{QuantifAI}. We validate our results against posterior samples from MCMC sampling algorithms and compute the posterior standard deviation. We found that \textsc{QuantifAI} produced more meaningful posterior standard deviations in comparison to the wavelet-based model. We also included numerical experiments with simulated MeerKAT ungridded visibilities, where we present \textsc{QuantifAI}'s performance and computing times going up to $\sim 10^{5}$ visibilities.

We explore several MAP-based UQ techniques that rely on the approximate HPD region. We carry out hypothesis tests of image structure to asses if some structures observed in the reconstructions are physical. We then computed local credible intervals (LCIs c.f. Bayesian error bars) to measure the pixel-wise uncertainty. These two approaches were proposed by \citet{cai2018_2}, and in this work, we validated them with MCMC posterior sampling results. Even if LCIs represent an already scalable alternative to sampling-based methods to provide pixel-wise UQ, they remain expensive for SKA-size data. Therefore, we proposed a novel pixel-wise UQ technique to approximate pixel errors at different scales that is three orders of magnitude faster than the LCIs. The new approach is based on thresholding the coefficients of a wavelet representation of the reconstruction until the HPD region saturates and is six orders of magnitude faster than sampling-based techniques.

\textsc{QuantifAI} proposed an approach with the potential to be highly scalable and performant to address UQ in RI imaging. In this work, we have compared \textsc{QuantifAI} to a wavelet-based model using numerical experiments and a variety of metrics. However, as both models rely on the Bayesian framework, we could make a Bayesian model comparison, a principled approach to model selection and determine which model the data favours. Recent developments in \citet{mcewen2023} extend the model comparison to the learnt setting, with data-driven priors. The focus of ongoing work is to implement the proposed methodology in existing RI imaging frameworks \textsc{purify}\footnote{\href{https://github.com/astro-informatics/purify}{https://github.com/astro-informatics/purify}} \citep{carrillo2014,pratley2018,pratley2019} and \textsc{sopt}\footnote{\href{https://github.com/astro-informatics/sopt}{https://github.com/astro-informatics/sopt}} \citep{carrillo2012,onose2016} to exploit massively parallelised computing environment \citep{pratley2019_2,pratley2019_3} and to realise the potential of scalability. In the near future, we plan to benchmark the speed and scalability of \textsc{QuantifAI} in a highly realistic setting.

A new perspective is to relax the convexity constraint of the prior by exploiting the fact that the posterior potential needs to be convex (rather than the prior) and that the RI imaging likelihood is already strongly convex. The relaxation of the CRR regulariser has been studied in a very recent work \citep{goujon2023_b}, where a weakly-convex-ridge-regulariser neural network (WCRR-NN) has been proposed. If the WCRR-NN is adopted, it could further enhance the expressiveness of the regulariser and the reconstruction performance of \textsc{QuantifAI} in the RI imaging problem.

\section*{Acknowledgements}

This work is supported by the UK Research and Innovation (UKRI) and Engineering and Physical Sciences Research Council (EPSRC) grant numbers EP/W007673/1 (LEXCI) and EP/T007346/1 (BOLT). MM is supported by the UCL Centre for Doctoral Training in Data Intensive Science (STFC Training grant ST/P006736/1).

We acknowledge the Python packages used in this work: IPython \citep{ipython}, Jupyter \citep{jupyter}, Matplotlib \citep{matplotlib}, Pytorch \citep{pytorch2019}, Numpy \citep{numpy}, Astropy \citep{astropy}, Scikit Image \citep{scikit-image}.

\section*{Data Availability}

We provide the \textsc{QuantifAI} PyTorch-based Python package in a publicly available repository\footnote{\href{https://github.com/astro-informatics/QuantifAI}{https://github.com/astro-informatics/QuantifAI}}. We are in favour of reproducible research, which is why the images, visibilities, scripts, trained CRR-NN model and code to reproduce this article's experiments can be found in the aforementioned repository.

\bibliographystyle{rasti}
\bibliography{convex_uq}

\appendix

\section{MAP calculation for the wavelet-based prior model}
\label{ap:wavelet_map}

The MAP estimation for the wavelet-based model can recasted as the following optimisation problem
\begin{equation}
	\hat{\bm{x}}_{\text{MAP}} = \argmin_{\bm{x} \in \mathbb{R}^{N}} \frac{1}{2 \sigma^{2}} \left\| \bm{y} - \Phi \bm{x} \right\|_{2}^{2} + \lambda_{\text{wav}} \big\| \Psi^{\dagger} \bm{x} \big\|_{1} + \iota_{\mathbb{R}^{N}}(\bm{x}) \,,
	\label{eq:optim_target_wavelet}
\end{equation}
where $\lambda_{\text{wav}}$ corresponds to the regularisation strength of the wavelet-based prior. The FISTA algorithm to estimate the MAP is presented in Algorithm \ref{al:optim_alg_wavelet} where we use the soft thresholding operator, $\text{soft}(\cdot)$, defined in Equation \ref{eq:soft_thresh_op}. The Lipschitz constant used to define the step size can be set as $L_{\text{wav}} = \| \Phi^{\dagger} \Phi \| / \sigma^{2}$.

\begin{algorithm}
	\caption{FISTA \citep{beck2009} tackling (\ref{eq:optim_target_wavelet})}
	\label{al:optim_alg_wavelet}
	\textbf{Input:}  $\Psi$, $\Phi$, $\sigma$, $\lambda_{\text{wav}}$, $\xi$, $a_{(1)}=1$, $\bm{z}_{(1)} = \bm{x}_{(0)} = \text{Re}(\Phi^{\dagger}\bm{y})$,  $\tau_{\text{wav}} = 0.98/L_{\text{wav}}$.\\
	\textbf{Output:} $\hat{\bm{x}}_{\text{MAP}}$ \vspace{0.05in} \\
	\For{$n=1, \ldots, N_{\text{max}}$}{
		$\tilde{\bm{z}}_{(n)} = \bm{z}_{(n)} - \frac{\tau_{\text{wav}}}{\sigma^{2}} \text{Re}(\Phi^{\dagger} (\Phi \bm{z}_{(n)} - \bm{y}))$ \\
		$\bm{x}_{(n)} = \tilde{\bm{z}}_{(n)} + \Psi \left( \text{soft}_{\lambda_{\text{wav}} \tau} (\Psi^{\dagger} \tilde{\bm{z}}_{(n)}) - \Psi^{\dagger} \tilde{\bm{z}}_{(n)} \right)$ \\
		$a_{(n+1)} = \frac{1}{2} (1 + \sqrt{4 a_{(n)}^{2} + 1})$  \\
		$\bm{z}_{(n+1)} = \bm{x}_{(n)} + \frac{a_{(n)} - 1}{a_{(n+1)}} (\bm{x}_{(n)} - \bm{x}_{(n-1)})$ \\
		\If{$\frac{\| \bm{x}_{(n)} - \bm{x}_{(n-1)} \|}{ \| \bm{x}_{(n-1)} \|} < \xi$}{
			break
		}
	} \vspace{0.05in}

	set $\hat{\bm{x}}_{\text{MAP}} = \bm{x}_{(n)}$
\end{algorithm}

\section{Wavelet-based Bayesian uncertainty quantification}

\subsection{Hypothesis testing of image structure}
\label{ap:hyp_test_structure}

\begin{figure}
	\centering
	\begin{tabular}{cc}
		\includegraphics[width=0.48\linewidth]{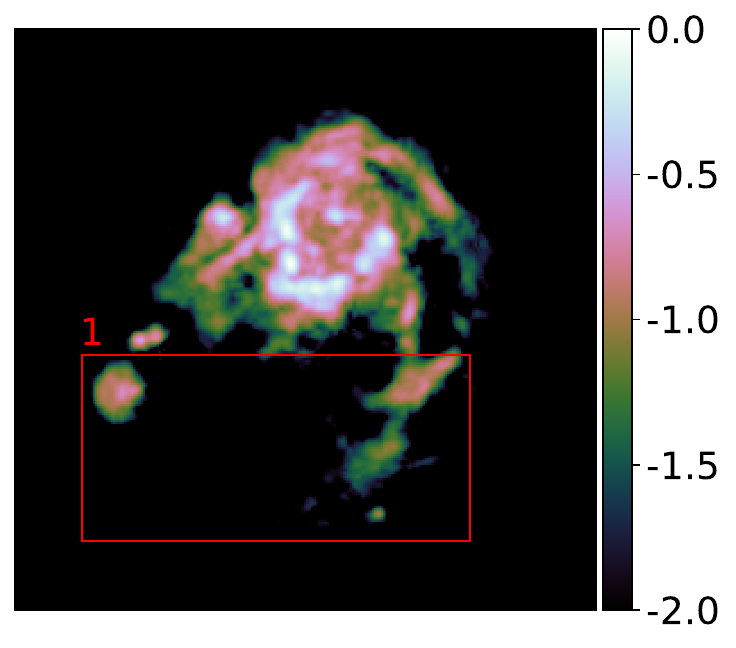}   &
		\includegraphics[width=0.48\linewidth]{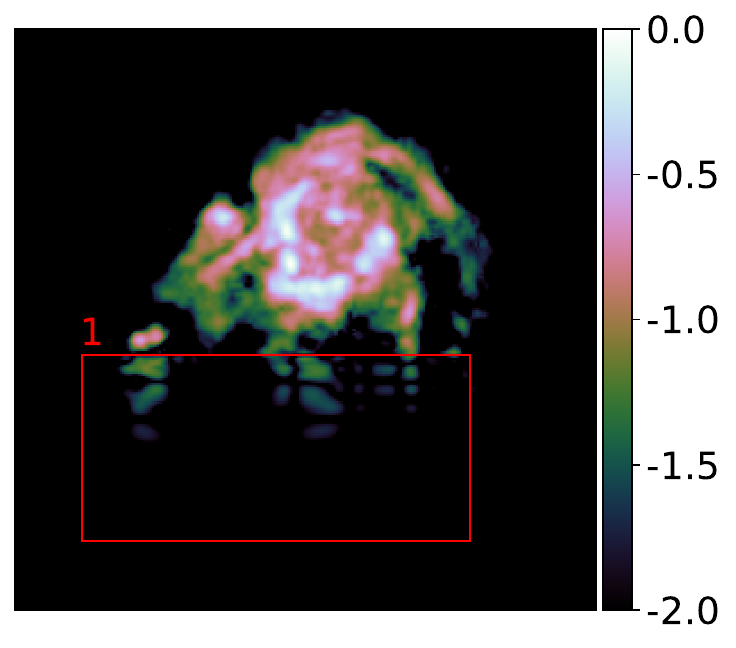}                         \\
		\includegraphics[width=0.48\linewidth]{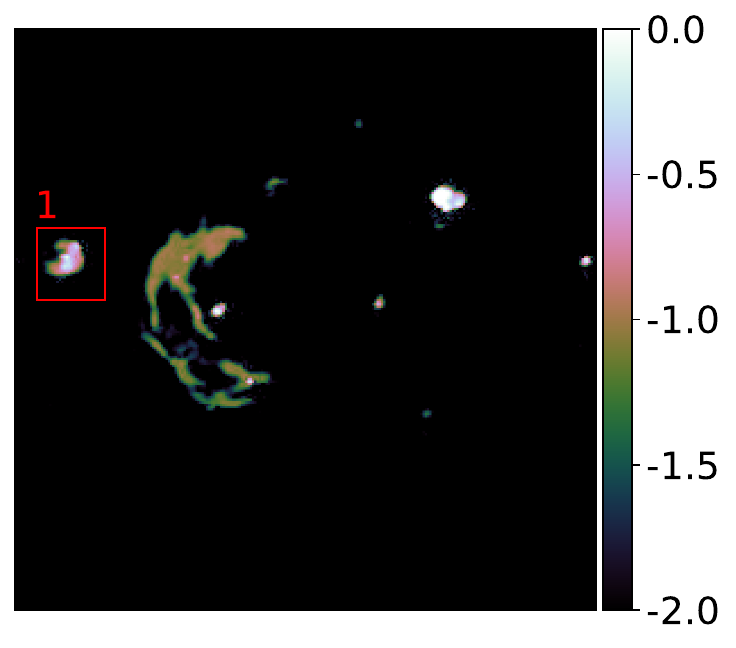}   &
		\includegraphics[width=0.48\linewidth]{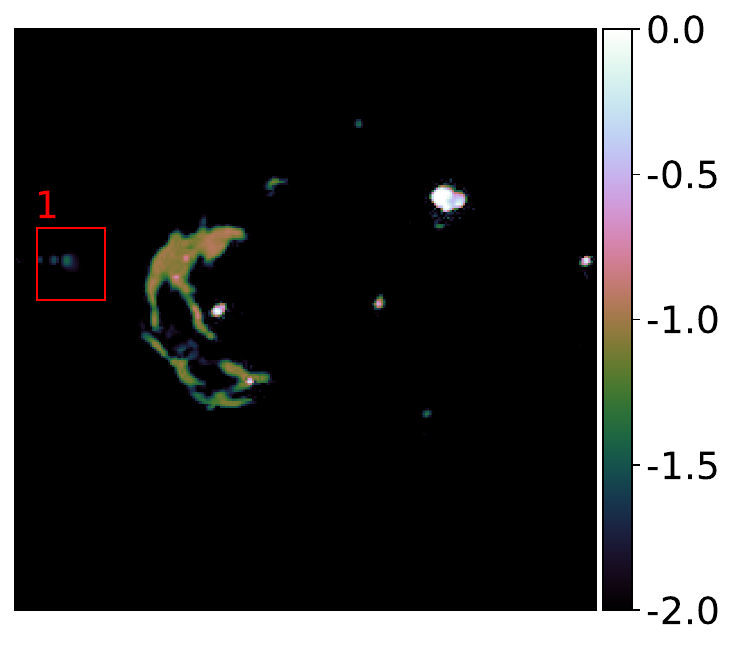}                         \\
		\includegraphics[width=0.48\linewidth]{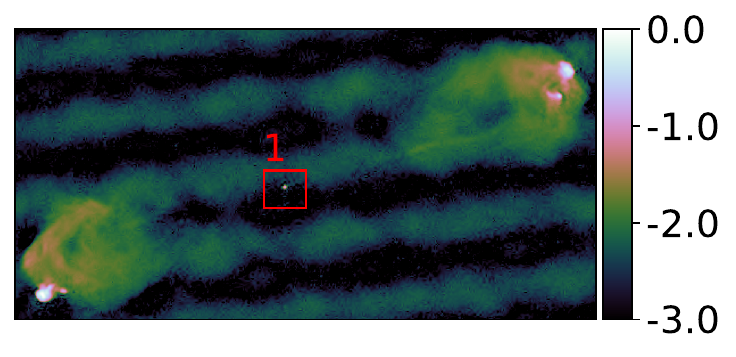}   &
		\includegraphics[width=0.48\linewidth]{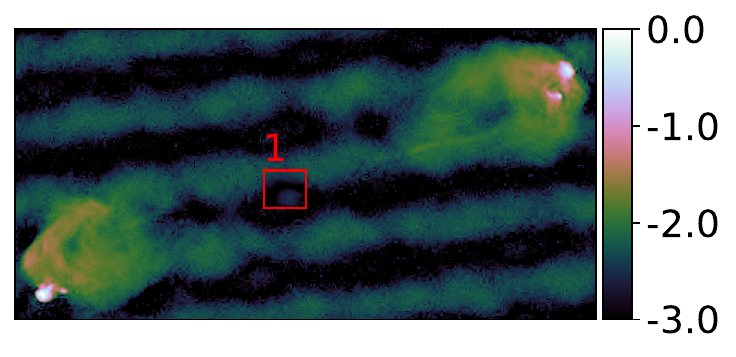}                         \\
		\includegraphics[width=0.48\linewidth]{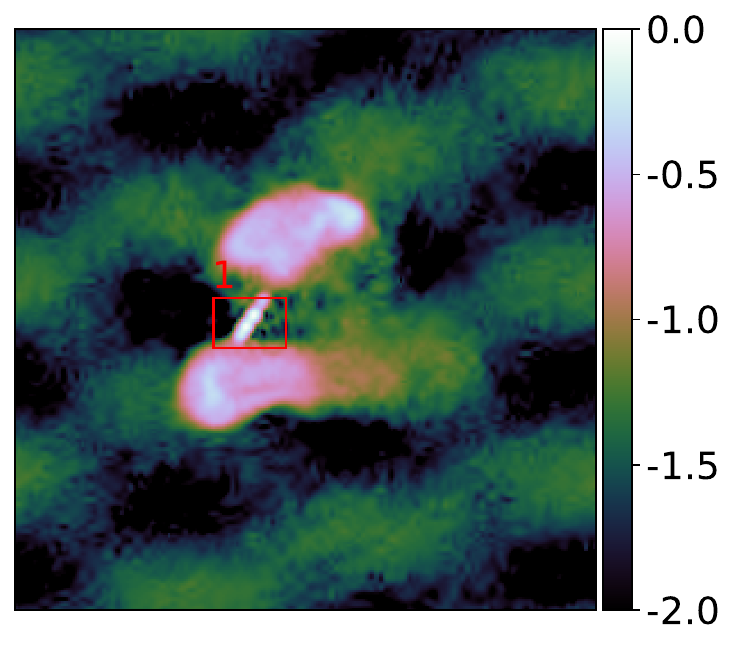} &
		\includegraphics[width=0.48\linewidth]{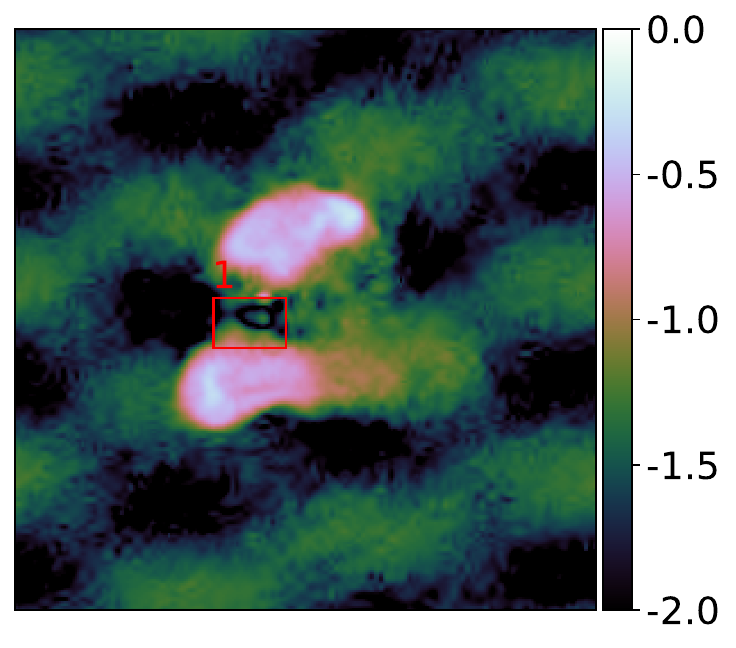}                       \\
		\includegraphics[width=0.48\linewidth]{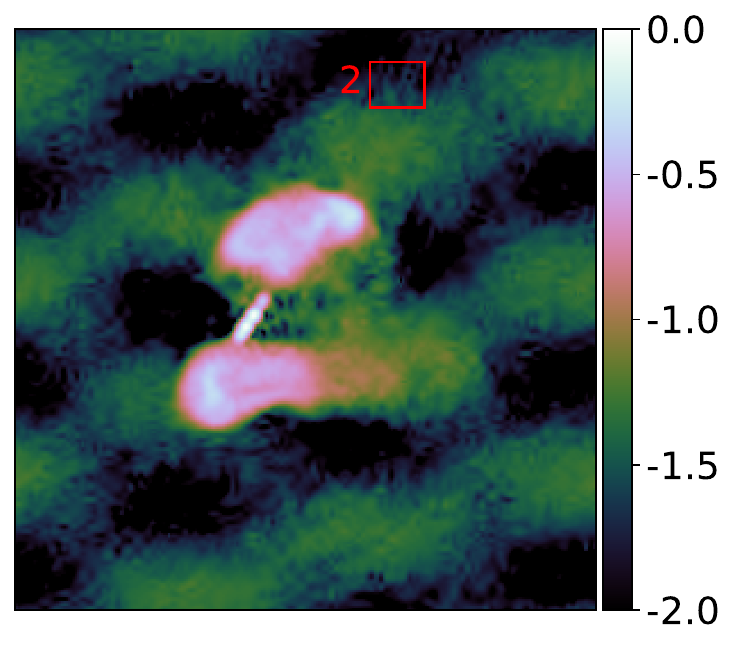} &
		\includegraphics[width=0.48\linewidth]{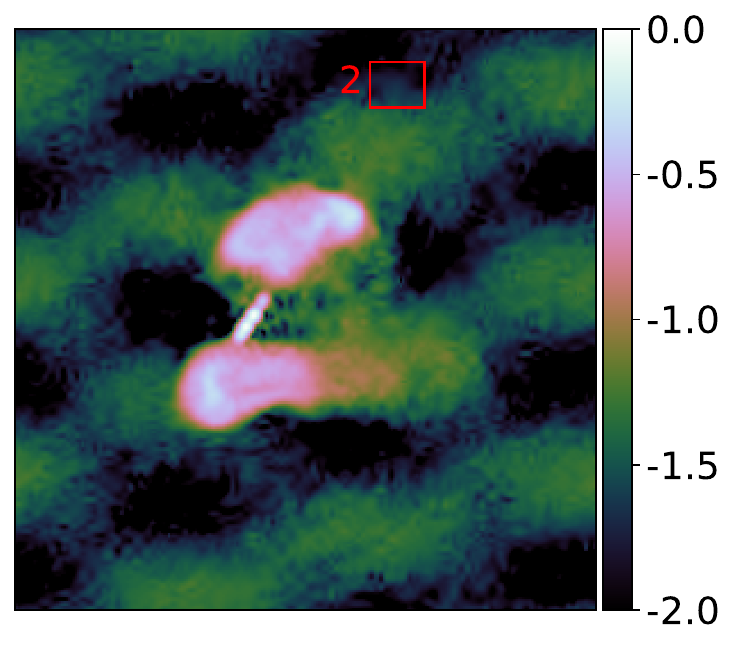}                       \\
		(a) MAP reconstruction                                                                            & (b) Inpainted surrogate
	\end{tabular}
	\caption{Hypothesis test of different regions of the four images, M31, W28, Cygnus A, and 3C288. All the images are shown in $\log_{10}$ scale. The figure is similar to Figure \ref{fi:hp_test_inpainting_CRR}, but the wavelet-based model has been used to generate the MAP. The wavelet prior was used to inpaint the surrogate image.}
	\label{fi:hp_test_inpainting_WAV}
\end{figure}

Figure \ref{fi:hp_test_inpainting_WAV} and Table \ref{tb:hp_test_inpainting_WAV} present the results of the hypothesis testing of structure for the model with the sparsity-promoting prior.

\begin{table*}
	\begin{center}
		\caption{Hypothesis test results for the inpainted surrogates from Figure \ref{fi:hp_test_inpainting_WAV} generated with the wavelet-based model. All values are in units $10^{4}$. The description of Table \ref{tb:hp_test_inpainting_CRR} applies in this table.}
		\label{tb:hp_test_inpainting_WAV}
		\begin{tabular}{cccccccc}
			\toprule
			\multirow{2}{*}{Images}   & Test               & Ground                   & \multirow{2}{*}{Method} & \multicolumn{1}{c}{Initial}                     & \multicolumn{1}{c}{Surrogate}                                & \multicolumn{1}{c}{Isocontour}            & Hypothesis \\
			                          & area               & truth                    &                         & \multicolumn{1}{c}{$(f + g)(\hat{\bm{x}}^{*})$} & \multicolumn{1}{c}{$(f + g)(\hat{\bm{x}}^{*,{\text{sgt}}})$} & \multicolumn{1}{c}{$\hat{\gamma}_{0.01}$} & test       \\
			\hline \hline
			\multirow{2}{*}{M31}      & \multirow{2}{*}{1} & \multirow{2}{*}{ \cmark} &
			SK-ROCK                   & $0.448$            & $\bf 1.396$              & $1.105$                 & \cmark                                                                                                                                                                  \\
			                          &                    &                          & MAP                     & $0.359$                                         & $\bf 1.335$                                                  & $1.039$                                   & \cmark     \\
			\midrule
			\multirow{2}{*}{Cygnus A} & \multirow{2}{*}{1} & \multirow{2}{*}{ \cmark} &
			SK-ROCK                   & $0.480$            & $0.533$                  & $\bf 1.639$             & \xmark                                                                                                                                                                  \\
			                          &                    &                          & MAP                     & $0.444$                                         & $0.514$                                                      & $\bf 1.789$                               & \xmark     \\
			\midrule
			\multirow{2}{*}{W28}      & \multirow{2}{*}{1} & \multirow{2}{*}{ \cmark} &
			SK-ROCK                   & $0.353$            & $\bf 5.190$              & $0.879$                 & \cmark                                                                                                                                                                  \\
			                          &                    &                          & MAP                     & $0.284$                                         & $\bf 5.204$                                                  & $0.964$                                   & \cmark     \\
			\midrule
			\multirow{4}{*}{3C288}    & \multirow{2}{*}{1} & \multirow{2}{*}{\cmark}  &
			SK-ROCK                   & $0.729$            & $\bf 2.487$              & $1.398$                 & \cmark                                                                                                                                                                  \\
			                          &                    &                          & MAP                     & $0.654$                                         & $\bf 2.409$                                                  & $1.333$                                   & \cmark     \\
			\cdashline{2-8}
			                          & \multirow{2}{*}{2} & \multirow{2}{*}{ \xmark} &
			SK-ROCK                   & $0.729$            & $0.729$                  & $\bf 1.398$             & \xmark                                                                                                                                                                  \\
			                          &                    &                          & MAP                     & $0.654$                                         & $0.654$                                                      & $\bf 1.333$                               & \xmark     \\
			\bottomrule
		\end{tabular}
	\end{center}
\end{table*}

\addtolength{\tabcolsep}{-\tabL}
\addtolength{\tabcolsep}{\tabL}

\section{More realistic experiment results}
\label{ap:more_realistic_results}

\begin{figure*}
	\centering
	\begin{tabular}{cccc}
		{\Large 1h} 	& {\Large 2h}	& {\Large 4h} & {\Large 8h} \\
		\includegraphics[width=0.18\linewidth]{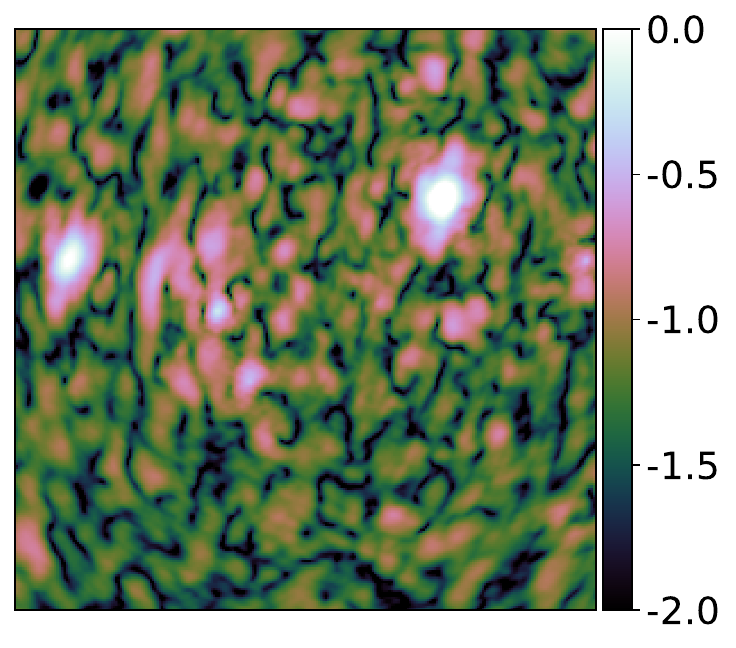} \put(-110,22){\rotatebox{90}{\large Dirty rec.}}  &
		\includegraphics[width=0.18\linewidth]{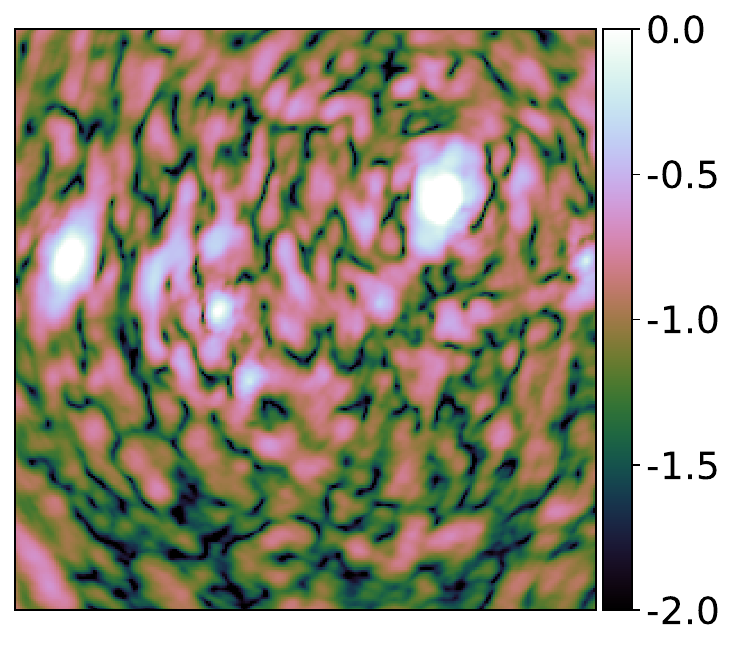}                                                                           &
		\includegraphics[width=0.18\linewidth]{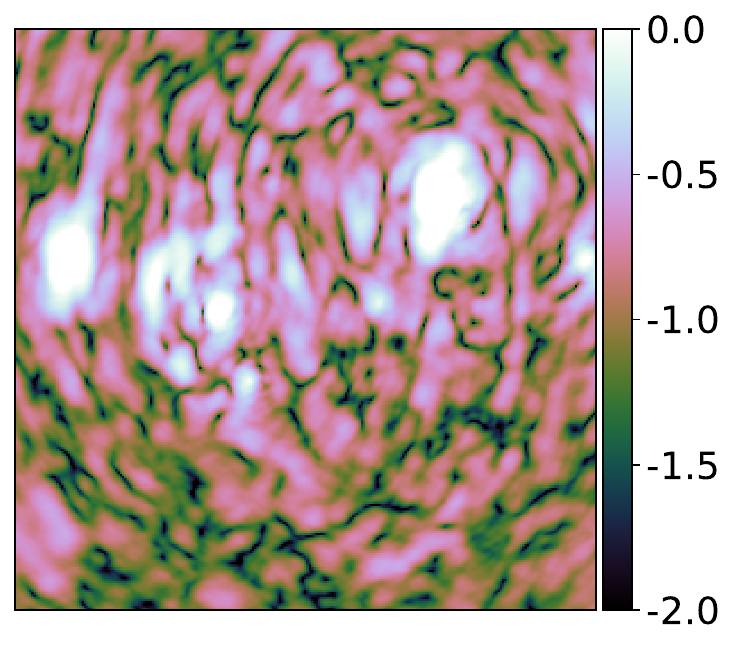}                                                              &
		\includegraphics[width=0.18\linewidth]{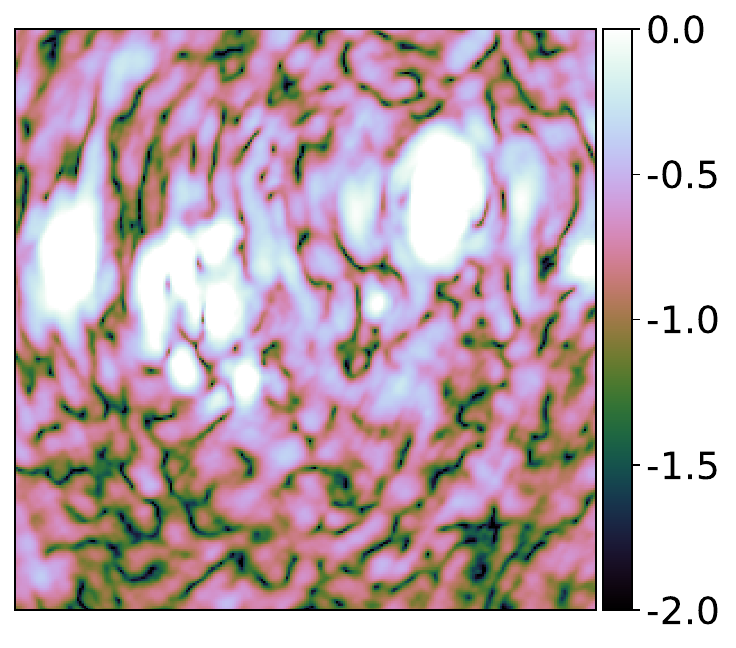}
		\\
		\includegraphics[width=0.18\linewidth]{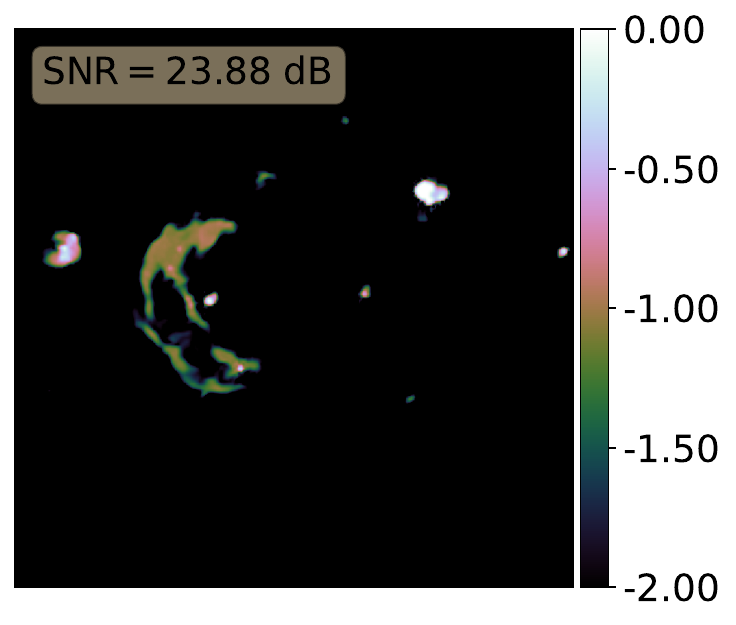} \put(-110,22){\rotatebox{90}{\large MAP rec.}} &
		\includegraphics[width=0.18\linewidth]{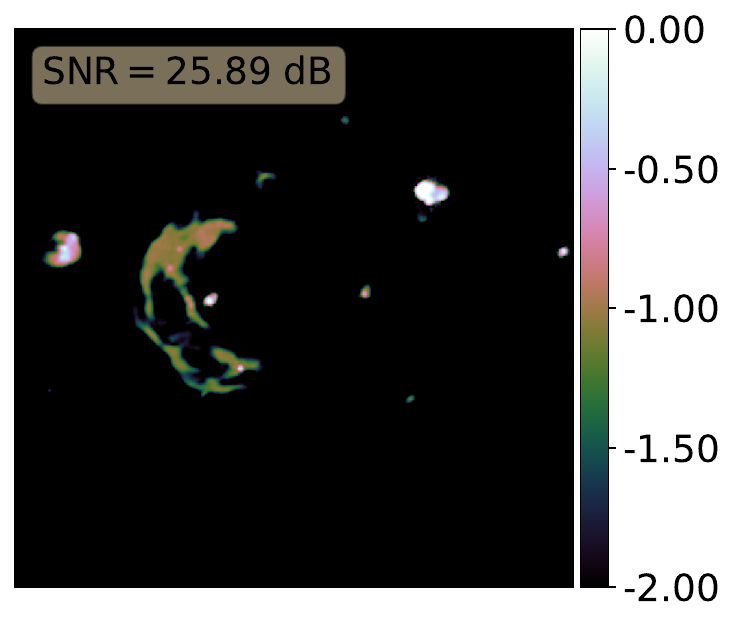}                                                                           &
		\includegraphics[width=0.18\linewidth]{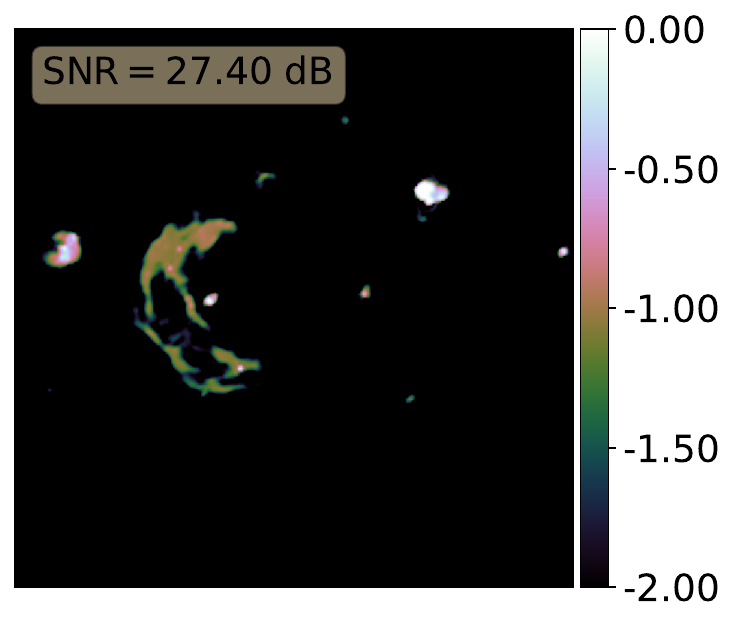}                                                              &
		\includegraphics[width=0.18\linewidth]{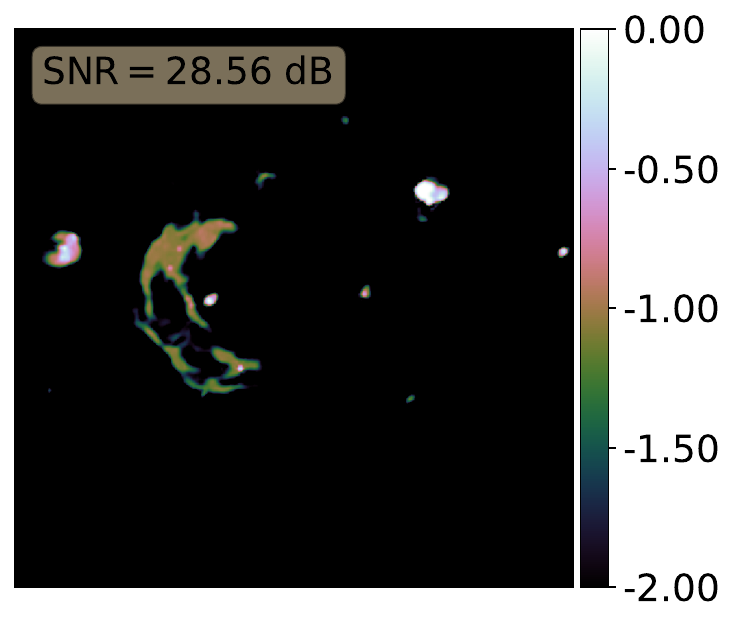}
		\\
		\includegraphics[width=0.18\linewidth]{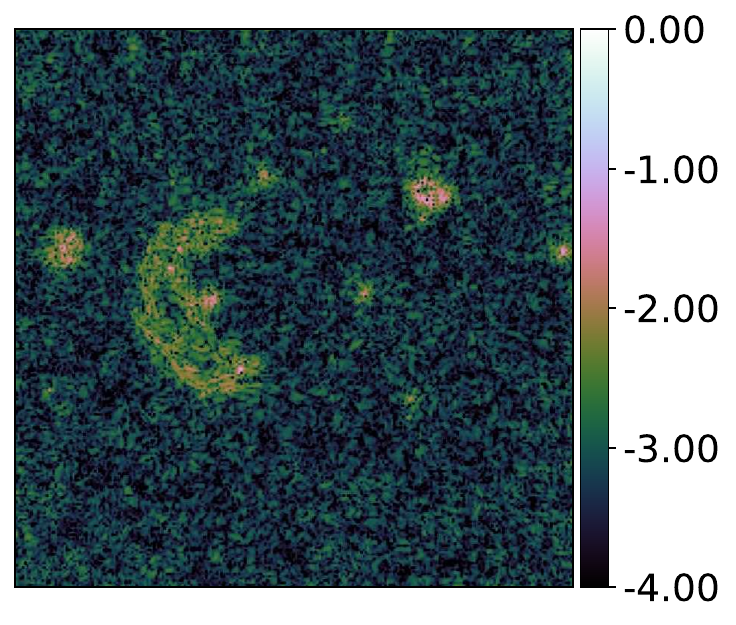} \put(-110,20){\rotatebox{90}{\large Oracle error}} &
		\includegraphics[width=0.18\linewidth]{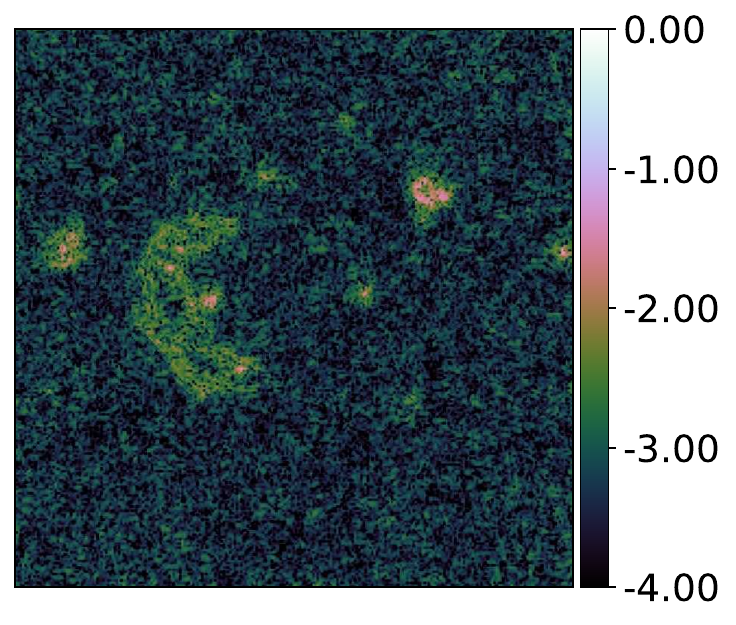}                                                                           &
		\includegraphics[width=0.18\linewidth]{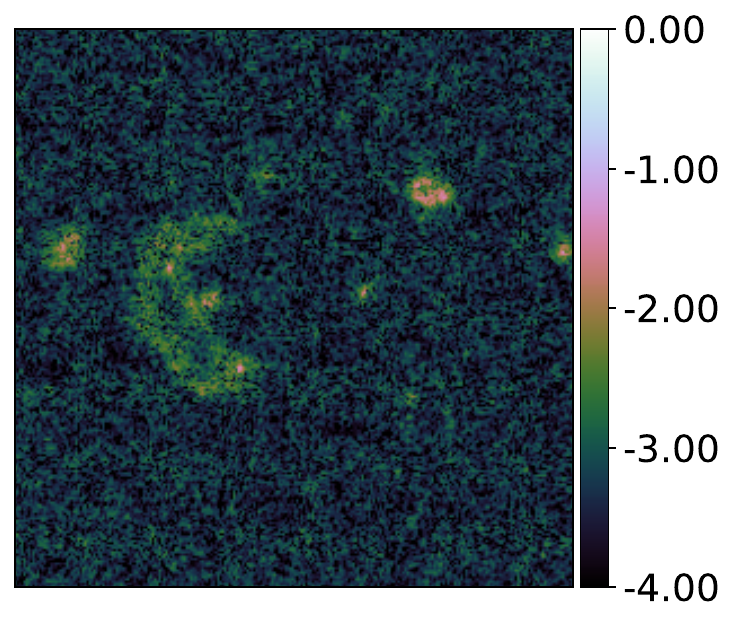}                                                              &
		\includegraphics[width=0.18\linewidth]{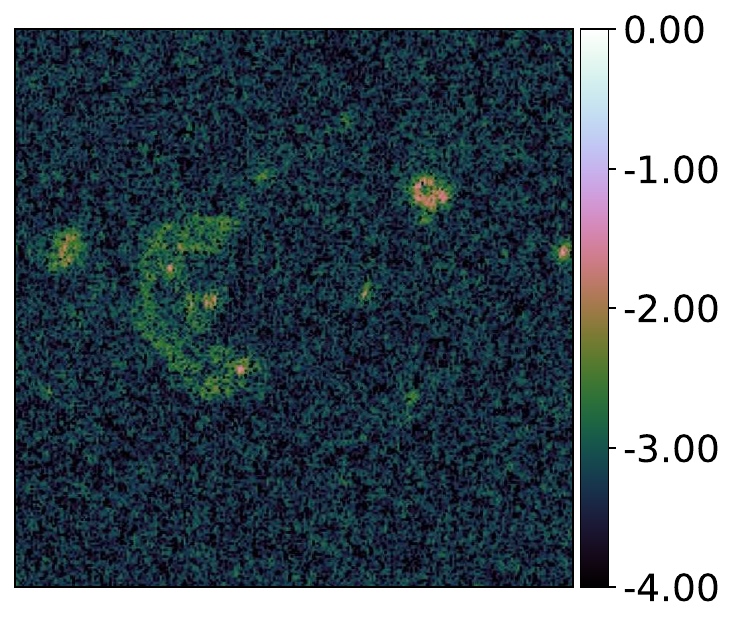}
		\\
		\multicolumn{4}{c}{\large Predicted error}  \\
		\includegraphics[width=0.18\linewidth]{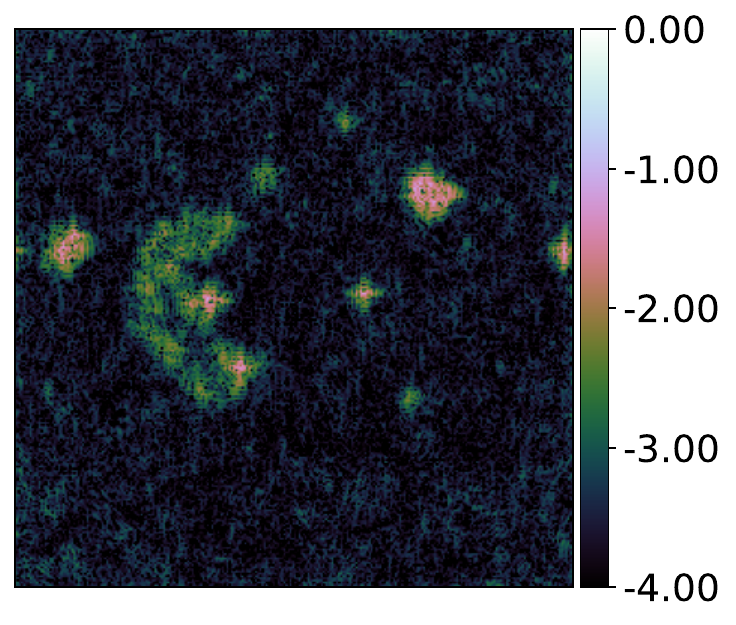} \put(-110,30){\rotatebox{90}{\large Level $4$}} &
		\includegraphics[width=0.18\linewidth]{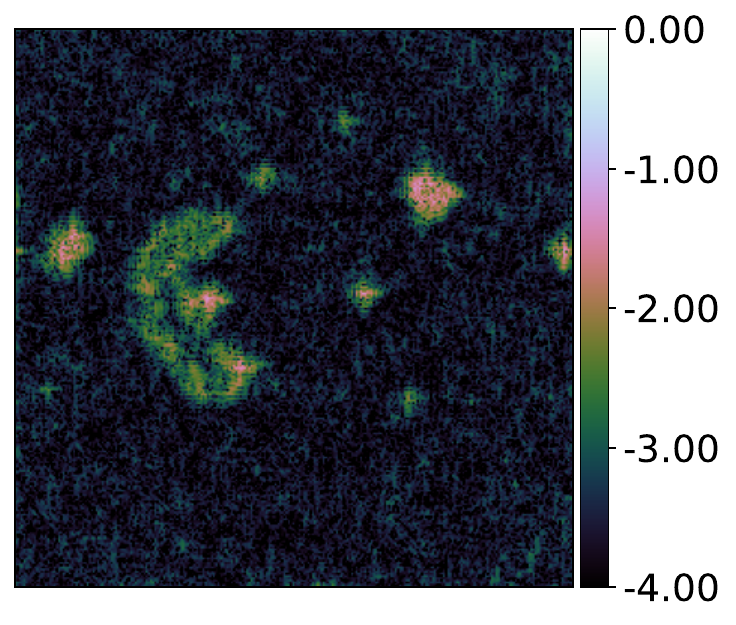}    &
		\includegraphics[width=0.18\linewidth]{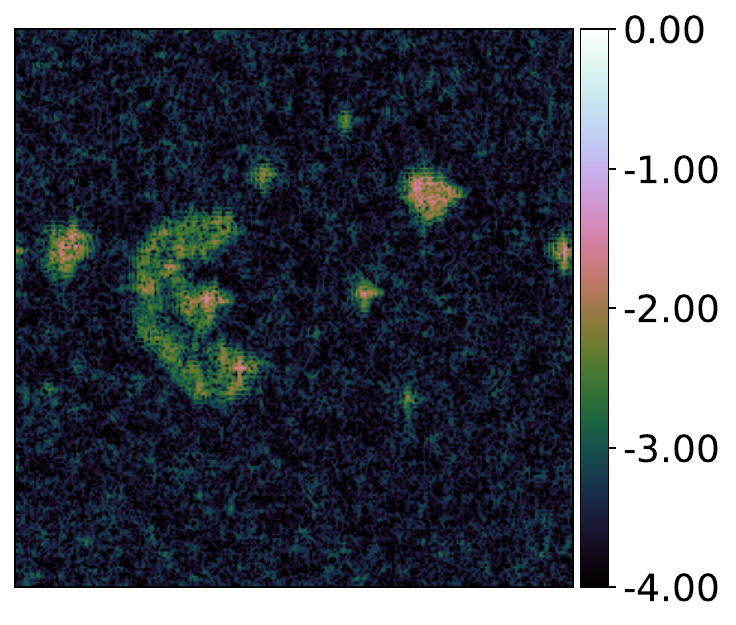}                                                 &
		\includegraphics[width=0.18\linewidth]{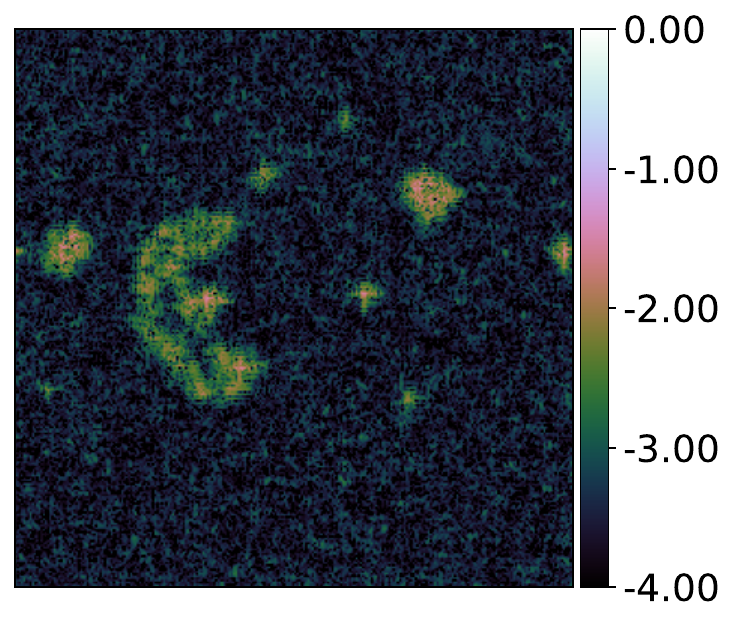}
		\\
		\includegraphics[width=0.18\linewidth]{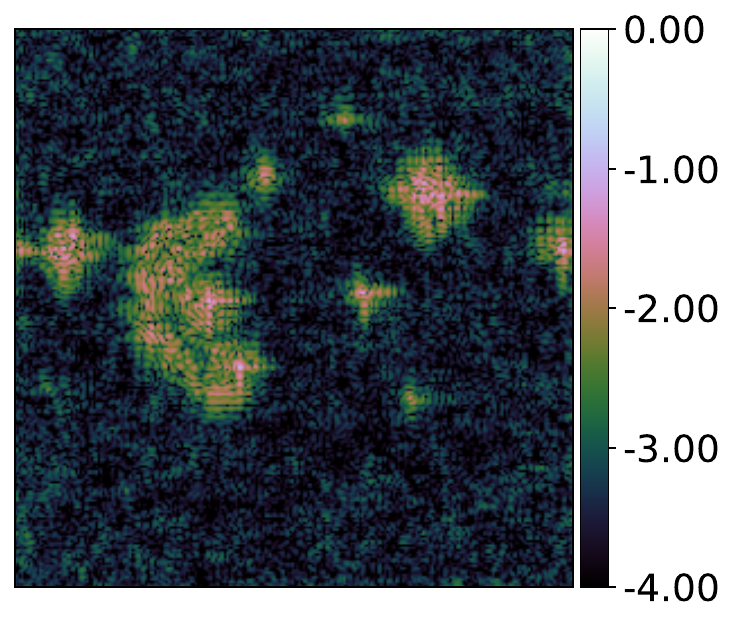} \put(-110,30){\rotatebox{90}{\large Level $3$}} &
		\includegraphics[width=0.18\linewidth]{fastUQ_ungridded/W28-CRR1h-newPixelUQ-MAP_thresholded_error_level_3.pdf}                                                          &
		\includegraphics[width=0.18\linewidth]{fastUQ_ungridded/W28-CRR1h-newPixelUQ-MAP_thresholded_error_level_3.pdf}                                                 &
		\includegraphics[width=0.18\linewidth]{fastUQ_ungridded/W28-CRR1h-newPixelUQ-MAP_thresholded_error_level_3.pdf}
		\\
		\includegraphics[width=0.18\linewidth]{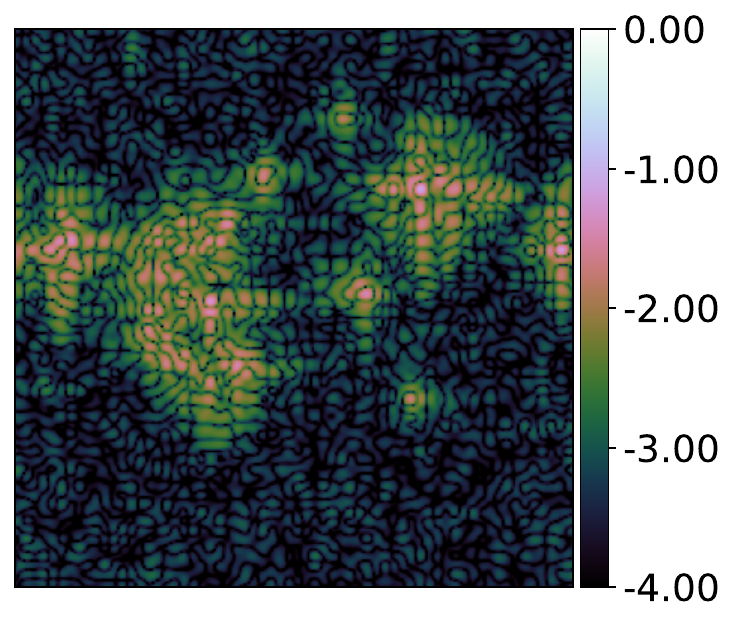} \put(-110,30){\rotatebox{90}{\large Level $2$}} &
		\includegraphics[width=0.18\linewidth]{fastUQ_ungridded/W28-CRR1h-newPixelUQ-MAP_thresholded_error_level_2.pdf}                                                          &
		\includegraphics[width=0.18\linewidth]{fastUQ_ungridded/W28-CRR1h-newPixelUQ-MAP_thresholded_error_level_2.pdf}                                                 &
		\includegraphics[width=0.18\linewidth]{fastUQ_ungridded/W28-CRR1h-newPixelUQ-MAP_thresholded_error_level_2.pdf}
		\\
		\includegraphics[width=0.18\linewidth]{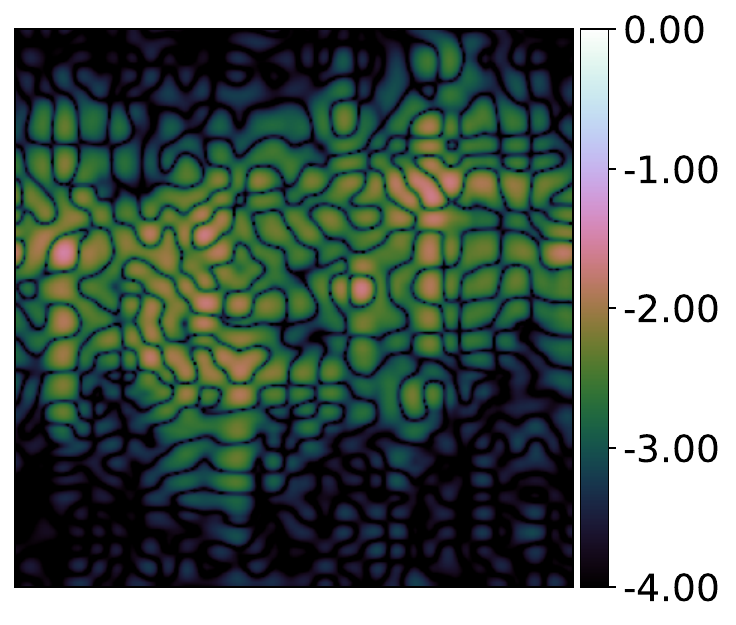} \put(-110,30){\rotatebox{90}{\large Level $1$}} &
		\includegraphics[width=0.18\linewidth]{fastUQ_ungridded/W28-CRR1h-newPixelUQ-MAP_thresholded_error_level_1.pdf}                                                          &
		\includegraphics[width=0.18\linewidth]{fastUQ_ungridded/W28-CRR1h-newPixelUQ-MAP_thresholded_error_level_1.pdf}                                                 &
		\includegraphics[width=0.18\linewidth]{fastUQ_ungridded/W28-CRR1h-newPixelUQ-MAP_thresholded_error_level_1.pdf}
		\\
	\end{tabular}
	\caption{Reconstructions and fast pixel uncertainty quantification (UQ) with the \textsc{QuantifAI} model for the W28 image with the four sets of simulated MeerKAT ungridded visibilities with a field of view of approximately $1$ deg$^2$. Each column corresponds to the four datasets with synthesis times of $1$, $2$, $4$ and $8$ hours. The first row represents the dirty reconstruction. The MAP reconstruction is presented in the second row, while the oracle error, which we do not have access to with real data, is shown in the third row. The different decomposition levels of pixel UQ are shown in the last four rows.}
	\label{fi:ungridded_results_W28}
\end{figure*}

\begin{figure*}
	\centering
	\begin{tabular}{cccc}
		{\Large 1h} 	& {\Large 2h}	& {\Large 4h} & {\Large 8h} \\
		\includegraphics[width=0.2\linewidth]{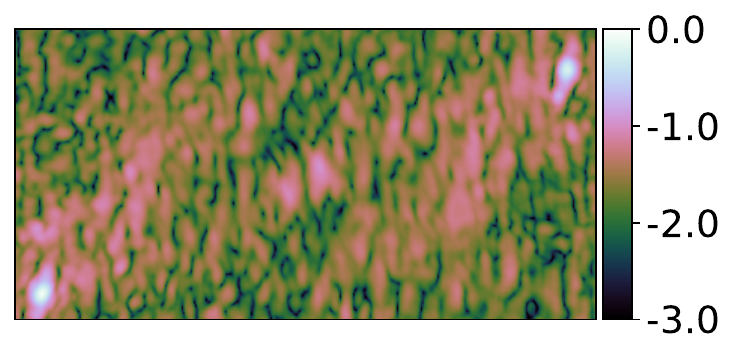} \put(-110,12){\rotatebox{90}{\large Dirty}}  &
		\includegraphics[width=0.2\linewidth]{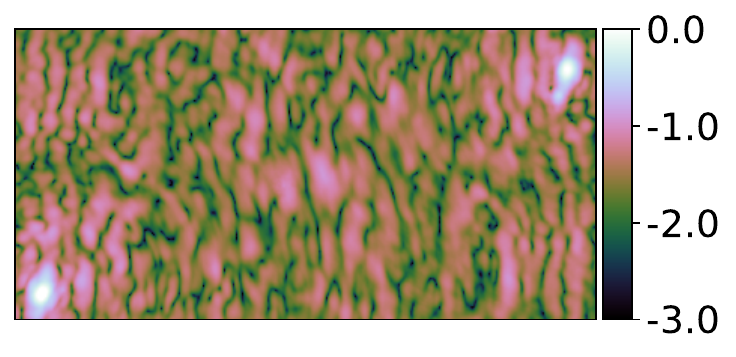}                                                                           &
		\includegraphics[width=0.2\linewidth]{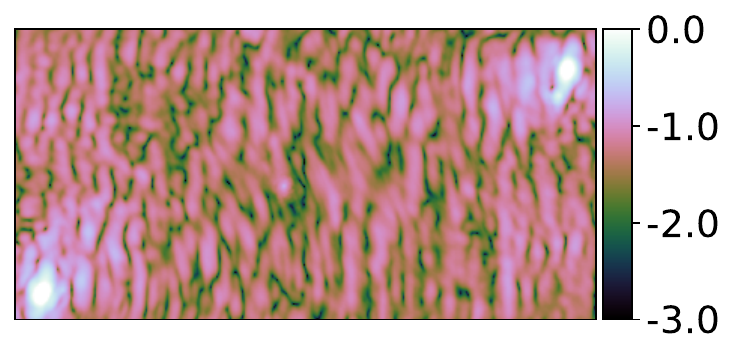}                                                              &
		\includegraphics[width=0.2\linewidth]{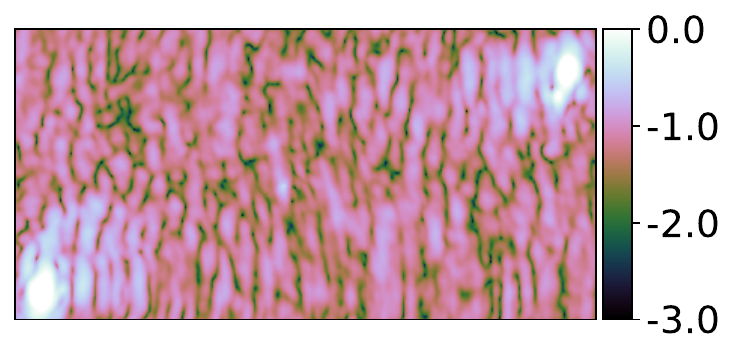}
		\\
		\includegraphics[width=0.2\linewidth]{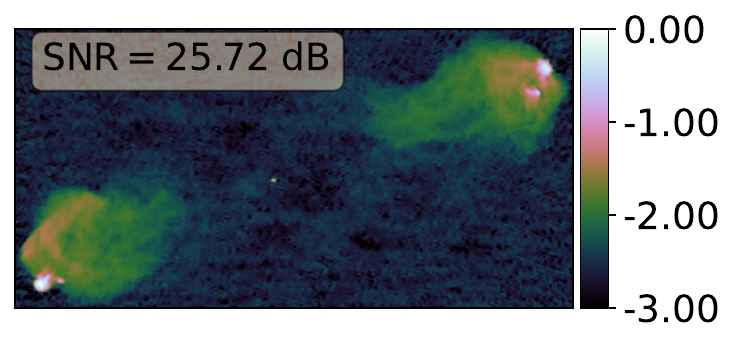} \put(-110,12){\rotatebox{90}{\large MAP}} &
		\includegraphics[width=0.2\linewidth]{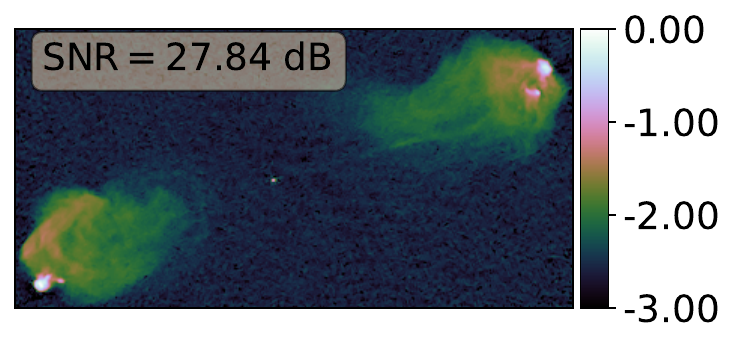}                                                                           &
		\includegraphics[width=0.2\linewidth]{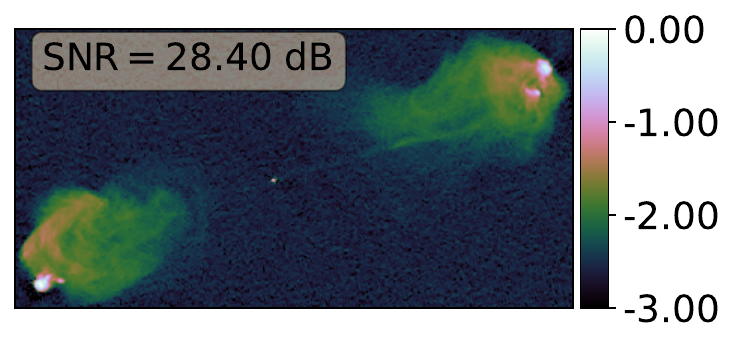}                                                              &
		\includegraphics[width=0.2\linewidth]{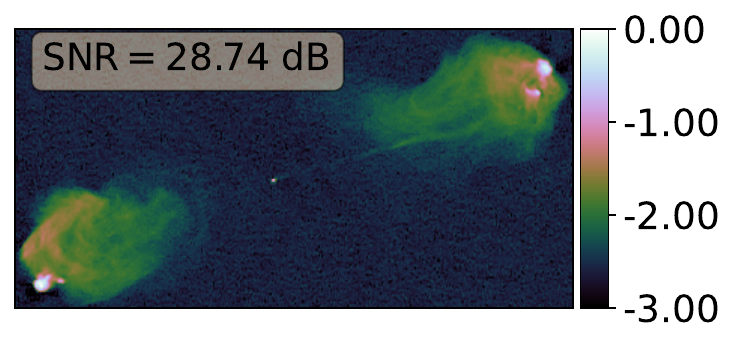}
		\\
		\includegraphics[width=0.2\linewidth]{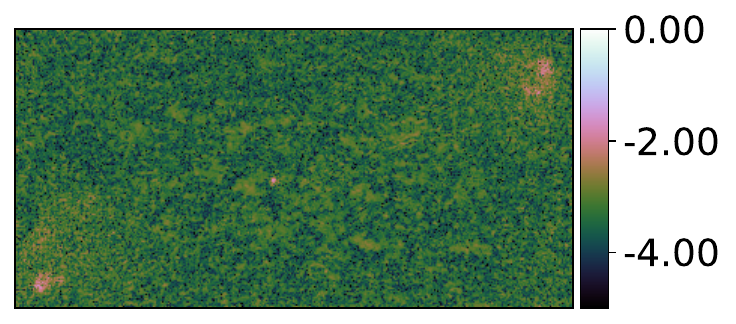} \put(-110,00){\rotatebox{90}{\large Oracle err.}} &
		\includegraphics[width=0.2\linewidth]{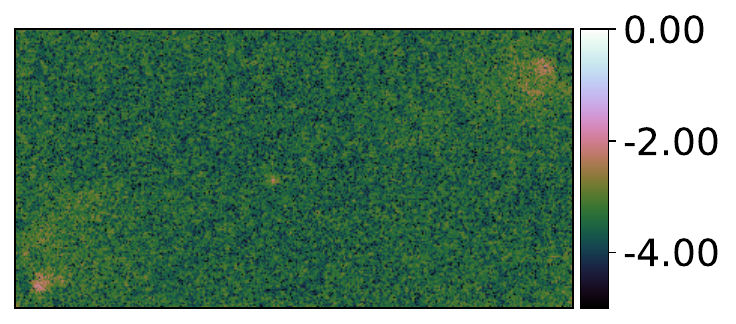}                                                                           &
		\includegraphics[width=0.2\linewidth]{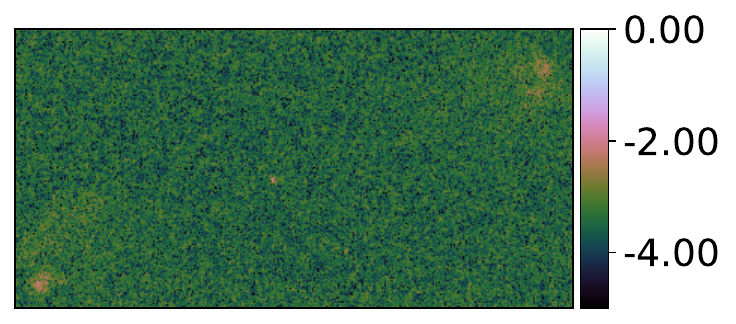}                                                              &
		\includegraphics[width=0.2\linewidth]{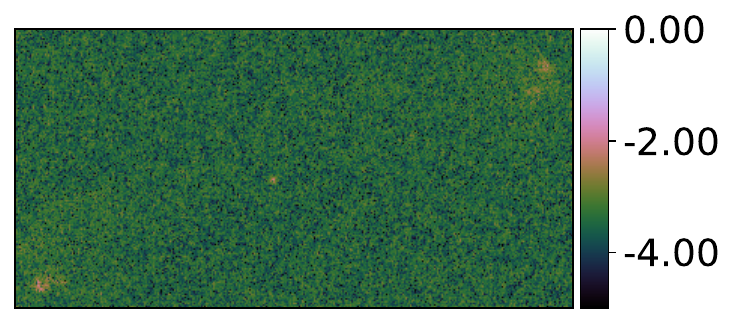}
		\\
		\multicolumn{4}{c}{\large Predicted error}  \\
		\includegraphics[width=0.2\linewidth]{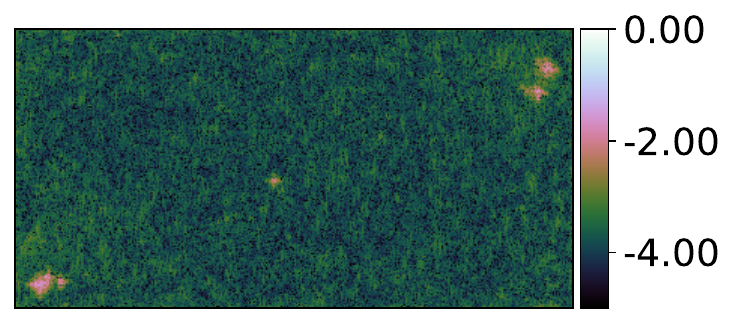} \put(-110,5){\rotatebox{90}{\large Level $4$}} &
		\includegraphics[width=0.2\linewidth]{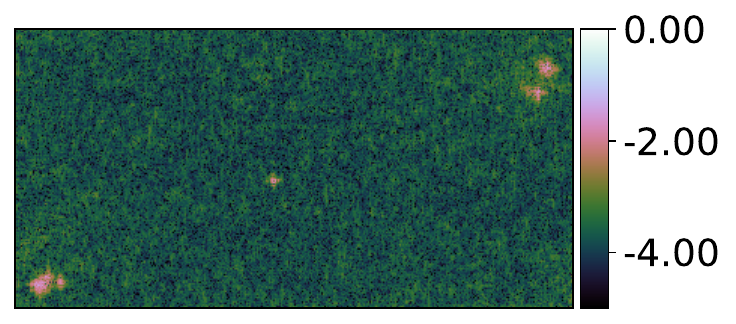}    &
		\includegraphics[width=0.2\linewidth]{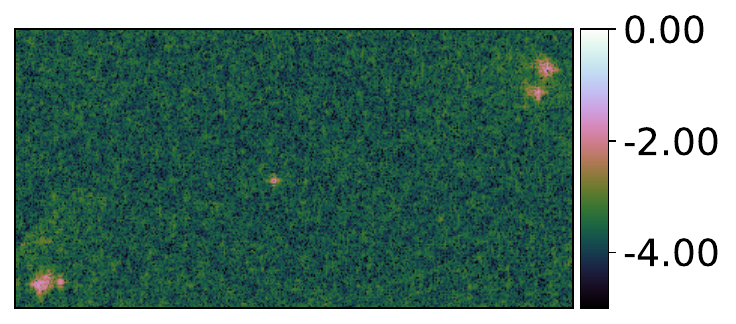}                                                 &
		\includegraphics[width=0.2\linewidth]{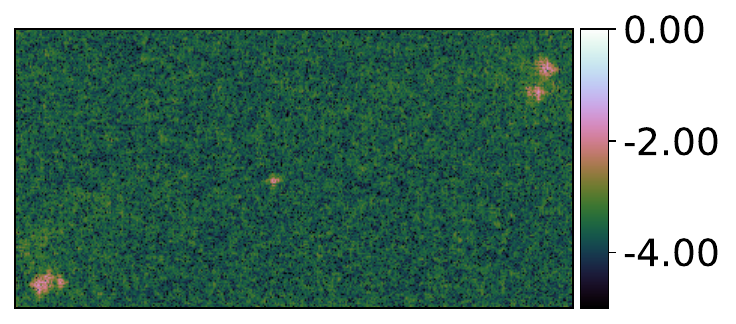}
		\\
		\includegraphics[width=0.2\linewidth]{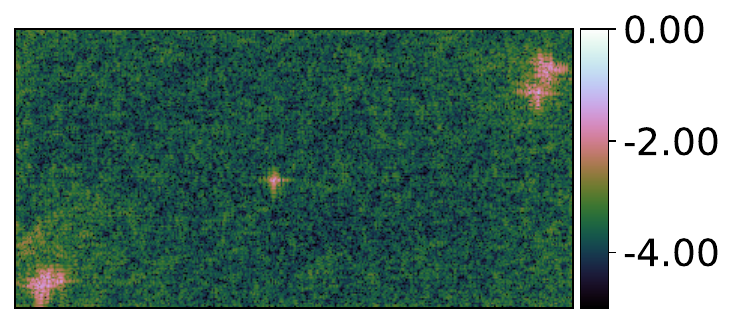} \put(-110,5){\rotatebox{90}{\large Level $3$}} &
		\includegraphics[width=0.2\linewidth]{fastUQ_ungridded/CYN-CRR1h-newPixelUQ-MAP_thresholded_error_level_3.pdf}                                                          &
		\includegraphics[width=0.2\linewidth]{fastUQ_ungridded/CYN-CRR1h-newPixelUQ-MAP_thresholded_error_level_3.pdf}                                                 &
		\includegraphics[width=0.2\linewidth]{fastUQ_ungridded/CYN-CRR1h-newPixelUQ-MAP_thresholded_error_level_3.pdf}
		\\
		\includegraphics[width=0.2\linewidth]{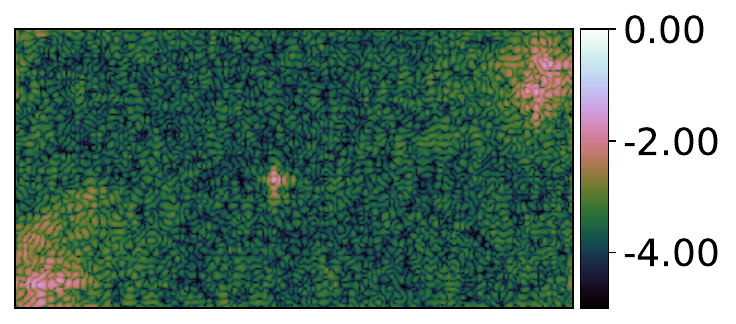} \put(-110,5){\rotatebox{90}{\large Level $2$}} &
		\includegraphics[width=0.2\linewidth]{fastUQ_ungridded/CYN-CRR1h-newPixelUQ-MAP_thresholded_error_level_2.pdf}                                                          &
		\includegraphics[width=0.2\linewidth]{fastUQ_ungridded/CYN-CRR1h-newPixelUQ-MAP_thresholded_error_level_2.pdf}                                                 &
		\includegraphics[width=0.2\linewidth]{fastUQ_ungridded/CYN-CRR1h-newPixelUQ-MAP_thresholded_error_level_2.pdf}
		\\
		\includegraphics[width=0.2\linewidth]{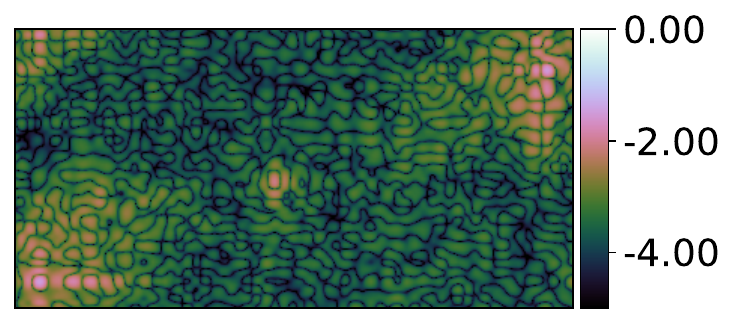} \put(-110,5){\rotatebox{90}{\large Level $1$}} &
		\includegraphics[width=0.2\linewidth]{fastUQ_ungridded/CYN-CRR1h-newPixelUQ-MAP_thresholded_error_level_1.pdf}                                                          &
		\includegraphics[width=0.2\linewidth]{fastUQ_ungridded/CYN-CRR1h-newPixelUQ-MAP_thresholded_error_level_1.pdf}                                                 &
		\includegraphics[width=0.2\linewidth]{fastUQ_ungridded/CYN-CRR1h-newPixelUQ-MAP_thresholded_error_level_1.pdf}
		\\
	\end{tabular}
	\caption{Reconstructions and fast pixel uncertainty quantification (UQ) with the \textsc{QuantifAI} model for the Cygnus A image with the four sets of simulated MeerKAT ungridded visibilities with a field of view of approximately $1$ deg$^2$. Each column corresponds to the four datasets with synthesis times of $1$, $2$, $4$ and $8$ hours. The first row represents the dirty reconstruction. The MAP reconstruction is presented in the second row, while the oracle error, which we do not have access to with real data, is shown in the third row. The different decomposition levels of pixel UQ are shown in the last four rows.}
	\label{fi:ungridded_results_CYN}
\end{figure*}

\begin{figure*}
	\centering
	\begin{tabular}{cccc}
		{\Large 1h} 	& {\Large 2h}	& {\Large 4h} & {\Large 8h} \\
		\includegraphics[width=0.18\linewidth]{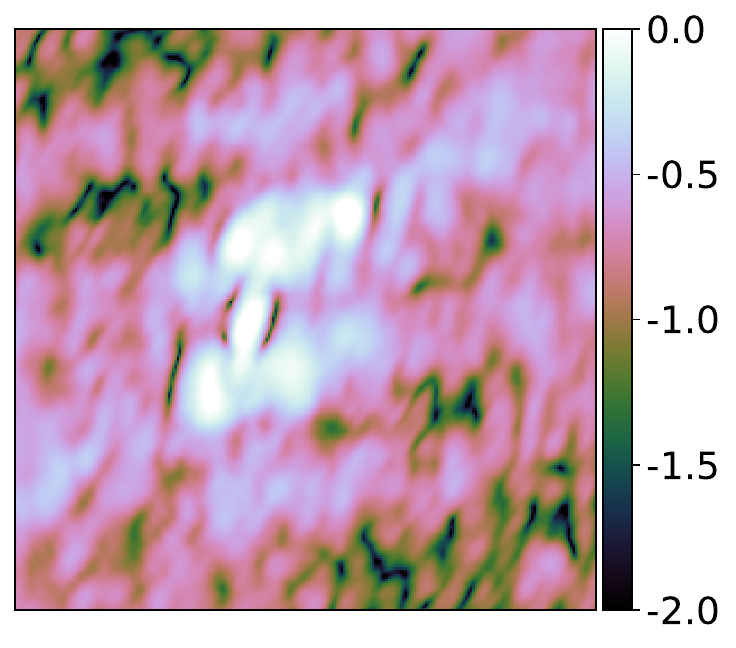} \put(-110,22){\rotatebox{90}{\large Dirty rec.}}  &
		\includegraphics[width=0.18\linewidth]{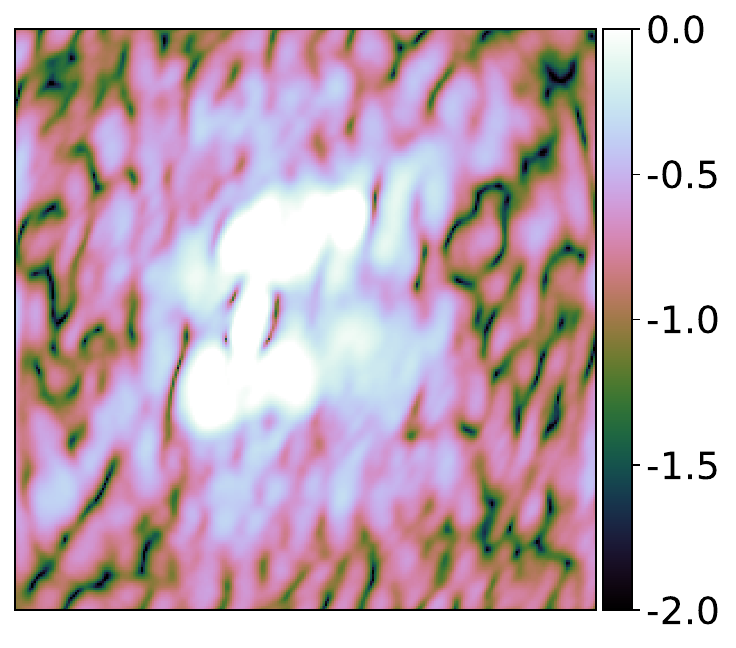}                                                                           &
		\includegraphics[width=0.18\linewidth]{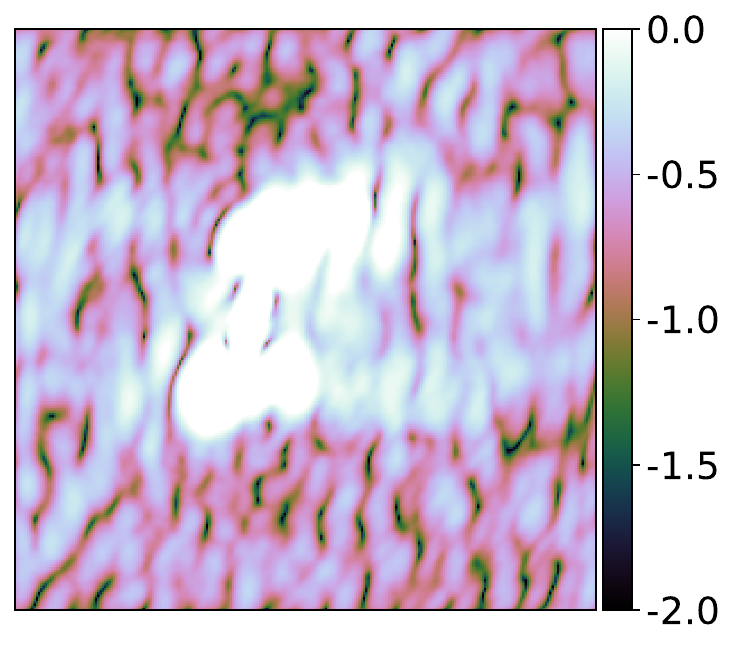}                                                              &
		\includegraphics[width=0.18\linewidth]{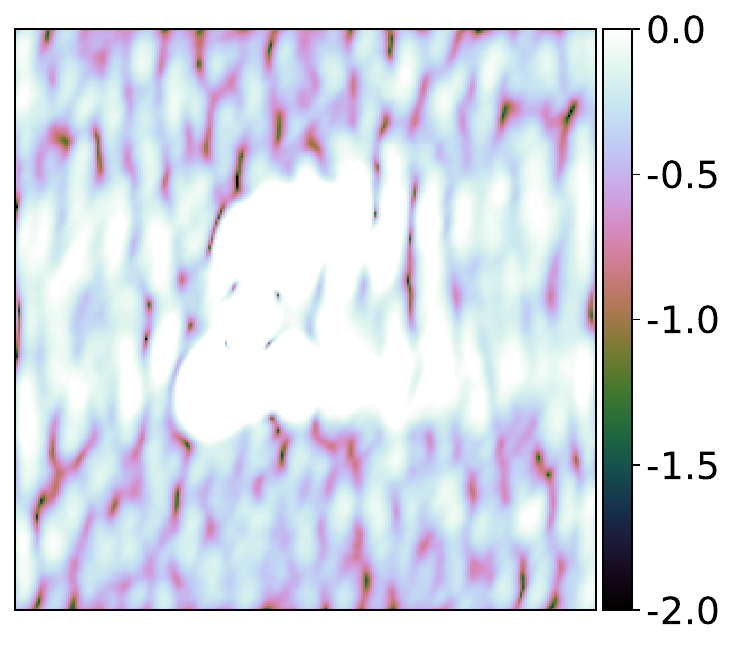}
		\\
		\includegraphics[width=0.18\linewidth]{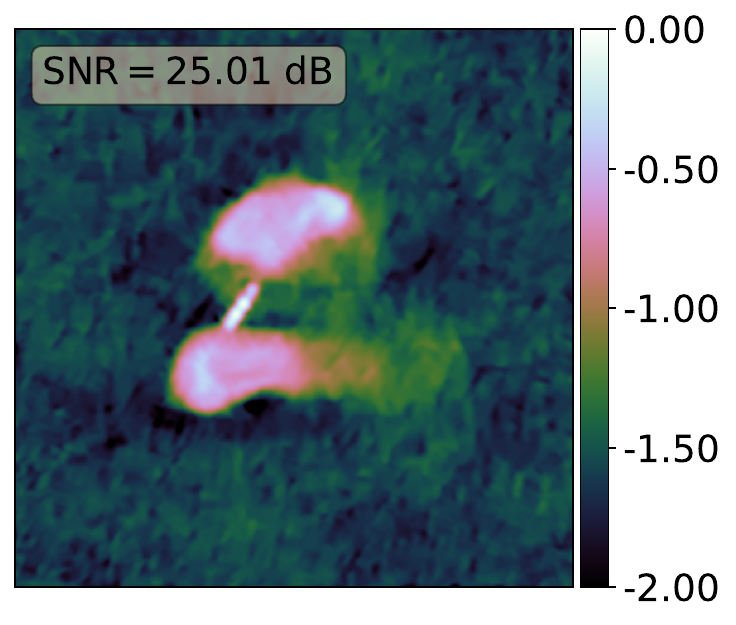} \put(-110,22){\rotatebox{90}{\large MAP rec.}} &
		\includegraphics[width=0.18\linewidth]{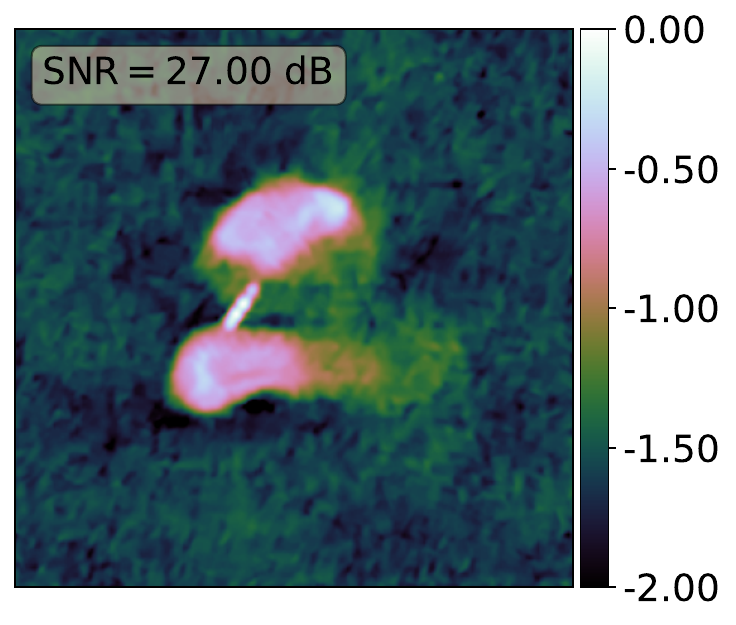}                                                                           &
		\includegraphics[width=0.18\linewidth]{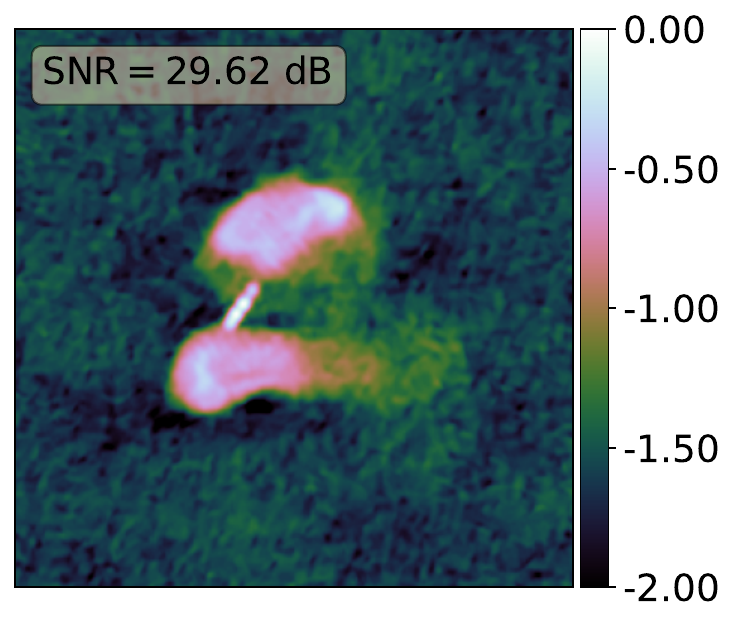}                                                              &
		\includegraphics[width=0.18\linewidth]{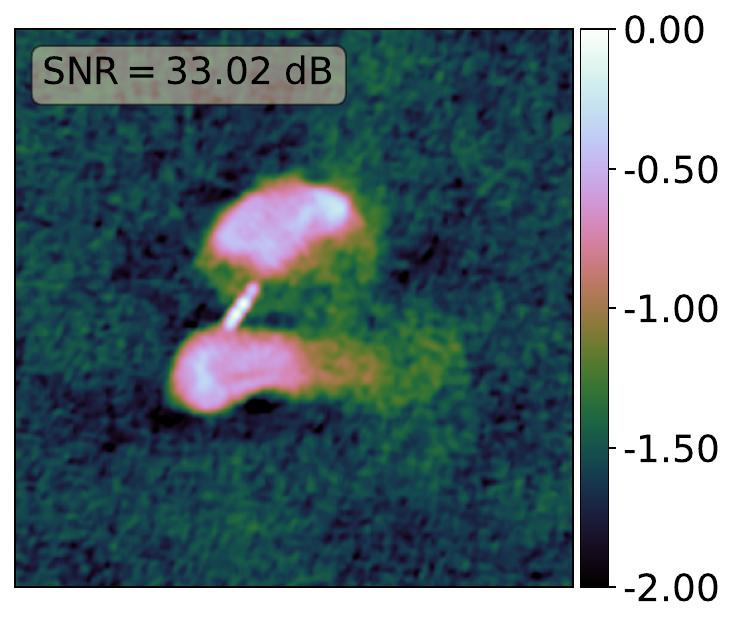}
		\\
		\includegraphics[width=0.18\linewidth]{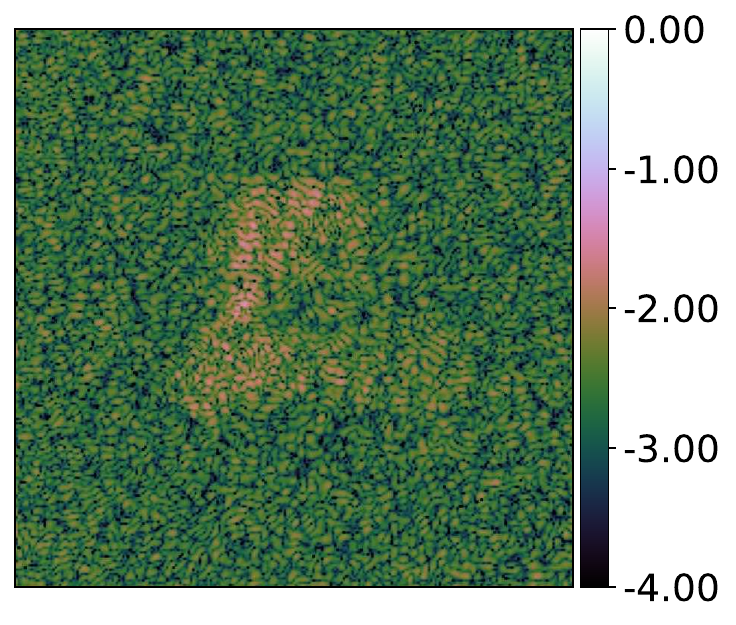} \put(-110,20){\rotatebox{90}{\large Oracle error}} &
		\includegraphics[width=0.18\linewidth]{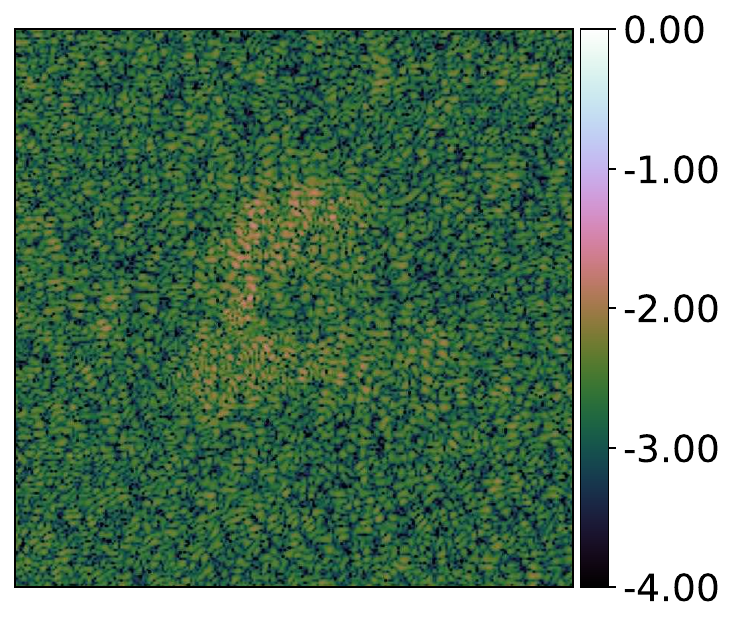}                                                                           &
		\includegraphics[width=0.18\linewidth]{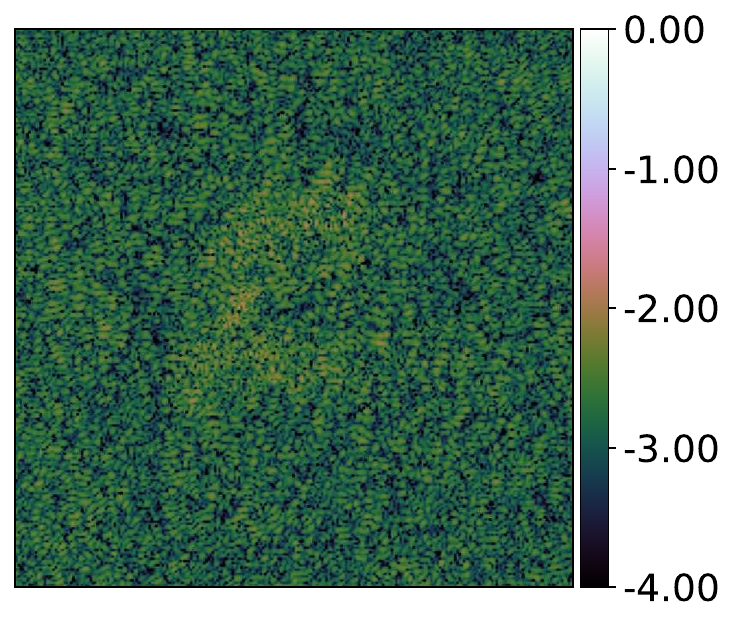}                                                              &
		\includegraphics[width=0.18\linewidth]{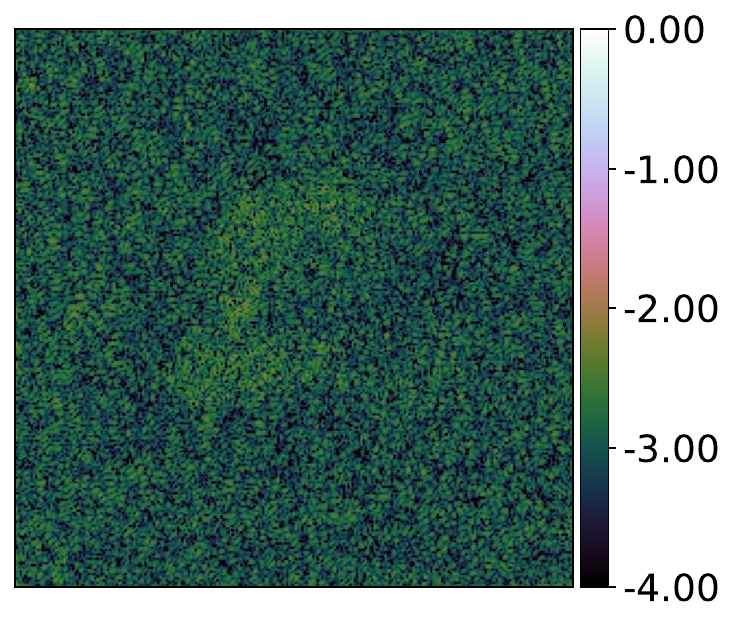}
		\\
		\multicolumn{4}{c}{\large Predicted error}  \\
		\includegraphics[width=0.18\linewidth]{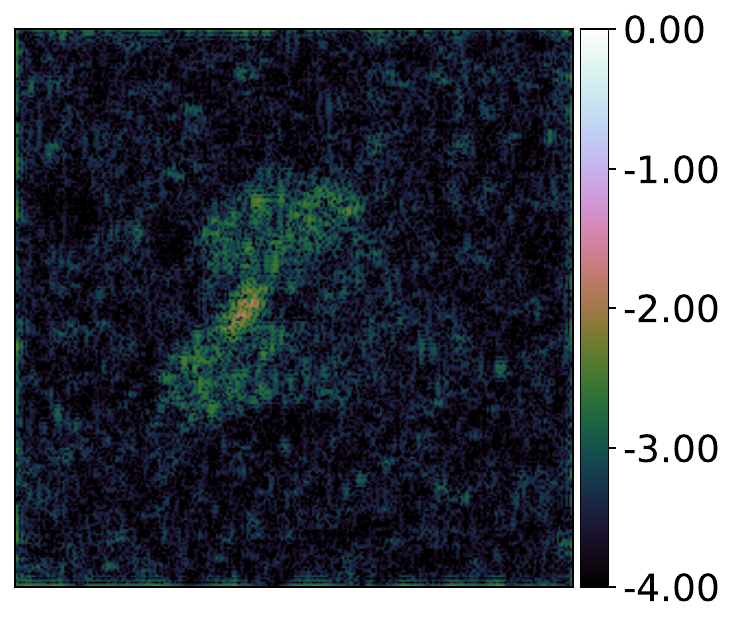} \put(-110,30){\rotatebox{90}{\large Level $4$}} &
		\includegraphics[width=0.18\linewidth]{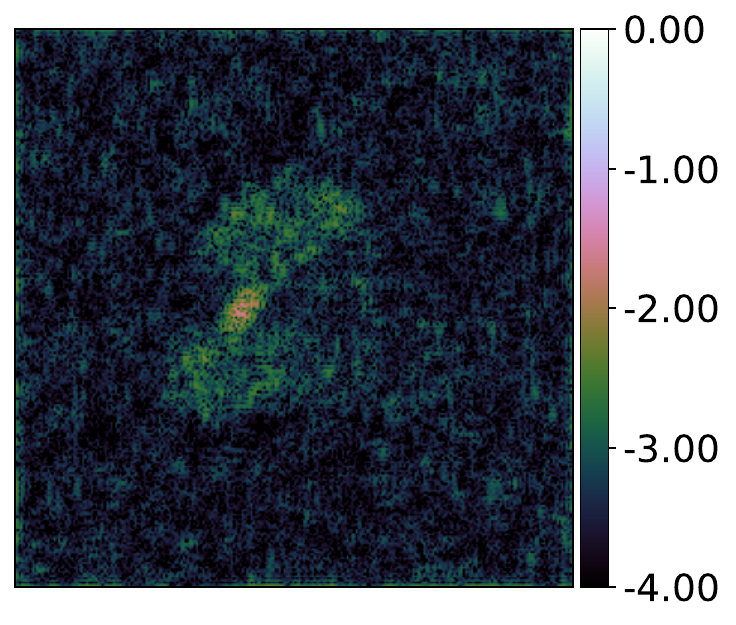}    &
		\includegraphics[width=0.18\linewidth]{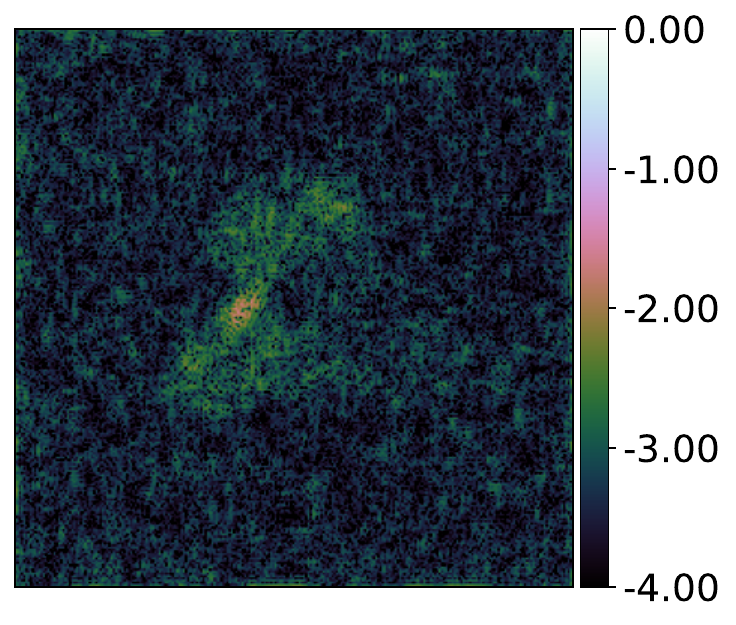}                                                 &
		\includegraphics[width=0.18\linewidth]{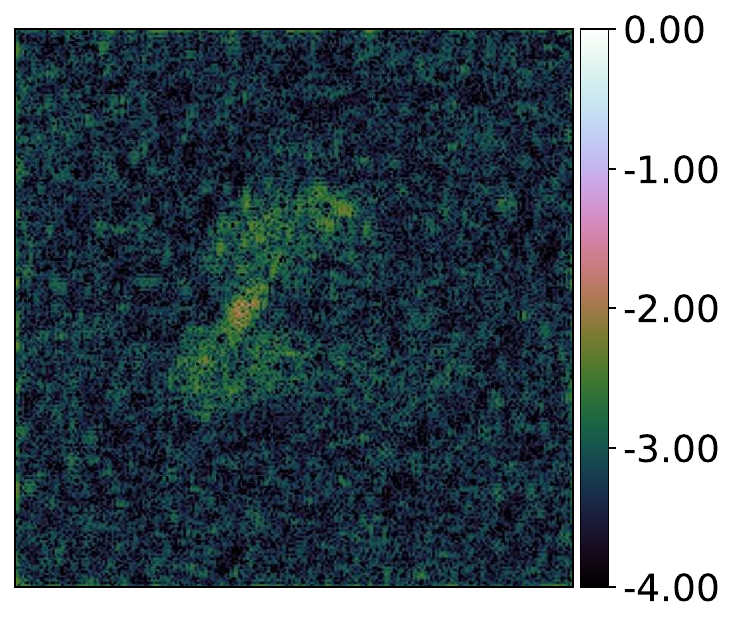}
		\\
		\includegraphics[width=0.18\linewidth]{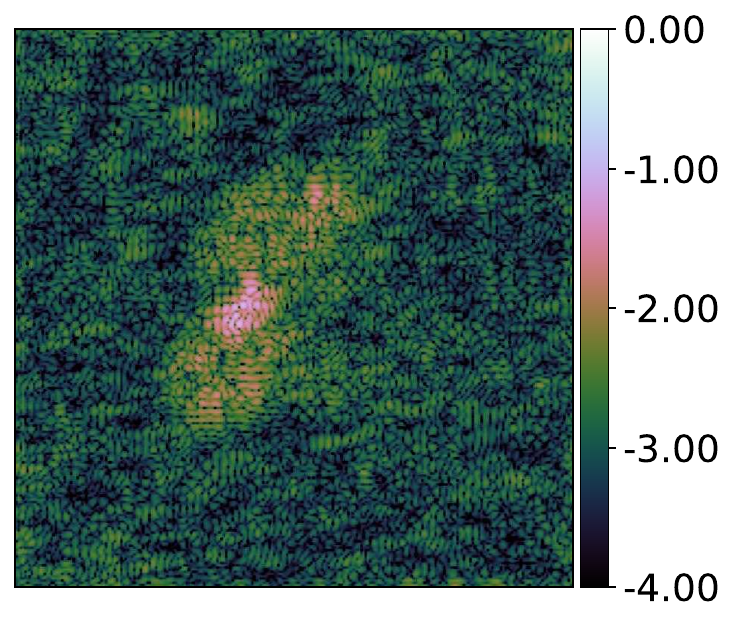} \put(-110,30){\rotatebox{90}{\large Level $3$}} &
		\includegraphics[width=0.18\linewidth]{fastUQ_ungridded/3c288-CRR1h-newPixelUQ-MAP_thresholded_error_level_3.pdf}                                                          &
		\includegraphics[width=0.18\linewidth]{fastUQ_ungridded/3c288-CRR1h-newPixelUQ-MAP_thresholded_error_level_3.pdf}                                                 &
		\includegraphics[width=0.18\linewidth]{fastUQ_ungridded/3c288-CRR1h-newPixelUQ-MAP_thresholded_error_level_3.pdf}
		\\
		\includegraphics[width=0.18\linewidth]{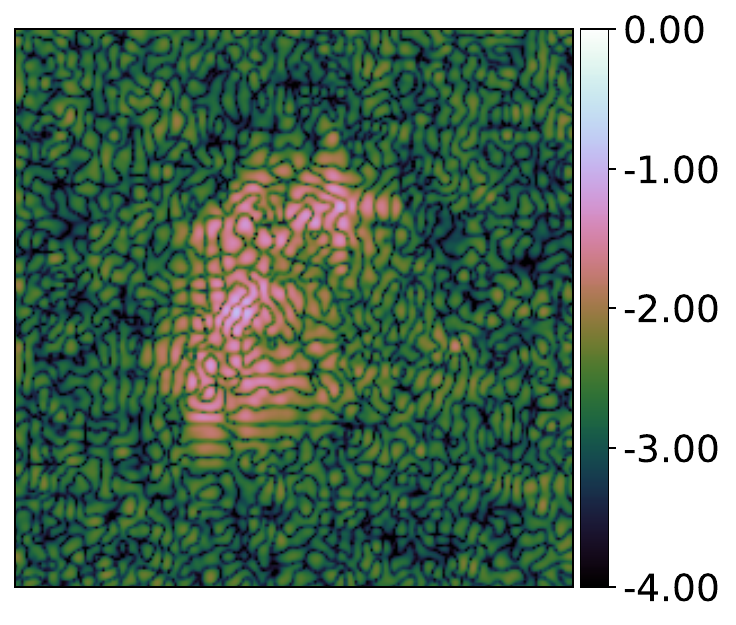} \put(-110,30){\rotatebox{90}{\large Level $2$}} &
		\includegraphics[width=0.18\linewidth]{fastUQ_ungridded/3c288-CRR1h-newPixelUQ-MAP_thresholded_error_level_2.pdf}                                                          &
		\includegraphics[width=0.18\linewidth]{fastUQ_ungridded/3c288-CRR1h-newPixelUQ-MAP_thresholded_error_level_2.pdf}                                                 &
		\includegraphics[width=0.18\linewidth]{fastUQ_ungridded/3c288-CRR1h-newPixelUQ-MAP_thresholded_error_level_2.pdf}
		\\
		\includegraphics[width=0.18\linewidth]{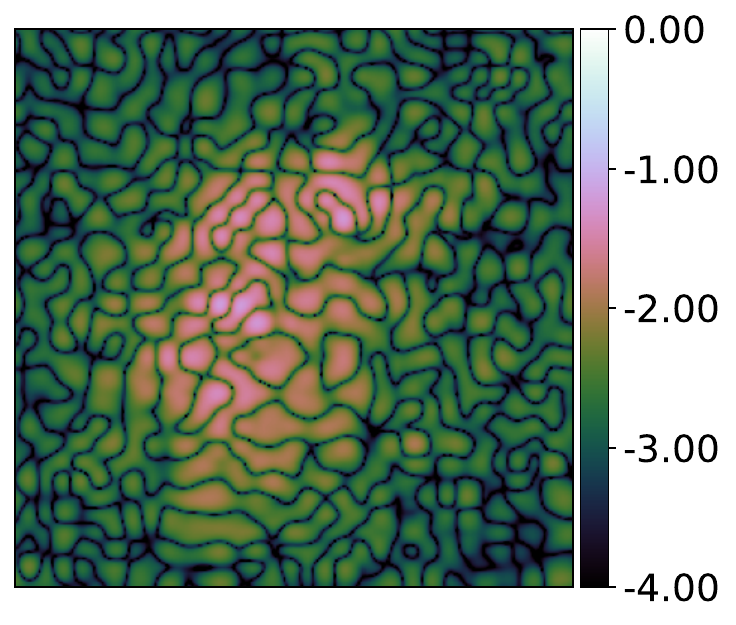} \put(-110,30){\rotatebox{90}{\large Level $1$}} &
		\includegraphics[width=0.18\linewidth]{fastUQ_ungridded/3c288-CRR1h-newPixelUQ-MAP_thresholded_error_level_1.pdf}                                                          &
		\includegraphics[width=0.18\linewidth]{fastUQ_ungridded/3c288-CRR1h-newPixelUQ-MAP_thresholded_error_level_1.pdf}                                                 &
		\includegraphics[width=0.18\linewidth]{fastUQ_ungridded/3c288-CRR1h-newPixelUQ-MAP_thresholded_error_level_1.pdf}
		\\
	\end{tabular}
	\caption{Reconstructions and fast pixel uncertainty quantification (UQ) with the \textsc{QuantifAI} model for the 3C288 image with the four sets of simulated MeerKAT ungridded visibilities with a field of view of approximately $1$ deg$^2$. Each column corresponds to the four datasets with synthesis times of $1$, $2$, $4$ and $8$ hours. The first row represents the dirty reconstruction. The MAP reconstruction is presented in the second row, while the oracle error, which we do not have access to with real data, is shown in the third row. The different decomposition levels of pixel UQ are shown in the last four rows.}
	\label{fi:ungridded_results_3c288}
\end{figure*}

Figures \ref{fi:ungridded_results_W28}, \ref{fi:ungridded_results_CYN} and \ref{fi:ungridded_results_3c288} present the results using the realistic MeerKAT ungridded visibility pattern of the images W28, Cygnus A, and 3C288, correspondingly. The Tables \ref{tb:realistic_results_W28}, \ref{tb:realistic_results_CYN} and \ref{tb:realistic_results_3c288} show the quantitative results of the experiment.

\addtolength{\tabcolsep}{-\tabL}
\addtolength{\tabcolsep}{\tabL}

\begin{table*}
	\begin{center}
		\caption{Main results of \textsc{QuantifAI} for the W28 image with the realistic ungridded MeerKAT visibility patterns with differing synthesis times. As the number of visibilities grows with the synthesis timer, so does the reconstruction SNR. The number of visibilities increases proportionally to the synthesis times.}
		\label{tb:realistic_results_W28}
		\begin{tabular}{cccccc}
			\toprule
			&\multirow{2}{*}{Metrics} & \multicolumn{4}{c}{Datasets} \\
			\cmidrule{3-6}
			& & 1h	& 2h	& 4h	& 8h \\
			\hline \hline
			& Number of visibilities			& $3 \times 10^{4}$  	& $6 \times 10^{4}$ 	& $1.2 \times 10^{5}$   	& $2.4 \times 10^{5}$  	\\
			& MAP reconstrucion SNR [dB]		& $23.88$  				& $25.89$ 				& $27.40$   				& $28.56$    			\\
			\cmidrule{3-6}
			\multirow{2}{*}{Reconstruction} 	& Measurement op. evaluations		& $6976$  	& $6124$  	& $5270$   		& $4062$ 	\\
												& Wall-clock time [s]				& $34.83$  	& $59.25$ 	& $93.61$   	& $137.2$ 	\\
			\cmidrule{3-6}
			\multirow{2}{*}{UQ} 				& Measurement op. evaluations		& $30$  	& $30$  	& $32$   		& $32$ 		\\
												& Wall-clock time [s]				& $0.31$  	& $0.47$ 	& $0.78$   		& $1.35$ 	\\
			\bottomrule
		\end{tabular}
	\end{center}
\end{table*}

\addtolength{\tabcolsep}{-\tabL}
\addtolength{\tabcolsep}{\tabL}

\begin{table*}
	\begin{center}
		\caption{Main results of \textsc{QuantifAI} for the Cygnus A image with the realistic ungridded MeerKAT visibility patterns with differing synthesis times. As the number of visibilities grows with the synthesis timer, so does the reconstruction SNR. The number of visibilities increases proportionally to the synthesis times.}
		\label{tb:realistic_results_CYN}
		\begin{tabular}{cccccc}
			\toprule
			&\multirow{2}{*}{Metrics} & \multicolumn{4}{c}{Datasets} \\
			\cmidrule{3-6}
			& & 1h	& 2h	& 4h	& 8h \\
			\hline \hline
			& Number of visibilities			& $3 \times 10^{4}$  	& $6 \times 10^{4}$ 	& $1.2 \times 10^{5}$   	& $2.4 \times 10^{5}$  	\\
			& MAP reconstrucion SNR [dB]		& $25.72$  				& $27.84$ 				& $28.40$   				& $28.74$    			\\
			\cmidrule{3-6}
			\multirow{2}{*}{Reconstruction} 	& Measurement op. evaluations		& $6482$  	& $5172$  	& $3686$   		& $2692$ 	\\
												& Wall-clock time [s]				& $36.17$  	& $51.71$ 	& $66.79$   	& $91.79$ 	\\
			\cmidrule{3-6}
			\multirow{2}{*}{UQ} 				& Measurement op. evaluations		& $30$  	& $30$  	& $30$   		& $32$ 		\\
												& Wall-clock time [s]				& $0.35$  	& $0.50$ 	& $0.77$   		& $1.36$ 	\\
			\bottomrule
		\end{tabular}
	\end{center}
\end{table*}

\addtolength{\tabcolsep}{-\tabL}
\addtolength{\tabcolsep}{\tabL}

\begin{table*}
	\begin{center}
		\caption{Main results of \textsc{QuantifAI} for the 3C288 image with the realistic ungridded MeerKAT visibility patterns with differing synthesis times. As the number of visibilities grows with the synthesis timer, so does the reconstruction SNR. The number of visibilities increases proportionally to the synthesis times.}
		\label{tb:realistic_results_3c288}
		\begin{tabular}{cccccc}
			\toprule
			&\multirow{2}{*}{Metrics} & \multicolumn{4}{c}{Datasets} \\
			\cmidrule{3-6}
			& & 1h	& 2h	& 4h	& 8h \\
			\hline \hline
			& Number of visibilities			& $3 \times 10^{4}$  	& $6 \times 10^{4}$ 	& $1.2 \times 10^{5}$   	& $2.4 \times 10^{5}$  	\\
			& MAP reconstrucion SNR [dB]		& $25.01$  				& $27.00$ 				& $29.62$   				& $33.02$    			\\
			\cmidrule{3-6}
			\multirow{2}{*}{Reconstruction} 	& Measurement op. evaluations		& $2730$  	& $2198$  	& $1976$   		& $1856$ 	\\
												& Wall-clock time [s]				& $13.93$  	& $21.22$ 	& $34.99$   	& $62.68$ 	\\
			\cmidrule{3-6}
			\multirow{2}{*}{UQ} 				& Measurement op. evaluations		& $26$  	& $28$  	& $32$   		& $32$ 		\\
												& Wall-clock time [s]				& $0.28$  	& $0.44$ 	& $0.78$   		& $1.33$ 	\\
			\bottomrule
		\end{tabular}
	\end{center}
\end{table*}


\bsp	
\label{lastpage}
\end{document}